\documentclass[aps,prd,twocolumn,showpacs,superscriptaddress,groupedaddress]{revtex4}  % for review and submission
\usepackage{graphicx}  % needed for figures
\usepackage{dcolumn}   % needed for some tables
\usepackage{bm}        % for math
\usepackage{amssymb}   % for math
\usepackage{comment}   % for commenting out entire sections
\usepackage{hyperref}
\usepackage{amsmath}
\usepackage{rotating}

\def\pythia{\textsc{pythia}}
\def\herwig{\textsc{herwig}}
\def\nlo{NLO}
\def\hej{\textsc{hej}}

\def\blackhat{\textsc{blackhat+sherpa}}
\def\alpgen{\textsc{alpgen}}
\def\alppyth{\textsc{alpgen+pythia}}

\def\sherpa{\textsc{sherpa}}

\def\wjet{\wnjet}
\def\wdijet{\ensuremath{W+\textrm{dijet}}}
\def\wjets{\wnjet}
\def\wnjet{\ensuremath{W+n\textrm{-jet}}}
\def\wenu{\ensuremath{W(\to e\nu)}}
\def\dphi12{\ensuremath{\Delta\phi(j_1,j_2)}}
\def\dphiFB{\ensuremath{\Delta\phi(j_F,j_B)}}
\def\dy12{\ensuremath{\Delta y(j_1,j_2)}}
\def\dyFB{\ensuremath{\Delta y(j_F,j_B)}}

\newcommand{\eqcd}{\ensuremath{\epsilon_\mathrm{MJ}}}

\let\esig=\ereal

\def\Guru{\textsc{guru}}
\def\guru{\textsc{\Guru}}

\def\MET{{\mbox{$p\kern-0.38em\raise0.18ex\hbox{/}_{T}$}}}
\def\met{{\mbox{$p\kern-0.38em\raise0.18ex\hbox{/}_{T}$}}}
\def\mex{{\mbox{$p\kern-0.38em\raise0.18ex\hbox{/}_{x}$}}}
\def\mey{{\mbox{$p\kern-0.38em\raise0.18ex\hbox{/}_{y}$}}}
\def\pt{$p_T$}

\def\Dzero{D0}
\def\dzero{D0}

\def\ifb{$\mathrm{fb}^{-1}$}

\def\tt{$t\bar{t}$}

\def\to{\rightarrow}

\def\lmet{$WH\rightarrow \ell\kern-0.45em\raise0.19ex\hbox{/} \nu b\bar{b}$}

\begin{document}

\hspace{5.2in} \mbox{FERMILAB-PUB-13-051-E}

\title{Studies of {\boldmath $W$} boson plus jets production in {\boldmath $p\bar{p}$} collisions at {\boldmath $\sqrt{s}=1.96$}~TeV}
\affiliation{LAFEX, Centro Brasileiro de Pesquisas F\'{i}sicas, Rio de Janeiro, Brazil}
\affiliation{Universidade do Estado do Rio de Janeiro, Rio de Janeiro, Brazil}
\affiliation{Universidade Federal do ABC, Santo Andr\'e, Brazil}
\affiliation{University of Science and Technology of China, Hefei, People's Republic of China}
\affiliation{Universidad de los Andes, Bogot\'a, Colombia}
\affiliation{Charles University, Faculty of Mathematics and Physics, Center for Particle Physics, Prague, Czech Republic}
\affiliation{Czech Technical University in Prague, Prague, Czech Republic}
\affiliation{Center for Particle Physics, Institute of Physics, Academy of Sciences of the Czech Republic, Prague, Czech Republic}
\affiliation{Universidad San Francisco de Quito, Quito, Ecuador}
\affiliation{LPC, Universit\'e Blaise Pascal, CNRS/IN2P3, Clermont, France}
\affiliation{LPSC, Universit\'e Joseph Fourier Grenoble 1, CNRS/IN2P3, Institut National Polytechnique de Grenoble, Grenoble, France}
\affiliation{CPPM, Aix-Marseille Universit\'e, CNRS/IN2P3, Marseille, France}
\affiliation{LAL, Universit\'e Paris-Sud, CNRS/IN2P3, Orsay, France}
\affiliation{LPNHE, Universit\'es Paris VI and VII, CNRS/IN2P3, Paris, France}
\affiliation{CEA, Irfu, SPP, Saclay, France}
\affiliation{IPHC, Universit\'e de Strasbourg, CNRS/IN2P3, Strasbourg, France}
\affiliation{IPNL, Universit\'e Lyon 1, CNRS/IN2P3, Villeurbanne, France and Universit\'e de Lyon, Lyon, France}
\affiliation{III. Physikalisches Institut A, RWTH Aachen University, Aachen, Germany}
\affiliation{Physikalisches Institut, Universit\"at Freiburg, Freiburg, Germany}
\affiliation{II. Physikalisches Institut, Georg-August-Universit\"at G\"ottingen, G\"ottingen, Germany}
\affiliation{Institut f\"ur Physik, Universit\"at Mainz, Mainz, Germany}
\affiliation{Ludwig-Maximilians-Universit\"at M\"unchen, M\"unchen, Germany}
\affiliation{Fachbereich Physik, Bergische Universit\"at Wuppertal, Wuppertal, Germany}
\affiliation{Panjab University, Chandigarh, India}
\affiliation{Delhi University, Delhi, India}
\affiliation{Tata Institute of Fundamental Research, Mumbai, India}
\affiliation{University College Dublin, Dublin, Ireland}
\affiliation{Korea Detector Laboratory, Korea University, Seoul, Korea}
\affiliation{CINVESTAV, Mexico City, Mexico}
\affiliation{Nikhef, Science Park, Amsterdam, the Netherlands}
\affiliation{Radboud University Nijmegen, Nijmegen, the Netherlands}
\affiliation{Joint Institute for Nuclear Research, Dubna, Russia}
\affiliation{Institute for Theoretical and Experimental Physics, Moscow, Russia}
\affiliation{Moscow State University, Moscow, Russia}
\affiliation{Institute for High Energy Physics, Protvino, Russia}
\affiliation{Petersburg Nuclear Physics Institute, St. Petersburg, Russia}
\affiliation{Instituci\'{o} Catalana de Recerca i Estudis Avan\c{c}ats (ICREA) and Institut de F\'{i}sica d'Altes Energies (IFAE), Barcelona, Spain}
\affiliation{Uppsala University, Uppsala, Sweden}
\affiliation{Lancaster University, Lancaster LA1 4YB, United Kingdom}
\affiliation{Imperial College London, London SW7 2AZ, United Kingdom}
\affiliation{The University of Manchester, Manchester M13 9PL, United Kingdom}
\affiliation{University of Arizona, Tucson, Arizona 85721, USA}
\affiliation{University of California Riverside, Riverside, California 92521, USA}
\affiliation{Florida State University, Tallahassee, Florida 32306, USA}
\affiliation{Fermi National Accelerator Laboratory, Batavia, Illinois 60510, USA}
\affiliation{University of Illinois at Chicago, Chicago, Illinois 60607, USA}
\affiliation{Northern Illinois University, DeKalb, Illinois 60115, USA}
\affiliation{Northwestern University, Evanston, Illinois 60208, USA}
\affiliation{Indiana University, Bloomington, Indiana 47405, USA}
\affiliation{Purdue University Calumet, Hammond, Indiana 46323, USA}
\affiliation{University of Notre Dame, Notre Dame, Indiana 46556, USA}
\affiliation{Iowa State University, Ames, Iowa 50011, USA}
\affiliation{University of Kansas, Lawrence, Kansas 66045, USA}
\affiliation{Louisiana Tech University, Ruston, Louisiana 71272, USA}
\affiliation{Northeastern University, Boston, Massachusetts 02115, USA}
\affiliation{University of Michigan, Ann Arbor, Michigan 48109, USA}
\affiliation{Michigan State University, East Lansing, Michigan 48824, USA}
\affiliation{University of Mississippi, University, Mississippi 38677, USA}
\affiliation{University of Nebraska, Lincoln, Nebraska 68588, USA}
\affiliation{Rutgers University, Piscataway, New Jersey 08855, USA}
\affiliation{Princeton University, Princeton, New Jersey 08544, USA}
\affiliation{State University of New York, Buffalo, New York 14260, USA}
\affiliation{University of Rochester, Rochester, New York 14627, USA}
\affiliation{State University of New York, Stony Brook, New York 11794, USA}
\affiliation{Brookhaven National Laboratory, Upton, New York 11973, USA}
\affiliation{Langston University, Langston, Oklahoma 73050, USA}
\affiliation{University of Oklahoma, Norman, Oklahoma 73019, USA}
\affiliation{Oklahoma State University, Stillwater, Oklahoma 74078, USA}
\affiliation{Brown University, Providence, Rhode Island 02912, USA}
\affiliation{University of Texas, Arlington, Texas 76019, USA}
\affiliation{Southern Methodist University, Dallas, Texas 75275, USA}
\affiliation{Rice University, Houston, Texas 77005, USA}
\affiliation{University of Virginia, Charlottesville, Virginia 22904, USA}
\affiliation{University of Washington, Seattle, Washington 98195, USA}
\author{V.M.~Abazov} \affiliation{Joint Institute for Nuclear Research, Dubna, Russia}
\author{B.~Abbott} \affiliation{University of Oklahoma, Norman, Oklahoma 73019, USA}
\author{B.S.~Acharya} \affiliation{Tata Institute of Fundamental Research, Mumbai, India}
\author{M.~Adams} \affiliation{University of Illinois at Chicago, Chicago, Illinois 60607, USA}
\author{T.~Adams} \affiliation{Florida State University, Tallahassee, Florida 32306, USA}
\author{G.D.~Alexeev} \affiliation{Joint Institute for Nuclear Research, Dubna, Russia}
\author{G.~Alkhazov} \affiliation{Petersburg Nuclear Physics Institute, St. Petersburg, Russia}
\author{A.~Alton$^{a}$} \affiliation{University of Michigan, Ann Arbor, Michigan 48109, USA}
\author{A.~Askew} \affiliation{Florida State University, Tallahassee, Florida 32306, USA}
\author{S.~Atkins} \affiliation{Louisiana Tech University, Ruston, Louisiana 71272, USA}
% ...
%
% visitor_addresses.tex                       11 January 2013 
%  available symbols are:
%  $\ast, \dag, \ddag, \S, \P, $\|$, $\ast\ast$, \dag\dag, \ddag\ddag ,\#
%
\collaboration{The D0 Collaboration\footnote{with visitors from
%{alton}
$^{a}$Augustana College, Sioux Falls, SD, USA,
%{burdin}
$^{b}$The University of Liverpool, Liverpool, UK,
%{garcia-guerra}
$^{c}$UPIITA-IPN, Mexico City, Mexico,
%{grohsjean}
$^{d}$DESY, Hamburg, Germany,
%{partridge}
$^{e}$SLAC, Menlo Park, CA, USA,
%{hesketh}
$^{f}$University College London, London, UK,
%{luna-garcia}
$^{g}$Centro de Investigacion en Computacion - IPN, Mexico City, Mexico,
%{podesta-lerma}
$^{h}$ECFM, Universidad Autonoma de Sinaloa, Culiac\'an, Mexico,
%{santos}
$^{i}$Universidade Estadual Paulista, S\~ao Paulo, Brazil,
%{meyer}
$^{j}$Karlsruher Institut f\"ur Technologie (KIT) - Steinbuch Centre for Computing (SCC)
and
%{patwa}
$^{k}$Office of Science, U.S. Department of Energy, Washington, D.C. 20585, USA.
%{falkowski}
%$^{?}$Laboratoire de Physique Theorique, Orsay, FR,
%{hooper}
%$^{?}$Visitor from Bradley University, Peoria, IL, USA.
%{kozminski}
%$^{?}$}Visitor from Lewis University, Romeoville, IL, USA.
%{weber}
%$^{?}$Universit{\"a}t Bern, Bern, Switzerland.
%{deceased}
%$^{\ddag}$Deceased.
}} \noaffiliation
\vskip 0.25cm
       
\date{February 26, 2013}

\begin{abstract}
We present a comprehensive analysis of inclusive $\wenu+n\textrm{-jet}$ ($n\geq1,2,3,4$) production in proton-antiproton collisions at a center-of-mass energy of 1.96~TeV
at the Tevatron collider using a 3.7~\ifb\ dataset collected by the D0 detector.
Differential cross sections are presented as a function of the jet rapidities ($y$),
lepton transverse momentum ($p_T$) and pseudorapidity ($\eta$), the scalar sum of the transverse energies of the $W$ boson and all jets ($H_T$), leading dijet $p_T$ and invariant mass, dijet rapidity
separations for a variety of jet pairings for $p_T$-ordered and angular-ordered jets, dijet opening angle, dijet azimuthal angular separations for $p_T$-ordered and angular-ordered jets,
and $W$ boson transverse momentum. 
The mean number of jets in an event containing a $W$ boson is measured as a function of $H_T$,
and as a function of the rapidity separations between the two highest-$p_T$ jets and between the most widely separated jets in rapidity.
Finally, the probability for third-jet emission in events containing a $W$ boson and at least two jets is studied by measuring the fraction of events in the 
inclusive $W+2\textrm{-jet}$ sample that contain a third jet over a $p_T$ threshold. 
The analysis employs a regularized singular value decomposition technique to accurately correct for detector effects and for the presence of backgrounds.
The corrected data are compared to particle level next-to-leading order perturbative QCD predictions, predictions from all-order resummation approaches, 
and a variety of leading-order and matrix-element plus parton-shower event generators.
Regions of the phase space where there is agreement or disagreement with the data are discussed for the different models tested.

\end{abstract}

\maketitle

\section{Introduction}
\label{sec:intro}

Measurements of vector boson production in association with jets are important tests of perturbative quantum chromodynamics (pQCD), 
the theory describing the strong interaction between quarks and gluons.
\wnjet\ production processes are also of interest because of the important role they play as backgrounds to beyond the standard model phenomena and as multi-scale QCD processes.
In the case of searches, \wjets\ production is a major background in several supersymmetry and Higgs boson decay channels. 
In the case of standard model processes with small cross sections, such as single top quark production and vector boson fusion (VBF) processes, \wjets\ processes often overwhelm the small signal.
Theoretical uncertainties on the production rates and kinematics of \wjets\ processes have large uncertainties and limit our ability to identify and characterize new phenomena.
Therefore, it is important to make \wjets\ measurements at the Fermilab Tevatron Collider and the CERN Large Hadron Collider (LHC) in order to constrain these backgrounds.
Measurements of \wjets\ production have also been performed by the CDF\,\cite{Aaltonen:2007ip}, \Dzero\,\cite{Abazov:2011rf}, 
ATLAS\,\cite{atlas2011,Aad:2012en}, and CMS\,\cite{Chatrchyan:2011ne} Collaborations.
We present here new measurements of \wjets\ production using a 3.7~\ifb\ data sample of proton-antiproton collisions collected with the \Dzero\ detector\,\cite{d0det,L1Cal,Layer0}
between 2002 and 2008.  
The measurements presented follow from earlier measurements of inclusive \wnjet\ production cross sections and differential cross sections as a 
function of the $n^\mathrm{th}$-jet $p_T$ up to $n=4$, using the same dataset\,\cite{Abazov:2011rf}, but providing further details and additional differential distributions.
Previous \wjets\ measurements have been used in testing and tuning theoretical models of $W$ boson production\,\cite{Berger:2009ep,Ellis:2009bu,Gleisberg:2008ta}.

In this article, we significantly expand on the number of measured observables in order to make a comprehensive study of \wjets\ production.   
These new measurements include differential cross sections of hadronic and leptonic variables, which will provide validation of new theoretical approaches and input for Monte Carlo (MC) tuning.  
We provide measurements of $n^\mathrm{th}$-jet rapidities to test the modeling of parton emission, which is difficult to predict accurately at large values of rapidity.
We measure the $W$ boson transverse momentum and the dijet invariant mass in inclusive $W+2\textrm{-jet}$ and $W+3\textrm{-jet}$ events, the latter being an important variable for electroweak 
VBF production. This observable is a useful validation tool for the reliability of background simulations for certain Higgs boson production and decay channels where the signal is 
extracted from the dijet mass distribution, and for the investigation of possible new phenomena.
All cross section measurements are normalized to the measured inclusive $W$ boson production cross section\,\cite{Abazov:2011rf}, allowing for the cancellation or reduction of several 
experimental systematic uncertainties.  

In addition to the single differential cross sections, we further probe QCD emissions in \wjets\ events through the study of observables such as the 
mean number of jets in an event as a function of the total hadronic and leptonic transverse energy in the event, $H_T$, and as a function of the rapidity span between jets in 
$W+\geq2\textrm{-jet}$ events.
The probability of additional jet emission as a function of dijet rapidity separation is also studied for the first time in inclusive $W+2\textrm{-jet}$ events, 
for both $p_T$-ordered and rapidity-ordered jets by measuring the fraction of events in the
inclusive $W+2\textrm{-jet}$ sample that contain a third jet above a $p_T$ threshold.
This variable has consequences for the design of jet vetoes in high jet multiplicity final states, which are particularly important for VBF Higgs and electroweak production.  
Such variables are also sensitive to BFKL-like dynamics\,\cite{BFKL,BFKL-HEJ} when the two jets are widely separated in rapidity. 

The methods employed for this measurement follow those used in the previous D0 $Z$+jet cross section\,\cite{Abazov:2008ez} and $Z$ boson $p_T$\,\cite{Abazov:2010kn} analyses,
as well as on the previous D0 \wjets\ analysis\,\cite{Abazov:2011rf}.  
We select a high purity sample of \wjets\ events in which the $W$ boson decays to an electron and a neutrino, while maintaining the bulk of the kinematic phase space.  
The measurements are corrected to the particle level, which includes energy from stable particles, the underlying event (partonic interactions from the same proton-antiproton scatter), muons, and neutrinos, as defined in Ref.~\cite{particle}.  
The unfolding uses a regularized singular value decomposition method\,\cite{Hocker:1995kb} as implemented in the program \guru.
This procedure corrects a measured observable back to the particle observable, deconvolving the effects of finite experimental resolution, detector response, acceptance, and efficiencies.

\section{The D0 Detector}
\label{sec:det}

The primary components of the \Dzero\ detector are a central tracking system, a calorimeter, a muon identification system, and a luminosity monitor.
To reconstruct the $W$ boson and the jets in this measurement, we use the central tracker to identify 
the location of the $p\bar{p}$ interaction vertex and the electron produced in the decay of the $W$ boson candidate, and use the liquid-argon and uranium 
calorimeter to identify electromagnetic and hadronic showers, as well as calculate the magnitude and direction of the missing transverse energy.
The luminosity monitor is employed to make a measurement of the integrated luminosity corresponding to the data collected, and the trigger system is 
used to make a basic selection of likely \wjets\ events.
A detailed description of the \Dzero\ detector can be found in Ref.~\cite{d0det,L1Cal,Layer0}.
Here we outline the most important elements of the \Dzero\ detector for performing the \wjets\ measurements presented in this paper.

\subsection{Tracking detectors}

The \Dzero\ central tracking system is made up of a silicon microstrip tracker and a fiber tracker.
The tracking detectors are primarily used to identify the charged track associated with the leptonic decay of the $W$ boson,
but are not used directly in jet reconstruction since the jet-finding algorithms in \Dzero\ use only energy deposits in the calorimeter towers. 

The tracking detectors are used to reconstruct the position of the primary vertex (PV) of the $p\bar{p}$ interaction, which is
necessary to measure the jet rapidity\,\cite{definitions} and transverse momentum.  The tracking system is also used to confirm that the jets originated from the 
PV in the event, thereby reducing the contamination from additional $p\bar{p}$ interactions.

The distribution of the PV along the beam axis follows a 20~cm wide Gaussian distribution function centered on the nominal interaction point at the center of the detector. 
We use a right-handed coordinate system in which the $z$-axis is along the proton beam direction, the $x$-axis points away from the center of the Tevatron ring, and the $y$-axis is upward. 
The inner tracking system, consisting of the silicon microstrip tracker, provides 35~$\mu$m vertex resolution along the beam line and $15~\mu$m resolution in 
the $r$-$\phi$ plane, where $\phi$ is the azimuthal angle, for tracks with a minimum $p_T$ of 10~GeV at $|\eta| = 0$. 
The outer tracking system, consisting of the central fiber tracker, includes eight axial and eight stereo doublet layers of 800 $\mu m$ diameter scintillating fibers to complement the silicon tracker. 
Both detectors are located inside the 1.9\,T magnetic field of the superconducting solenoidal magnet to allow measurements of the momentum of charged particles.   

\subsection{Calorimeter}

The calorimeter system consists of a uranium/liquid-argon calorimeter, divided into a central (CC) and two end (EC) sections, and a 
plastic scintillator inter-cryostat detector. Both the CC and EC are segmented longitudinally into electromagnetic (EM),
fine hadronic, and coarse hadronic sections.

The calorimeter is transversely segmented into cells along the polar and azimuthal axes in a projective tower geometry.  
The CC covers detector pseudorapidity $|\eta|<1.2$, where $\eta = -\ln\tan(\theta/2)$ and $\theta$ is the polar angle defined with 
respect to the beamline, and the two ECs extend the range up to $|\eta| = 4.2$.
Both the electromagnetic and fine hadronic calorimeters are sampling calorimeters with an active medium of liquid argon and absorber plates of nearly pure depleted uranium. 
Incoming particles traversing the uranium absorber plates initiate showers of secondary particles that ionize the argon in the gaps between the absorber plates.
Due to a high-voltage electric field, the free electrons collect on resistively-coated copper pads that act as signal boards. The outer part of the calorimeter, 
the coarse hadronic section, uses copper in the CC and stainless steel in the EC for the absorber plates. 
The calorimeter is transversely segmented into cells in $\Delta\eta \times \Delta\phi$
of $0.1 \times 0.1$ ($0.05 \times 0.05$ in the third layer of the EM calorimeter for $|\eta|<3.2$ to allow for a precise location of EM shower centroids).
At $|\eta| > 3.2$, the cell size grows to 0.2 or more in both $\eta$ and $\phi$.  The energy resolution of jets reconstructed beyond $|\eta| > 3.2$ is therefore degraded.
The total depth of the EM calorimeter is about 20~radiation lengths, and the combined thickness of the electromagnetic and hadronic 
calorimeters is about seven nuclear interaction lengths.

\subsection{Trigger system}
The \wjets\ events are selected by triggering on single electron or electron-plus-jet signatures with a three-level trigger system.  
The electron trigger signature is similar to the electron reconstruction signature (including electromagnetic shower shape and a track matched to the EM shower), 
albeit more loosely-defined to facilitate a fast enough trigger decision.
Several single electron triggers are used in a logical~\texttt{OR} to maximize the trigger efficiency.
The $p_T$ threshold on the electron triggers varies between 15 and 35~GeV as different triggers are activated at different instantaneous luminosities.

To further maximize the trigger efficiency, data collected with $\mathrm{electron+jet}$ triggers are included.  In the bulk of the dataset considered, 
the \Dzero\ calorimeter trigger performs clustering of the trigger towers, which are $\Delta\eta\times\Delta\phi = 0.2 \times 0.2$ sums of the calorimeter cells, 
using a sliding windows algorithm\,\cite{L1Cal}.  
This clustering improves the energy resolution of the trigger jet objects, which allows triggering on relatively low-$p_T$ jets.   
As for the electron triggers, the $p_T$ threshold defined for the jet trigger objects varies between 15 and 25~GeV, 
depending on the instantaneous luminosity of the data delivered by the Tevatron.   

\subsection{Luminosity detector}
The measurement of the \Dzero\ luminosity is made by the luminosity monitor (LM).   The LM consists of scintillating tiles on either side of the interaction point, 
which measure the particles created in inelastic collisions.   The luminosity is determined as
\begin{equation}
\mathcal{L} = \frac{f\cdot N_{\mathrm{LM}}}{\sigma_{\mathrm{LM}}},
\end{equation}
where $f$ is the $p\bar{p}$ bunch crossing frequency, $N_{\mathrm{LM}}$ is the average number of observed interactions, and $\sigma_{\mathrm{LM}}$ is the 
effective cross section for inelastic collisions measured by the LM that takes into account event losses due to inefficiencies and 
geometric acceptance\,\cite{d0lumi}.   The uncertainty on the luminosity determination is estimated to be 6.1\%.   
The uncertainty is dominated by a 4.2\% uncertainty coming from the determination of $\sigma_{\mathrm{LM}}$\,\cite{Casey:2012rr}.  

\section{Event Reconstruction and Selection}
\label{sec:evtsel}

Our measurements use a sample of $\wenu+n\textrm{-jet}$ candidate events 
corresponding to 3.7~\ifb\ of data collected with the \Dzero\ detector in Run~II of the Fermilab Tevatron Collider.  
The analysis techniques used are identical to those described in Ref.~\cite{Abazov:2011rf}, although 
we quote an integrated luminosity using a convention which now includes the loss due to data quality corrections.
The data are grouped into two time periods. Run~IIa refers to the data collected prior to 2006 (1.1~\ifb\ of data), when two major upgrades were installed in the \Dzero\ detector.   
An additional layer of silicon was added to the inner tracker to improve track position resolution\,\cite{Layer0}, and the Level~1 calorimeter trigger was 
replaced with a system that performed electron, jet, and tau identification\,\cite{L1Cal}.  
Because of these changes, the data collected after the summer of 2006 are triggered and reconstructed in a different manner and are referred to as Run~IIb data 
(a total integrated luminosity of 2.6~\ifb). The measurements presented here are limited by systematic uncertainties and the inclusion of additional data would neither
improve the overall precision of the measurements, nor appreciably enhance their kinematic reach.

The events are then processed through the \Dzero\
reconstruction program, which identifies jet and $W$ boson candidates.  
Jets are identified with the \Dzero\ midpoint cone algorithm\,\cite{d0jets}, with a split-merge fraction of 0.5 using a cone of radius 
$R=\sqrt{(\Delta y)^2+(\Delta\phi)^2}=0.5$ to cluster calorimeter energy into jets.  
Jets are corrected for calorimeter response, instrumental and out-of-cone showering effects, additional energy deposits in the calorimeter that 
arise from detector noise and underlying event energy, and for pile-up arising from multiple $p\bar{p}$ interactions and previous bunch crossings. 
These jet energy scale corrections\,\cite{Abazov:2011vi} are determined using transverse momentum imbalance in $\gamma+\mathrm{jet}$
events, where the electromagnetic response is calibrated using $Z/\gamma^* \rightarrow e^+e^-$ events.
Jets are required to have at least two tracks that point to their associated PV (jet-vertex confirmation) to improve
the identification of the jets and ensure they are associated with the same proton-antiproton collision as the $W$ boson under consideration.
These tracks must have $p_T>0.5$~GeV, at least one hit in the SMT detector, and a distance of closest approach with respect to the PV of less than 0.5~cm 
in the transverse plane and less than 1~cm along the beam axis ($z$).
Jets are ordered in decreasing transverse momentum, and we denote the jet with the highest transverse momentum the ``leading'' jet.

Electrons are identified as clusters of calorimeter cells in which at least 95\% of the energy in the shower is deposited in the EM section. 
Electron candidates must be isolated from other calorimeter energy deposits, have spatial distributions consistent with those expected for EM showers, 
and contain a reconstructed track pointing to the PV and matched to an EM shower that is isolated from other tracks.  
The energy in an isolation cone around the electron track must not exceed 15\% of the electron $p_T$.
The extrapolated electron track must lie within 1~cm of the primary vertex along the $z$ direction.
Events with a second isolated electron are removed to suppress the background from $Z$ boson and Drell-Yan production.  
The missing transverse energy in the event is calculated as the vector sum of all the electromagnetic and fine hadronic cell energies, 
and the coarse hadronic cell energies that are contained in jets provided they have an energy greater than four standard deviations of the electronic noise
or are neighbors of an energetic cell\,\cite{Vlimant}.
A correction for the presence of any muons is applied to the missing transverse energy calculation.   
All energy corrections which are applied to electrons and jets in the event are also propogated to the missing transverse energy calculation.  
Because the longitudinal component of the momentum of the neutrino is not measured, the calculated properties of 
each $W$ boson candidate is limited to its transverse energy, $E_T^W$, and transverse mass, defined as $M_T^W = \sqrt{(\met+p_T^e)^2 - (\mex + p_x^e)^2 - (\mey + p_y^e)^2}$, 
where \met\ is the magnitude of the missing transverse energy vector,
$p_T^e$ is the transverse momentum of the electron, and the remaining variables are the associated $x$ and $y$ components.   

The following event selections are used to suppress background while maintaining high efficiency for events in which a $W$ boson was produced: 
$p_T^e \ge 15$ GeV and electron pseudorapidity $|\eta^e|<1.1$, $\met > 20$~GeV, $M_T^W \ge 40$~GeV, for all jets $p_T^{\text{jet}} \ge 20$~GeV 
and rapidity $|y^{\text{jet}}|<  3.2$, $\Delta R = \sqrt{(\Delta \phi)^2 + (\Delta \eta)^2}>0.5$ between the electron and the nearest jet, 
and the $z$ position of the interaction vertex is restricted to $|z_{\text{vtx}}|<60$~cm\,\cite{definitions}.
Events must have a reconstructed $p\bar{p}$ interaction vertex with at least three associated tracks,
with $\Delta z <$ 1~cm between the PV and the electron track extrapolated to the beam axis.  
Finally, events with excessive calorimeter noise are removed from the sample.   

In this paper, we refer to the samples containing a $W$ boson and at least $n$ jets as inclusive \wnjet\ events, or simply as \wnjet\ events.
The inclusive $W$ boson sample contains $2\,184\,821$ events, and there are $265\,713$, $39\,805$, $5\,962$, and $1\,028$ events in the inclusive
$W$+1-jet, $W$+2-jet, $W$+3-jet, and $W$+4-jet samples, respectively.
The majority of these events are true $W(\rightarrow e\nu)$+jets events, but there are background processes contaminating this dataset.
These background processes include $W(\rightarrow \tau \nu)$+jets events, QCD multijet events in which a jet is reconstructed as an electron, 
$Z\rightarrow e^+e^-$ events in which one electron is not reconstructed, and diboson and top quark processes.

\subsection{QCD multijet background}

In QCD multijet events there is a small but non-negligible probability
of instrumental backgrounds or decays to electrons/photons in or near jets that may create a fake-electron signature.
A jet composed primarily of neutral particles with a high electromagnetic fraction may pass electron candidate identification criteria, 
or a photon might be misidentified as an electron.
Since the QCD multijet cross section is large, the contribution from such instances of fake-electron events to the measured distributions must be taken into account. 
These backgrounds are difficult to accurately model in simulation;
therefore, we estimate our background using a data-driven approach with a \dzero\ dataset that is orthogonal to that used for the main measurement.

To estimate this background contribution, we first define two samples of events, one with ``loose'' selection criteria 
and one with ``tight'' selection criteria corresponding to our standard event selection, where the latter is a subset of the former. 
The loose sample (containing $N_\mathrm{loose}$ events) consists of $N^{\mathrm{signal}}_\mathrm{loose}$ events with a real electron candidate originating from $W/Z+\mathrm{jets}$,
diboson, or top quark production sources, and $N^{\mathrm{MJ}}_\mathrm{loose}$ multijet background events with a fake-electron signature.
In the case of our tight data selection (containing $N_\mathrm{tight}$ events), we have a similar relation:
\begin{equation}
N_\mathrm{tight}=N^{\mathrm{signal}}_\mathrm{tight}+N^{\mathrm{MJ}}_\mathrm{tight}.
\label{eqn:tight}
\end{equation}

To determine the shape and overall normalization of the QCD multijet distributions, 
we then define a ``loose-not-tight'' (LNT) data sample that is orthogonal to our standard selection (containing $N_\mathrm{LNT}$ events),
requiring that an electron candidate pass the loose selection but fail the tight.

This LNT sample is composed of events with a real electron ($N^{\mathrm{signal}}_\mathrm{LNT}$)
and events where a jet is misidentified as an electron ($N^{\mathrm{MJ}}_\mathrm{LNT}$):
\begin{equation}
N_\mathrm{LNT}=N^{\mathrm{signal}}_\mathrm{LNT}+N^{\mathrm{MJ}}_\mathrm{LNT}.
\label{eqn:LNT}
\end{equation}

By combining Eqs.~\ref{eqn:tight} and~\ref{eqn:LNT}, the number of events in our loose selection can be written as:
\begin{equation}
N_\mathrm{LNT}+N_\mathrm{tight}=N_\mathrm{loose}=N^{\mathrm{signal}}_\mathrm{loose}+N^{\mathrm{MJ}}_\mathrm{loose},
\end{equation}
with the relationship between the numbers of real electrons and misidentified jets passing the loose selection that also pass the 
tight selection being defined by:
\begin{eqnarray}
N^\mathrm{signal}_\mathrm{tight}&=&\esig\cdot N^\mathrm{signal}_\mathrm{loose} \label{eqn:eqcd4}\\
N^\mathrm{MJ}_\mathrm{tight}&=&\eqcd\cdot N^\mathrm{MJ}_\mathrm{loose}, \label{eqn:eqcd5}
\end{eqnarray}
where \esig\ and \eqcd\ then represent the efficiencies for a real electron and for a misidentified jet passing the loose selection to also pass the tight selection.

From these relations, the number of multijet events with tight electron requirements in a given bin
can be determined as follows:

\begin{align}
\begin{split}
N^{\mathrm{MJ}}_\mathrm{tight}&=\left(\frac{\eqcd}{1-\eqcd}\right)\cdot N_\mathrm{LNT}\\
&\qquad -\left(\frac{\eqcd}{1-\eqcd}\right)\cdot \left(\frac{1-\esig}{\esig}\right)\cdot N^\mathrm{signal}_\mathrm{tight}.
\end{split}
\end{align}
That is, we estimate the shape of the multijet component from the loose-not-tight sample,
with overall normalization determined from the relative efficiency for a misidentified jet passing the loose selection to also pass the tight selection, 
\eqcd, and a small correction derived from the tight sample to account for the presence of real electrons in
the loose-not-tight sample. To determine the QCD multijet component we first need to calculate the values of \esig\ and \eqcd, as well as $N^\mathrm{signal}_\mathrm{tight}$.

\begin{figure}[htbp]
  \begin{center}
    \includegraphics[width=\columnwidth]{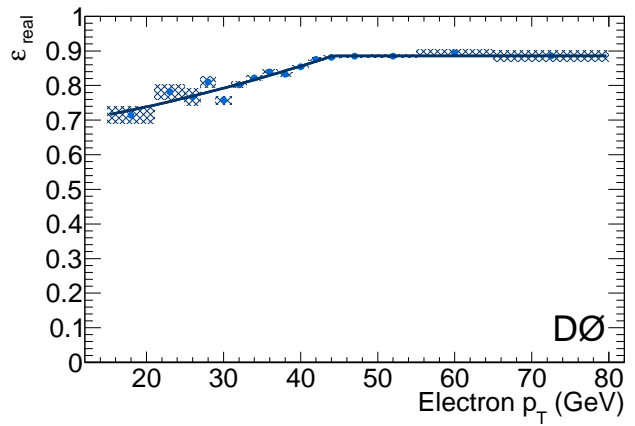}
    \caption{Probability that a real electron candidate passing the loose electron identification requirements also passes
      the tight electron identification requirements. 
      The shaded band represents the systematic uncertainty originating from the determination of the tight and loose electron efficiencies.
      \label{fig:signalefficiency}
    }
  \end{center}
\end{figure}

The probability that a true electron candidate in the loose sample also passes the tight criteria, 
\esig, is calculated from the ratio of electron identification efficiencies (calculated using $Z\to e^+e^-$ events) derived under the loose and tight selection criteria.
Significant variation in \esig\ is observed as a function of electron \pt\ (shown in Fig.~\ref{fig:signalefficiency})
with a plateau of $(88.6\pm 0.3)\%$ reached at a \pt\ of approximately 45~GeV, and thus events in the tight and loose-not-tight samples are assigned 
a value of \esig\ based on the electron candidate \pt.

To determine \eqcd, we define a multijet-enriched data sample with selection criteria as for the standard event 
selection except that the loose electron identification criteria are used, the $W$ boson transverse mass requirement is removed, and the $\met>20$~GeV
requirement is reversed and lowered to $\met<10$~GeV.
This selection is then applied to the data and to the MC signal and background samples (see Sec.~\ref{sec:mcsamples}). \eqcd\ is determined from the data sample by
calculating the fraction of events in this loose multijet-enriched sample that are also in the multijet-enriched selection once tight requirements are applied.
A small component of true electrons can still contaminate these multijet-enriched samples for both the loose and tight selections.
MC samples with multijet-enriched selection criteria are used to model and remove this signal contamination in both selection samples
before the determination of \eqcd.

\begin{figure}[htbp]
  \begin{center}
    \includegraphics[width=\columnwidth]{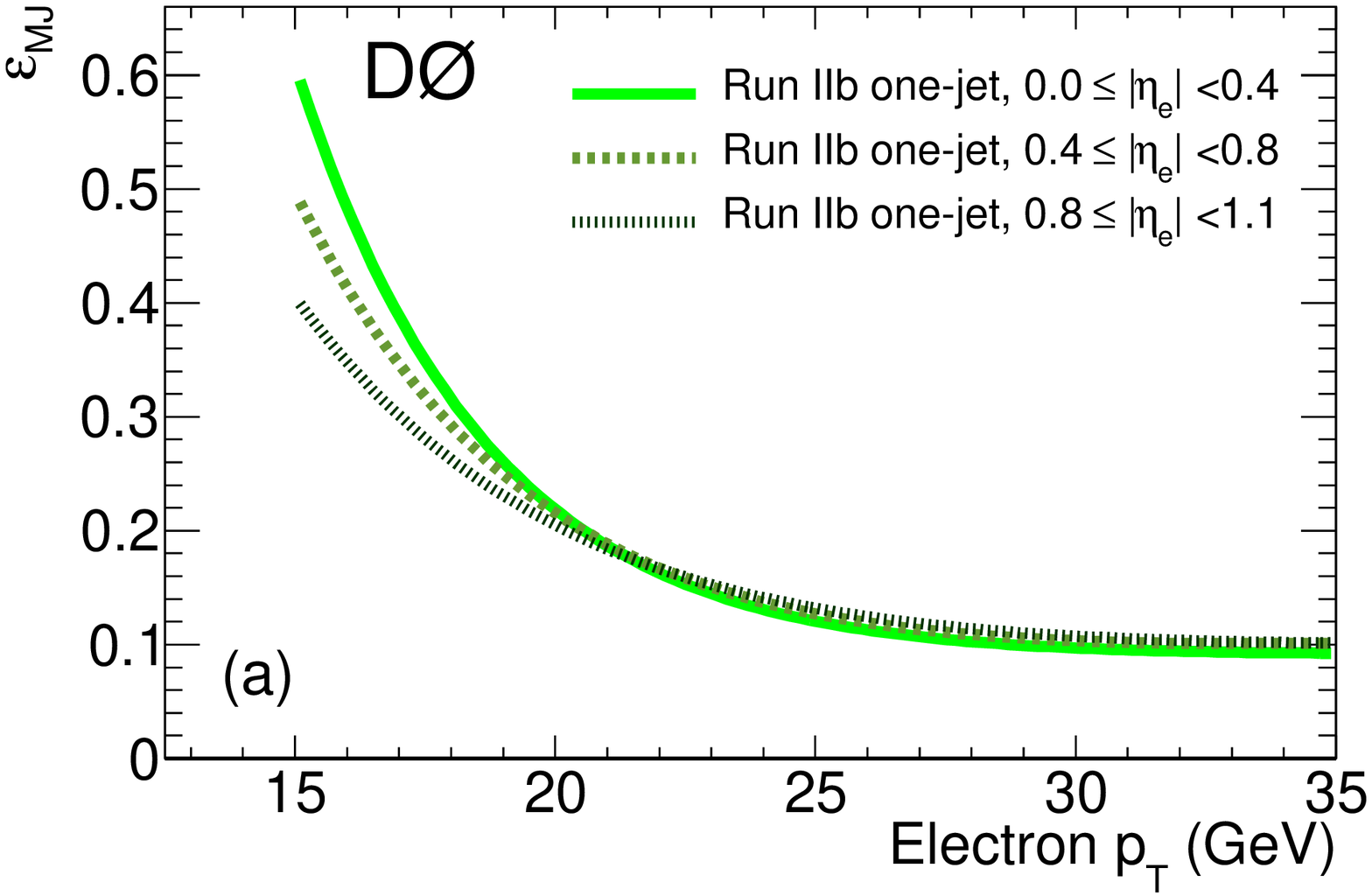}
    \includegraphics[width=\columnwidth]{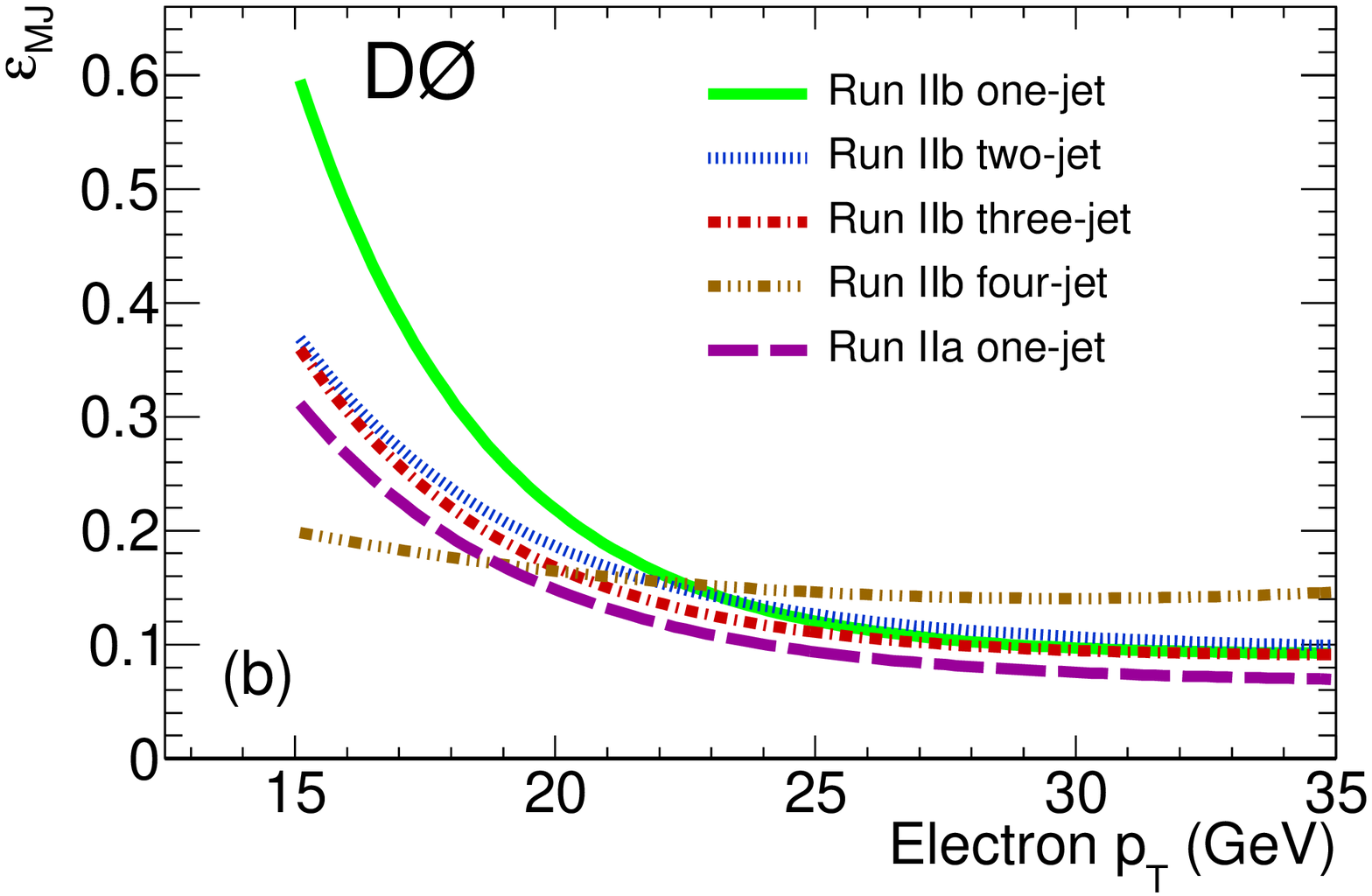}
    \caption{(color online) Parametrized \eqcd\ (as defined in Eq.~\ref{eqn:eqcd5}), used for the determination of the multijet component of reconstructed data distributions.
      \eqcd\ is parametrized as a function of electron $p_T$, electron pseudorapidity, inclusive jet multiplicity, and determined separately for Run~IIa and Run~IIb data.
      Fig.~\ref{fig:QCDFakeRates}(a) shows the variation of \eqcd\ as a function of electron $p_T$ for three electron pseudorapidity intervals in Run~IIb.
      Fig.~\ref{fig:QCDFakeRates}(b) shows the variation of \eqcd\ with jet multiplicity for Run~IIa and Run~IIb data. Similar variations with respect to electron 
      pseudorapidity and jet multiplicity are also observed for Run~IIa.
      \label{fig:QCDFakeRates}
    }
  \end{center}
\end{figure}
Dependencies of \eqcd\ on jet multiplicity, electron transverse momentum, and, to a lesser extent, electron pseudorapidity are observed and taken into account.
The rate at which fake electron candidates passing the loose selection criteria also pass the tight selection (\eqcd, as defined in Eq.~\ref{eqn:eqcd5}) 
is determined as a function of electron $p_T$, electron $|\eta|$, and jet multiplicity, and is shown in Fig.~\ref{fig:QCDFakeRates}.
Absolute uncertainties on \eqcd\ are approximately $(2-3)\%$ and are dominated by statistical uncertainties on the multijet-enriched data sample.
We find that the value of \eqcd, particularly at low misidentified electron $p_T$, increases in the Run~IIb data-taking period
compared to Run~IIa, which can be attributed to tighter electron shower-shape requirements applied in trigger selections in Run~IIb that reduce 
the differences between jets misidentified as electrons and real electrons entering the loose sample.

\subsection{Background and signal process simulation}
\label{sec:mcsamples}

\wjets\ events dominate the inclusive data sample, but there are backgrounds from $Z$+jets, $t\bar{t}$, diboson, single top quark, and multijet events.  
With the exception of multijet production, all processes are simulated using MC event generators.
All simulated samples are processed through the full \textsc{geant3}-based\,\cite{geant} simulation of the \Dzero\ detector.
Data events from random bunch crossings are overlaid on the simulated events to mimic the effects of detector noise and the presence of additional concurrent
$p\bar{p}$ interactions. 
The simulated events are weighted such that the instantaneous luminosity profile in the simulation matches the distribution observed in data.
These events are then reconstructed using the same software that is used on data.
The impact of the trigger efficiency turn-on curves is simulated by the application of trigger turn-on curves measured in data.   
Independent electron and jet samples are used to measure electron and jet trigger object efficiencies using tag-and-probe techniques.   
The overall trigger efficiency for the logical~\texttt{OR} of electron and electron+jet triggers is then calculated, taking into account all correlations, and is applied to the MC as an event weight.

We simulate the $W/Z+\mathrm{jets}$ and \tt\ processes with \alpgen\ v2.11\,\cite{Mangano:2002ea} interfaced with \pythia\ v6.403\,\cite{pythia} 
for the simulation of initial and final-state radiation and for jet hadronization, with the underlying event parameter settings tuned using ``Tune-A''\,\cite{TuneA}. 
A factorization and renormalization scale choice of $Q^2=M^2_V+\sum p^2_{Tj}$ is used for vector boson plus jets processes (where $M_V$ is the vector boson mass, and $p_{Tj}$ is
the transverse momentum of a jet in the event).
The normalization of \tt\ backgrounds is determined from NNLO calculations\,\cite{Moch:2008qy}.
The \pythia\ generator is used to simulate diboson production, with next-to-leading order (NLO) cross sections\,\cite{Campbell:1999ah} derived from the \textsc{mcfm} program\,\cite{Campbell:2010ff}, 
while production of single top quarks is simulated using the \textsc{comphep}-based NLO event generator \textsc{singletop}\,\cite{Boos:2004kh}. 

The $W/Z+\mathrm{jets}$ normalization is corrected by a multiplicative factor to match the inclusive $W/Z+\mathrm{jets}$
cross sections calculated at NLO\,\cite{MCFM}. Kinematic distributions are weighted to match 
existing $Z$ boson transverse momentum measurements in inclusive $Z$ boson events\,\cite{Zpt}, with corresponding corrections for $W$ boson events derived
through the application of $W$ to $Z$ $p_T$ distribution ratios from NLO predictions.
The heavy-flavor fractions are further corrected by the ratio of heavy-to-light NLO multiplicative factors as discussed in Ref.~\cite{Abazov:2011mi},
determined from NLO pQCD calculations from \textsc{mcfm}.

The proportion of the data that is attributed to each of these background processes and to the signal process can be seen in Fig.~\ref{fig:njetmult}.   
\begin{figure}[htbp]
  \begin{center}
    \includegraphics[width=\columnwidth]{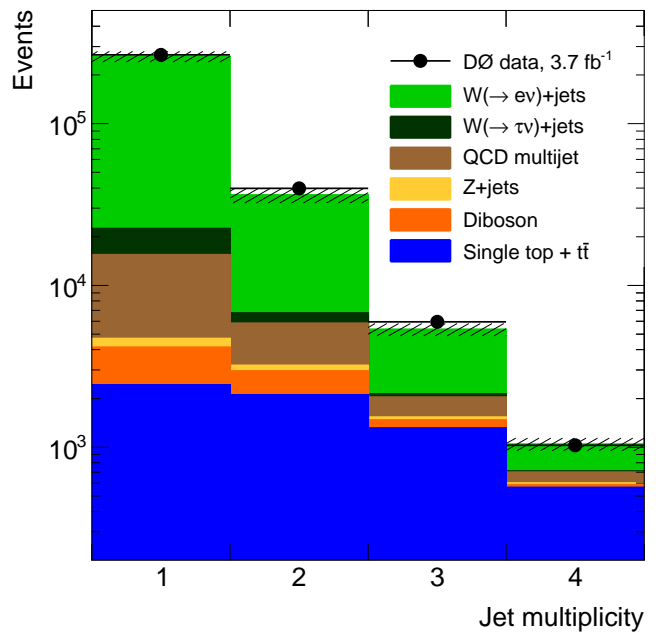}
    \caption{(color online) Uncorrected inclusive jet multiplicity distributions in the $\wenu+\mathrm{jet}$ event selection. 
      Hatched regions indicate normalization and shape uncertainties on the sum of the predicted contributions.
      All signal and background sources are derived from MC simulations with the exception of the QCD multijet component which is estimated from data. 
     \label{fig:njetmult}
    }
  \end{center}
\end{figure}

\begin{figure}[htbp]
  \begin{center}
    \includegraphics[width=\columnwidth]{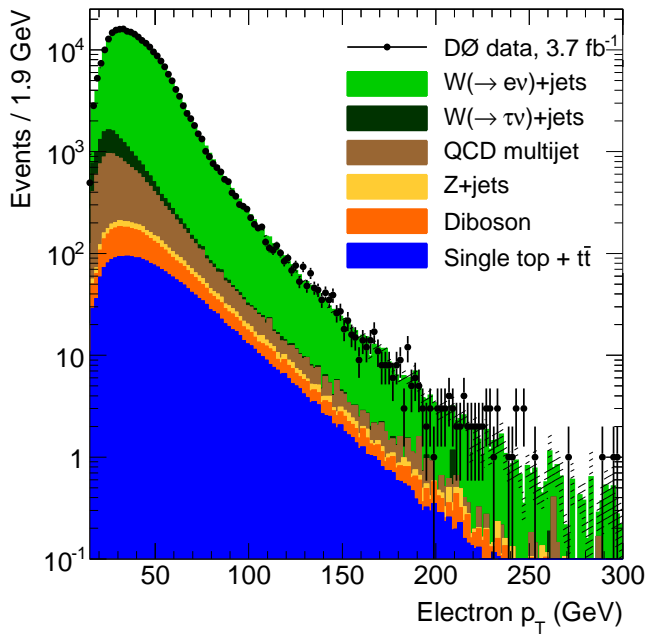}
    \caption{(color online) Uncorrected electron $p_T$ distribution
      for events with a $W$ boson candidate and one or more jets.
      Hatched regions indicate normalization and shape uncertainties on the predicted distributions.
      \label{fig:controlkinematics}
    }
  \end{center}
\end{figure}

Figures~\ref{fig:controlkinematics} and~\ref{fig:controlkinematics2} illustrate kinematic distributions for selected data events
and MC simulations plus data-driven multijet background contributions for some representative observables.
The estimated fraction of the data sample that is due to background processes ranges from (2--40)\% as a function of the measured observables
and the fraction of background due to top quark production ranges from (0--20)\%, with the larger contributions at higher jet multiplicities in both cases.

\begin{figure*}[htbp]
  \begin{center}
    \includegraphics[width=0.49\textwidth]{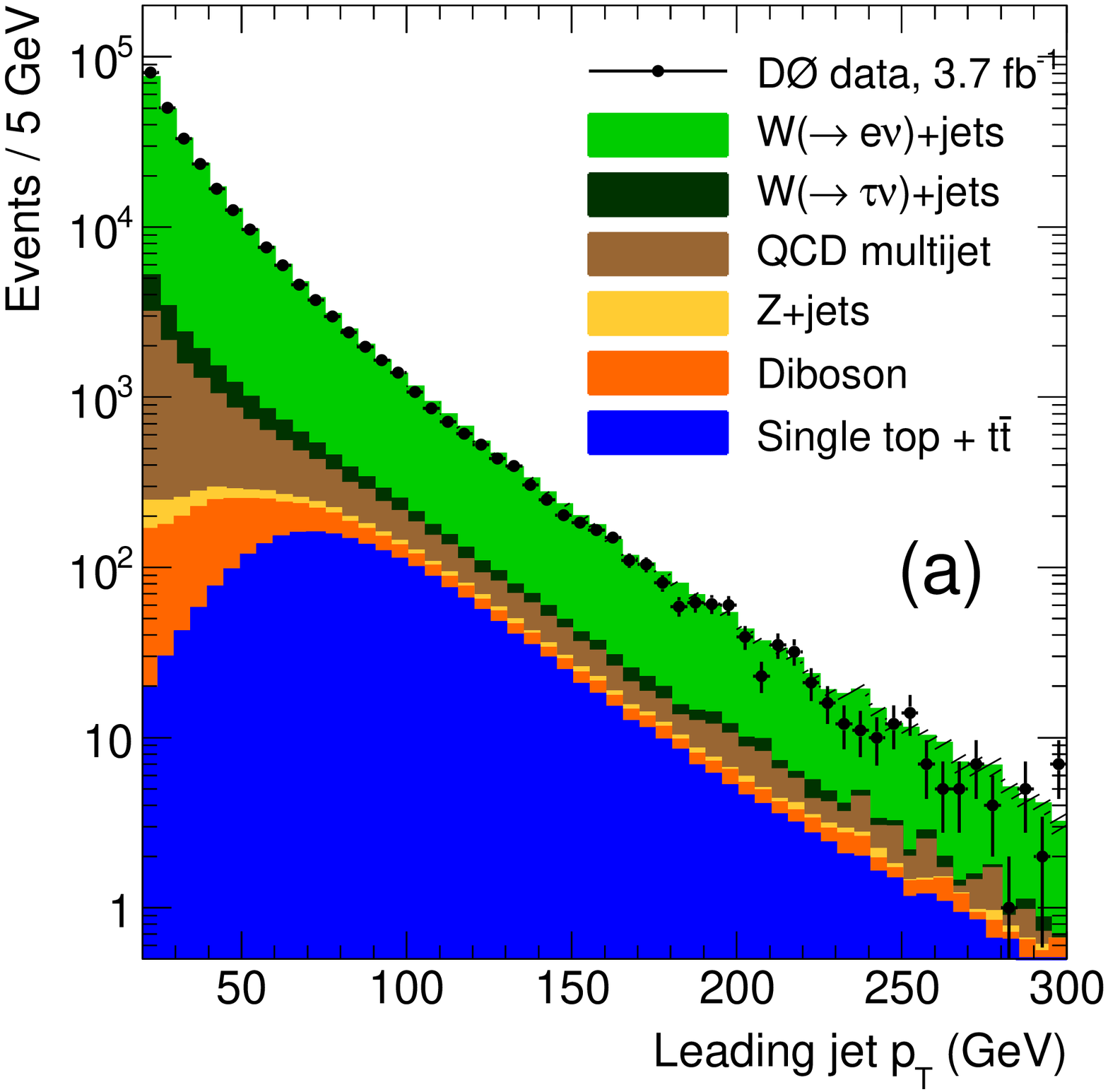}
    \includegraphics[width=0.49\textwidth]{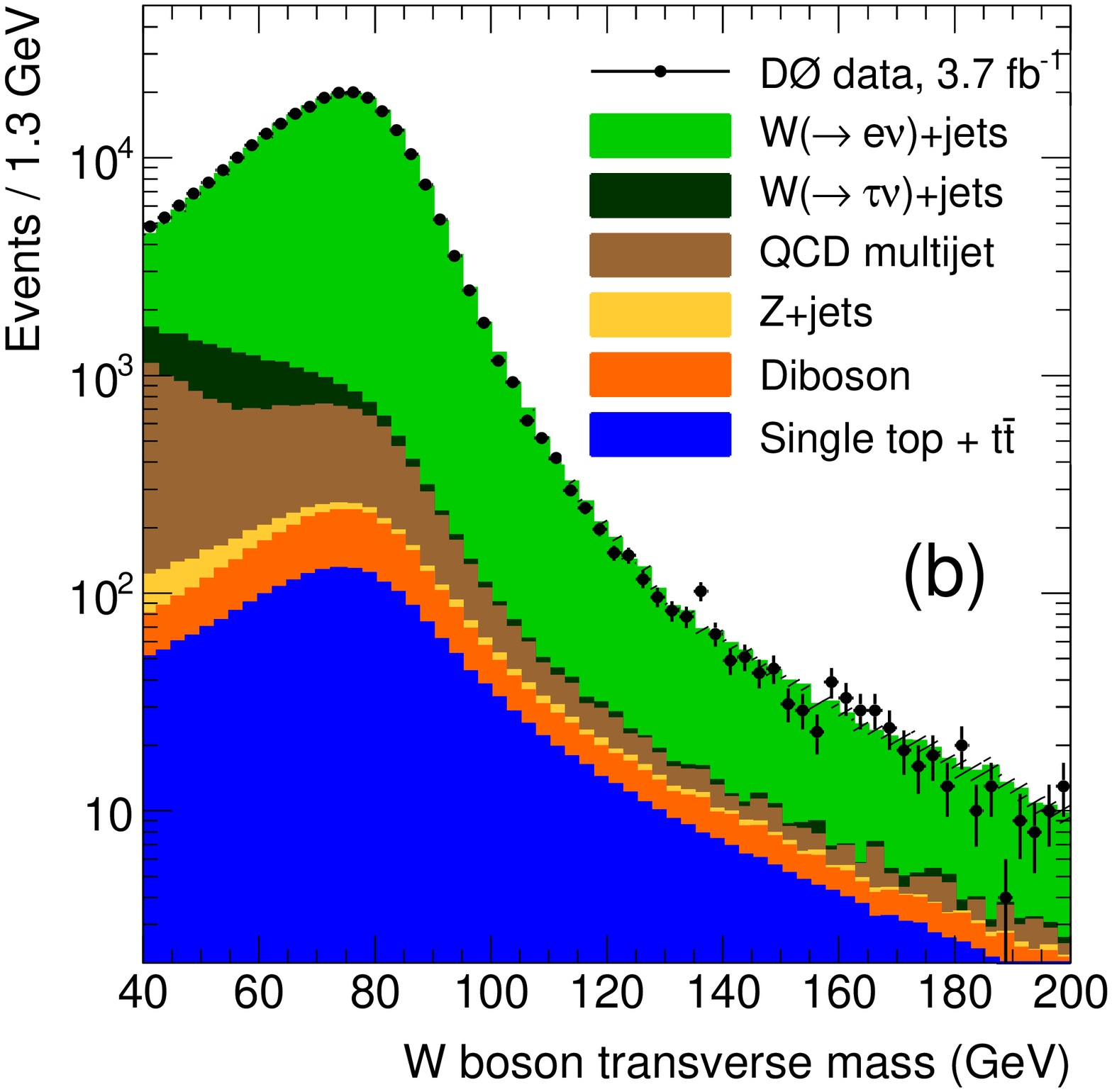}
    \caption{(color online) Uncorrected kinematic distributions of (a) leading jet $p_T$, (b) $W$ boson transverse mass
      for events with a $W$ boson candidate and one or more jets.
      Hatched regions indicate normalization and shape uncertainties on the predicted distributions.
      \label{fig:controlkinematics2}
    }
  \end{center}
\end{figure*}

\section{Correction of experimental data for detector effects}
\label{sec:unfold}

The background-subtracted yields of \wjets\ signal candidates are corrected back to the particle level taking into account corrections for
detector acceptance and efficiencies, as well as detector resolution effects. These corrections are performed using a singular value decomposition
regularized unfolding approach as implemented in \guru\,\cite{Hocker:1995kb}.
We define the kinematic phase space into which we unfold our final results by the selection in Table~\ref{tab:phasespace}
(the same selection criteria are applied at the reconstruction level).
\begin{table}[htbp]
\caption{\label{tab:phasespace} Unfolded phase space of the measurement.}
\begin{ruledtabular}
\begin{tabular}{lr}
 Jet transverse momentum & $p^\mathrm{jet}_T>20$~GeV \\
 Jet rapidity            & $|y_\mathrm{jet}|<3.2$ \\
 Electron transverse momentum & $p^e_T>15$~GeV \\
 Electron pseudorapidity  & $|\eta^e|<1.1$ \\
 Sum of all neutrino transverse energies & \met$>20$~GeV \\
 Transverse $W$ boson mass requirement & $M_T^W>40$~GeV \\
\end{tabular}
\end{ruledtabular}
\end{table}

Electron candidates are defined at the particle level to have the electron four-momentum modified to include all 
collinear radiation within a cone of radius $R=0.2$ to account for final-state radiation.
At the particle level we define \met\ as the magnitude of the neutrino transverse momenta.
Particle level jets are constructed using the \Dzero\ Run II midpoint cone algorithm running at particle level.
The $W$ boson decay products (including collinear emissions from the electron) are removed from the list of stable particles 
before constructing jets with a cone radius $R=0.5$. 

Bin boundaries in the unfolded observables are chosen based on detector resolution 
(chosen to define bin widths to be significantly larger than the corresponding resolution for measurements within each bin)
and available data statistics, while allowing for sensitivity to the shape of the unfolded observable.

\subsection{Regularized unfolding using GURU}

Due to the limited resolution of the detector, a significant fraction of events
may be measured to be in a different kinematic interval than they were
at the particle level, so a simple bin-by-bin correction for acceptance and efficiencies is not adequate.
The aim of unfolding is to correct a measured observable back to the particle level observable, accounting both for the effect
of finite experimental resolution, and for the detector response and acceptances. The relationship of the
true particle level distribution $T(x^\mathrm{true})$ to the reconstructed distribution $R(x)$ for an observable $x$ can be written as follows:
\begin{equation}
R(x)=\int^{x^\mathrm{true}_\mathrm{max}}_{x^\mathrm{true}_\mathrm{min}}dx^\mathrm{true}{\cal A}(x^\mathrm{true}){\cal M}(x^\mathrm{true},x)T(x^\mathrm{true})
\end{equation}
where the limits $x^\mathrm{true}_\mathrm{min}$ to $x^\mathrm{true}_\mathrm{max}$ reflects the range of the variable we wish to measure,
${\cal A}(x^\mathrm{true})$ represents the probability for a given observable to be seen at reconstruction level as a function of its particle level value
(which takes into account acceptance, efficiencies, and analysis requirements), and ${\cal M}(x^\mathrm{true},x)$ is the migration matrix.

Experimental resolution affects the relationship between the reconstruction level and particle level objects so that corrections, ${\cal A}$, need to be applied
and are derived using \wjets\ \alpgen+\pythia\ MC simulation.
There are events passing the reconstruction level selection requirements that are not within the phase space defined at the particle level. 
There are also events that pass both reconstruction and particle level
selections, but due to jet energy resolution, the jet $p_T$-ordering (or rapidity-ordering in the case of those observables dependent on selecting the most forward-rapidity jets) 
is not consistent between particle and reconstructed jets.   These effects need to be corrected to ensure a well-defined relationship
between the reconstructed and particle level objects before bin migration corrections are applied. To account for this, we correct the measured distribution
for those events that originate outside the particle level phase space or where jet re-ordering (up to the jet multiplicity concerned) has occurred.
With this correction we also account for the presence of jets originating from additional $p\bar{p}$ interactions in the same and neighboring bunch crossings.
A further acceptance correction is applied as part of the unfolding procedure itself, correcting for those particle level events that would fail detector-level selection 
requirements and thus not be reconstructed.

The migration matrix, obtained from the same MC simulation used to build the acceptance corrections,
accounts for the probability of an event in a given particle level bin entering into various reconstruction level bins (or vice-versa) through the relation:
\begin{equation}
\vec{x}_\mathrm{reco}={\cal M}\cdot\vec{x}_\mathrm{true}.
\end{equation}
Ideally then, simply applying the inverse of the matrix ${\cal M}^{-1}$ to the measured reconstructed observables would provide us with the unfolded distributions. 
However, low-significance bins can introduce numerical instabilities and give rise to large, rapidly oscillating fluctuations that contain little 
meaningful information about the particle level distributions. 

The program \Guru\ counters this problem using a singular value decomposition technique that
allows for a regularized inversion of the migration matrix\,\cite{Hocker:1995kb}.
This regularization imposes the requirement that the second derivatives of the distributions be small, equivalent to the condition 
that the unfolded distribution should be smooth. 
While this suppresses fluctuations in the unfolded data that might otherwise be observed (and statistical correlations 
accounted for in \guru\ between bins can be large), the statistical uncertainties assigned in each bin of the 
unfolded spectrum still accurately reflect the possible statistical spread of the data that could be caused by such fluctuations.
Using a regularized unfolding approach allows for reduced dependency on MC inputs to the unfolding procedure and
reduced uncertainties on the final results compared to a bin-by-bin correction.

\begin{figure}[htbp]
  \begin{center}
    \includegraphics[width=0.98\columnwidth]{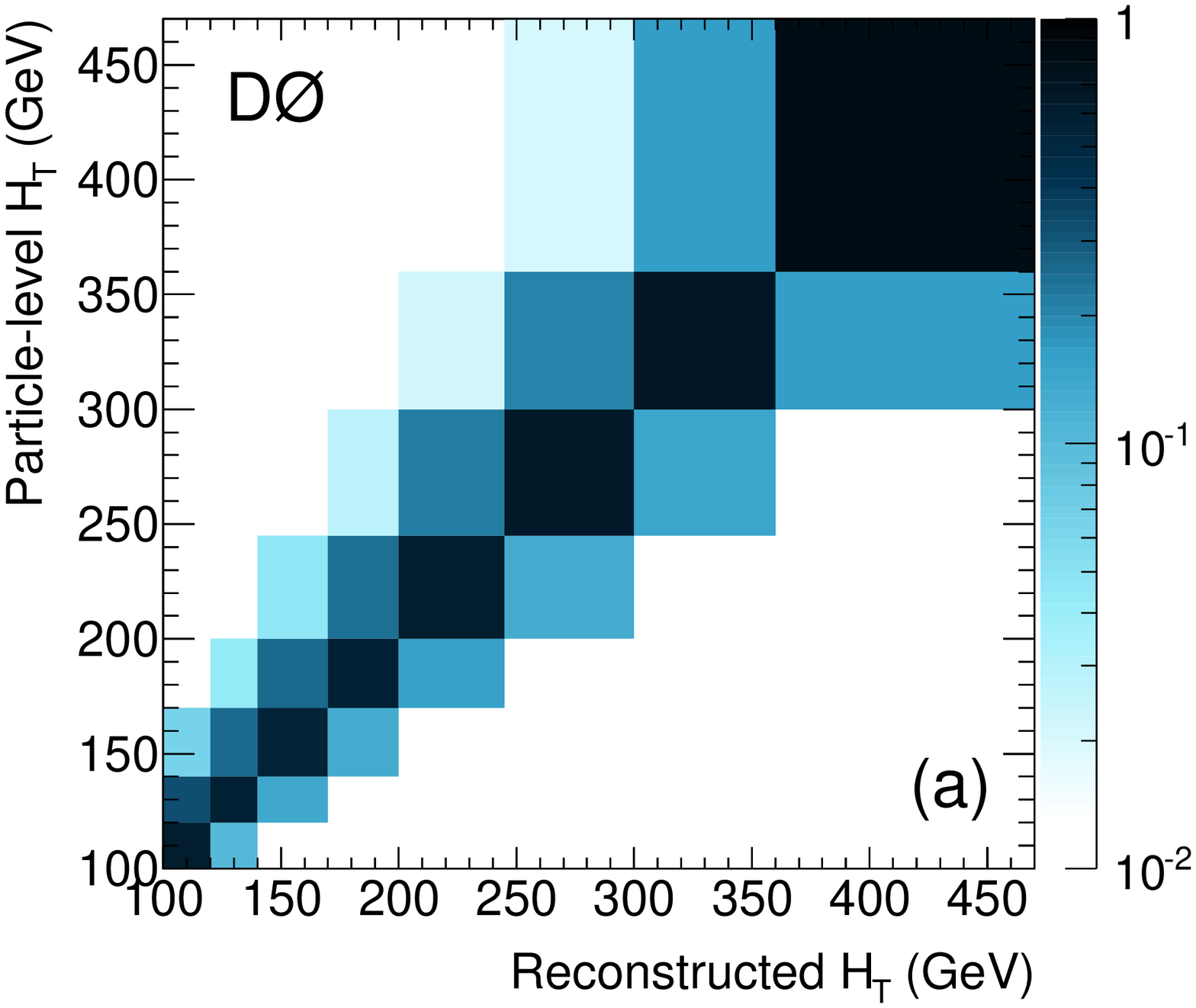}
    \includegraphics[width=0.98\columnwidth]{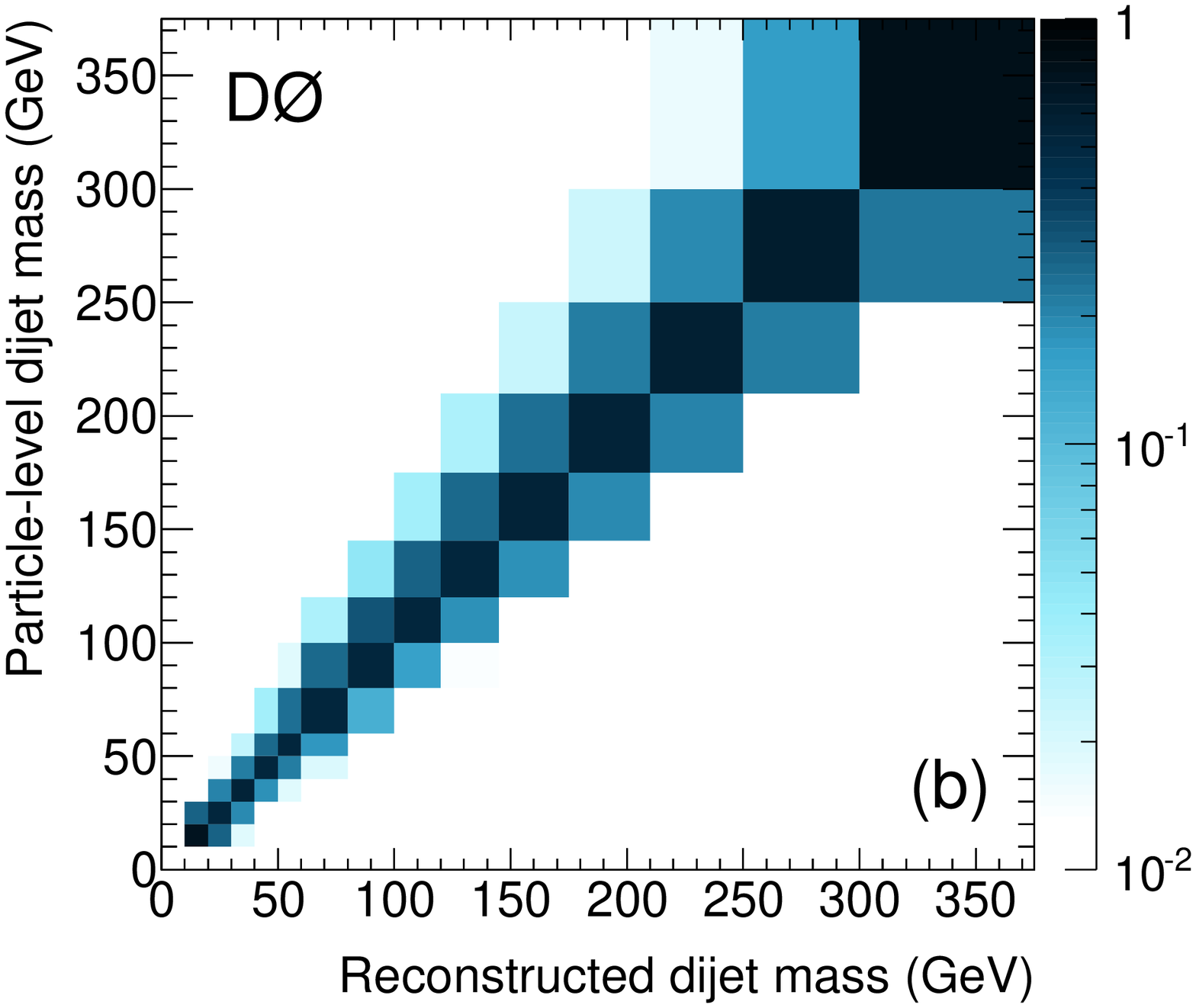}
    \includegraphics[width=0.98\columnwidth]{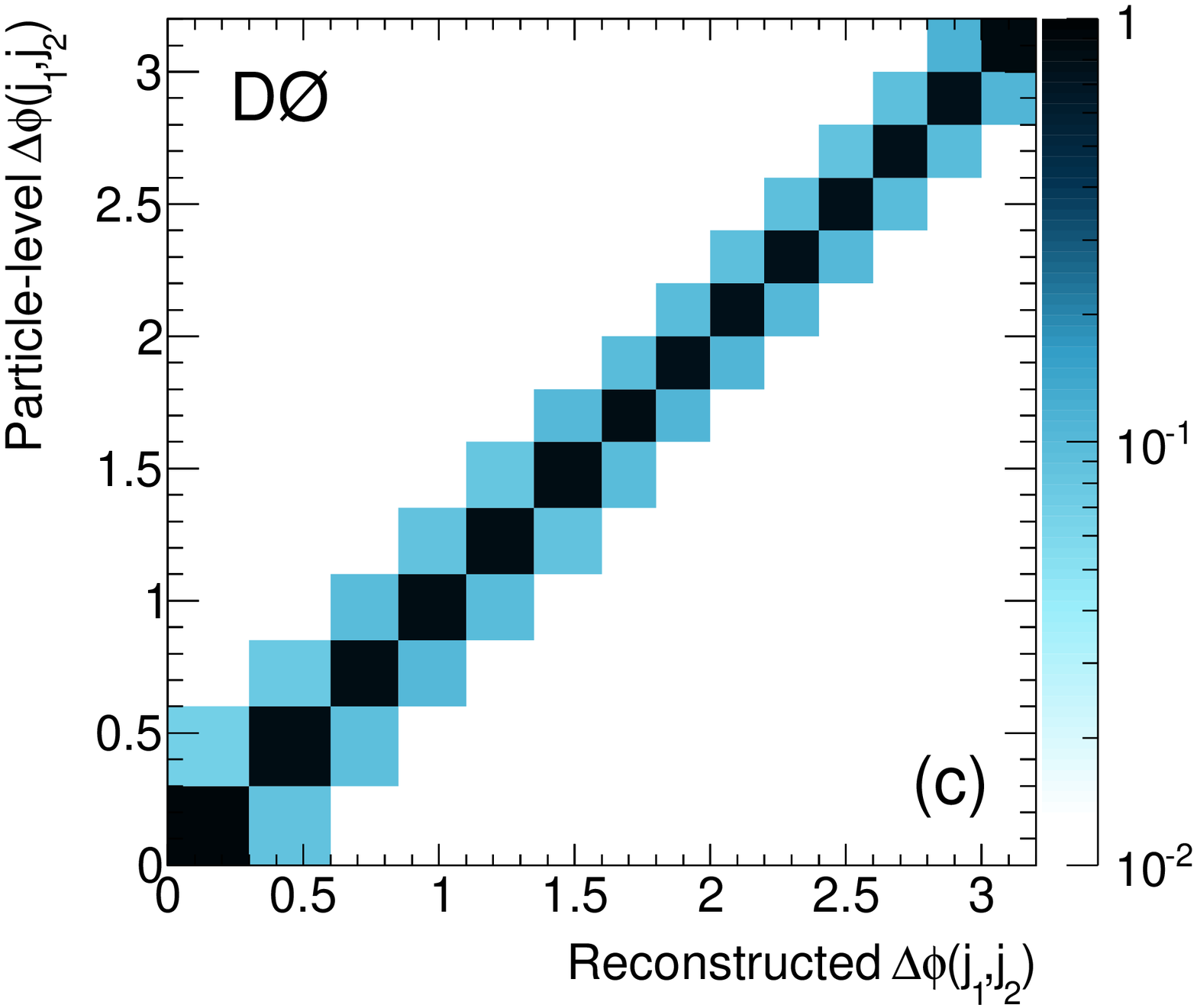}
    \caption{(color online) Migration matrices for (a) $H_T$ in the one-jet inclusive bin, (b) dijet mass in the two-jet inclusive bin, and (c) \dphi12\ in the two-jet inclusive bin.
      Element $(i,j)$ represents the probability for the particle level observable in bin $i$ to be reconstructed in observable bin $j$, and is represented by the axis on the right.
      \label{fig:migmatrix}
    }
  \end{center}
\end{figure}
Figure~\ref{fig:migmatrix} highlights three examples of the migration matrices used for observables measured
in this paper, expressed as a probability that a particle level observable in a given bin is reconstructed in the same or different bin.
Events entering the migration matrix must pass the selection at both reconstruction and particle levels,
and the reconstructed jets relevant to the observable in question must retain their $p_T$-ordering (or rapidity-ordering, where applicable)
between particle and reconstruction levels.

Following unfolding, the resultant distributions are normalized to the inclusive $\wenu$ production cross section to reduce experimental uncertainties.  
This value of $\sigma_W = 1097 {}^{+79}_{-89}$~pb
was measured\,\cite{Abazov:2011rf} in the same phase space, without any jet requirements, but with all other selection criteria as described in this analysis, 
and makes use of the same dataset. Total inclusive $n$-jet cross sections were also previously measured\,\cite{Abazov:2011rf} to be:
\begin{itemize}
\item $\sigma_{W+1\textrm{-jet}} = 119.5 {}^{+9.4}_{-8.3}$~pb,
\item $\sigma_{W+2\textrm{-jet}} = 19.0 {}^{+2.4}_{-1.9}$~pb,
\item $\sigma_{W+3\textrm{-jet}} = 2.9 {}^{+0.4}_{-0.4}$~pb,
\item $\sigma_{W+4\textrm{-jet}} = 0.39 {}^{+0.09}_{-0.07}$~pb.
\end{itemize}

\subsection{Evaluation of unfolding biases and\protect\\ statistical uncertainties}

To assess and correct for any bias that might have been introduced into the unfolded distribution by our acceptance and unfolding corrections,
and to determine the statistical and systematic uncertainties on the final results, we perform pseudo-experiments using ensembles constructed to mimic the measured
and corrected data distributions that replicate the statistical fluctuations present in the data.
A large sample of \alpgen+\pythia\ MC \wjets\ events first receives an ad-hoc correction so as to describe data at both the reconstruction level and after unfolding.
This correction is performed independently for each distribution under study, with the aim of creating distributions that mimic the data distributions.

Five hundred ensembles per distribution are then drawn from this corrected \alpgen+\pythia\ MC sample. The probability of an
event entering a given bin in a given distribution is chosen such that the ensembles not only reproduce the data distributions at particle and reconstruction levels
but also have statistical fluctuations, both bin-to-bin and in overall yield, as are observed in the data.
For each distribution, a particle level and reconstruction level equivalent is constructed, and these ensembles reflect the results that 
would be expected from repeated independent experimental measurements.

Each of the ensembles in turn then receives the same acceptance corrections as are applied to the measured data distribution, and are unfolded using \guru\ under the same conditions.
Unlike in the data, however, for each of these unfolded distributions we may compare the results to the corresponding ensemble's particle level distribution.
For all ensembles in turn, for each distribution, for each bin, the residual
\begin{equation}
r = \frac{\textrm{particle level result} - \textrm{unfolded result}}{\textrm{unfolded result}}
\end{equation}
is calculated and determines the fractional shift in the unfolded distribution from its true value.
In a given bin, over all ensembles, $r$ follows a Gaussian distribution. The mean value of $r$
is the fractional bias due to the unfolding procedure and the width 
of the Gaussian fit representing the statistical uncertainty on the measured unfolded result in that bin.

The unfolding bias for a given bin is applied as a correction to the unfolded data distribution in that bin.
This correction is small, generally a few percent in magnitude, and always much smaller than the statistical uncertainty in the bin
except in the case of the $p_T$-ordered third jet emission probability results, where the bias becomes larger than the statistical uncertainty at wide opening angles.
After the initial correction for the unfolding bias is applied to the data, the \alpgen+\pythia\ MC samples used to obtain the acceptance corrections and migration matrices 
are corrected to the data.   
After the \guru\ unfolding and bias assessment procedures are applied to these new MC inputs, no further biases are observed.
The statistical uncertainty on the fitted mean and any difference between the extracted bias from a Gaussian fit and the arithmetic mean 
of the distribution are taken as systematic uncertainties on the value of this bias.

\section{Systematic Uncertainties}
\label{sec:systs}

The two dominant sources of uncertainty on the majority of the unfolded distributions come from uncertainties in the jet energy scale (JES) and in the jet energy resolution (JER).
At large jet rapidities and large opening angles between jets, uncertainties on the jet-vertex confirmation requirement also contribute a significant amount to the overall uncertainties,
while at very low and at high electron transverse momenta, trigger efficiency uncertainties become one of the dominant systematics (due to limited statistics in the $Z/\gamma^*\to e^+e^-$
data samples used to calculate the efficiencies).
At large dijet opening angles, $p_T$, and invariant mass systematic uncertainties from multijet backgrounds and the unfolding procedure become important
due to limited statistics in the multijet-enriched data samples and relatively large unfolding corrections.

Systematic uncertainties from JES, JER, jet-vertex confirmation, trigger efficiency, and jet identification efficiency are assessed with ensemble tests. 
The same MC ensembles used in determination of the statistical uncertainties and the unfolding bias are once again unfolded, but this time 
using acceptance corrections and migration matrices derived using fully-reconstructed \alpgen+\pythia\ events produced with detector responses shifted 
by one standard deviation from the nominal values, for each of the sources of systematic uncertainty separately.
Unfolding biases are then re-determined for these systematically shifted unfolded ensembles.

As the input ensembles to the unfolding procedure are identical to those used for the nominal unfolding, any change in the unfolding bias must be due 
to the modified detector response and acceptance correction inputs to \guru. Any change in the unfolding bias from the nominal to the systematically shifted 
ensembles is therefore attributed to the effect of the shifted detector response and assigned as the corresponding systematic uncertainty due to that detector effect.
This method enables the impact of a shifted detector response to be translated into the unfolded cross section result while also accounting for possible 
changes in the bin migrations due to the change in the response.

The uncertainties derived using this method are compared with the
ratios of systematically shifted to nominal response \wjets\ MC samples at reconstruction level. The shifts seen were found to be in
good agreement with those found at the particle level.
The impact of these sources of systematic uncertainty on the background contributions (that are subtracted from the reconstruction level
data before unfolding) is determined using this method, and the uncertainties on the signal due to the background subtraction are added in quadrature to give the total systematic uncertainty.

\begin{figure*}[htbp]
  \begin{center}
    \includegraphics[width=0.49\textwidth]{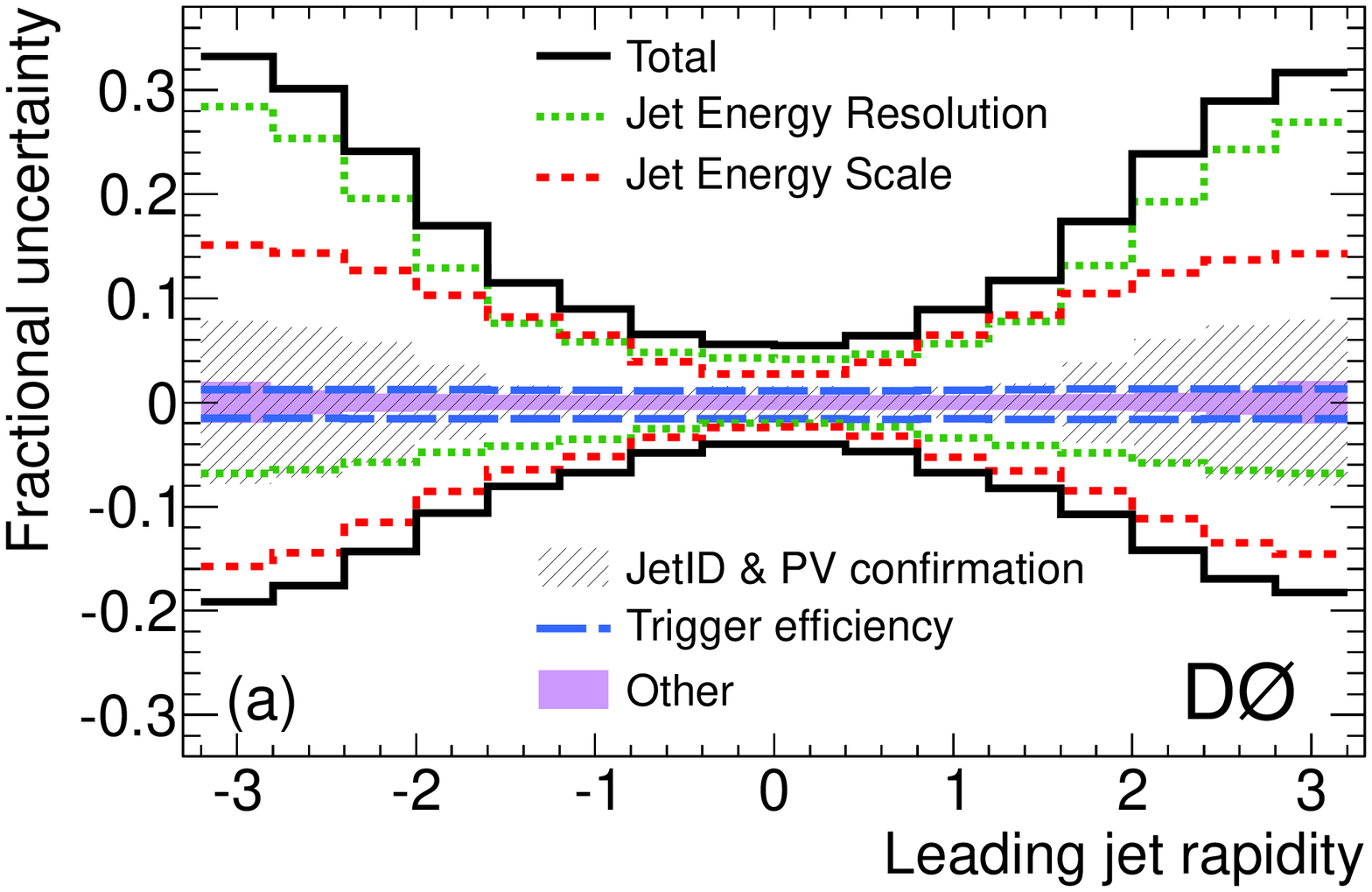}
    \includegraphics[width=0.49\textwidth]{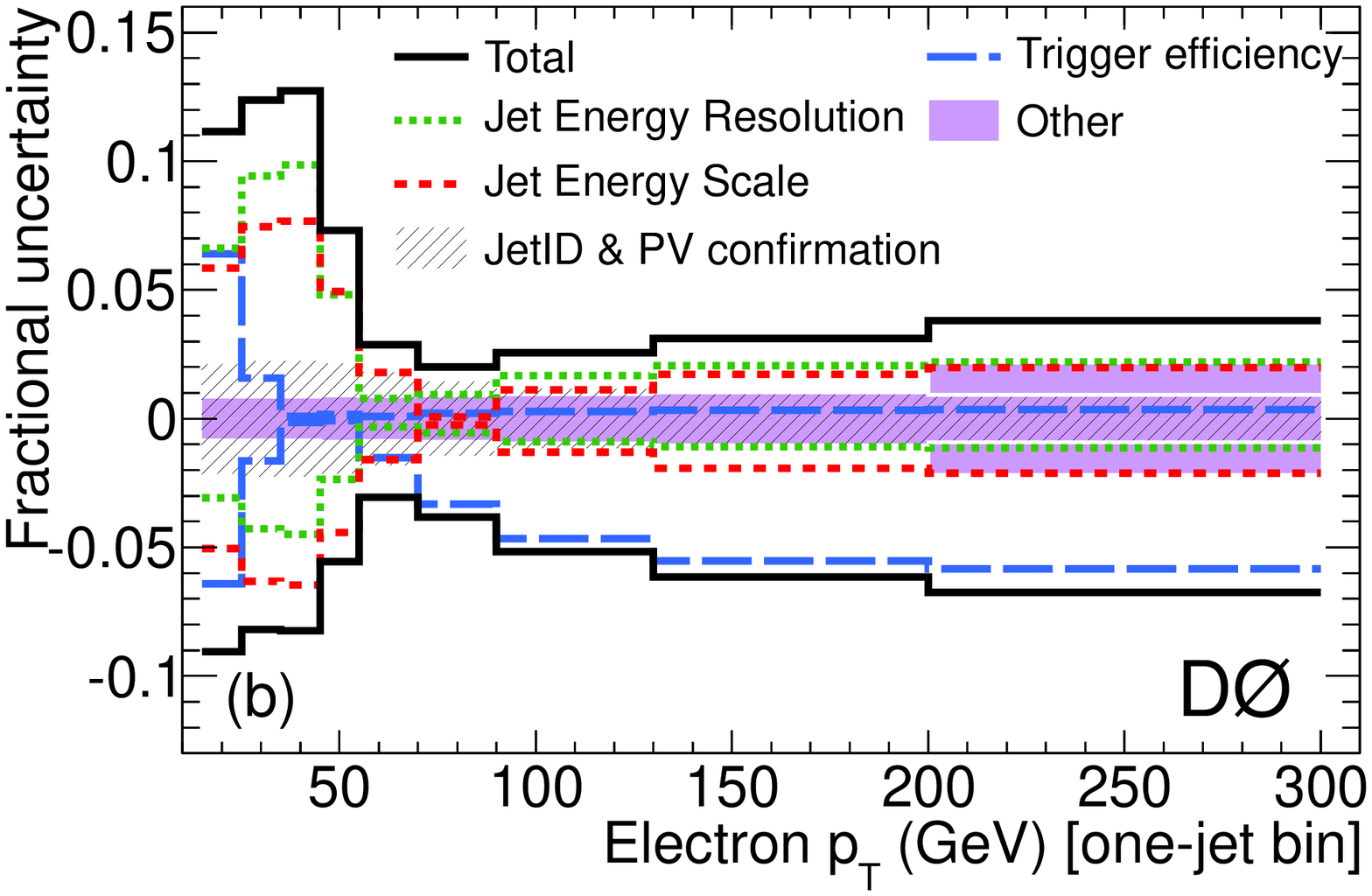}
    \includegraphics[width=0.49\textwidth]{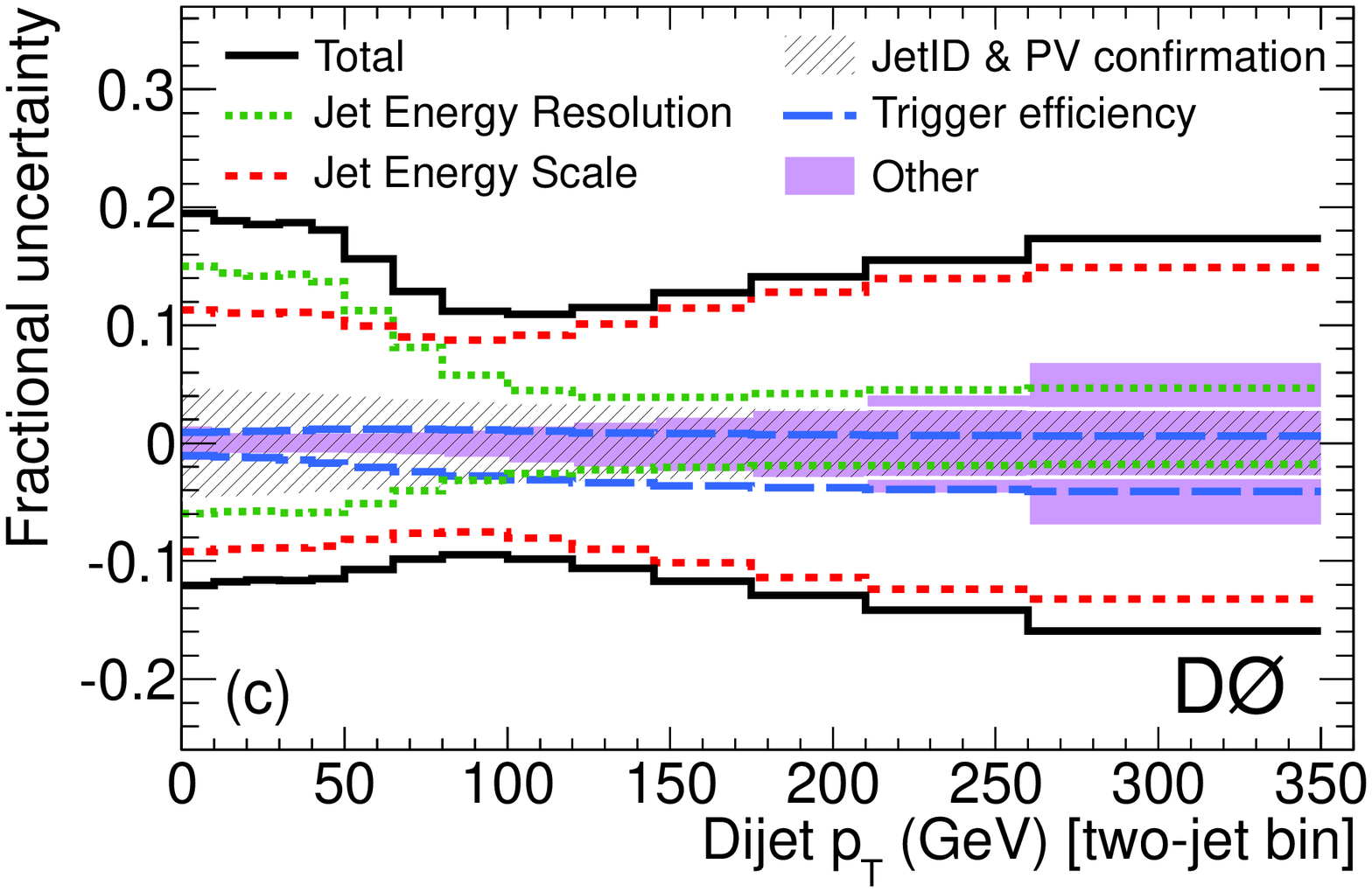}
    \includegraphics[width=0.49\textwidth]{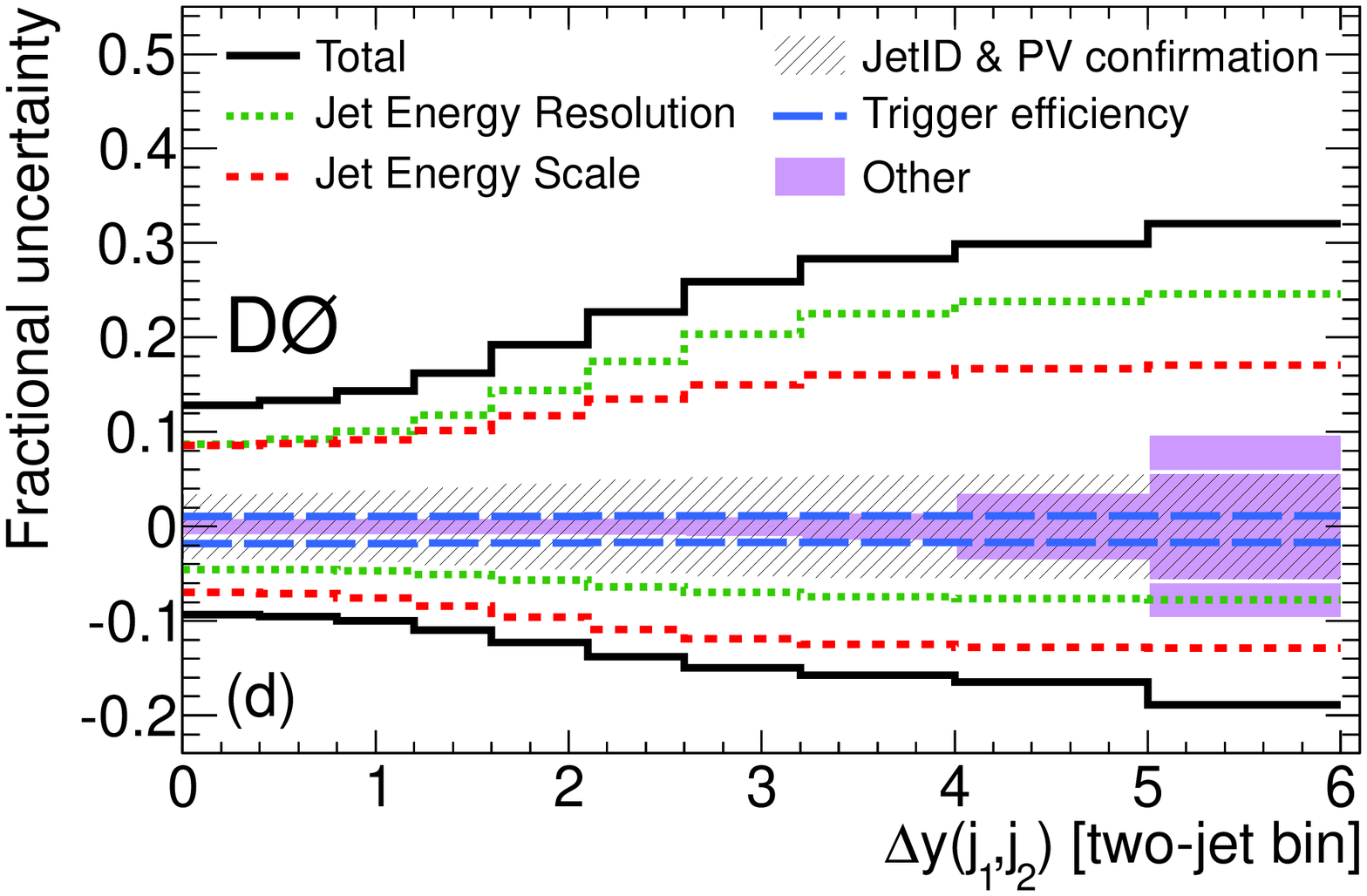}
    \caption{(color online) 
      Summary of systematic uncertainties on the normalized cross sections of (a) leading jet rapidity and (b) electron $p_T$ in the inclusive one-jet bin, and of the (c) dijet $p_T$ 
      and (d) dijet rapidity separation of the two highest-$p_T$ jets in the inclusive two-jet bin. The most significant sources of uncertainty are shown separately. 
      Additional sources of uncertainty due to background modeling, electron identification, and the unfolding procedure are grouped under ``Other.''
      \label{fig:systs}
    }
  \end{center}
\end{figure*}

Uncertainties on the normalization and shape of the data-driven multijet background can arise from uncertainties in the electron efficiency, in the \eqcd\ determination, 
and from the statistical uncertainties on the data samples used for the background modeling. The dominant contribution to the total uncertainty on the multijet background is the \eqcd\ uncertainty except in the tails of distributions where control sample data statistics also play a role.
The uncertainty in \eqcd\ is determined for Run~IIa and Run~IIb independently, as a function of each of the observables measured. The two uncertainties are combined
in each analysis bin by taking the corresponding systematic uncertainty from each data period and 
scaling its relative contribution according to the fractional multijet content originating from Run~IIa and Run~IIb data in a given bin.
Despite being the largest background contribution to the low jet multiplicity events (see Fig.~\ref{fig:njetmult}), 
total uncertainties due to multijet backgrounds in the inclusive one-jet bin are less than 1\% on average.
For high jet multiplicities this uncertainty can rise to 8\% at high electron $p_T$ and 20\% for the largest jet rapidities in the inclusive four-jet bin.

The systematic uncertainty associated with the $t\bar{t}$ production cross section is determined by varying the assumed $t\bar{t}$ production rate within its theoretical uncertainty\,\cite{Moch:2008qy} ($+6,-9\%$), and determining the associated uncertainty on the \wjets\ contribution by 
assuming the additional (reduced) top quark contribution reduces (increases) the \wjets\ contribution directly.
This theoretical uncertainty translates into an uncertainty on the $t\bar{t}$-subtracted \wjets\ signal
of $(10-15)\%$ or larger where top quark backgrounds dominate the event selection, in high jet multiplicity events, and at high $W$ boson $p_T$ or at high $H_T$.

Electron identification uncertainties can originate from uncertainties in the background subtraction and fits to the efficiency turn-on curves. We also
consider uncertainties on the $p_T$, instantaneous luminosity, and jet multiplicity (including electron-jet spatial separation) dependence of the efficiencies.
We benefit from a reduction of the electron identification systematic uncertainty in the final measurements through normalization of the differential cross sections
to the total measured $W$ boson production cross section. After cancellations, the residual electron identification systematic uncertainty is approximately 1\%.

The systematic uncertainty on the measurement due to the unfolding procedure includes uncertainties from the derivation of the unfolding bias value (both the statistical
uncertainty on the mean and the difference between the correction determined from a Gaussian fit or from the arithmetic mean) and from statistical
uncertainties on the acceptance corrections used by \guru. To check the dependence of an imperfect MC modeling of the kinematic spectra on the inputs to the 
unfolding procedure, the unfolding is repeated with a data-derived correction of the MC samples used to generate the acceptance and migration matrices
so as to provide the best description of the observed data and the shift in the final results with these new inputs assigned as a systematic uncertainty. 
The uncertainties on the unfolding procedure are small $(\lesssim1\%)$ in most analysis bins, but can become more significant, most notably at 
large dijet rapidity separations $(6-9)\%$ and large jet rapidity $(2-4)\%$.

The jet spatial matching criterion used in the acceptance and bin migration corrections is set to half and twice the size of the cone radius $R=0.5$ of a reconstructed jet 
to test the dependence of the corrections on the matching choice. The impact on the final cross sections is found to be well below 1\% for most distributions, but reaching up
to 2\% for high jet multiplicity events with high $H_T$ and in events with wide dijet rapidity separations.

All sources of systematic uncertainty on the theoretical modeling, detector response-based systematics, background subtraction, and the unfolding procedure are added in quadrature to
arrive at a total systematic uncertainty on the unfolded distributions. 
Figure~\ref{fig:systs} illustrates the total systematic uncertainty as a function of four representative unfolded observables.
The contribution to the total uncertainty from jet energy scale, jet energy resolution, trigger efficiency, jet identification efficiency and jet-vertex confirmation
are shown separately. Smaller sources of uncertainty including electron identification efficiency, background shape/normalization, and unfolding procedure uncertainties,
are shown as a combined contribution.

\section{Theoretical Predictions}
\label{sec:theory}

We compare the data after correction for detector efficiencies and resolution effects to several theoretical models.
Comparisons are made to a range of widely used parton shower and matrix element plus parton shower matched MC programs,
to all-order resummation predictions, and to next-to-leading order (NLO) pQCD calculations.

\subsection{Monte Carlo programs}

We compare our results to \pythia~6.425 with the \textsc{perugia2011} underlying event tune and the CTEQ5l parton density function (PDF) set\,\cite{cteq5}. We also compare to
\herwig~6.520\,\cite{herwig} at leading-order (LO) in $\alpha_s$, using the CTEQ6ll PDF set\,\cite{cteq6l}, 
and interfaced to \textsc{jimmy}~4.31\,\cite{jimmy} for modeling of multiple parton interactions (MPI).
To assess the impact of the inclusion of additional matrix elements provided in the \alpgen\ MC program, we compare predictions from \alpgen~2.414 hadronized in two ways,
using either \pythia\ or \herwig(+\textsc{jimmy}) with the same program version, underlying event tune, and PDF set as the stand-alone predictions.

Comparisons are also made to leading-order matrix element plus parton shower matched MC produced with \sherpa\ v.1.4.0\,\cite{Sherpa} using
the CT10\,\cite{CT10} PDF set and with the factorization and renormalization scales chosen as discussed in Ref.~\cite{SherpaMerging}.
The \sherpa\ default tuning parameters are used, with the exception of the MPI cutoff scale, which was tuned by the \sherpa\ authors to fit the
CDF underlying event data in Drell-Yan production\,\cite{CDFrivet}. Hadronization is conducted using the \sherpa\ internal cluster fragmentation approach\,\cite{Sherpa}.

For these comparisons, results are provided using the D0 Run II midpoint cone algorithm at the particle level, with cone radius $R=0.5$,
and differential distributions are normalized to the inclusive $W$ boson production cross section determined from the same MC 
program in the same measurement phase space (as summarized in Table~\ref{tab:phasespace}).

\subsection{All-order resummation}

\textsc{High Energy Jets} (\hej)\,\cite{HEJ,BFKL-HEJ} is an implementation in a parton level MC generator of an exclusive, 
all-order resummation of the perturbative contributions to production of wide angle emissions at hadron colliders. 
Such predictions are particularly suited for description of events containing two jets with a large rapidity separation.

Predictions are produced using the \Dzero\ Run II midpoint cone algorithm at the particle level with $R=0.5$. 
The factorization and renormalization scale is chosen to be $\mu=\mu_F=\mu_R=2\times \mathrm{max}\{p^j_T\}$, with scale uncertainties
estimated by varying this central scale choice by a factor of two or one-half.
As \hej\ is capable only of describing two-jet and higher multiplicity events, differential cross sections are normalized by the measured inclusive $W$ boson cross section\,\cite{Abazov:2011rf}
to allow a like-to-like comparison of the distributions with the data.
Scale uncertainties from \hej\ are generally larger than those in NLO pQCD calculations, due in part to \hej\ not including all running coupling terms
and also from not being able to benefit from cancellations in the inclusive $W$ boson cross section normalization.

\subsection{Next-to-leading order pQCD}

Next-to-leading order pQCD predictions of the production of \wjets\ in $p\bar{p}$ collisions at $\sqrt{s}=1.96$~TeV have recently become available for up to four partons
in the final state\,\cite{Berger:2010zx} from the Blackhat Collaboration\,\cite{Berger:2009ba}.
NLO \blackhat\ predictions are obtained using \textsc{blackhat} for calculation of the virtual terms, interfaced to \sherpa\ for
calculation of all real emissions.
Previous comparisons\,\cite{Abazov:2011rf} of the predictions from \blackhat\ and another NLO pQCD calculation approach, 
\textsc{rocket+mcfm}\,\cite{MCFM,Rocket}, at Tevatron energies have been found to be in good agreement with each other for numerically similar choices 
of renormalization and factorization scale in their prediction of jet transverse momenta and total cross sections for up to four jets, 
so only comparison of experimental data with \blackhat\ predictions are made in this paper.

The \blackhat\ predictions employ the renormalization and factorization scale choice $\mu=\mu_F=\mu_R=\frac{1}{2}{H}^\prime_T$ where 
${H}^\prime_T$ is defined as the scalar sum of the parton and $W$ boson transverse energies.
Scale uncertainties are estimated by varying $\mu$ by a factor of two or one-half.
The calculations use the MSTW2008nlo68cl PDF set\,\cite{MSTW}, with values of $\alpha_s(\mu)$ set consistently with the PDF choice, using a two-loop running at \nlo.
Uncertainties arising from the choice of PDF are studied\,\cite{MSTW} using the Hessian method 
by the Blackhat Collaboration and found to be negligible ($\lesssim2\%$) in comparison with scale uncertainties.

Predictions are generated with the \textsc{siscone}\,\cite{Salam:2007xv} jet algorithm applied at the particle level with split-merging fraction $f=0.5$ and cone radius $R=0.5$, 
rather than with the \Dzero\ Run II midpoint cone jet algorithm.
The predicted distributions are then corrected for the effect of using \textsc{siscone} rather than the \Dzero\ Run II midpoint cone jet algorithm using \sherpa\ (described below).
All cross section results are normalized to the theoretical inclusive $W$ boson production cross section prediction at NLO\,\cite{Berger:2009ba} 
determined to be $1153^{+17}_{-7}$~pb in the unfolded phase space.

\subsubsection*{Non-perturbative corrections}
\label{sec:NPCs}

\begin{figure*}[hbtp]
  \begin{center}
    \includegraphics[width=0.32\textwidth]{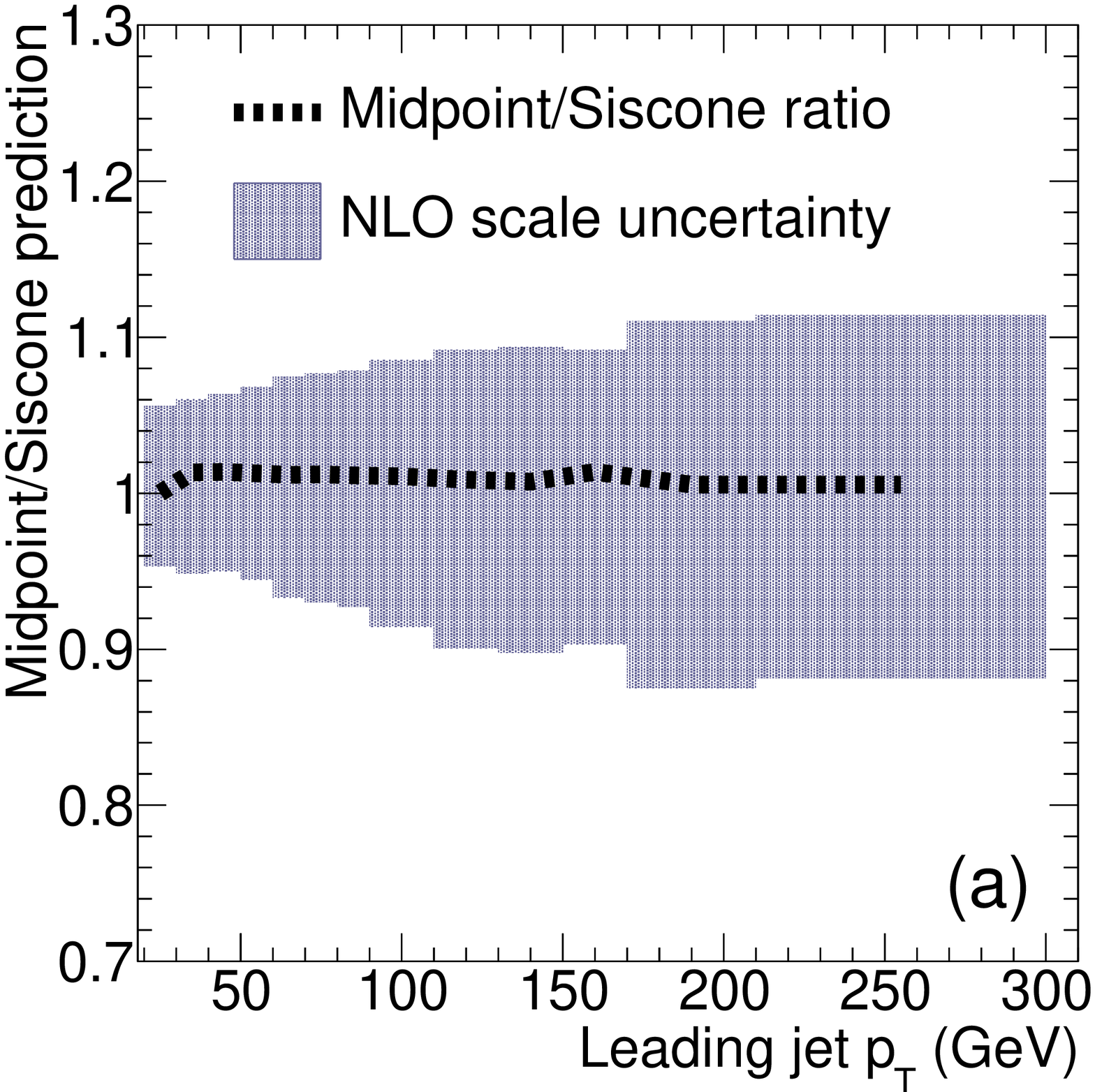}
    \includegraphics[width=0.32\textwidth]{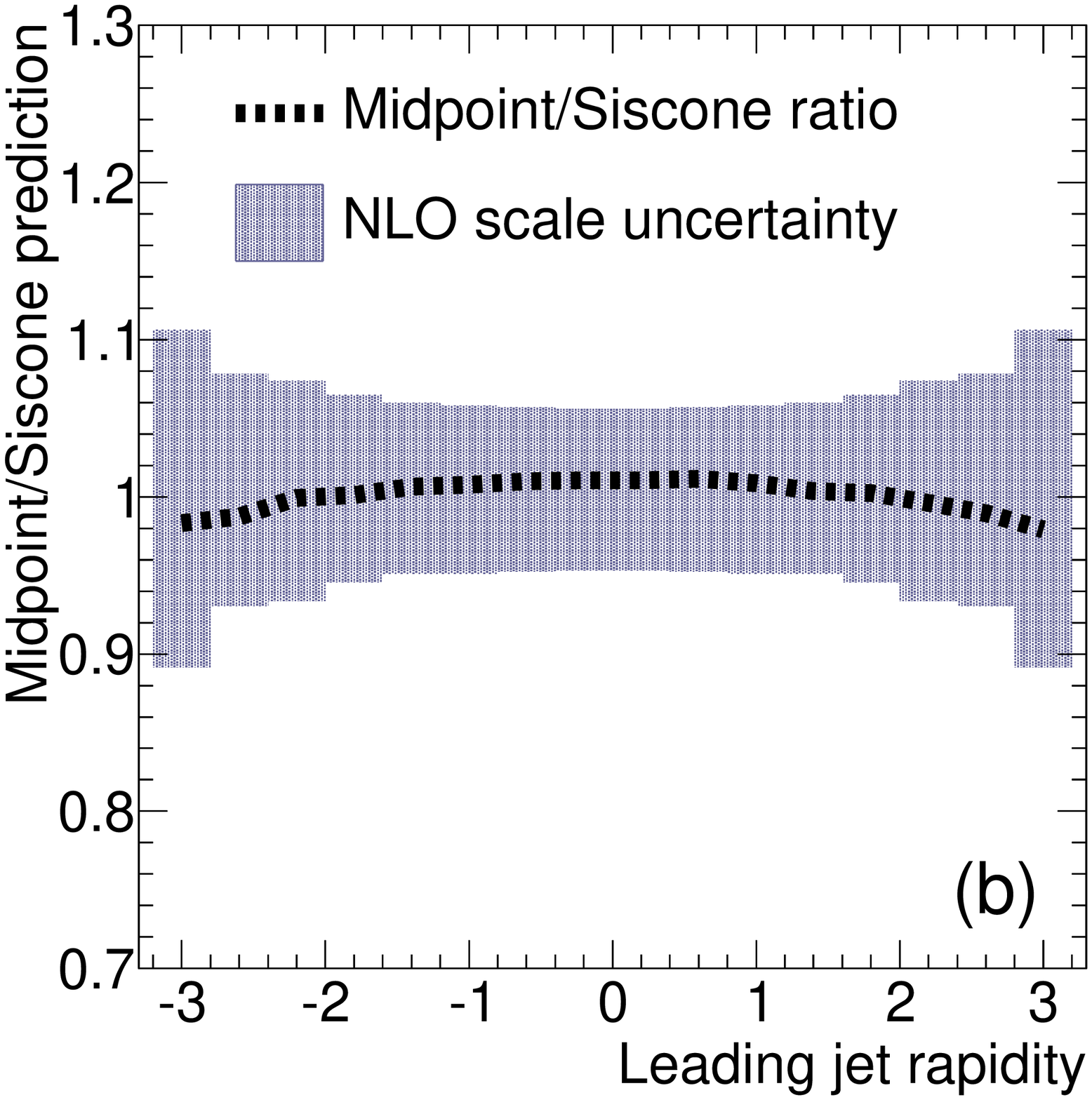}
    \includegraphics[width=0.32\textwidth]{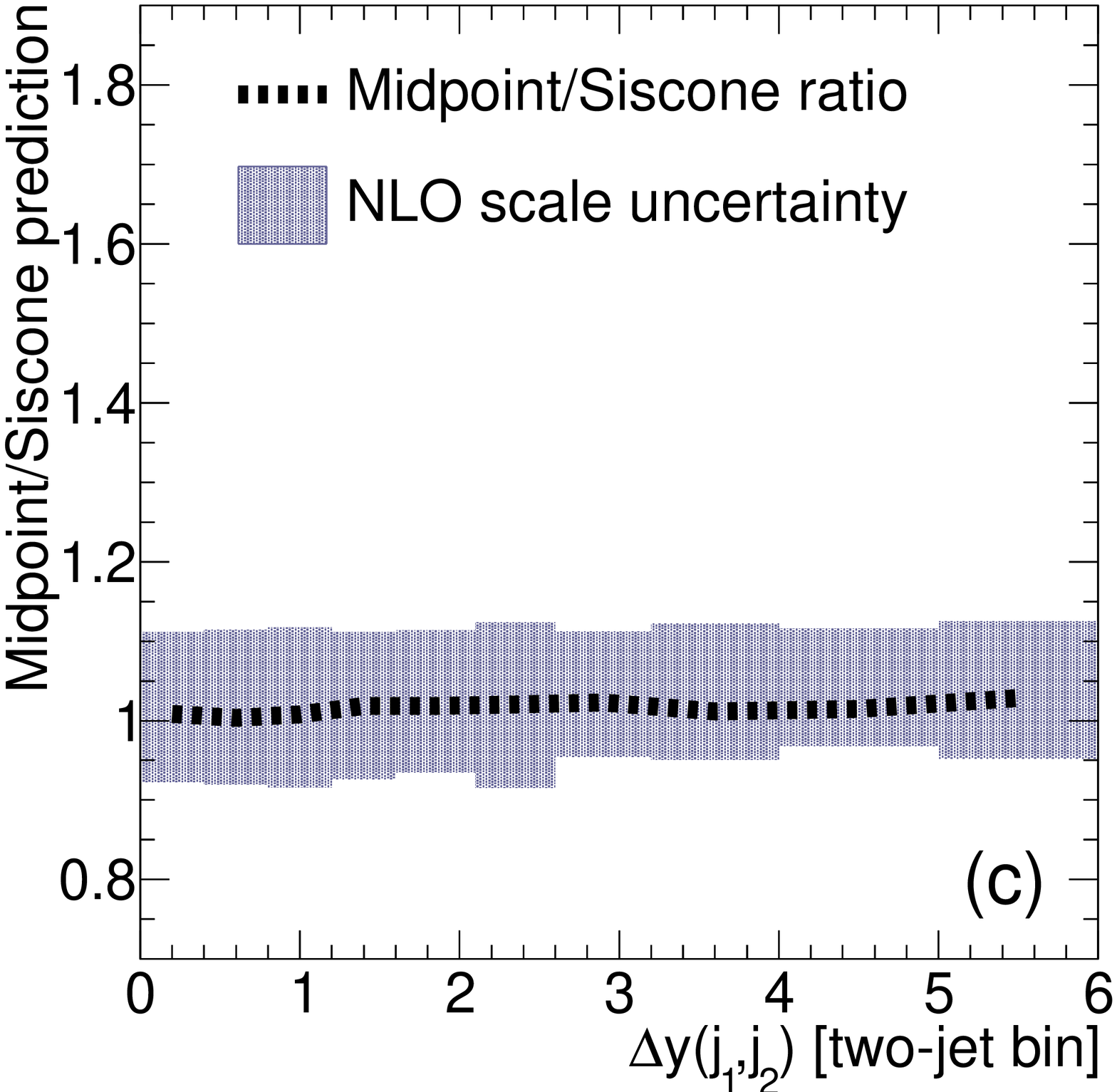}
    \includegraphics[width=0.32\textwidth]{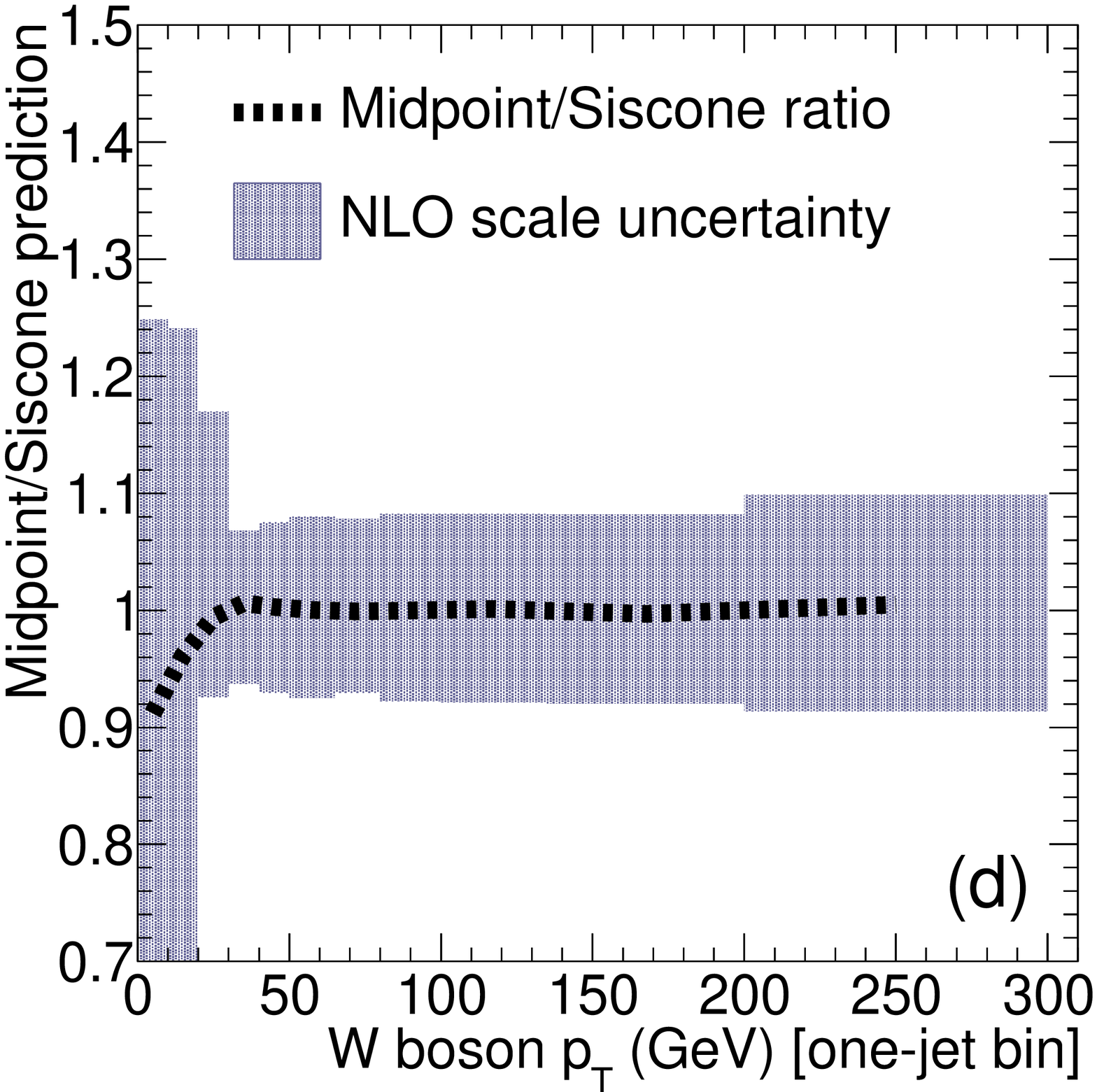}
    \includegraphics[width=0.32\textwidth]{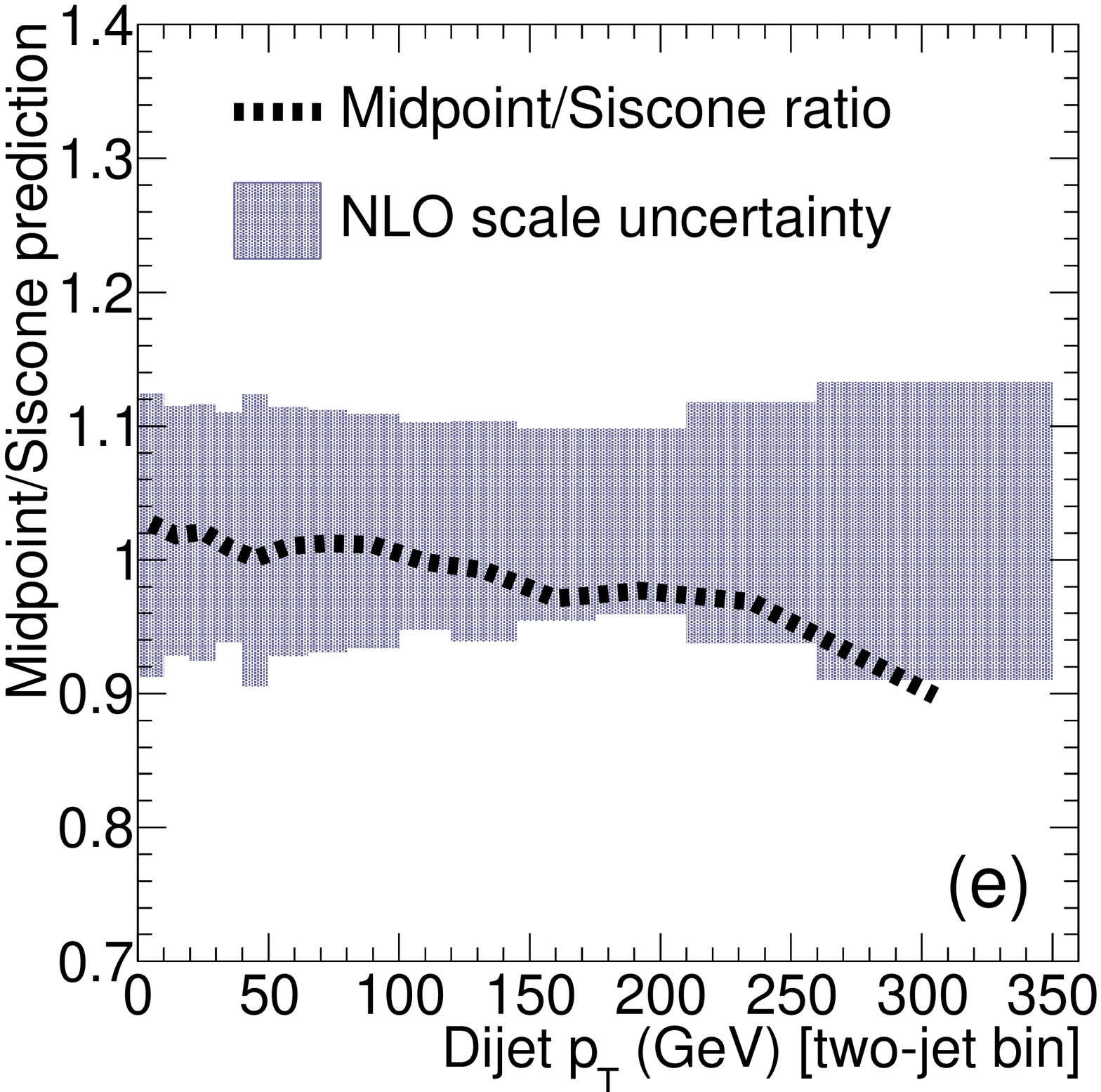}
    \includegraphics[width=0.32\textwidth]{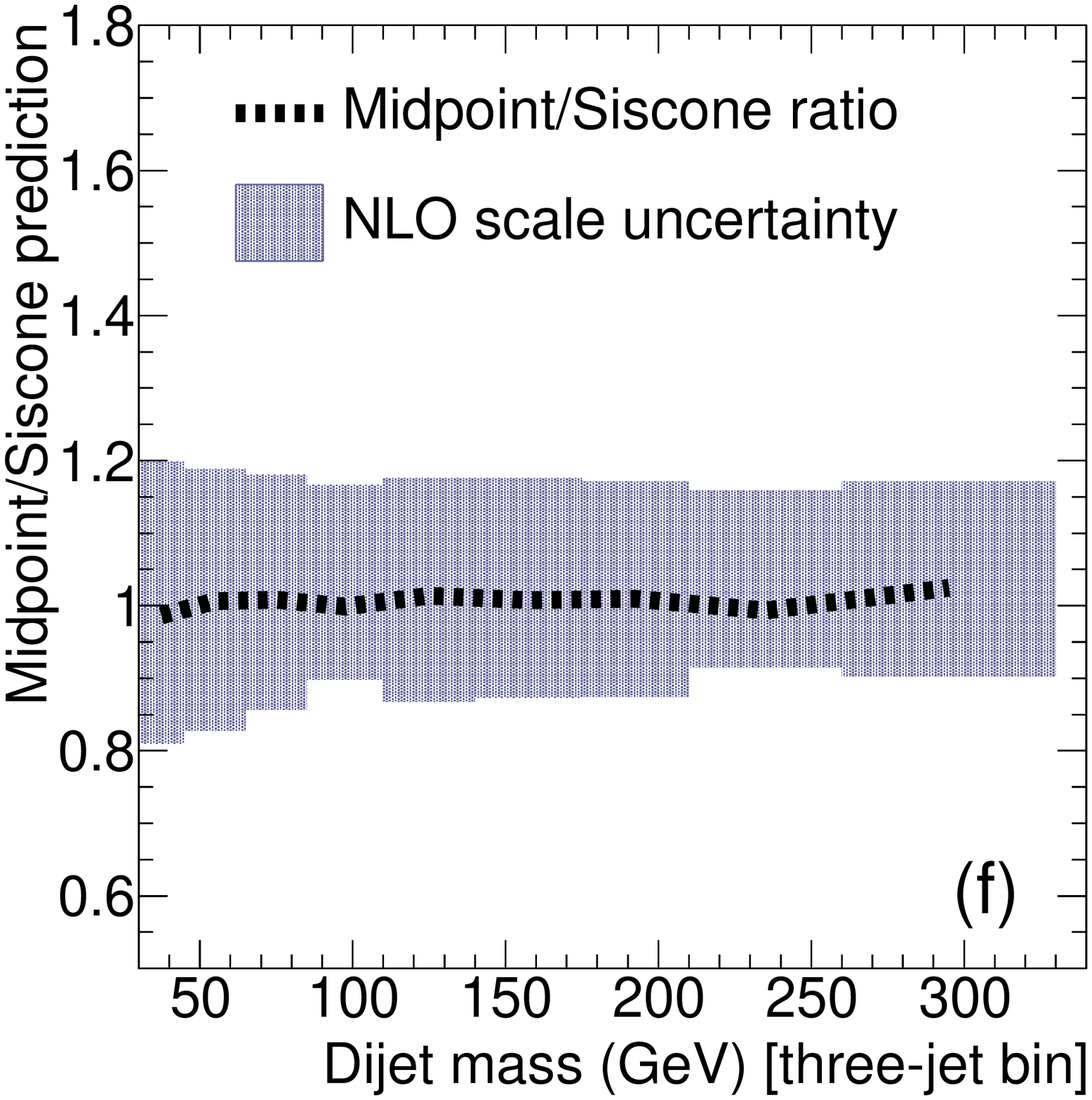}
    \caption{Examples of correction factors (derived from \sherpa) accounting for differences between the
      \textsc{siscone} and \Dzero\ midpoint cone jet algorithm, applied to NLO \blackhat\ predictions for (a) leading jet $p_T$ in inclusive $W+1\textrm{-jet}$ events,
      (b) leading jet rapidity in inclusive $W+1\textrm{-jet}$ events, (c) dijet rapidity separation between the leading two jets in inclusive $W+2\textrm{-jet}$ events,
      (d) $W$ boson $p_T$ in inclusive $W+1\textrm{-jet}$ events, (e) dijet $p_T$ in inclusive $W+2\textrm{-jet}$ events, (f) dijet invariant mass in inclusive $W+2\textrm{-jet}$ events.
      NLO renormalization and factorization scale uncertainties are shown for reference purposes.
      \label{fig:jetAlg}
    }
  \end{center}
\end{figure*}
As a fixed-order pQCD prediction, \blackhat\ provides a parton level prediction that is not immediately comparable to the unfolded experimental data.
Bin-by-bin corrections for non-perturbative QCD effects, due to hadronization and the underlying event, must be derived to compare the NLO predictions
with data. Non-perturbative corrections were produced using \sherpa~1.4.0 with the CT10 PDF set.

These corrections are calculated by taking the ratio of the observed differential cross section derived from \sherpa\ at the particle level 
(using \sherpa's internal cluster fragmentation model) to the differential cross section from \sherpa\ at the parton level, which includes parton showering 
but without hadronization or MPI and with electron final state emission disabled.
Uncertainties on these non-perturbative corrections are obtained\,\cite{Hoeche:2011fd} by recalculating the particle level \sherpa\ results as described above 
using the Lund string fragmentation model\,\cite{Andersson:1983ia}, taking the difference between the two as a symmetric systematic uncertainty on this correction.

Within the non-perturbative correction, we also apply an additional correction to account for the impact of the jet algorithm differences between experiment and
NLO \blackhat\ by computing the particle level cross sections with the \textsc{siscone} jet algorithm, using the \Dzero\ midpoint cone algorithm for the particle level predictions.
The total correction applied is then a term to account for hadronization and underlying event effects, and a further correction to account for the jet algorithm mismatch:
\begin{equation}
\textrm{total~correction}=\frac{\sigma^\mathrm{midpoint}_\mathrm{particle}}{\sigma^\textsc{siscone}_\mathrm{particle}} 
         \times \frac{\sigma^\textsc{siscone}_\mathrm{particle}}{\sigma_\mathrm{parton}} 
         = \frac{\sigma^\mathrm{midpoint}_\mathrm{particle}}{\sigma_\mathrm{parton}}.
\label{eq:npcs}
\end{equation}

Figure~\ref{fig:jetAlg} presents some examples of the \textsc{siscone} to \Dzero\ midpoint cone jet algorithm correction factors 
as a function of a representative sub-sample of the unfolded observables presented in this paper, and as computed with \sherpa. 
These jet algorithm corrections are of smaller magnitude than the corrections for underlying event and hadronization effects.

For most distributions the overall correction is small (with respect to the scale uncertainties on the theory). 
Where total non-perturbative corrections become large ($\geq50\%$), \blackhat\ predictions are not displayed in comparison to data.
Examples of the combined correction applied to the data for those distributions that do exhibit notable shape dependencies are shown in Fig.~\ref{fig:NPCs}.
\begin{figure}[htbp]
  \begin{center}
    \includegraphics[width=\columnwidth]{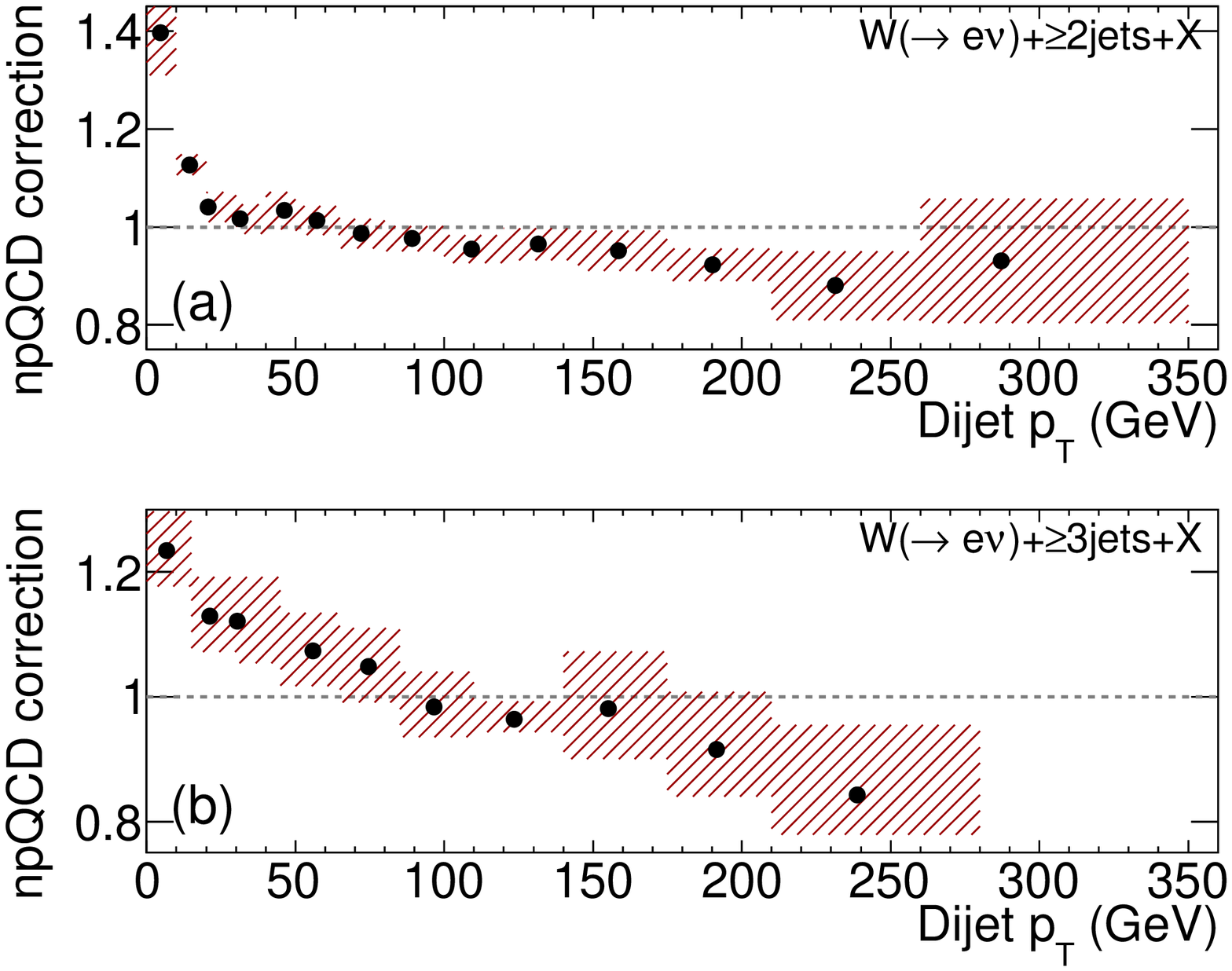}
    \includegraphics[width=\columnwidth]{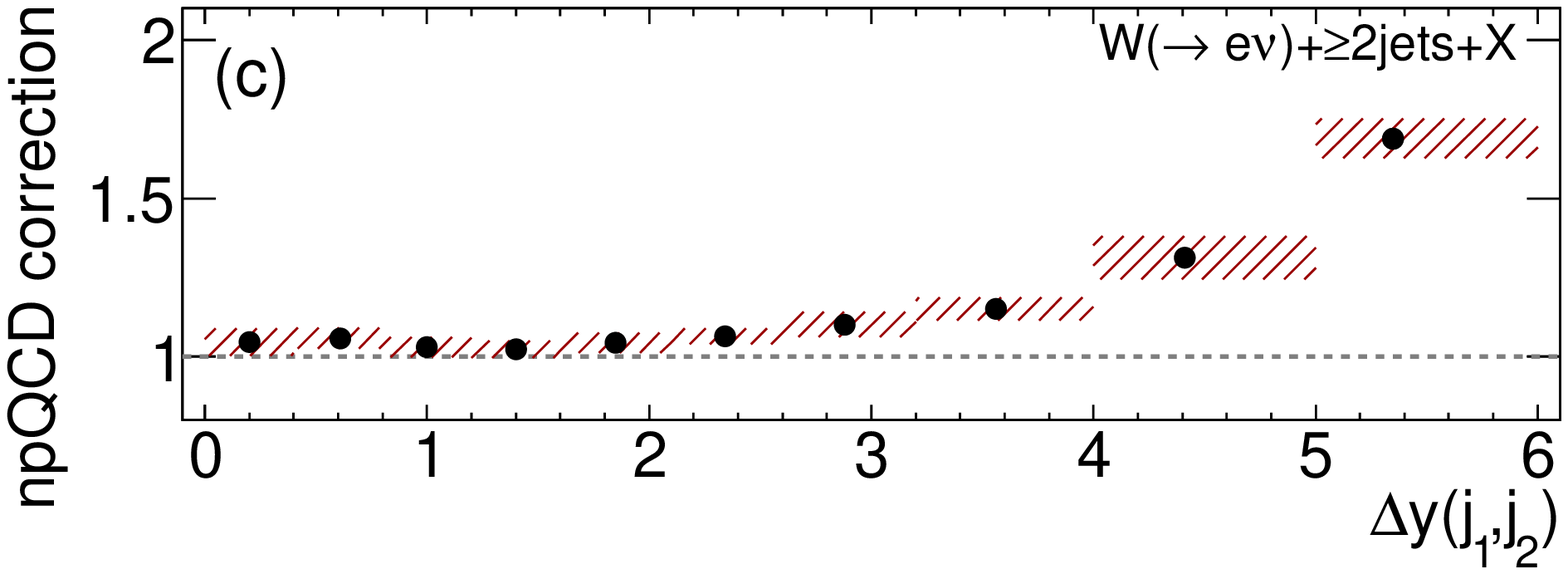}
    \caption{Non-perturbative QCD (npQCD) correction factors used to correct NLO \blackhat\ theoretical predictions from the parton to the particle level.
Uncertainties are estimated by combining the statistical uncertainty with the systematic uncertainty arising from different hadronization models, and are shown as the shaded band.  
      Examples shown are for the dijet $p_T$ distributions in the (a) inclusive two-jet and (b) three-jet multiplicity bins, 
      and for (c) dijet \dy12\ in the inclusive two-jet multiplicity bin, and are among the largest. Corrections are derived using \sherpa.
      \label{fig:NPCs}
    }
  \end{center}
\end{figure}

Full tables of the non-perturbative corrections applied to the NLO \blackhat\ predictions, including the jet algorithm correction terms, are available in HEPDATA\,\cite{HEPDATA}
and documented in Appendix~\ref{appendix} to facilitate comparison between future pQCD calculations and the experimental data presented here.

\section{Experimental Results}

Measurements of forty observables are documented in this paper, measured in the phase space defined in Table~\ref{tab:phasespace}.
These consist of thirty-three differential cross section measurements, normalized to the inclusive $W$ boson cross section in the same phase space,
four measurements of average jet activity as a function of dijet rapidity separations and the scalar sum of $W$ boson and jet transverse energies, 
and three measurements of the probability of subsequent jet emission in $W+\mathrm{dijet}$ events as a function of dijet rapidity separation under various conditions.
Unless otherwise noted, all jets are ordered by $p_T$ and all \wnjet\ distributions specify $n$-jet inclusive multiplicities.
Figures~\ref{fig:result_jetrap} to~\ref{result:gapfrac} present the results in comparison with various theoretical predictions. 
Data points are placed at the bin average, defined as the value where the theoretical differential cross section is equal to the mean cross section within the bin,
following the prescription detailed in Ref.~\cite{Lafferty:1994cj}. 
Error bars on data points represent statistical and systematic uncertainties added in quadrature.

All results presented here are available in tabulated form in HEPDATA\,\cite{HEPDATA} and in Appendix~\ref{appendix},
along with correlation matrices for the statistical and systematic uncertainties.

\subsection{Differential cross sections}

Measurement of relative rates and shapes of jet rapidities in \wjets\ events allow us to compare different theoretical approaches to jet emissions
and are key to understanding searches for new physics characterized by forward jet emission, as well as standard model measurements including vector boson fusion and vector boson scattering.
For all observables, experimental uncertainties are smaller or of similar magnitude to corresponding theoretical uncertainties on predictions 
and have the potential to discriminate between different theoretical approaches.

Measured normalized \wjets\ cross sections are shown as a function of the $n^\mathrm{th}$-jet rapidity in inclusive \wnjet\ events in Fig.~\ref{fig:result_jetrap},
highlighting the wide range of predicted differential spectra between the models considered.
\begin{figure}[htbp]
  \begin{center}
    \includegraphics[width=\columnwidth]{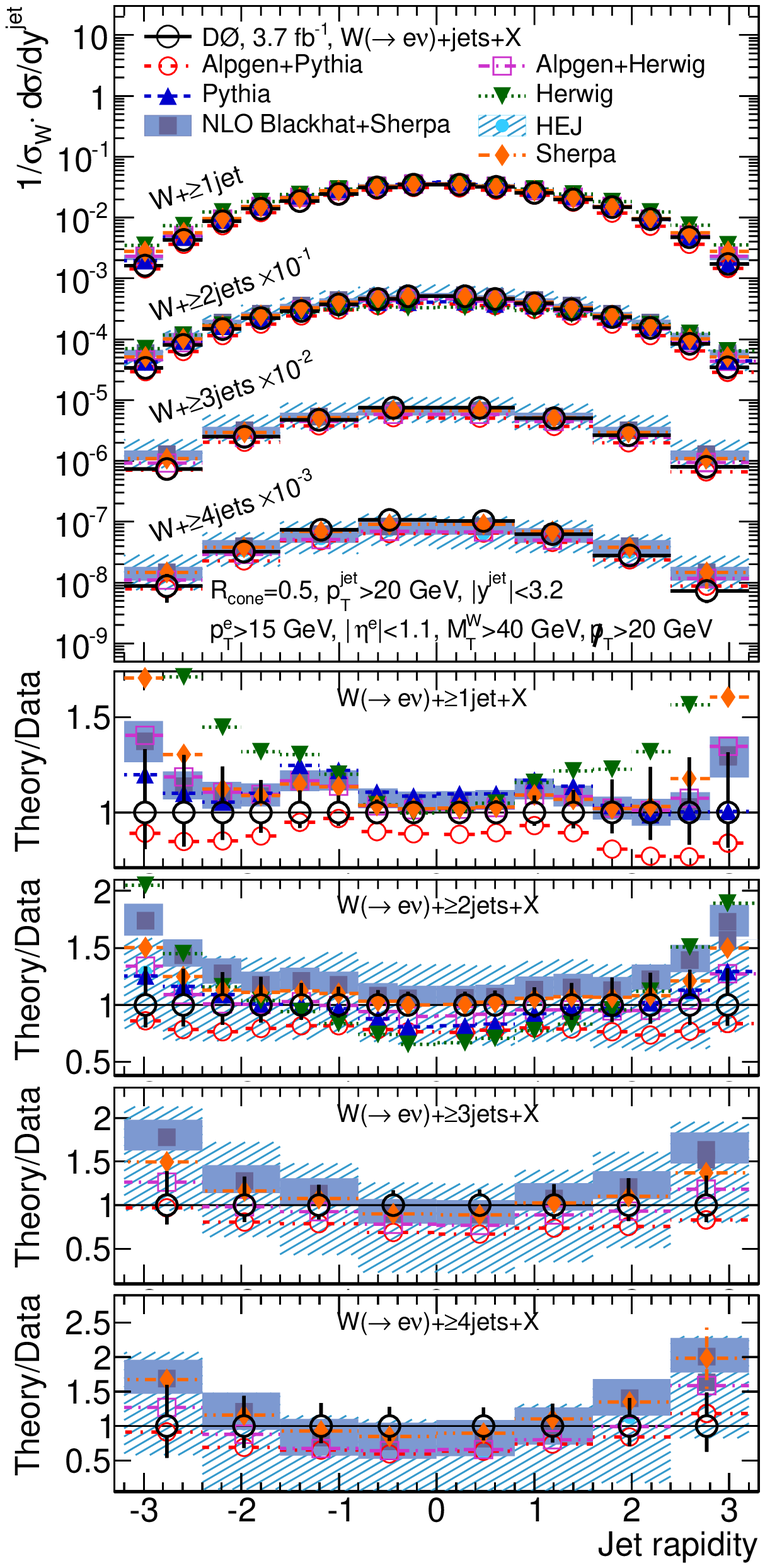}
    \caption{(color online) Measurement of the $n^\mathrm{th}$-jet rapidity distributions in inclusive \wnjet\ events for $n=\mathrm{1-4}$ and comparison to various theoretical predictions.
      Lower panes show theory/data comparisons for each of the $n$-jet multiplicity bin results separately.
      \label{fig:result_jetrap}
    }
  \end{center}
\end{figure}
While all predictions largely agree in shape at central rapidities, for rapidities $|y|>1$, discrepancies with the data begin to emerge, 
resulting in large differences at forward rapidities. 
NLO pQCD, \hej, \sherpa, and \herwig\ predictions are found to slightly overpredict the forward jet rate, while
\pythia\ and \alppyth\ give predictions approximately in agreement with the data.
Bjorken $x$ values of gluon PDFs probed by typical \wjets\ events at large rapidity where predictions begin to diverge from data are $x\approx\mathcal{O}(10^{-3})$.
Similar values of $x$ are probed by the ATLAS Collaboration in \wjets\ events at higher center-of-mass energies, where discrepancies are also observed\,\cite{Aad:2012en}.
These observations may suggest some tension with current determinations of the small $x$ gluon.

\begin{figure}[htbp]
  \begin{center}
    \includegraphics[width=\columnwidth]{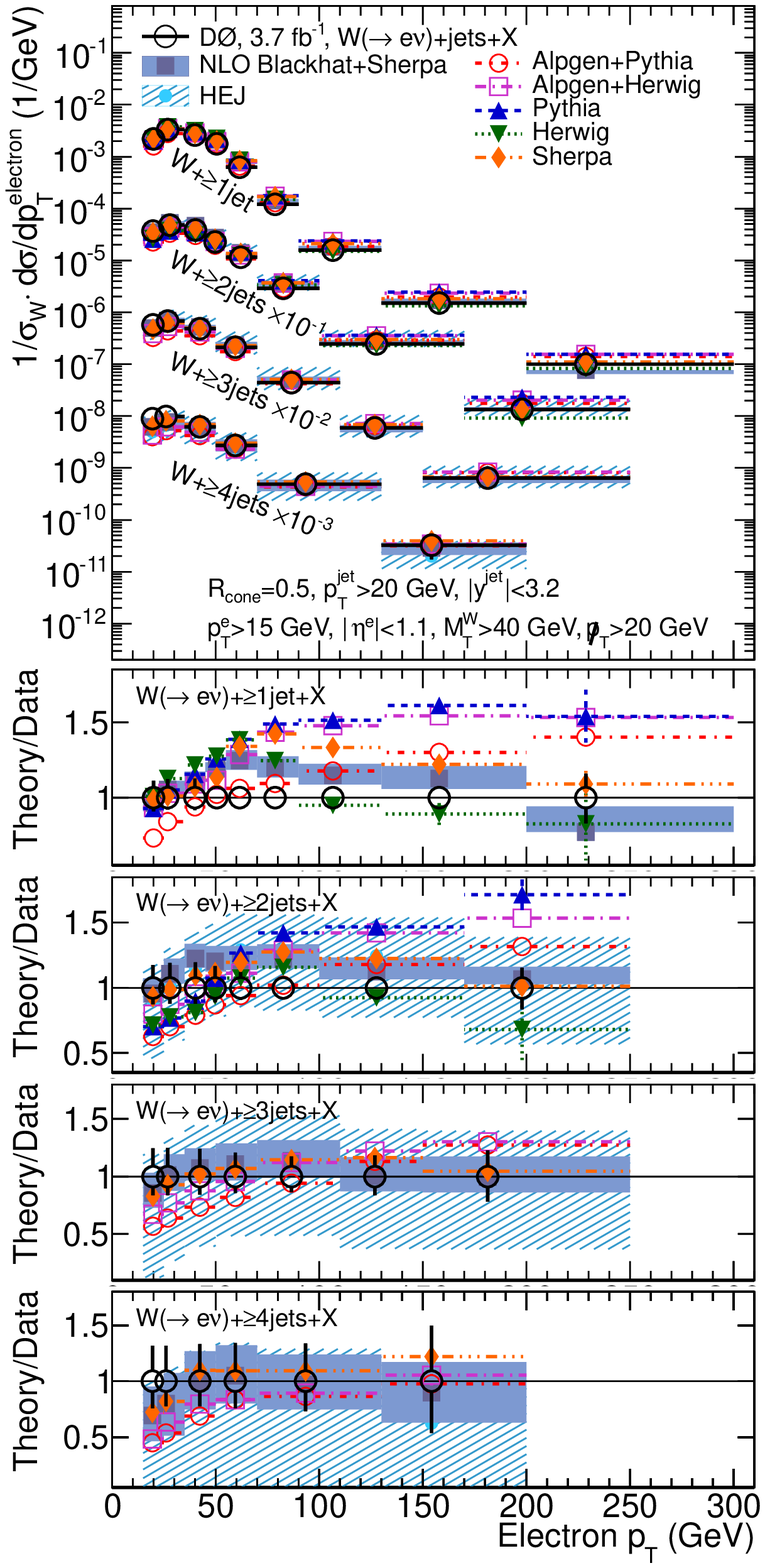}
    \caption{(color online) Measurement of the electron transverse momentum distributions in inclusive \wnjet\ events for $n=\mathrm{1-4}$ and comparison to various theoretical predictions.
      Lower panes show theory/data comparisons for each of the $n$-jet multiplicity bin results separately.
      \label{fig:result_leptonpt}
    }
  \end{center}
\end{figure}
Figure~\ref{fig:result_leptonpt} presents normalized cross sections as a function of the electron transverse momentum for inclusive one- to four-jet events.
The study of electron $p_T$ provides kinematic information that is complementary to the previously measured jet $p_T$ distributions
as the electron $p_T$ is only partly correlated to the $p_T$ of any jet in the event.
We observe a trend for predictions to slightly underestimate the data at low ($<50$~GeV) $p_T$ in higher $n$-jet channels. NLO pQCD predictions are expected to be unreliable in this region
as events are dominantly $W+1\textrm{-jet}$ configurations where the $W$ boson and jet are back-to-back.

At higher electron transverse momenta, \blackhat, \hej, and \sherpa\ predictions describe the spectrum well, although the one-jet rate is slightly overpredicted by \blackhat\
and \sherpa\ in this region. \herwig\ does not describe the low-$p_T$ shape, and a distinct change in slope is observed above $\approx60$~GeV 
related to the transition from a pure parton shower to a matching of the parton shower to the $W+1\textrm{-jet}$ matrix element in \herwig.
\pythia\ predicts a harder $p_T$ spectrum than observed in data, overpredicting the rate at high $p_T$ by about 50\%.
Interfacing the \pythia\ parton shower to the \alpgen\ matrix element calculation improves the description significantly, bringing \alppyth\
predictions in closer agreement to those from NLO pQCD. Similar improvements are not observed when the \herwig\ parton shower is matched with the \alpgen\ matrix element calculation.

The corresponding electron pseudorapidity spectra for each jet multiplicity are presented in Fig.~\ref{fig:result_leptoneta}.
\begin{figure}[htbp]
  \begin{center}
    \includegraphics[width=\columnwidth]{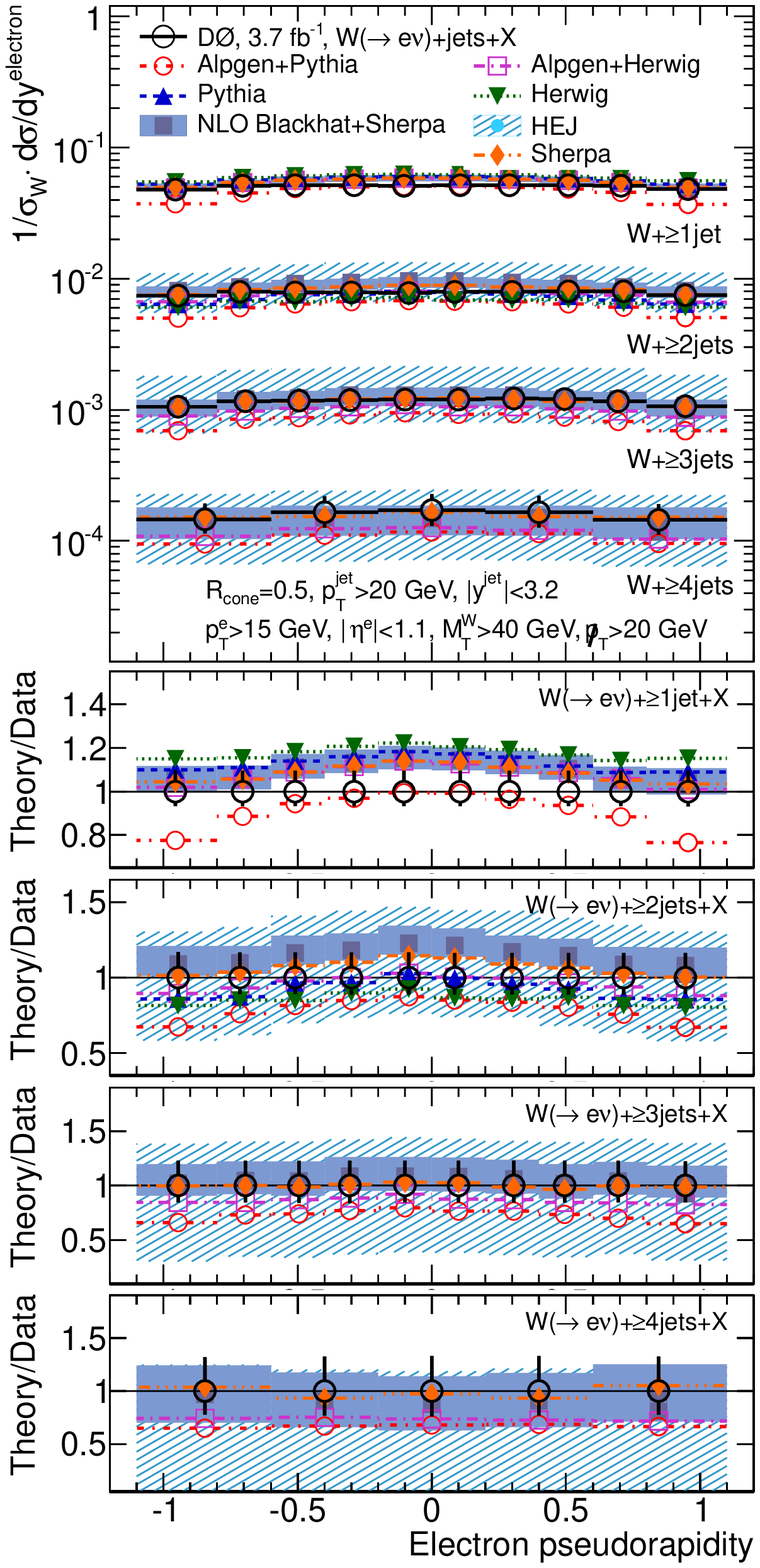}
    \caption{(color online) Measurement of the electron pseudorapidity distributions in inclusive \wnjet\ events for $n=\mathrm{1-4}$ and comparison to various theoretical predictions.
      Lower panes show theory/data comparisons for each of the $n$-jet multiplicity bin results separately.
      \label{fig:result_leptoneta}
    }
  \end{center}
\end{figure}
\alppyth\ predicts a narrower electron pseudorapidity spectrum than observed in data at low jet multiplicities,
underpredicting the rate at $|\eta|=1.0$ by over 20\%. Other MC generators and \blackhat\ predictions do not exhibit the same behavior although
there is some indication of a shape difference between theory and data at the 10\% level across the measured interval.

In Fig.~\ref{fig:result_DelRap}, normalized cross sections are presented as a function of dijet rapidity separation in inclusive two-jet and three-jet events,
for two distinct jet pairings. The first configuration defines the dijet rapidity separation between the two highest-$p_T$ jets in the event [\dy12],
the second defines the separation between the two most rapidity-separated jets (generally with one forward jet, $j_F$, and one backward jet, $j_B$, in rapidity and both jets
with $p^\mathrm{jet}_T>20$~GeV) in the event [\dyFB].

\begin{figure*}[htbp]
  \begin{center}
    \includegraphics[width=\columnwidth]{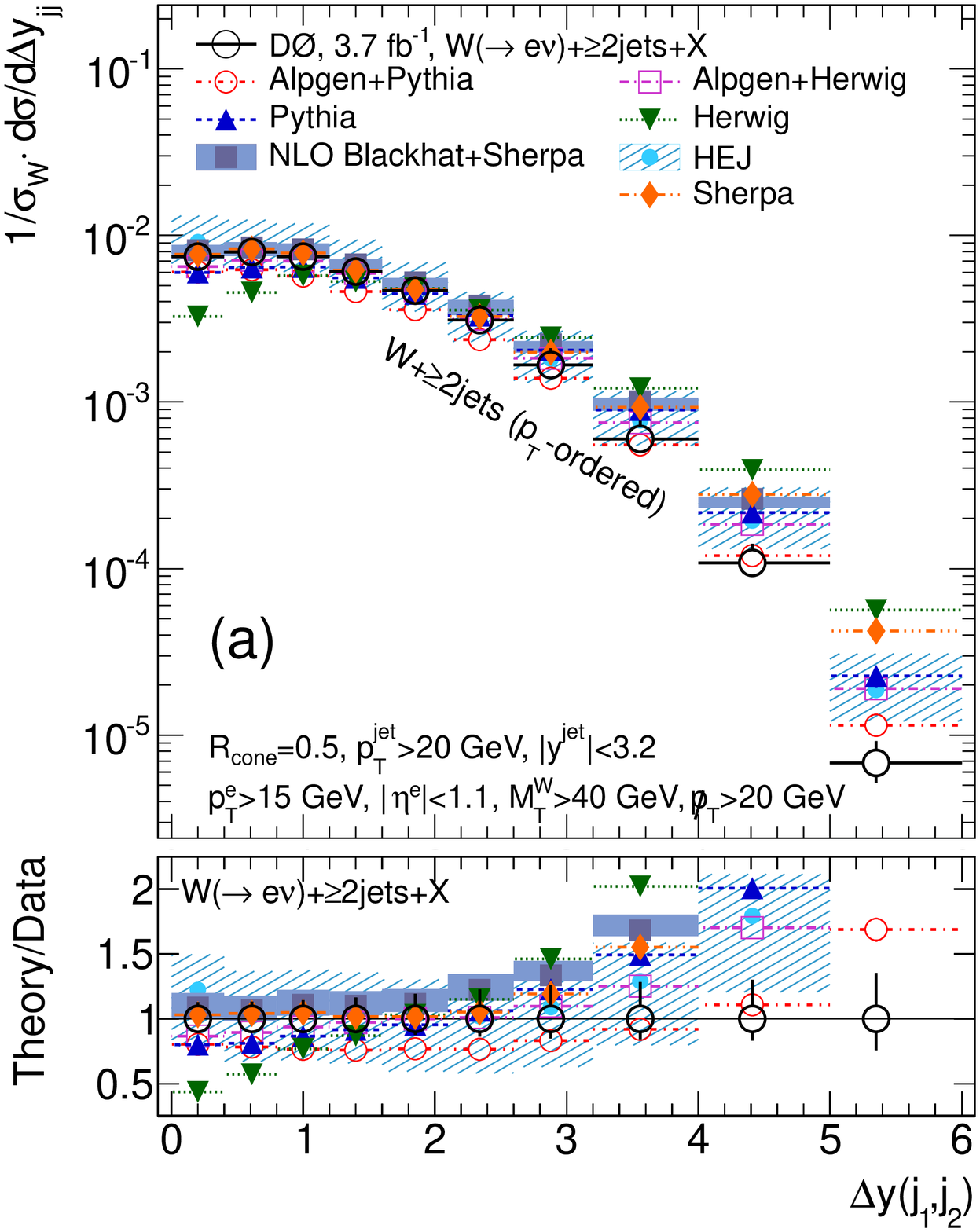}
    \includegraphics[width=\columnwidth]{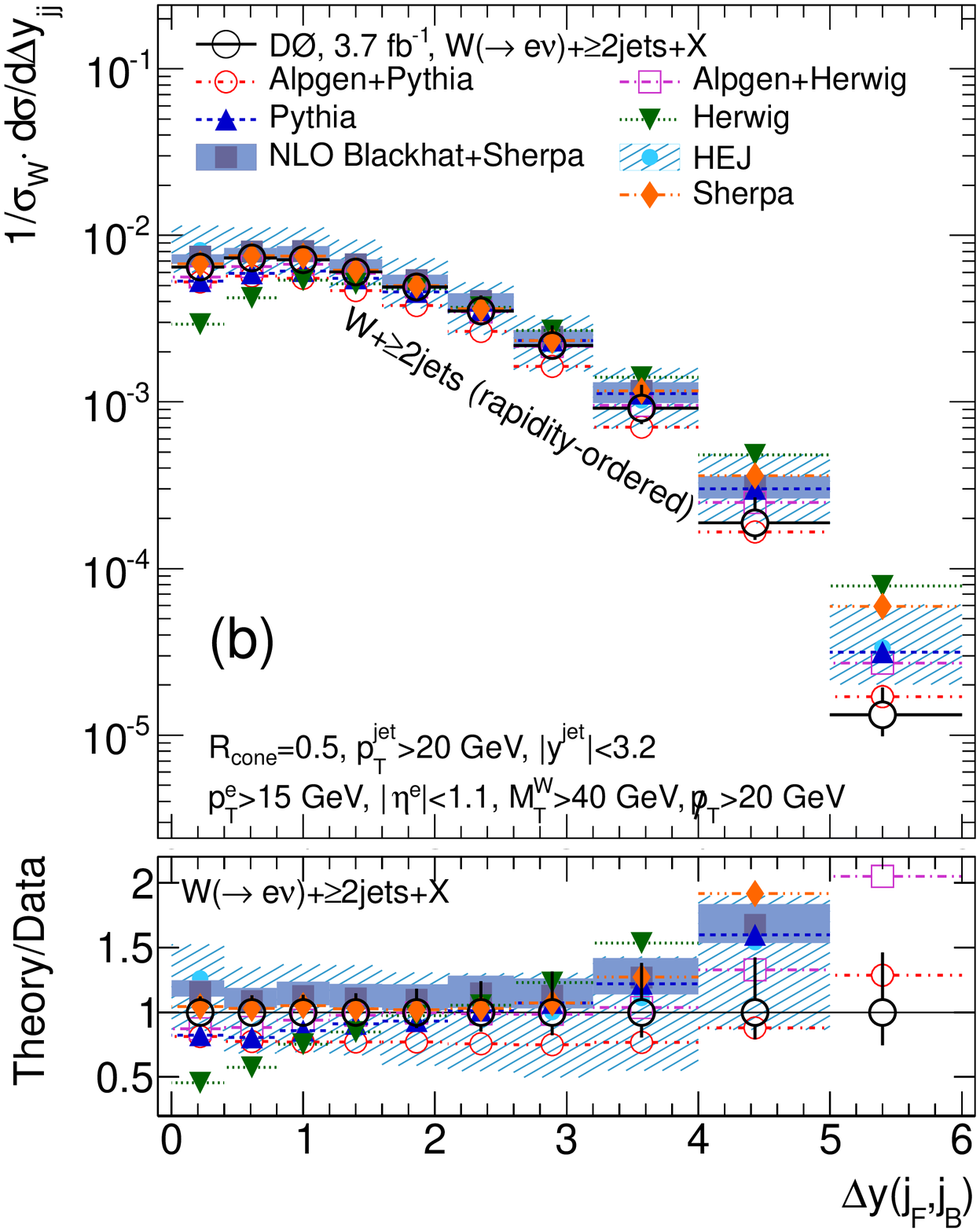}
    \includegraphics[width=\columnwidth]{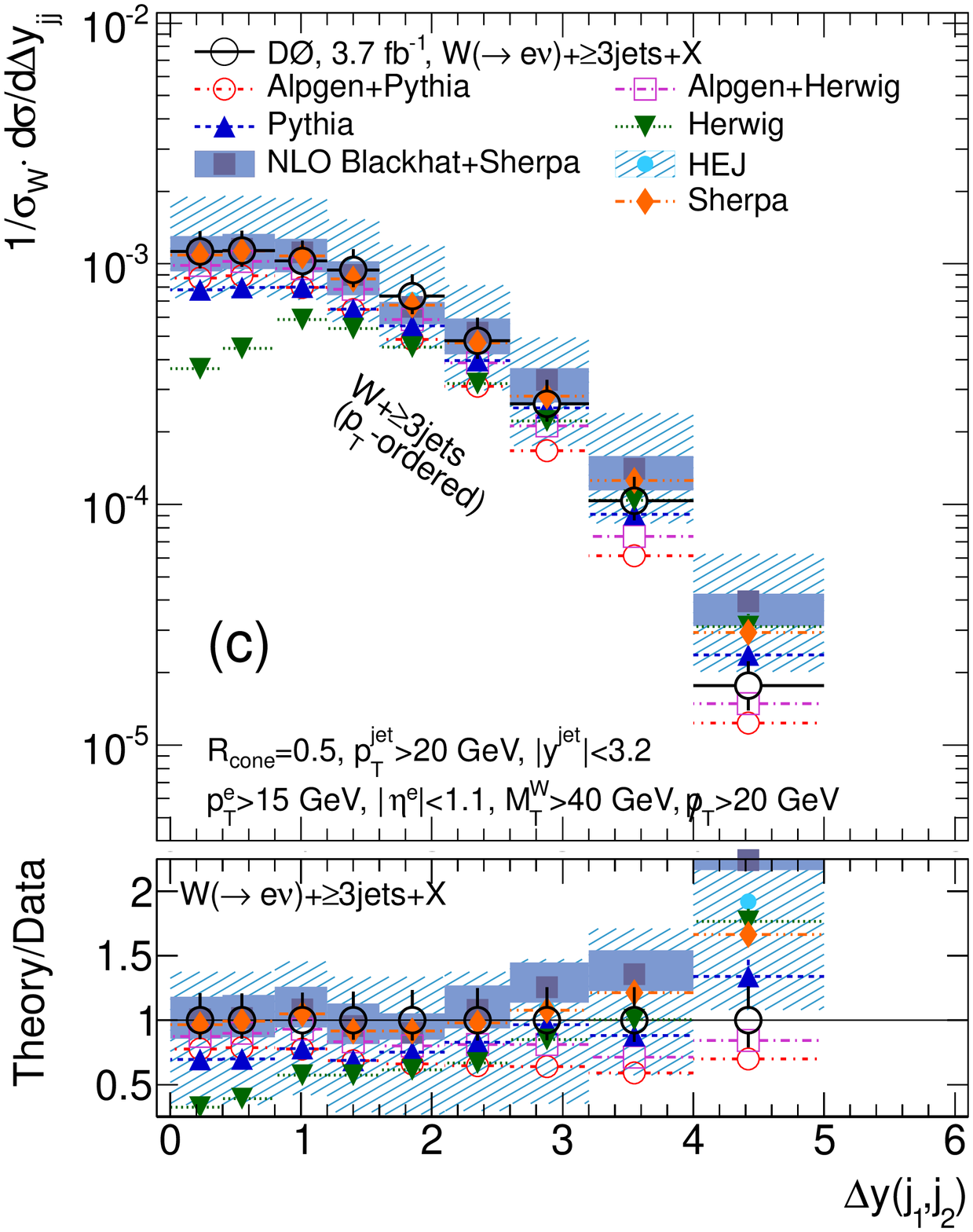}
    \includegraphics[width=\columnwidth]{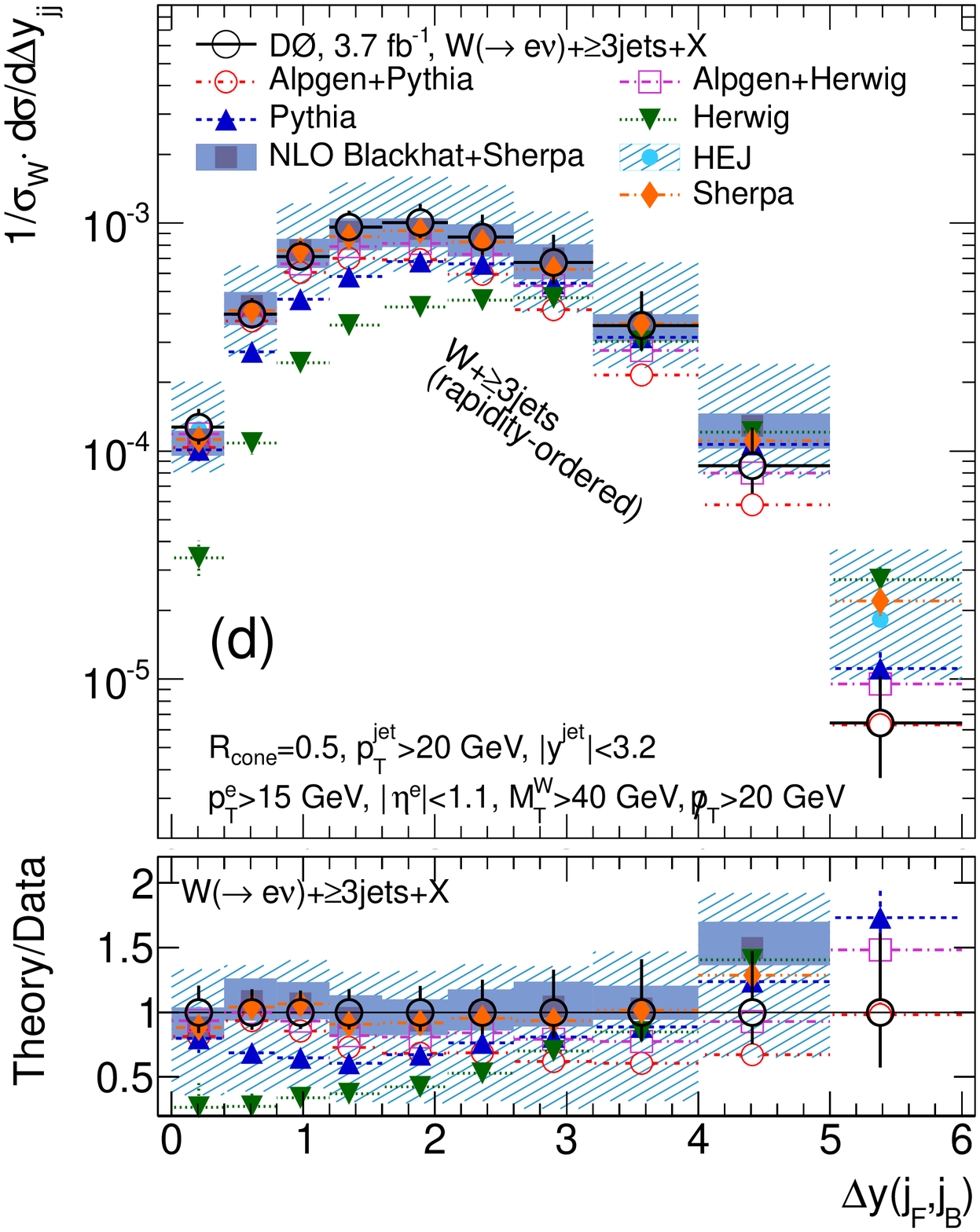}
    \caption{(color online) Measured dijet rapidity separation in inclusive $W+2\textrm{-jet}$ events for (a) the two leading $p_T$ jets and (b) the two most rapidity-separated jets 
      and in inclusive $W+3\textrm{-jet}$ events for (c) the two leading $p_T$ jets and (d) the two most rapidity-separated jets with comparison to various theoretical predictions.
      Lower panes show theory/data comparisons.
      \label{fig:result_DelRap}
    }
  \end{center}
\end{figure*}

Study of the rapidity separation of the two leading (highest-$p_T$) jets in the event in \wdijet\ (and $W+3\textrm{-jet}$) events is a test
of wide-angle soft parton radiation and matrix element plus parton shower matching schemes. Understanding the distribution of this variable
is important to distinguish VBF processes from the larger $W+textrm{jets}$ contributions and is key for background modeling for future searches and measurement of the Higgs boson
in the vector boson fusion and vector boson scattering modes. 

Measuring the rapidity separation of the two most rapidity-separated jets in inclusive \wdijet\ events provides sensitive information on additional QCD radiation in the event, as
does the measurement of the same variable as a function of the two leading jets. However, the rapidity-ordered configuration is sensitive to BFKL-like dynamics
because, in this case, the dijet invariant mass is much larger than the transverse momentum of the jets, allowing tests of the advanced modeling of high-$p_T$ wide-angle emissions
missed by a standard parton shower approach.

\begin{figure*}[htbp]
  \begin{center}
    \includegraphics[width=\columnwidth]{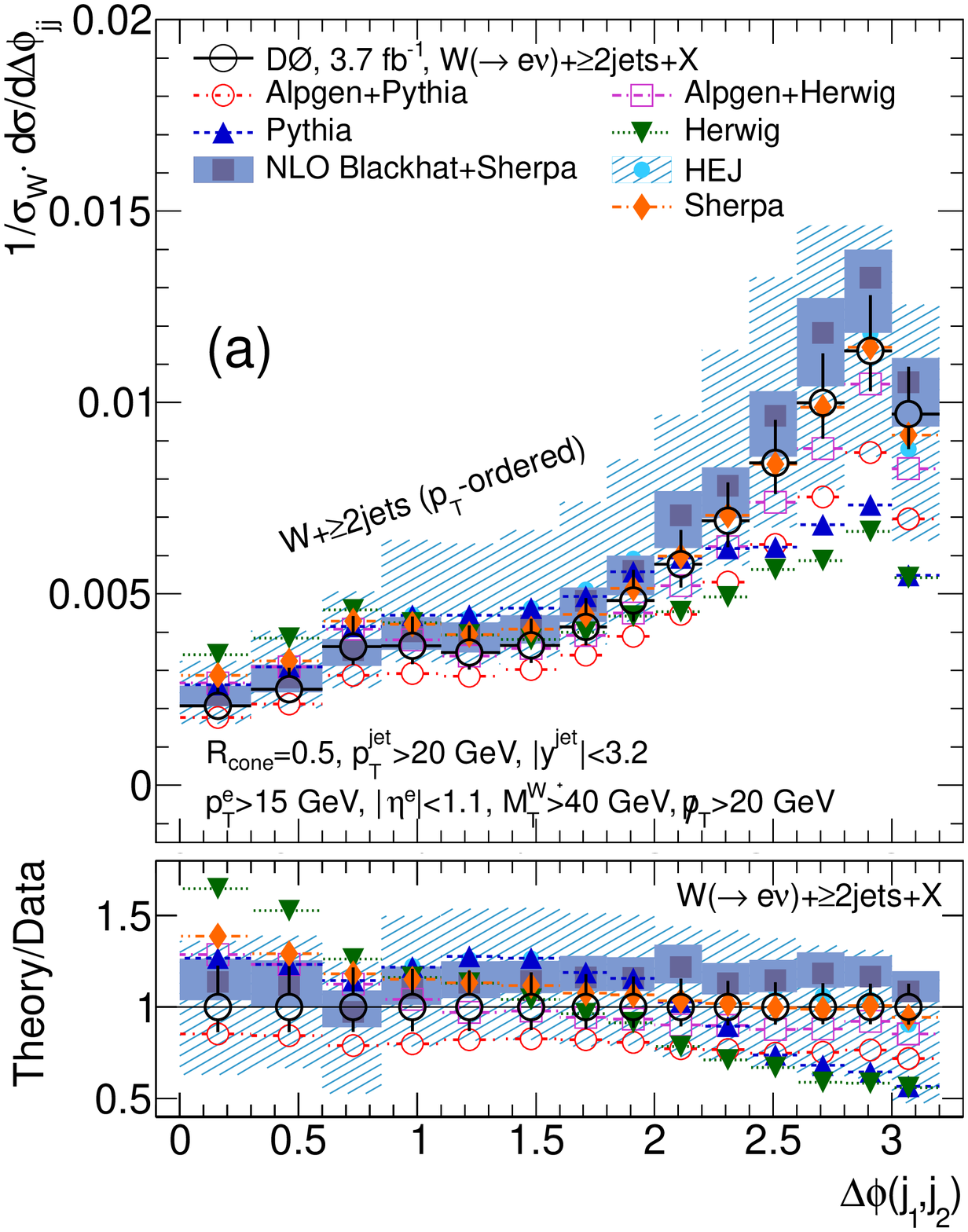}
    \includegraphics[width=\columnwidth]{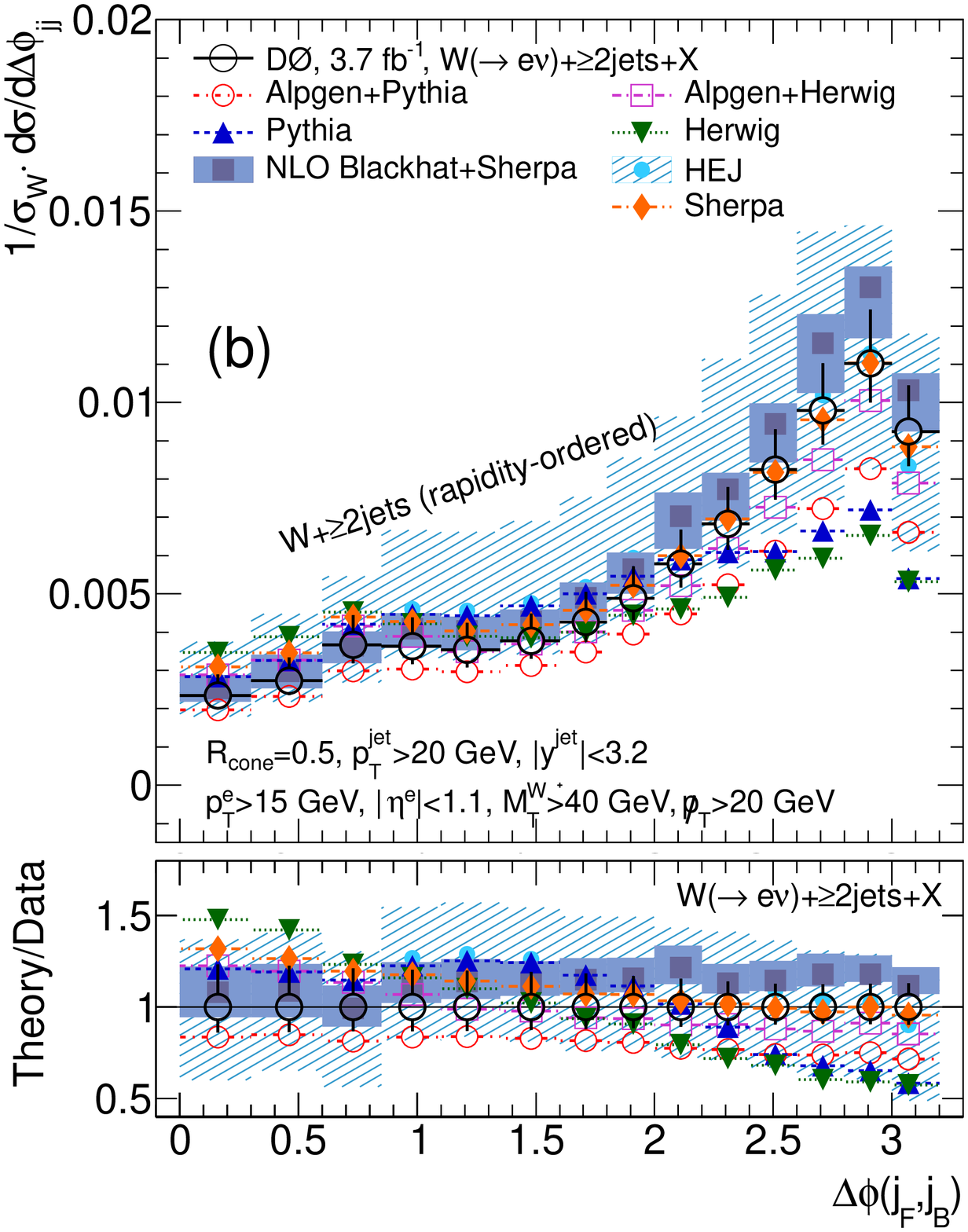}
    \caption{(color online) Measured differential cross sections and various theoretical predictions for dijet $\Delta\phi$ for
      (a) the two highest-$p_T$ jets and (b) the two most rapidity-separated jets in $W+2\textrm{-jet}$ events.
      Lower panes show theory/data comparisons.
      \label{fig:result_dPhi}
    }
  \end{center}
\end{figure*}

Predictions from \herwig\ consistently favor wider separations between both the highest-$p_T$ and most rapidity-separated jets
but underpredicts the measured rate at small rapidity intervals. This mismodelling increases in the three-jet events and in rapidity-separated configurations.
NLO pQCD predictions are able to describe small rapidity intervals well, but increasingly overpredict the rate seen in data as the rapidity separation grows.
The distribution is dominated by contributions from non-perturbative processes at large rapidity separations. Typical corrections for non-perturbative QCD effects 
of $\approx$40\% for $4\leq\dy12<5$ and $\approx$75\% for $5\leq\dy12<6$ thus limit the applicability of NLO pQCD predictions in this region.
The shape in inclusive three-jet events is well-described by \textsc{alpgen+(pythia/herwig)}, \sherpa, and \pythia, all of which
have better performance than NLO pQCD, suggesting that the contributions of soft emissions from the parton shower are necessary and well-tuned.

The azimuthal angle separation between the two leading jets or most rapidity-separated jets
in \wdijet\ events is a sensitive test of modeling of higher-order corrections in theoretical calculations.
Some theoretical and experimental analyses prefer to study $p_T$-ordered jets, while others use rapidity-ordered jets. 
These azimuthal correlations are therefore studied as a function of both the two leading [\dphi12] and two most rapidity-separated [\dphiFB] jets with the results
presented in Fig.~\ref{fig:result_dPhi}. The two corrected observables [\dphi12\ and \dphiFB] are similar in shape.

Jet pairings with large (close to $\pi$) separation are generally modeled via matrix element calculations while small separations are modeled
mainly via the parton shower. Hard radiative corrections from all-order resummation approaches can modify the predicted spectrum for this observable.
The spectra are well described by the all-order resummation (\hej) and NLO (\blackhat) approaches, although the latter is a little high in overall rate.
\sherpa\ and \alpgen\ provide a reasonable description of the data within experimental uncertainties, 
although when interfaced to \pythia\ parton showering \alpgen\ slightly underestimates the rate at large $\Delta\phi$.
Both the parton shower MC generators, \pythia\ and \herwig, predict significantly reduced emissions at large $\Delta \phi$ than are observed in data and more collinear emissions
from the parton shower, the modeling of which can be improved with these data.

Figure~\ref{fig:result_dR} shows the $\Delta R = \sqrt{(\Delta y)^2 + (\Delta\phi)^2}$ spectrum
between the two leading jets in the inclusive \wdijet\ sample.
Study of the opening angle between the two highest-$p_T$ jets in the event (in $y$-$\phi$ space) allows for testing the final-state radiation
modeling in theoretical calculations. This is also an important experimental variable to properly model dijet correlations
in backgrounds for precision measurement and searches for new physics. Both \hej\ and NLO \blackhat\ calculations do not accurately
model the shape of this distribution at large and small opening angles. 
Both \sherpa\ and \alppyth\ provide a good description of the shape observed in data except at the largest $\Delta R$.

\begin{figure}[htbp]
  \begin{center}
    \includegraphics[width=\columnwidth]{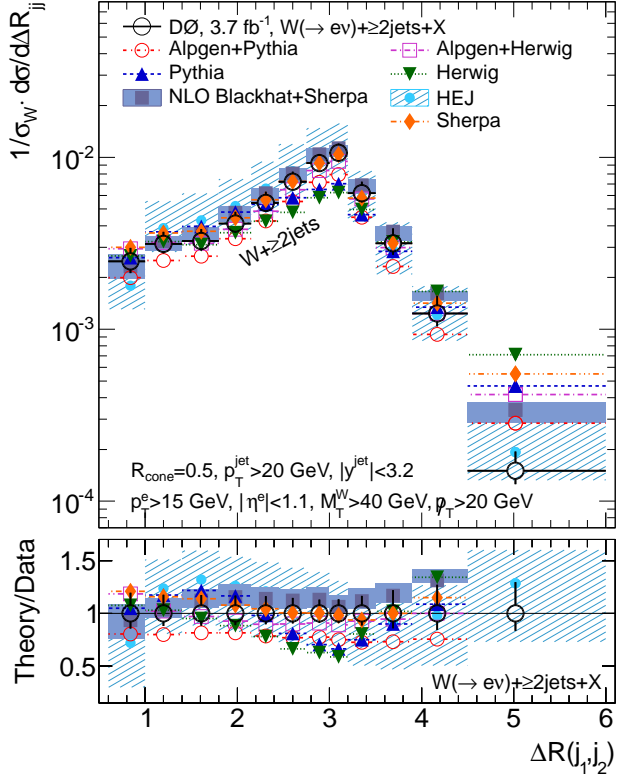}
    \caption{(color online) Measured differential cross sections and various theoretical predictions for dijet $\Delta R = \sqrt{(\Delta y)^2 + (\Delta\phi)^2}$ 
      in inclusive $W+2\textrm{-jet}$ events. The lower pane shows theory/data comparisons.
      \label{fig:result_dR}
    }
  \end{center}
\end{figure}

Figure~\ref{fig:result_DeltaRap3} presents measurements of the rapidity separation between the third-hardest jet in inclusive $W+3\textrm{-jet}$ events and either the 
leading or sub-leading jet in the event. Measurement of the angular correlations between various jet pairings gives us further information to constrain QCD radiation modeling.
In particular, this variable is of interest as a test of initial state radiation modeling. 
\sherpa, \hej, and NLO \blackhat\ provide a good description of the shapes of these distributions, with some tension 
again observed at the very largest rapidity separations. \pythia, \herwig, and \alpgen\ matrix element matched approaches show deviations from the data, particularly at low
rapidity separation.

\begin{figure}[htbp]
  \begin{center}
    \includegraphics[width=\columnwidth]{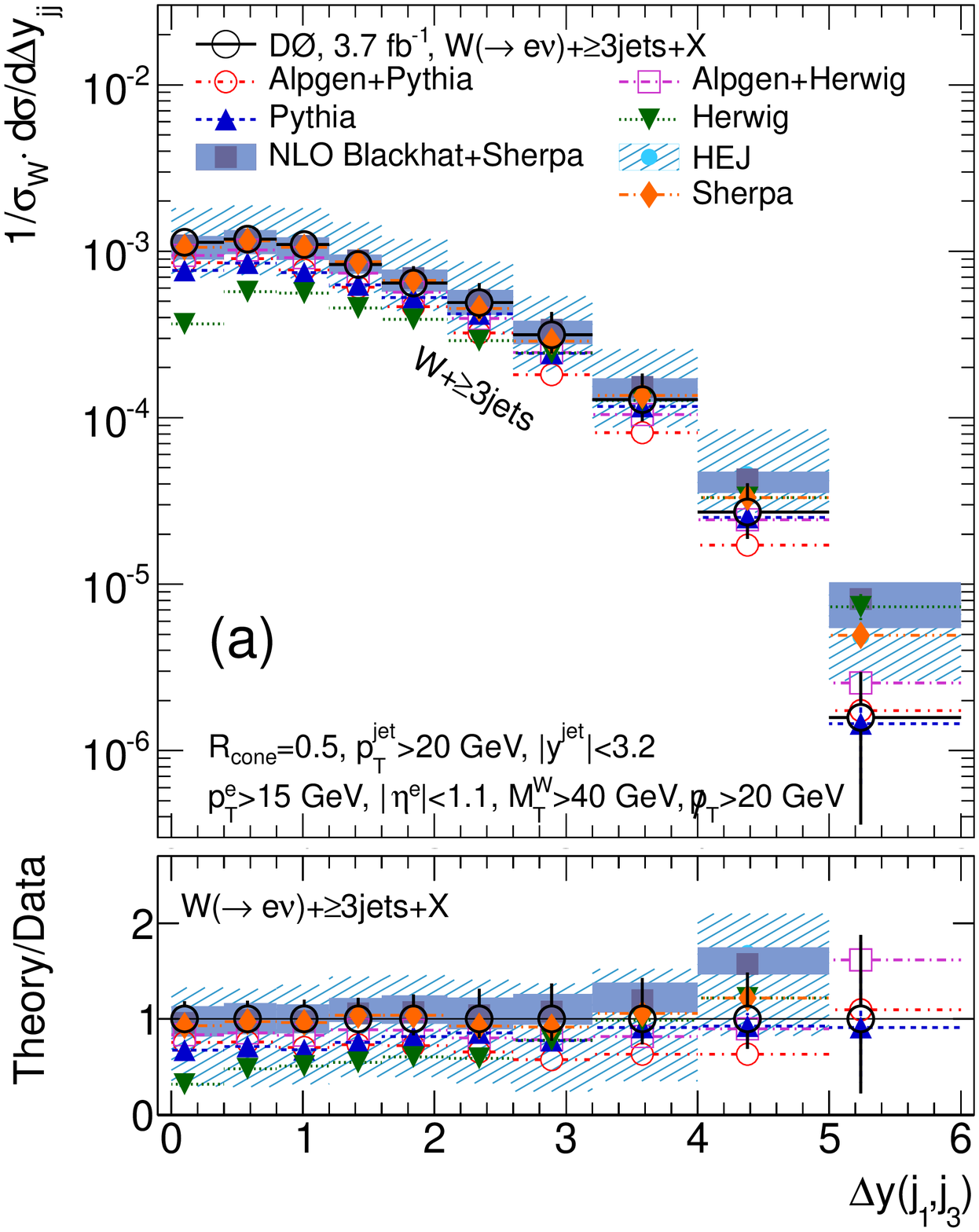}
    \includegraphics[width=\columnwidth]{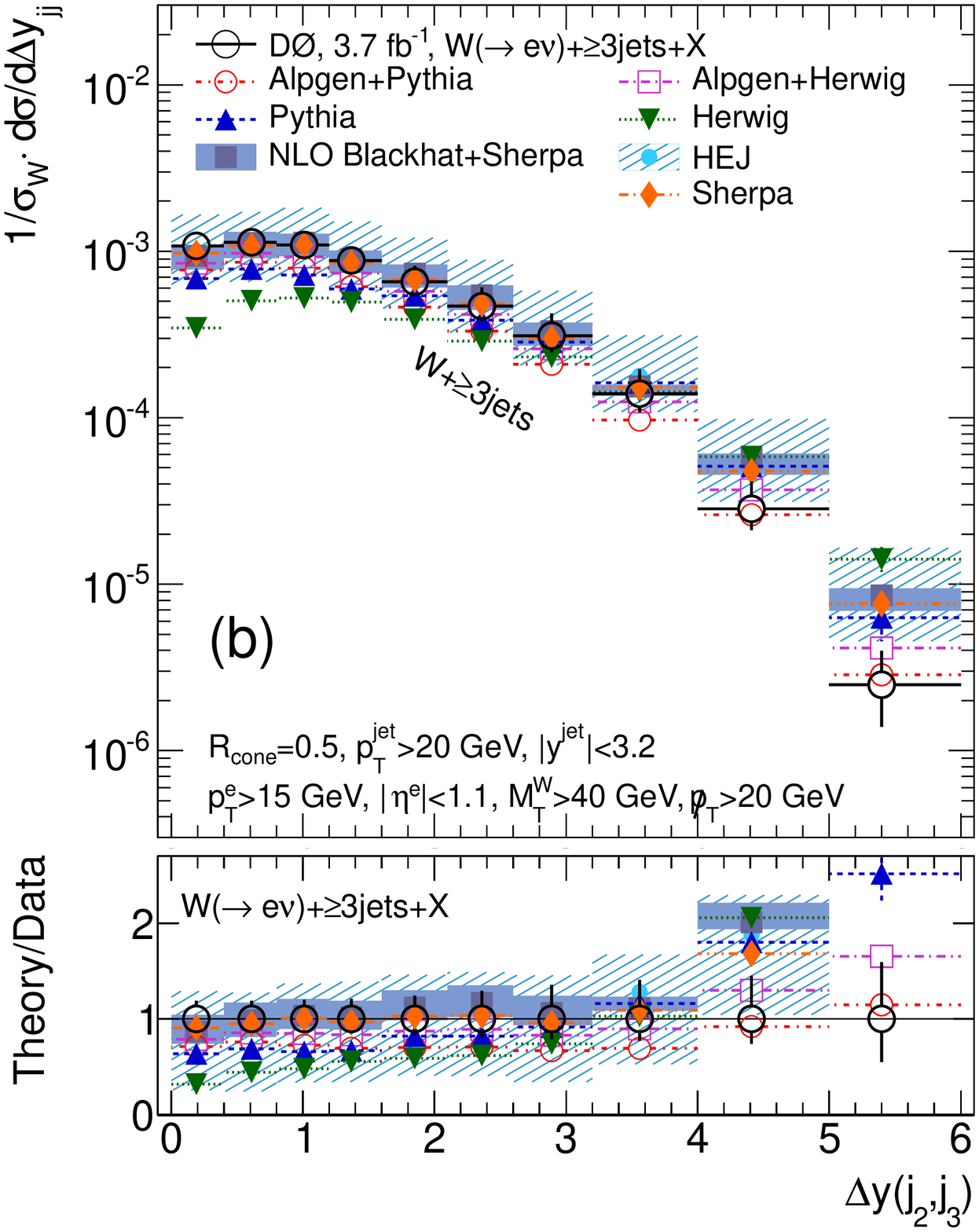}
    \caption{(color online) Measurement of the spectrum of dijet rapidity separation of (a) the first and third and, (b) the second and third $p_T$-ordered jets 
      in inclusive $W+3\textrm{-jet}$ events and comparison to various theoretical predictions.       Lower panes show theory/data comparisons.}
      \label{fig:result_DeltaRap3}
  \end{center}
\end{figure}

The $W$ boson transverse momentum in inclusive $n=\mathrm{1-4}$-jet multiplicity bins is shown in Fig.~\ref{fig:result_Wpt}.
Good agreement between the data and \blackhat\ and \hej\ are observed for all jet multiplicities.
\begin{figure}[htbp]
  \begin{center}
    \includegraphics[width=\columnwidth]{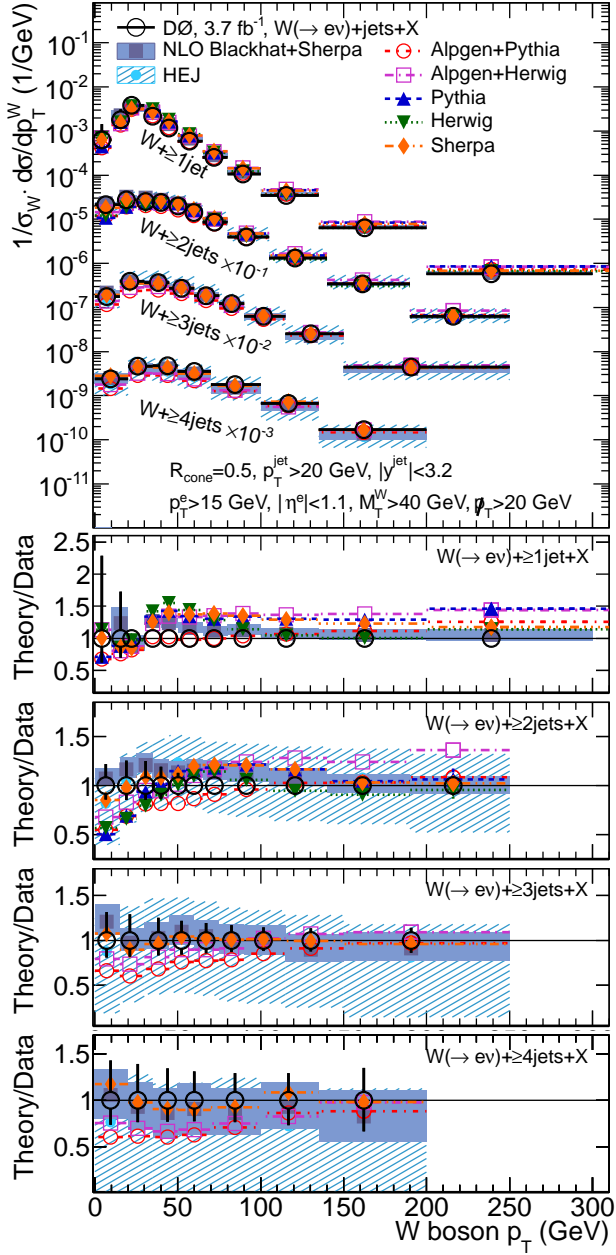}
    \caption{(color online) Measurement of the $W$ boson transverse momentum distributions in inclusive \wnjet\ events for $n=\mathrm{1-4}$ and comparison to various theoretical predictions.
      Lower panes show theory/data comparisons for each of the $n$-jet multiplicity bin results separately.
      \label{fig:result_Wpt}
    }
  \end{center}
\end{figure}
A change in behavior in NLO pQCD/data is observed at $W$ boson transverse momenta near the jet $p_T$ threshold of 20~GeV,
most notably in the inclusive $W+1\textrm{-jet}$ sample. As the boson $p_T$ approaches this threshold, the boson and jet are produced back-to-back.
At transverse momenta below this threshold, non-perturbative effects dominate and the fixed-order calculations are expected to be unreliable.

The dijet transverse momentum and invariant mass spectra in the inclusive two-jet and three-jet multiplicity bins are shown in Figs.~\ref{fig:result_dijetpt}
and~\ref{fig:result_dijetmass}. Dijet quantities in this article are calculated from the highest and second highest-$p_T$ jets in the event.
As well as testing the modeling of correlations between the two highest-$p_T$ jets in the event in MC generators, validation of theoretical 
modeling of the $p_T$ distribution of the dijet system in $W+2\textrm{-jet}$ events and accurate accounting for the kinematic correlations of the jets
is important for searches for beyond the standard model physics.
We provide measurements in this variable to allow the study of modeling differences between theoretical approaches.

Agreement in the shape of the dijet $p_T$ distribution (Fig.~\ref{fig:result_dijetpt}) is observed between data and predictions from NLO \blackhat, \hej, and \sherpa.
Notable discrepancies in the \alpgen, \pythia, and \herwig\  modeling are observed at low $p_T$.

\begin{figure}[htbp]
  \begin{center}
    \includegraphics[width=\columnwidth]{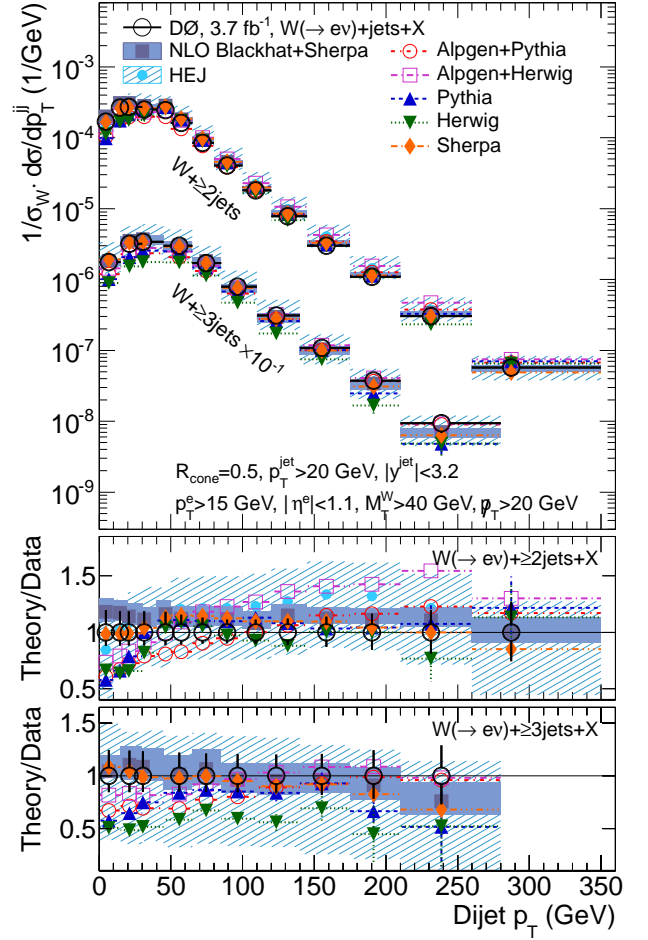}
    \caption{(color online) Measurement of the dijet transverse momentum spectrum of the dijet system in inclusive $W+2\textrm{-jet}$ and $W+3\textrm{-jet}$ events 
      and comparison to various theoretical predictions.
      Lower panes show theory/data comparisons for each of the $n$-jet multiplicity bin results separately.
      \label{fig:result_dijetpt}
    }
  \end{center}
\end{figure}

As a function of dijet invariant mass (Fig.~\ref{fig:result_dijetmass}), \hej\ and \sherpa\ predictions
model the shape well, but NLO pQCD predictions increasingly overestimate the high mass rate, particularly in the inclusive two-jet bin.
\begin{figure}[htbp]
  \begin{center}
    \includegraphics[width=\columnwidth]{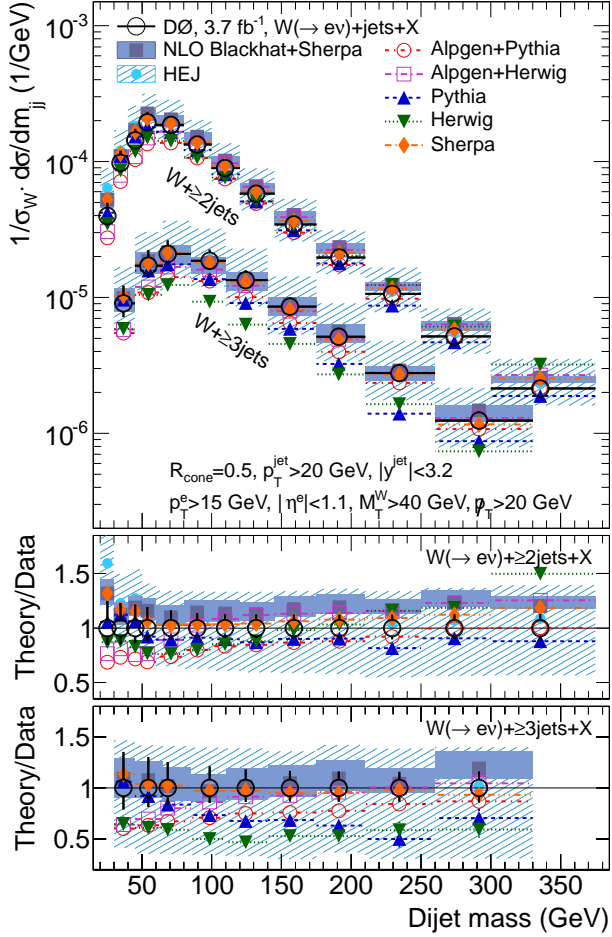}
    \caption{(color online) Measurement of the invariant mass spectrum of the dijet system in inclusive $W+2\textrm{-jet}$ and $W+3\textrm{-jet}$ events and comparison to various theoretical predictions.
      Lower panes show theory/data comparisons for each of the $n$-jet multiplicity bin results separately.
      \label{fig:result_dijetmass}
    }
  \end{center}
\end{figure}

Figure~\ref{fig:result_jetHT} shows the differential distributions of \wjets\ events as a function of $H_T$, 
the scalar sum of the transverse energies of the $W$ boson and the partons in the event.
Accurate prediction of the distribution of the scalar sum of the transverse energies of the $W$ boson and all high-$p_T$ ($p_T>20$~GeV) jets in \wjets\ events is important as this 
variable is often used as the preferred renormalization and factorization scale choice for theoretical predictions of 
vector boson plus jet events at the Tevatron and the LHC. 
In addition, this variable is often chosen as a discriminant in searches for signals of physics beyond the standard model at hadron colliders.
Calculation of high $H_T$ events is sensitive to higher-order corrections and so high $H_T$ data provides discrimination power between 
the various theoretical approaches for accounting for these contributions.

\begin{figure}[htbp]
  \begin{center}
    \includegraphics[width=\columnwidth]{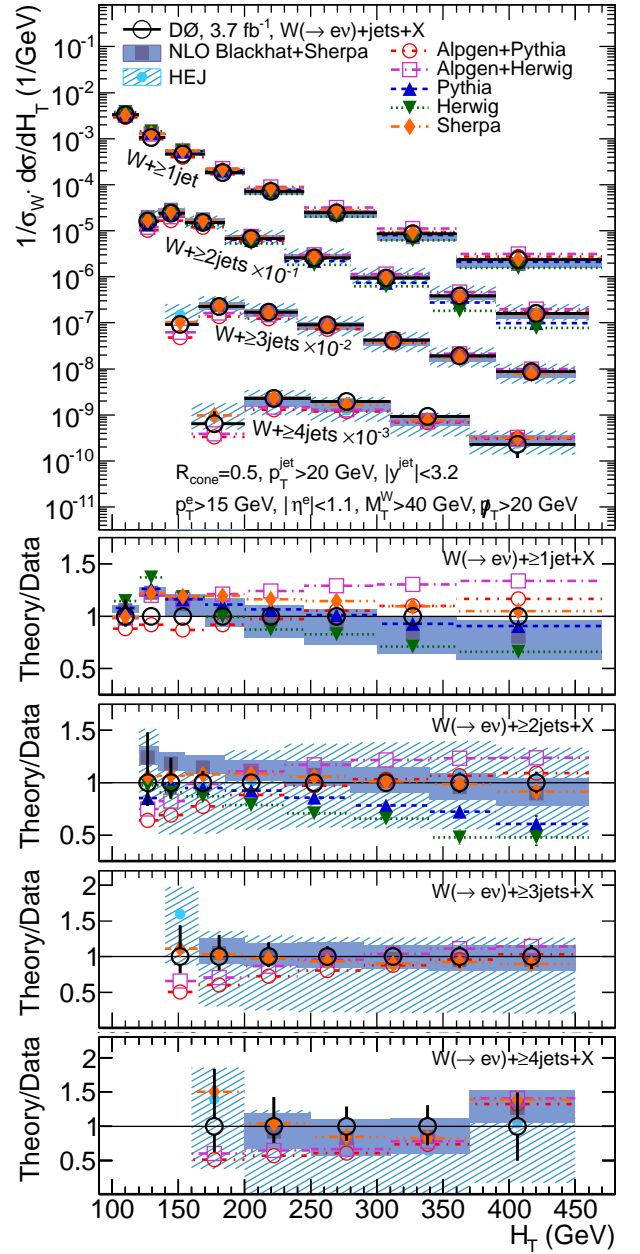}
    \caption{(color online) Measurement of the distribution of the scalar sum of transverse energies of the $W$ boson and all jets in the event for inclusive \wnjet\ 
      events for $n=\mathrm{1-4}$ and comparison to various theoretical predictions.       Lower panes show theory/data comparisons for each of the $n$-jet multiplicity bin results separately.
      \label{fig:result_jetHT}
    }
  \end{center}
\end{figure}

We observe significant variation in the predicted shapes of the $H_T$ spectrum from the various theoretical approaches.
\sherpa, \pythia, \herwig, and \alpgen\ show discrepancies in shape by $\pm 25\%$ in the one-jet bin, and up to $\pm 50\%$ in the four-jet bin.
The data are significantly more precise than the spread of these predictions and can be used to improve the modeling.
\hej\ exhibits a good description of the data, albeit with large scale uncertainties, but the trend for NLO \blackhat\ (particularly noticeable in the one-jet bin)
is for predictions to progressively underestimate the data as $H_T$ increases. 
NLO $W+n\textrm{-jet}$ calculations include $n$ and $n+1$ parton emissions and this limitation becomes apparent when studying observables
such as $H_T$ that are sensitive to higher-order contributions at high $H_T$, where the omission of matrix elements with three or more real emissions in the NLO 
calculation becomes apparent. Similar behavior was also observed in ATLAS \wjets\ data\,\cite{Aad:2012en}.
\alppyth, which includes LO matrix elements with up to five partons in the final state, gives the best description of the inclusive one-jet spectrum in data.

\subsection{Mean jet multiplicities}

Figures~\ref{fig:result_njet_HT}(a) and~(b) show the mean number of high-$p_T$ jets produced in \wjets\ events as a function of $H_T$
in inclusive $W+1\textrm{-jet}$ and $W+2\textrm{-jet}$ events, respectively,
allowing us to investigate how the jet multiplicity in \wjets\ events correlates with increasing boson and parton transverse energy.

\begin{figure}[hbtp]
  \begin{center}
    \includegraphics[width=0.98\columnwidth]{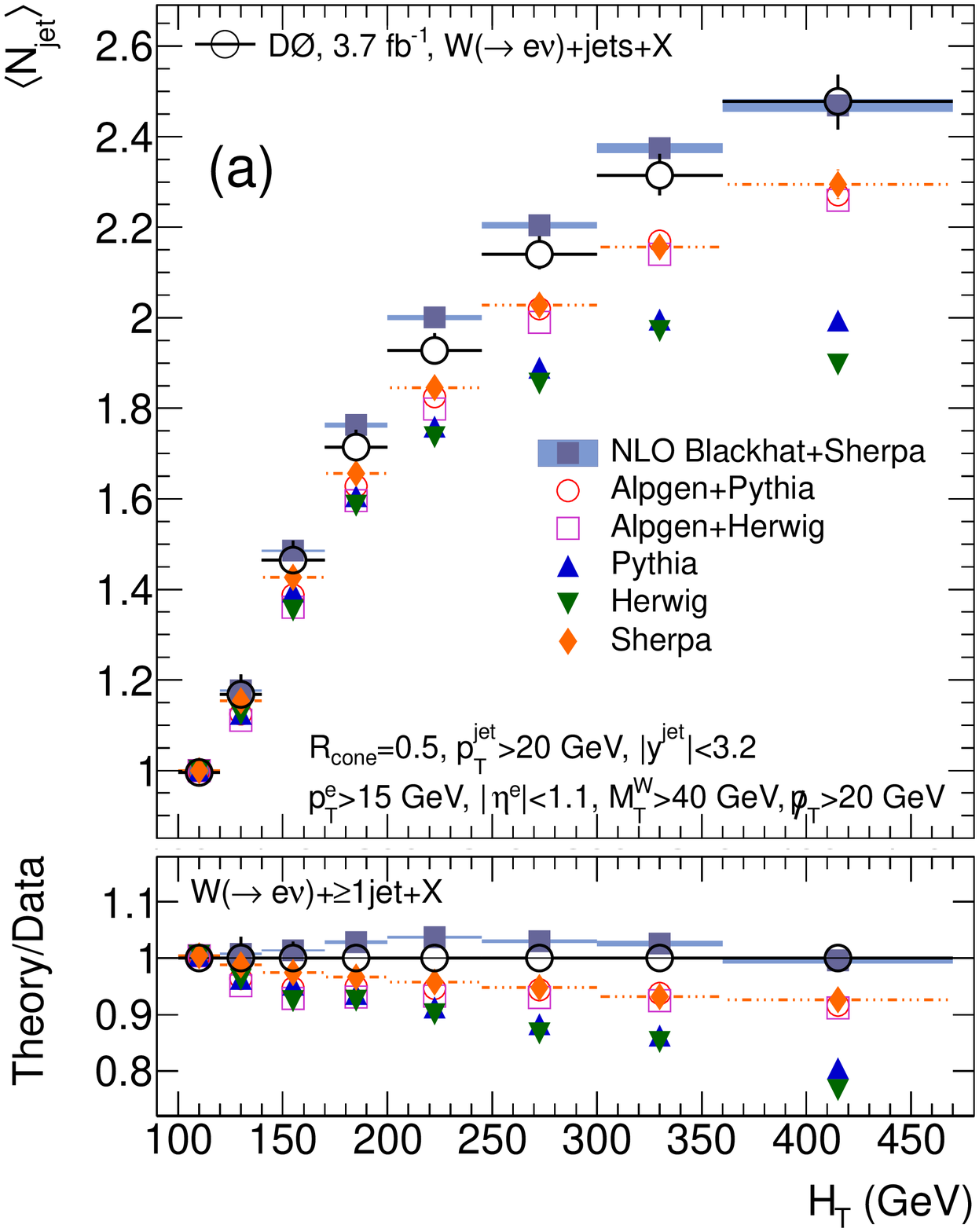}
    \includegraphics[width=0.98\columnwidth]{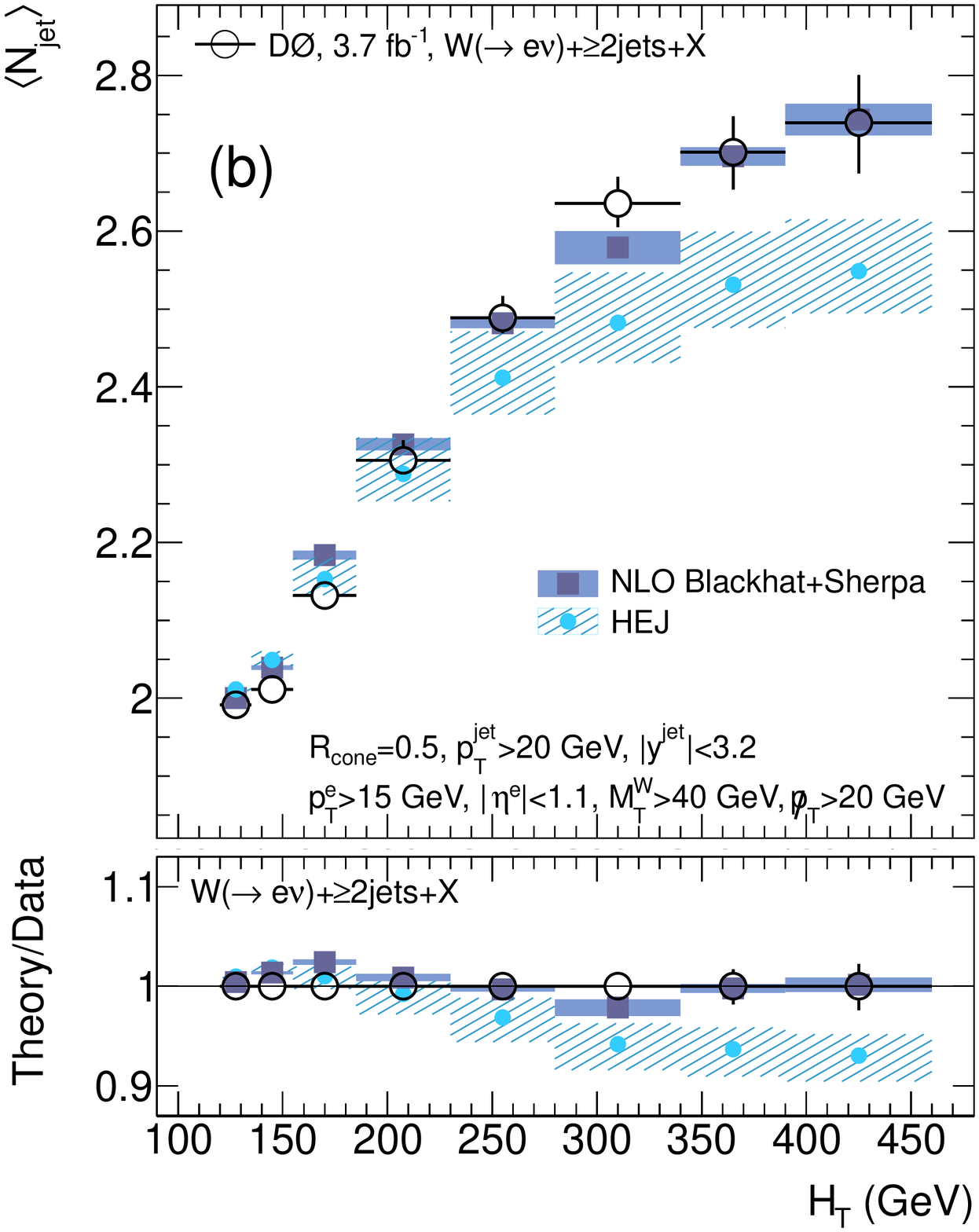}
    \caption{(color online) Measurement of the mean number of jets in (a) inclusive $W+1\textrm{-jet}$ events and (b) inclusive $W+2\textrm{-jet}$ events
      as a function of the scalar sum of transverse energies of the $W$ boson and all jets in the event, with comparison to various theoretical predictions. 
      The lower pane shows theory/data comparisons.
      \label{fig:result_njet_HT}
    }
  \end{center}
\end{figure}

The data display a sharp rise in the mean number of jets versus $H_T$. 
We observe that high $H_T$ events are typically high jet multiplicity events with moderate jet $p_T$ rather than low multiplicity events with high $p_T$.
The high $H_T$ region is therefore particularly sensitive to higher-order corrections
and proper modeling of the jet emissions in such a region will be necessary to permit discrimination between standard model vector boson plus jets 
production and indications of new physics with different high $H_T$ properties.

The \textsc{blackhat} collaboration use the following prescription for calculating the expected mean number of jets within a given kinematic interval 
in an inclusive $W+n\textrm{-jet}$ event to improve the description beyond the standard NLO pQCD calculation: 
\begin{equation}
\langle N_\mathrm{jet}\rangle = n+\left(d\sigma^\mathrm{NLO}_{n+1} + d\sigma^\mathrm{LO}_{n+2}\right)/d\sigma^\mathrm{NLO}_n.
\end{equation}
Such a definition includes all NLO corrections, plus some higher-order terms in $\alpha_s$, but essentially becomes a leading-order calculation
where $\langle N_\mathrm{jet}\rangle\to n+1$ in an inclusive \wnjet\ event, leading to reduced reliability in the predictions.

\begin{figure*}[h!tbp]
  \begin{center}
    \includegraphics[width=\columnwidth]{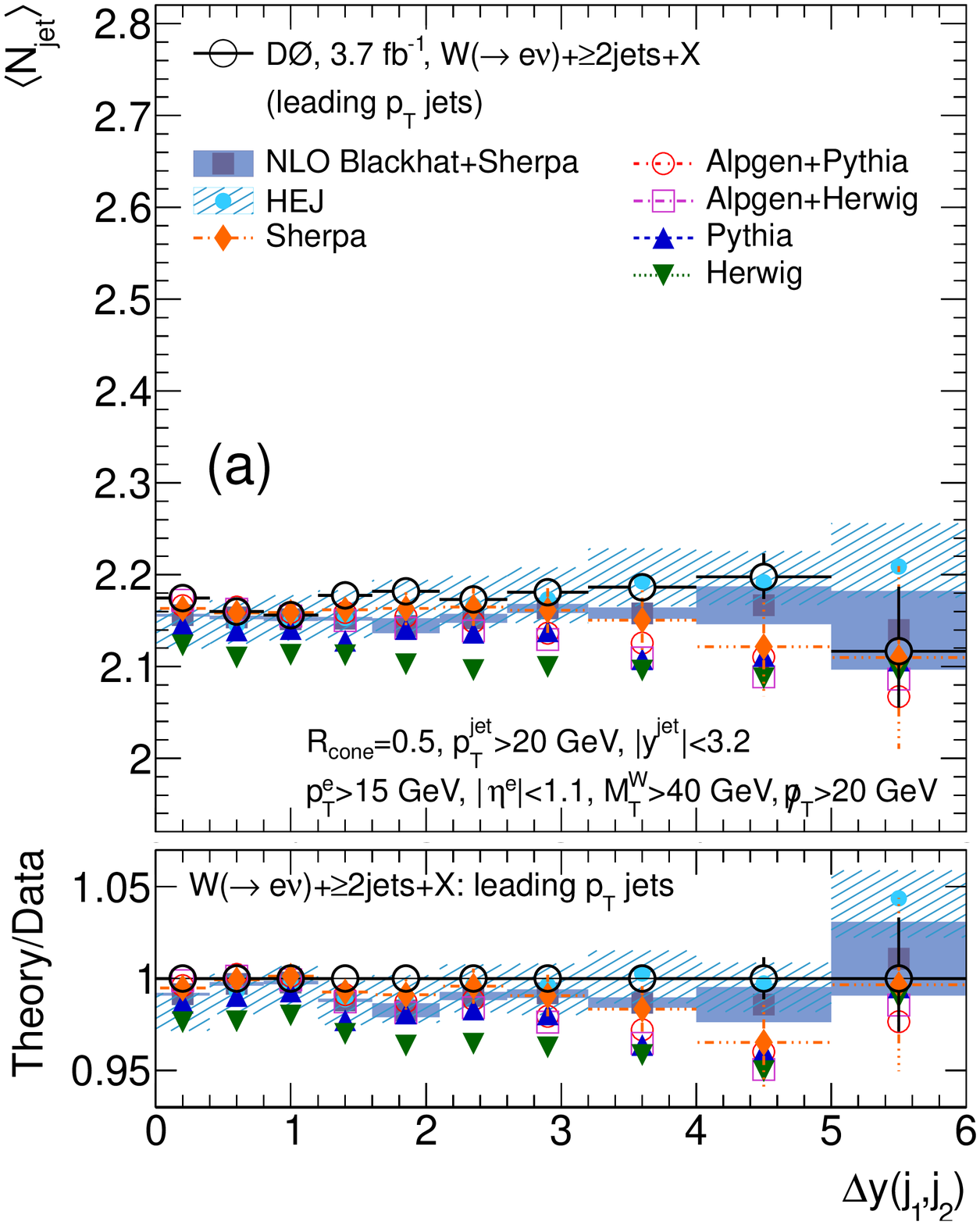}
    \includegraphics[width=\columnwidth]{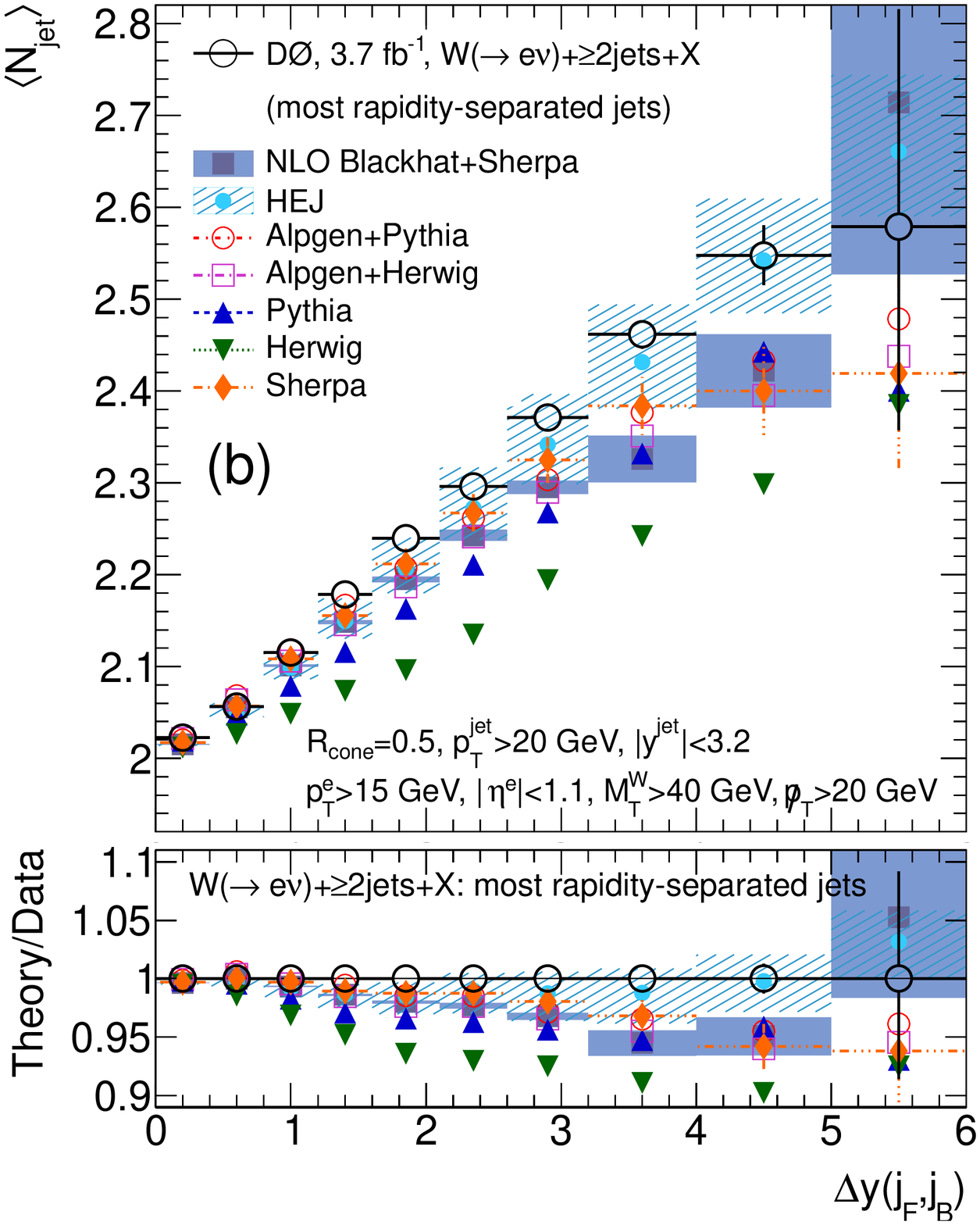}
    \caption{(color online) Measurement of the mean number of jets in inclusive $W+2\textrm{-jet}$ events as a function of the dijet rapidity separation of the two leading jets in both
      (a) $p_T$-ordered and (b) rapidity-ordered scenarios, along with comparison to various theoretical predictions.
      Lower panes show theory/data comparisons for each of the configurations.
      \label{fig:result_njet_DeltaRap}
    }
  \end{center}
\end{figure*}

In inclusive one-jet events (Fig.~\ref{fig:result_njet_HT}(a)), parton shower approaches are unable to describe the jet emission dependence, diverging from the data even at the lowest $H_T$ accessible.
Such predictions plateau below $\langle N_\mathrm{jet}\rangle = 2$ due to the limitations of the $W+1\textrm{-jet}$ matrix element.
Both matrix element plus matched parton shower approaches from \textsc{alpgen+(pythia/herwig)} and \sherpa\ do a somewhat better job of describing the 
jet multiplicity increase but again reach a maximum at $\langle N_\mathrm{jet}\rangle = 2.2$, well below the data which reaches a maximum jet multiplicity of 2.5.
In contrast, the NLO \blackhat\ approach is successful in describing the $\langle N_\mathrm{jet}\rangle$ spectrum across the entire $H_T$ range in inclusive one-jet events with good accuracy.

For inclusive two-jet events (Fig.~\ref{fig:result_njet_HT}(b)), we focus on resummation and NLO pQCD approaches,
due to parton shower simulations, which contain LO matrix elements only, having reduced accuracy and less predictive power for high jet-multiplicity final states.
Here, the NLO prediction again describes the data well over the full range of measured $H_T$.
Calculations from \hej\ perform well at low $H_T$, but begin to underestimate the amount of high-$p_T$ jet emission above $H_T>250$~GeV.

We also measure the mean jet multiplicity as a function of the rapidity separation between the two highest-$p_T$ jets 
and between the two most rapidity-separated jets in inclusive \wdijet\ events, with the results shown in Fig.~\ref{fig:result_njet_DeltaRap}.
The mean number of jets as a function of dijet rapidity separation provides a sensitive test of high-$p_T$ jet emission in \wjets\ events.
As a function of the $\Delta y$ between the two highest-$p_T$ jets, the mean jet multiplicity is approximately constant up
to rapidity spans of six units of rapidity with $\langle N_\mathrm{jet}\rangle \approx2.17$. 
Parton shower and matrix element matched theoretical approaches are able to describe the shape of the rapidity separation dependence (if not
the overall jet emission rate) until $\Delta y>3$, where these approaches consistently predict a 5\% drop in mean jet multiplicity not observed in the data.
Predictions from NLO \blackhat\ and \hej\ accurately predict the uniform jet multiplicity distribution seen in the data.

In the case of the most rapidity-separated jet configuration, a strong $\langle N_\mathrm{jet}\rangle$ dependence is observed with rapidity separation,
in contrast to the $p_T$-ordered configuration, varying from $\langle N_\mathrm{jet}\rangle = 2.0$ jets at small separation (where there is limited phase
space for emission of a third jet with $p_T>20$~GeV between the two forward jets) increasing steadily with rapidity separation to approximately $2.6$ jets at the widest spans
as shown in Fig.~\ref{fig:result_njet_DeltaRap}(b).
This is a particularly important probe for validation of theoretical understanding of wide angle gluon emission in vector boson plus jet processes.

Unlike the \dy12\ configurations, both parton shower and matrix element plus matched parton shower generators underpredict the rate of increase in 
the number of jets as a function of \dyFB. Predictions from \blackhat\ also show a trend for NLO pQCD to underestimate the jet multiplicity in a similar
manner to \alpgen\ and \sherpa. Resummation predictions from \hej\ are able to accurately describe the jet multiplicity dependence on jet rapidity separation
across the full interval studied, with high precision.

\subsection{Jet emission probabilities / gap fraction}

Figures~\ref{result:jetprob_dy12}--\ref{result:jetprob_dy12restrict} present measurements of the probability for a third high-$p_T$ jet to be emitted in inclusive \wdijet\ events 
calculated as the fraction of events in the inclusive $W+2\textrm{-jet}$ sample that contain a third jet over a $p_T>20$~GeV threshold
as a function of dijet rapidity separation using:
\begin{enumerate}
\item the two highest-$p_T$ jets,
\item the two most rapidity-separated jets ($p_T>20$~GeV), and
\item the two highest-$p_T$ jets with the requirement that the third jet be emitted into the rapidity interval between the two highest-$p_T$ jets.
\end{enumerate}
The probability of emission of a third jet in inclusive \wdijet\ events is strongly correlated with the mean number of jets in the event
presented in Figs.~\ref{fig:result_njet_HT} and~\ref{fig:result_njet_DeltaRap}.
However, with the probability observable, we specifically focus on the emission of a single additional jet beyond the two used to define the dijet rapidity interval.

\begin{figure}[htbp]
  \begin{center}
    \includegraphics[width=\columnwidth]{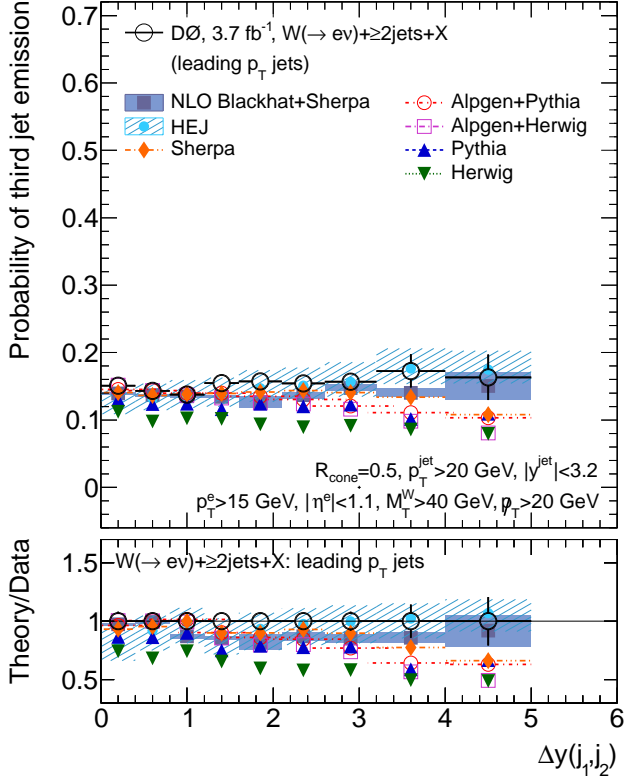}
    \caption{(color online) Measurement of the probability of emission of a third jet in inclusive $W+2\textrm{-jet}$ events as a 
      function of the dijet rapidity separation of the two highest-$p_T$ jets.
      Comparison is made to predictions from various theoretical approaches.       
      The lower pane shows theory/data comparisons.
      \label{result:jetprob_dy12}
    }
  \end{center}
\end{figure}

The probability of third jet emission as a function of the rapidity span between the two leading jets is approximately $15\%$
and is shown in Fig.~\ref{result:jetprob_dy12} in comparison to a variety of theoretical predictions.
Both parton shower and matrix-element plus parton shower matched MC programs underpredict the overall emission rate, particularly
at large rapidity separations where these programs predict a drop in jet emission not supported by the data. Unlike the MC
predictions that underestimate the high-$p_T$ radiation at large rapidity separations, \hej\ and NLO \blackhat\ approaches are able to model
the constant jet emission dependence well.

As a function of the most rapidity-separated jets, a significant variation in third jet emission probability is observed in the data.
At the smallest rapidity separations, emission probabilities are $\approx0\%$, but at the largest rapidity spans, half of all inclusive \wdijet\
events are found to have a third high-$p_T$ jet present. 
This measurement is shown in comparison to a variety of theoretical models in Fig.~\ref{result:jetprob_dyFB}.
\begin{figure}[htbp]
  \begin{center}
    \includegraphics[width=\columnwidth]{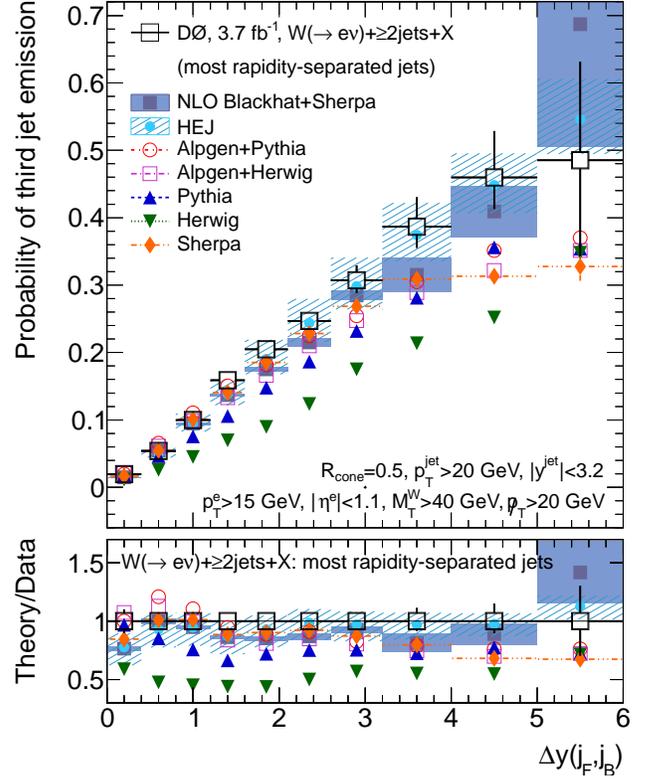}
    \caption{(color online) Measurement of the probability of emission of a third jet in inclusive $W+2\textrm{-jet}$ events as a function of the dijet rapidity separation of the two 
      most rapidity-separated jets (with $p_T>20$~GeV).
      Comparison is made to predictions from various theoretical approaches.       
      The lower pane shows theory/data comparisons.
      \label{result:jetprob_dyFB}
    }
  \end{center}
\end{figure}

The exact correlation of jet emission probability with rapidity interval is dependent on the interplay between two effects: the increasing 
phase space for high-$p_T$ emission between jets versus the probability to actually emit into that rapidity interval (which decreases at large
rapidity separations due to steeply falling PDFs as Bjorken $x\to1$). There is some evidence that as we approach the largest separations studied,
PDF suppression may be beginning to dominate over the increased phase space (in both the highest-$p_T$ jet and most rapidity-separated jet configurations). 
Proper modeling of \wjets\ behavior, particularly in the most rapidity-separated jet case,
will be important for understanding central jet vetoes in future VBF Higgs studies.

As for $\langle N_\mathrm{jet}\rangle$ in Fig.~\ref{fig:result_njet_DeltaRap}(b), parton shower and matrix-element plus parton shower matched 
predictions underestimate the rise in jet emission probability 
with increased rapidity separation and plateau at a maximum probability of around 35\% as shown in Fig.~\ref{result:jetprob_dyFB}.
NLO pQCD and \hej\ resummation approaches are able to describe the emission probability across the full range of study.

Third jet emission probabilities are also presented as a function of dijet rapidity separation with an additional requirement that the third high-$p_T$ jet be emitted 
into the rapidity interval between the leading two jets and compared with theoretical models in Fig.~\ref{result:jetprob_dy12restrict}.
\begin{figure}[htbp]
  \begin{center}
    \includegraphics[width=\columnwidth]{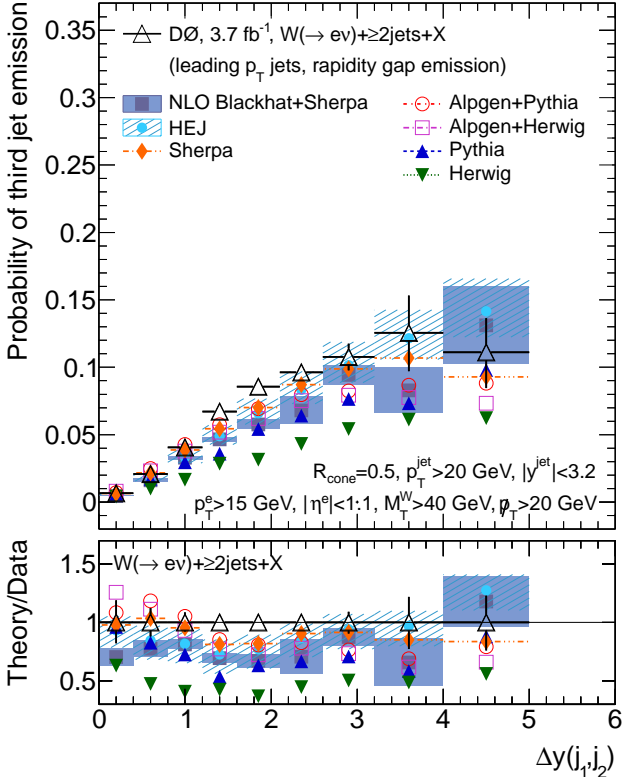}
    \caption{(color online) Measurement of the probability of emission of a third jet in inclusive $W+2\textrm{-jet}$ events as a function of the dijet rapidity separation of the two highest-$p_T$ jets
      with an additional requirement that the third jet be emitted into the rapidity interval defined by the two leading jets.
      Comparison is made to predictions from various theoretical approaches.       
      The lower pane shows theory/data comparisons.
      \label{result:jetprob_dy12restrict}
    }
  \end{center}
\end{figure}

This configuration represents a conceptual hybrid between the rapidity-ordered configuration (which has the requirement that the third jet be emitted 
between the two most rapidity-separated jets by construction) 
and a $p_T$-ordered jet configuration where it is the highest-$p_T$, rather than most rapidity-separated, jets that are probed.

\begin{figure}[h!tbp]
  \begin{center}
    \includegraphics[width=\columnwidth]{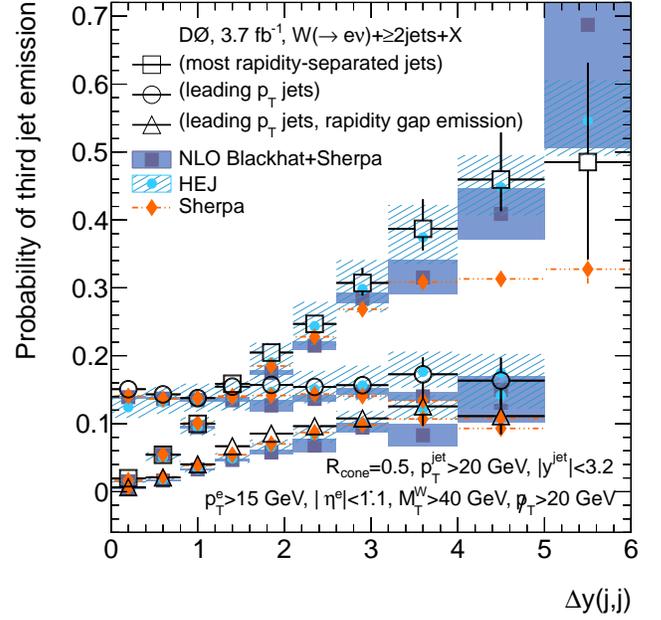}
    \caption{(color online) Measurement of the probability of emission of a third jet in inclusive $W+2\textrm{-jet}$ events 
      as a function of the dijet rapidity separation of the two leading jets in both
      $p_T$-ordered and rapidity-ordered scenarios, and a hybrid scenario where the rapidity separation is built from the two highest-$p_T$ jets but the third jet is required to be
      emitted into the rapidity interval defined by the two leading jets. 
      Comparison is made to predictions from NLO \blackhat, \hej, and \sherpa.
      \label{result:jetprob}
    }
  \end{center}
\end{figure}

The three measurements of third jet emission probability are summarized in Fig.~\ref{result:jetprob} in comparison to predictions from NLO \blackhat, \hej, and \sherpa.
The probability (${\mathbb P}$) ratio 
\begin{equation}
{\cal R}_{\mathbb P} = {\mathbb P}(\text{third jet in rapidity interval})/{\mathbb P}(\text{third jet}) 
\end{equation}
provides information on the probability that the third-highest $p_T$ jet in inclusive $W+3\textrm{-jet}$ events is emitted within the rapidity interval defined by the two highest-$p_T$ jets.
For $\Delta y\to 0$, ${\cal R}_{\mathbb P}\to 0$ as the available phase space for emission is reduced. As the rapidity interval widens, ${\cal R}_{\mathbb P}\to 1$ 
as the phase space for third jet emission at larger rapidities than the two leading jets decreases.
In the limit of $\Delta y\to 0$, the leading $p_T$ jet rapidity interval configuration is bounded by that of the rapidity-ordered jet results, and in the 
wide-angle limit is bounded by the emission probability of the $p_T$-ordered jet configuration without a rapidity interval requirement. 
As such, emission probabilities again start at $\approx0\%$ at small jet separation and rise quickly with increasing jet spans, but are 
limited to a plateau of around 15\% at the largest rapidity spans.
\pythia\ and \herwig\ in particular have trouble modeling this observable, both in overall jet emission rate and in the dependence with $\Delta y$.
\alpgen\ and \sherpa\ provide improved descriptions of the jet emission probability in this configuration, but still predict a lower emission rate at 
larger rapidity spans than observed in data.
NLO pQCD predictions systematically underestimate the total emission rate by about 30\%, but otherwise describe the emission rate dependence on rapidity 
interval well across the full range. Resummation predictions from \hej\ are best able to describe both the rate and shape across the full rapidity range.

The probability, $\mathbb{P}$, of additional jet emission as a function of dijet rapidity separation can be re-interpreted as 
a ``gap-fraction'' $\mathbb{F}$, where $\mathbb{F}=1-\mathbb{P}$. This gap fraction is defined as the fraction of inclusive $W+\textrm{dijet}$ events that do 
not have an additional jet with a transverse momentum larger than a given veto scale (in this analysis, $20$~GeV) within the rapidity interval defined by 
the rapidities of the two highest-$p_T$ (or most rapidity-separated) jets in the event.
Figure~\ref{result:gapfrac} shows the gap fraction dependence on rapidity-separation in both the $p_T$-ordered and rapidity-ordered jet configurations.
\begin{figure}[h!tbp]
  \begin{center}
    \includegraphics[width=\columnwidth]{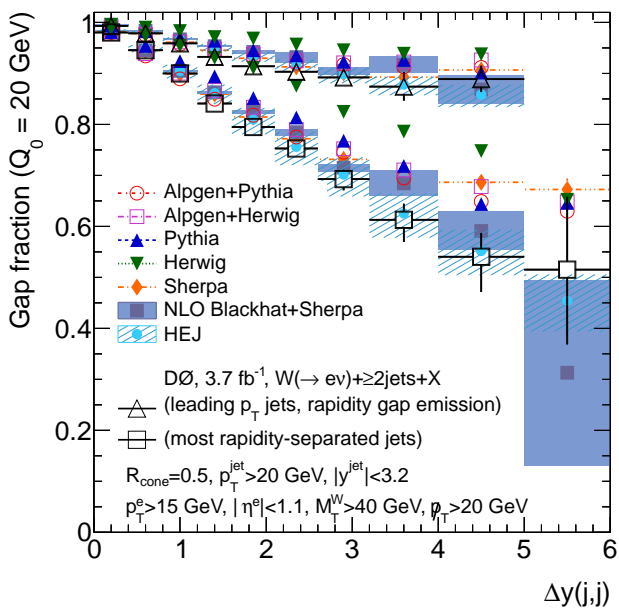}
    \caption{(color online) Gap fraction of inclusive $W+\mathrm{dijet}$ events presented as a function of the rapidity separation between the two highest-$p_T$ jets 
      or two most rapidity-separated jets in the event. The veto scale, $Q_0$, for the lower transverse momentum threshold for additional jets emitted
      into the rapidity interval is $Q_0=20$~GeV. Comparison is made to predictions from various theoretical approaches.
      Theory/data ratios are as for the results presented in Figs.~\ref{result:jetprob_dyFB} and~\ref{result:jetprob_dy12restrict}.
      \label{result:gapfrac}
    }
  \end{center}
\end{figure}

\section{Summary}

This paper presents measurements of the production rates and properties of $\wenu+n\mathrm{-jet}$ production (for $n\geq1,2,3,4$) in $p\bar{p}$ collisions 
at a center-of-mass energy of 1.96~TeV using 3.7~\ifb\ of \Dzero\ experiment data.
These measurements provide the most comprehensive set of measurements of the $W+textrm{jets}$ processes made to date.  They are important in their own right as precise 
studies of important QCD processes that will guide future theoretical refinements, and are essential for establishing backgrounds for searches for new rare processes.

Differential cross sections, normalized to the total $W$ boson cross section and fully corrected for detector effects, 
are presented for various inclusive jet multiplicities as a function of the jet rapidities; electron $p_T$ and pseudorapidity; 
rapidity separations of the first, second, and third-highest $p_T$ jets in the event; 
rapidity separations of the most rapidity-separated jets in the event (above $p_T>20$~GeV); azimuthal angle separations between jets; 
the angular separation between the two leading jets in $y$-$\phi$ space; $W$ boson $p_T$;
dijet system $p_T$ and invariant mass; and $H_T$ (the scalar sum of the jet and $W$ transverse energies).
Many of these observables are studied here for the first time in \wjets\ events or substantially improve on the precision of existing measurements.
These measurements complement previous measurements\,\cite{Abazov:2011rf} of the total inclusive \wnjet\ production cross sections (for $n\geq1,2,3,4$) and differential
cross sections of the $n^\mathrm{th}$-jet transverse momenta performed in the same phase space as measurements presented here.

Additionally, we present measurements of the evolution of mean jet multiplicities of \wjets\ events as a function of $H_T$ in the inclusive one- and two-jet multiplicity bins;
and within the region bounded by the dijet rapidity defined by the two jets that are the highest-$p_T$ jets, and also between the most rapidity-separated jets above a $p_T>20$~GeV threshold.
The probability of third high-$p_T$ jet emission in $W+\mathrm{dijet}$ events is also measured as a function of the dijet rapidity separation in three configurations: 
the first, where the rapidity separation is defined by the two highest-$p_T$ jets in the event; the second, as previously but where any third jet is required to be emitted into 
the rapidity interval defined by the other two jets; and finally, where the rapidity separation is defined by the two most rapidity-separated jets.
These results can be recast as a measure of the gap fraction in \wjets\ events, with a veto on additional emissions with $p_T>20$~GeV.

Presented measurements can be used for constraining the modeling of QCD radiation between the two jets, 
for understanding the efficacy of a central jet veto\,\cite{Rainwater} used for discriminating Higgs boson plus dijet events produced through vector boson fusion 
from standard model backgrounds and subsequent study of the Higgs boson properties\,\cite{Couplings}. In addition, they can contribute to improved understanding
of vector boson plus jet contributions to diverse topics such as studies of $W_L W_L \to WW$ scattering and searches for MSSM signatures through VBF production\,\cite{Dutta:2012xe}.
Measurements of the gap fraction in vector boson plus dijet events complement existing measurements in inclusive dijet\,\cite{ATLASdijetveto} events.

Comparisons of the experimental data are made to predictions from a variety of theoretical approaches.
Over most of the phase space in which the measurements are presented, experimental uncertainties are smaller than the theoretical uncertainties on NLO \blackhat,
and on \hej\ resummation predictions. The predictions from various Monte Carlo programs are found to have significant variations between each other, greater in magnitude than the 
experimental uncertainties, and thus these data can be used to improve the modeling of \wjets\ production and the emission of QCD radiation in such event generators.

The authors would like to thank the Blackhat Collaboration, in particular Fernando Febres Cordero, for providing 
NLO \blackhat\ predictions, and also Daniel Maitre and Lance Dixon for valuable conversations.
We also thank Stefan H\"{oche} for \sherpa\ predictions; Tuomas Hapola, Jeppe Andersen, and Jennifer Smillie for \hej\ predictions;
and Vato Kartvelishvili for useful discussions and advice on use of the \guru\ unfolding program.

% acknowledgement.tex                            11 January 2013 
%
We thank the staffs at Fermilab and collaborating institutions,
and acknowledge support from the
DOE and NSF (USA);
CEA and CNRS/IN2P3 (France);
MON, NRC KI and RFBR (Russia);
CNPq, FAPERJ, FAPESP and FUNDUNESP (Brazil);
DAE and DST (India);
Colciencias (Colombia);
CONACyT (Mexico);
NRF (Korea);
FOM (The Netherlands);
STFC and the Royal Society (United Kingdom);
MSMT and GACR (Czech Republic);
BMBF and DFG (Germany);
SFI (Ireland);
The Swedish Research Council (Sweden);
and
CAS and CNSF (China).

\clearpage
\newpage
\appendix
\widetext

\section{Tables of Measurements, Theory Calculations and Non-Perturbative Hadronization Corrections}
\label{appendix}

\begin{center}
{\bf To appear as an Electronic Physics Auxiliary Publication (EPAPS)}
\end{center}

In this appendix, we provide tables of the measured differential cross sections, theory predictions, and hadronization corrections described in this paper.   
The region that defines the kinematic phase space of the measurement at particle level is given by the electron transverse momentum, $p_T^e \ge 15$~GeV, 
and pseudorapidity $|\eta^e|< 1.1$, total missing transverse energy $\met>20$~GeV, $W$ boson transverse mass $M_T^W > 40$~GeV, and jet transverse 
momentum $p_T^\mathrm{jet}\ge 20$~GeV and rapidity $|y^\mathrm{jet}|<3.2$.  

\begin{table}[htbp]
\caption{\label{tab:results_jet1eta}The measured differential cross section, normalized to the measured inclusive $W$ boson cross section, as a function of leading jet rapidity for events with one or more jets produced in association with a $W$ boson. The first data uncertainty is statistical, the second is systematic. Also shown are predictions from NLO pQCD after hadronization corrections have been applied (also shown).}
\begin{ruledtabular}
\begin{tabular}{rc r@{$\pm$\hspace{-9mm}}l@{${}$\hspace{-9mm}}l ll}
\multicolumn{7}{c}{$1/\sigma_W\cdot  d\sigma/dy^\mathrm{jet}~\textrm{[one-jet bin]}~(1/\Delta y)$} \\
Bin interval  & Bin center  & \multicolumn{3}{c}{\hspace{-10mm}Measurement} & npQCD correction & NLO \textsc{blackhat+sherpa} \\[-0.5ex]
\multicolumn{1}{c}{$y$}  & $y$ & \multicolumn{3}{c}{\hspace{-6mm}$(\times 10^{-3})$} & & $(\times 10^{-3})$ \\[1ex]
\hline
$-3.2$~--~$-2.8$ & $ -3.0$ & $ 1.62$ & $0.03$ & ${}^{+0.54}_{-0.31}$ & $0.99\pm 0.02$ & $2.23{}^{+0.24}_{-0.24}$ \\
$-2.8$~--~$-2.4$ & $ -2.6$ & $ 4.3$ & $0.1$ & ${}^{+1.3}_{-0.8}$ & $1.01\pm 0.02$ & $4.9{}^{+0.3}_{-0.4}$ \\
$-2.4$~--~$-2.0$ & $ -2.2$ & $ 8.6$ & $0.1$ & ${}^{+2.1}_{-1.2}$ & $1.01\pm 0.01$ & $9.3{}^{+0.6}_{-0.7}$ \\
$-2.0$~--~$-1.6$ & $ -1.8$ & $ 13.8$ & $0.1$ & ${}^{+2.4}_{-1.5}$ & $1.01\pm 0.01$ & $15.0{}^{+0.8}_{-1.0}$ \\
$-1.6$~--~$-1.2$ & $ -1.4$ & $ 18.5$ & $0.1$ & ${}^{+2.1}_{-1.5}$ & $1.01\pm 0.01$ & $21.5{}^{+1.1}_{-1.3}$ \\
\end{tabular}
\end{ruledtabular}
\end{table}

\begin{table}[htbp]
\caption{\label{tab:results_jet2eta}The measured differential cross section, normalized to the measured inclusive $W$ boson cross section, as a function of second jet rapidity for events with two or more jets produced in association with a $W$ boson. The first data uncertainty is statistical, the second is systematic. Also shown are predictions from NLO pQCD after hadronization corrections have been applied (also shown) and \textsc{hej} resummation predictions.}
\begin{ruledtabular}
\begin{tabular}{rc r@{$\pm$\hspace{-3mm}}l@{${}$\hspace{-3mm}}l rll}
\multicolumn{8}{c}{$1/\sigma_W\cdot  d\sigma/dy^\mathrm{jet}~\textrm{[two-jet bin]}~(1/\Delta y)$} \\
Bin interval  & Bin center  & \multicolumn{3}{c}{\hspace{-6mm}Measurement} & npQCD correction & NLO \textsc{blackhat+sherpa} & \textsc{hej}\\[-0.5ex]
\multicolumn{1}{c}{$y$}  & $y$ & \multicolumn{3}{c}{\hspace{-6mm}$(\times 10^{-3})$} & & $(\times 10^{-3})$ & $(\times 10^{-3})$  \\[1ex]
\hline
$-3.2$~--~$-2.8$ & $ -3.0$ & $ 0.34$ & $0.02$ & ${}^{+0.11}_{-0.07}$ & $1.10\pm 0.04$ & $0.59{}^{+0.08}_{-0.09}$ & $0.44{}^{+0.22}_{-0.13}$ \\
$-2.8$~--~$-2.4$ & $ -2.6$ & $ 0.81$ & $0.02$ & ${}^{+0.25}_{-0.15}$ & $1.09\pm 0.03$ & $1.16{}^{+0.12}_{-0.16}$ & $0.94{}^{+0.45}_{-0.27}$ \\
$-2.4$~--~$-2.0$ & $ -2.2$ & $ 1.48$ & $0.03$ & ${}^{+0.42}_{-0.25}$ & $1.08\pm 0.03$ & $1.89{}^{+0.18}_{-0.25}$ & $1.69{}^{+0.78}_{-0.48}$ \\
$-2.0$~--~$-1.6$ & $ -1.8$ & $ 2.21$ & $0.04$ & ${}^{+0.53}_{-0.33}$ & $1.05\pm 0.02$ & $2.60{}^{+0.17}_{-0.30}$ & $2.51{}^{+1.13}_{-0.70}$ \\
$-1.6$~--~$-1.2$ & $ -1.4$ & $ 2.94$ & $0.05$ & ${}^{+0.57}_{-0.38}$ & $1.05\pm 0.02$ & $3.56{}^{+0.27}_{-0.41}$ & $3.35{}^{+1.47}_{-0.93}$ \\
\end{tabular}
\end{ruledtabular}
\end{table}

\begin{table}[htbp]
\caption{\label{tab:results_jet3eta}The measured differential cross section, normalized to the measured inclusive $W$ boson cross section, as a function of third jet rapidity for events with three or more jets produced in association with a $W$ boson. The first data uncertainty is statistical, the second is systematic. Also shown are predictions from NLO pQCD after hadronization corrections have been applied (also shown) and \textsc{hej} resummation predictions.}
\begin{ruledtabular}
\begin{tabular}{rc r@{$\pm$\hspace{-3mm}}l@{${}$\hspace{-3mm}}l rll}
\multicolumn{8}{c}{$1/\sigma_W\cdot  d\sigma/dy^\mathrm{jet}~\textrm{[three-jet bin]}~(1/\Delta y)$} \\
Bin interval  & Bin center  & \multicolumn{3}{c}{\hspace{-6mm}Measurement} & npQCD correction & NLO \textsc{blackhat+sherpa} & \textsc{hej}\\[-0.5ex]
\multicolumn{1}{c}{$y$}  & $y$ & \multicolumn{3}{c}{\hspace{-6mm}$(\times 10^{-3})$} & & $(\times 10^{-3})$ & $(\times 10^{-3})$  \\[1ex]
\hline
$-3.2$~--~$-2.4$ & $ -2.8$ & $ 0.073$ & $0.005$ & ${}^{+0.028}_{-0.015}$ & $1.07\pm 0.05$ & $0.130{}^{+0.020}_{-0.026}$ & $0.126{}^{+0.101}_{-0.050}$ \\
$-2.4$~--~$-1.6$ & $ -2.0$ & $ 0.25$ & $0.01$ & ${}^{+0.08}_{-0.05}$ & $1.09\pm 0.04$ & $0.32{}^{+0.04}_{-0.06}$ & $0.34{}^{+0.24}_{-0.13}$ \\
$-1.6$~--~$-0.8$ & $ -1.2$ & $ 0.48$ & $0.02$ & ${}^{+0.11}_{-0.07}$ & $1.10\pm 0.04$ & $0.54{}^{+0.07}_{-0.10}$ & $0.53{}^{+0.36}_{-0.20}$ \\
$-0.8$~--~$0.0$ & $ -0.4$ & $ 0.75$ & $0.02$ & ${}^{+0.12}_{-0.09}$ & $1.09\pm 0.05$ & $0.67{}^{+0.08}_{-0.11}$ & $0.67{}^{+0.44}_{-0.24}$ \\
$0.0$~--~$0.8$ & $ 0.4$ & $ 0.75$ & $0.02$ & ${}^{+0.13}_{-0.09}$ & $1.09\pm 0.05$ & $0.67{}^{+0.08}_{-0.11}$ & $0.67{}^{+0.44}_{-0.24}$ \\
$0.8$~--~$1.6$ & $ 1.2$ & $ 0.50$ & $0.02$ & ${}^{+0.12}_{-0.08}$ & $1.10\pm 0.04$ & $0.54{}^{+0.07}_{-0.10}$ & $0.52{}^{+0.35}_{-0.19}$ \\
$1.6$~--~$2.4$ & $ 2.0$ & $ 0.26$ & $0.01$ & ${}^{+0.08}_{-0.05}$ & $1.09\pm 0.04$ & $0.32{}^{+0.04}_{-0.06}$ & $0.33{}^{+0.23}_{-0.12}$ \\
$2.4$~--~$3.2$ & $ 2.8$ & $ 0.079$ & $0.005$ & ${}^{+0.026}_{-0.015}$ & $1.07\pm 0.05$ & $0.130{}^{+0.020}_{-0.026}$ & $0.124{}^{+0.094}_{-0.048}$ \\
\end{tabular}
\end{ruledtabular}
\end{table}

\begin{table}[htbp]
\caption{\label{tab:results_jet4eta}The measured differential cross section, normalized to the measured inclusive $W$ boson cross section, as a function of fourth jet rapidity for events with four or more jets produced in association with a $W$ boson. The first data uncertainty is statistical, the second is systematic. Also shown are predictions from NLO pQCD after hadronization corrections have been applied (also shown) and \textsc{hej} resummation predictions.}
\begin{ruledtabular}
\begin{tabular}{rc r@{$\pm$\hspace{-3mm}}l@{${}$\hspace{-3mm}}l rll}
\multicolumn{8}{c}{$1/\sigma_W\cdot  d\sigma/dy^\mathrm{jet}~\textrm{[four-jet bin]}~(1/\Delta y)$} \\
Bin interval  & Bin center  & \multicolumn{3}{c}{\hspace{-6mm}Measurement} & npQCD correction & NLO \textsc{blackhat+sherpa} & \textsc{hej}\\[-0.5ex]
\multicolumn{1}{c}{$y$}  & $y$ & \multicolumn{3}{c}{\hspace{-6mm}$(\times 10^{-3})$} & & $(\times 10^{-3})$ & $(\times 10^{-3})$  \\[1ex]
\hline
$-3.2$~--~$-2.4$ & $ -2.8$ & $ 0.009$ & $0.003$ & ${}^{+0.005}_{-0.003}$ & $1.23\pm 0.15$ & $0.015{}^{+0.003}_{-0.004}$ & $0.014{}^{+0.014}_{-0.006}$ \\
$-2.4$~--~$-1.6$ & $ -2.0$ & $ 0.033$ & $0.006$ & ${}^{+0.013}_{-0.009}$ & $1.22\pm 0.09$ & $0.039{}^{+0.010}_{-0.011}$ & $0.034{}^{+0.032}_{-0.015}$ \\
$-1.6$~--~$-0.8$ & $ -1.2$ & $ 0.074$ & $0.007$ & ${}^{+0.024}_{-0.016}$ & $1.13\pm 0.07$ & $0.062{}^{+0.017}_{-0.017}$ & $0.053{}^{+0.048}_{-0.023}$ \\
$-0.8$~--~$0.0$ & $ -0.5$ & $ 0.107$ & $0.009$ & ${}^{+0.028}_{-0.021}$ & $1.23\pm 0.13$ & $0.084{}^{+0.021}_{-0.022}$ & $0.072{}^{+0.064}_{-0.031}$ \\
$0.0$~--~$0.8$ & $ 0.5$ & $ 0.102$ & $0.007$ & ${}^{+0.027}_{-0.019}$ & $1.23\pm 0.13$ & $0.084{}^{+0.021}_{-0.022}$ & $0.064{}^{+0.058}_{-0.028}$ \\
$0.8$~--~$1.6$ & $ 1.2$ & $ 0.062$ & $0.007$ & ${}^{+0.019}_{-0.013}$ & $1.13\pm 0.07$ & $0.062{}^{+0.017}_{-0.017}$ & $0.055{}^{+0.050}_{-0.024}$ \\
$1.6$~--~$2.4$ & $ 2.0$ & $ 0.028$ & $0.004$ & ${}^{+0.011}_{-0.007}$ & $1.22\pm 0.09$ & $0.039{}^{+0.010}_{-0.011}$ & $0.032{}^{+0.030}_{-0.014}$ \\
$2.4$~--~$3.2$ & $ 2.8$ & $ 0.007$ & $0.002$ & ${}^{+0.003}_{-0.002}$ & $1.23\pm 0.15$ & $0.015{}^{+0.003}_{-0.004}$ & $0.014{}^{+0.014}_{-0.006}$ \\
\end{tabular}
\end{ruledtabular}
\end{table}

\clearpage
\newpage

\begin{table}[htbp]
\caption{\label{tab:results_lepton1pt}The measured differential cross section, normalized to the measured inclusive $W$ boson cross section, as a function of electron transverse momentum for events with one or more jets produced in association with a $W$ boson. The first data uncertainty is statistical, the second is systematic. Also shown are predictions from NLO pQCD after hadronization corrections have been applied (also shown).}
\begin{ruledtabular}
\begin{tabular}{rc r@{$\pm$\hspace{-9mm}}l@{${}$\hspace{-9mm}}l ll}
\multicolumn{7}{c}{$1/\sigma_W\cdot  d\sigma/dp_{T}^\mathrm{electron}~\textrm{[one-jet bin]}$ (GeV${}^{-1}$)} \\
Bin interval  & Bin center  & \multicolumn{3}{c}{\hspace{-10mm}Measurement} & npQCD correction & NLO \textsc{blackhat+sherpa} \\[-0.5ex]
\multicolumn{1}{c}{$\textrm{(GeV)}$}              &$\textrm{(GeV)}$             & \multicolumn{3}{c}{\hspace{-6mm}$(\times 10^{-6})$} & & $(\times 10^{-6})$ \\[1ex]
\hline
$15.0-25.0$ & $ 20.0$ & $ 2209$ & $9$ & ${}^{+246}_{-200}$ & $1.04\pm 0.01$ & $2128{}^{+117}_{-136}$ \\
$25.0-35.0$ & $ 26.8$ & $ 3355$ & $9$ & ${}^{+415}_{-275}$ & $1.02\pm 0.02$ & $3529{}^{+179}_{-209}$ \\
$35.0-45.0$ & $ 40.1$ & $ 2602$ & $8$ & ${}^{+331}_{-215}$ & $1.05\pm 0.03$ & $2879{}^{+188}_{-196}$ \\
$45.0-55.0$ & $ 50.4$ & $ 1753$ & $6$ & ${}^{+128}_{-97}$ & $0.96\pm 0.01$ & $2092{}^{+48}_{-90}$ \\
$55.0-70.0$ & $ 61.6$ & $ 617$ & $3$ & ${}^{+17}_{-19}$ & $0.98\pm 0.00$ & $768{}^{+47}_{-53}$ \\
$70.0-90.0$ & $ 78.6$ & $ 119.7$ & $1.0$ & ${}^{+2.4}_{-4.6}$ & $0.99\pm 0.00$ & $143.6{}^{+9.7}_{-11.0}$ \\
$90.0-130.0$ & $ 106.5$ & $ 15.82$ & $0.29$ & ${}^{+0.40}_{-0.82}$ & $0.98\pm 0.00$ & $18.18{}^{+1.12}_{-1.36}$ \\
$130.0-200.0$ & $ 157.8$ & $ 1.50$ & $0.07$ & ${}^{+0.05}_{-0.09}$ & $0.96\pm 0.01$ & $1.69{}^{+0.12}_{-0.14}$ \\
$200.0-300.0$ & $ 228.7$ & $ 0.10$ & $0.02$ & ${}^{+0.00}_{-0.01}$ & $1.01\pm 0.03$ & $0.08{}^{+0.00}_{-0.01}$ \\
\end{tabular}
\end{ruledtabular}
\end{table}

\begin{table}[htbp]
\caption{\label{tab:results_lepton2pt}The measured differential cross section, normalized to the measured inclusive $W$ boson cross section, as a function of electron transverse momentum for events with two or more jets produced in association with a $W$ boson. The first data uncertainty is statistical, the second is systematic. Also shown are predictions from NLO pQCD after hadronization corrections have been applied (also shown) and \textsc{hej} resummation predictions.}
\begin{ruledtabular}
\begin{tabular}{rc r@{$\pm$\hspace{-3mm}}l@{${}$\hspace{-3mm}}l rll}
\multicolumn{8}{c}{$1/\sigma_W\cdot  d\sigma/dp_{T}^\mathrm{electron}~\textrm{[two-jet bin]}$ (GeV${}^{-1}$)} \\
Bin interval  & Bin center  & \multicolumn{3}{c}{\hspace{-6mm}Measurement} & npQCD correction & NLO \textsc{blackhat+sherpa} & \textsc{hej}\\[-0.5ex]
\multicolumn{1}{c}{$\textrm{(GeV)}$}               &$\textrm{(GeV)}$             & \multicolumn{3}{c}{\hspace{-6mm}$(\times 10^{-6})$} & & $(\times 10^{-6})$ & $(\times 10^{-6})$  \\[1ex]
\hline
$15.0-25.0$ & $ 19.7$ & $ 365$ & $4$ & ${}^{+63}_{-45}$ & $1.07\pm 0.03$ & $350{}^{+26}_{-40}$ & $329{}^{+146}_{-91}$ \\
$25.0-35.0$ & $ 27.9$ & $ 481$ & $4$ & ${}^{+93}_{-59}$ & $1.08\pm 0.03$ & $533{}^{+40}_{-62}$ & $495{}^{+214}_{-136}$ \\
$35.0-45.0$ & $ 40.3$ & $ 383$ & $4$ & ${}^{+75}_{-47}$ & $1.10\pm 0.03$ & $470{}^{+34}_{-54}$ & $443{}^{+190}_{-121}$ \\
$45.0-55.0$ & $ 49.7$ & $ 228$ & $3$ & ${}^{+38}_{-25}$ & $1.02\pm 0.02$ & $274{}^{+21}_{-32}$ & $277{}^{+120}_{-76}$ \\
$55.0-70.0$ & $ 62.1$ & $ 115$ & $1$ & ${}^{+14}_{-10}$ & $0.99\pm 0.02$ & $138{}^{+10}_{-16}$ & $148{}^{+65}_{-41}$ \\
$70.0-100.0$ & $ 82.5$ & $ 28.6$ & $0.5$ & ${}^{+2.3}_{-2.2}$ & $0.99\pm 0.02$ & $35.0{}^{+2.6}_{-4.0}$ & $36.6{}^{+16.5}_{-10.3}$ \\
$100.0-170.0$ & $ 127.6$ & $ 2.46$ & $0.08$ & ${}^{+0.15}_{-0.18}$ & $0.97\pm 0.02$ & $2.77{}^{+0.17}_{-0.30}$ & $3.06{}^{+1.44}_{-0.89}$ \\
$170.0-250.0$ & $ 197.9$ & $ 0.134$ & $0.019$ & ${}^{+0.007}_{-0.010}$ & $0.96\pm 0.02$ & $0.143{}^{+0.005}_{-0.013}$ & $0.144{}^{+0.074}_{-0.045}$ \\
\end{tabular}
\end{ruledtabular}
\end{table}

\begin{table}[htbp]
\caption{\label{tab:results_lepton3pt}The measured differential cross section, normalized to the measured inclusive $W$ boson cross section, as a function of electron transverse momentum for events with three or more jets produced in association with a $W$ boson. The first data uncertainty is statistical, the second is systematic. Also shown are predictions from NLO pQCD after hadronization corrections have been applied (also shown) and \textsc{hej} resummation predictions.}
\begin{ruledtabular}
\begin{tabular}{rc r@{$\pm$\hspace{-3mm}}l@{${}$\hspace{-3mm}}l rll}
\multicolumn{8}{c}{$1/\sigma_W\cdot  d\sigma/dp_{T}^\mathrm{electron}~\textrm{[three-jet bin]}$ (GeV${}^{-1}$)} \\
Bin interval  & Bin center  & \multicolumn{3}{c}{\hspace{-6mm}Measurement} & npQCD correction & NLO \textsc{blackhat+sherpa} & \textsc{hej}\\[-0.5ex]
\multicolumn{1}{c}{$\textrm{(GeV)}$}               &$\textrm{(GeV)}$             & \multicolumn{3}{c}{\hspace{-6mm}$(\times 10^{-6})$} & & $(\times 10^{-6})$ & $(\times 10^{-6})$  \\[1ex]
\hline
$15.0-25.0$ & $ 19.8$ & $ 57$ & $2$ & ${}^{+14}_{-9}$ & $1.12\pm 0.04$ & $49{}^{+6}_{-9}$ & $47{}^{+33}_{-18}$ \\
$25.0-35.0$ & $ 26.9$ & $ 68$ & $2$ & ${}^{+17}_{-11}$ & $1.13\pm 0.05$ & $69{}^{+10}_{-13}$ & $66{}^{+46}_{-24}$ \\
$35.0-50.0$ & $ 42.4$ & $ 48$ & $1$ & ${}^{+11}_{-7}$ & $1.11\pm 0.05$ & $52{}^{+6}_{-9}$ & $52{}^{+35}_{-19}$ \\
$50.0-70.0$ & $ 59.4$ & $ 21.2$ & $0.6$ & ${}^{+4.4}_{-3.0}$ & $1.03\pm 0.03$ & $23.5{}^{+3.4}_{-4.2}$ & $24.7{}^{+16.9}_{-9.2}$ \\
$70.0-110.0$ & $ 86.5$ & $ 4.44$ & $0.19$ & ${}^{+0.75}_{-0.57}$ & $1.01\pm 0.02$ & $5.03{}^{+0.74}_{-0.90}$ & $5.14{}^{+3.47}_{-1.89}$ \\
$110.0-150.0$ & $ 126.9$ & $ 0.589$ & $0.064$ & ${}^{+0.085}_{-0.072}$ & $0.97\pm 0.04$ & $0.589{}^{+0.078}_{-0.100}$ & $0.613{}^{+0.411}_{-0.226}$ \\
$150.0-250.0$ & $ 181.4$ & $ 0.063$ & $0.012$ & ${}^{+0.008}_{-0.007}$ & $0.98\pm 0.04$ & $0.063{}^{+0.008}_{-0.011}$ & $0.065{}^{+0.043}_{-0.024}$ \\
\end{tabular}
\end{ruledtabular}
\end{table}

\begin{table}[htbp]
\caption{\label{tab:results_lepton4pt}The measured differential cross section, normalized to the measured inclusive $W$ boson cross section, as a function of electron transverse momentum for events with four or more jets produced in association with a $W$ boson. The first data uncertainty is statistical, the second is systematic. Also shown are predictions from NLO pQCD after hadronization corrections have been applied (also shown) and \textsc{hej} resummation predictions.}
\begin{ruledtabular}
\begin{tabular}{rc r@{$\pm$\hspace{-3mm}}l@{${}$\hspace{-3mm}}l rll}
\multicolumn{8}{c}{$1/\sigma_W\cdot  d\sigma/dp_{T}^\mathrm{electron}~\textrm{[four-jet bin]}$ (GeV${}^{-1}$)} \\
Bin interval  & Bin center  & \multicolumn{3}{c}{\hspace{-6mm}Measurement} & npQCD correction & NLO \textsc{blackhat+sherpa} & \textsc{hej}\\[-0.5ex]
\multicolumn{1}{c}{$\textrm{(GeV)}$}               &$\textrm{(GeV)}$             & \multicolumn{3}{c}{\hspace{-6mm}$(\times 10^{-6})$} & & $(\times 10^{-6})$ & $(\times 10^{-6})$  \\[1ex]
\hline
$15.0-25.0$ & $ 19.5$ & $ 8.9$ & $0.8$ & ${}^{+2.7}_{-2.00}$ & $1.31\pm 0.08$ & $6.2{}^{+1.4}_{-1.6}$ & $5.0{}^{+4.7}_{-2.2}$ \\
$25.0-35.0$ & $ 26.0$ & $ 9.8$ & $0.7$ & ${}^{+3.0}_{-2.08}$ & $1.18\pm 0.12$ & $7.8{}^{+2.2}_{-2.2}$ & $6.7{}^{+6.1}_{-2.9}$ \\
$35.0-50.0$ & $ 42.2$ & $ 6.2$ & $0.4$ & ${}^{+2.0}_{-1.31}$ & $1.24\pm 0.17$ & $6.2{}^{+1.6}_{-1.7}$ & $5.1{}^{+4.7}_{-2.3}$ \\
$50.0-70.0$ & $ 59.5$ & $ 2.73$ & $0.25$ & ${}^{+0.90}_{-0.61}$ & $1.12\pm 0.04$ & $2.86{}^{+0.72}_{-0.77}$ & $2.46{}^{+2.27}_{-1.09}$ \\
$70.0-130.0$ & $ 93.3$ & $ 0.49$ & $0.07$ & ${}^{+0.15}_{-0.11}$ & $1.06\pm 0.03$ & $0.48{}^{+0.11}_{-0.13}$ & $0.42{}^{+0.38}_{-0.18}$ \\
$130.0-200.0$ & $ 154.2$ & $ 0.032$ & $0.013$ & ${}^{+0.009}_{-0.007}$ & $1.06\pm 0.07$ & $0.029{}^{+0.008}_{-0.008}$ & $0.020{}^{+0.018}_{-0.009}$ \\
\end{tabular}
\end{ruledtabular}
\end{table}

\begin{table}[htbp]
\caption{\label{tab:results_lepton1eta}The measured differential cross section, normalized to the measured inclusive $W$ boson cross section, as a function of electron pseudorapidity for events with one or more jets produced in association with a $W$ boson. The first data uncertainty is statistical, the second is systematic. Also shown are predictions from NLO pQCD after hadronization corrections have been applied (also shown).}
\begin{ruledtabular}
\begin{tabular}{rc r@{$\pm$\hspace{-9mm}}l@{${}$\hspace{-9mm}}l ll}
\multicolumn{7}{c}{$1/\sigma_W\cdot  d\sigma/d\eta^\mathrm{electron}~\textrm{[one-jet bin]}~(1/\Delta y)$} \\
Bin interval  & Bin center  & \multicolumn{3}{c}{\hspace{-10mm}Measurement} & npQCD correction & NLO \textsc{blackhat+sherpa} \\[-0.5ex]
\multicolumn{1}{c}{$\eta^\mathrm{e}$}  & $\eta^\mathrm{e}$  & \multicolumn{3}{c}{\hspace{-6mm}$(\times 10^{-3})$} & & $(\times 10^{-3})$ \\[1ex]
\hline
$-1.1$~--~$-0.8$ & $ -1.0$ & $ 47.8$ & $0.2$ & ${}^{+4.6}_{-3.2}$ & $1.02\pm 0.02$ & $50.3{}^{+2.8}_{-3.2}$ \\
$-0.8$~--~$-0.6$ & $ -0.7$ & $ 51.1$ & $0.2$ & ${}^{+4.9}_{-3.4}$ & $1.01\pm 0.02$ & $54.0{}^{+2.7}_{-3.2}$ \\
$-0.6$~--~$-0.4$ & $ -0.5$ & $ 51.5$ & $0.2$ & ${}^{+4.9}_{-3.4}$ & $1.01\pm 0.02$ & $57.2{}^{+3.0}_{-3.5}$ \\
$-0.4$~--~$-0.2$ & $ -0.3$ & $ 51.5$ & $0.2$ & ${}^{+4.9}_{-3.4}$ & $1.01\pm 0.02$ & $58.3{}^{+2.8}_{-3.5}$ \\
$-0.2$~--~$0.0$ & $ -0.1$ & $ 51.1$ & $0.2$ & ${}^{+4.8}_{-3.3}$ & $1.01\pm 0.02$ & $58.8{}^{+2.9}_{-3.4}$ \\
$0.0$~--~$0.2$ & $ 0.1$ & $ 51.4$ & $0.2$ & ${}^{+4.8}_{-3.5}$ & $1.01\pm 0.02$ & $58.8{}^{+2.9}_{-3.4}$ \\
$0.2$~--~$0.4$ & $ 0.3$ & $ 51.7$ & $0.2$ & ${}^{+4.9}_{-3.4}$ & $1.01\pm 0.02$ & $58.3{}^{+2.8}_{-3.5}$ \\
$0.4$~--~$0.6$ & $ 0.5$ & $ 51.7$ & $0.2$ & ${}^{+4.9}_{-3.4}$ & $1.01\pm 0.02$ & $57.2{}^{+3.0}_{-3.5}$ \\
$0.6$~--~$0.8$ & $ 0.7$ & $ 51.3$ & $0.2$ & ${}^{+4.9}_{-3.5}$ & $1.01\pm 0.02$ & $54.1{}^{+2.7}_{-3.2}$ \\
$0.8$~--~$1.1$ & $ 1.0$ & $ 48.3$ & $0.2$ & ${}^{+4.6}_{-3.3}$ & $1.02\pm 0.02$ & $50.3{}^{+2.8}_{-3.2}$ \\
\end{tabular}
\end{ruledtabular}
\end{table}

\begin{table}[htbp]
\caption{\label{tab:results_lepton2eta}The measured differential cross section, normalized to the measured inclusive $W$ boson cross section, as a function of electron pseudorapidity for events with two or more jets produced in association with a $W$ boson. The first data uncertainty is statistical, the second is systematic. Also shown are predictions from NLO pQCD after hadronization corrections have been applied (also shown) and \textsc{hej} resummation predictions.}
\begin{ruledtabular}
\begin{tabular}{rc r@{$\pm$\hspace{-3mm}}l@{${}$\hspace{-3mm}}l rll}
\multicolumn{8}{c}{$1/\sigma_W\cdot  d\sigma/d\eta^\mathrm{electron}~\textrm{[two-jet bin]}~(1/\Delta y)$} \\
Bin interval  & Bin center  & \multicolumn{3}{c}{\hspace{-6mm}Measurement} & npQCD correction & NLO \textsc{blackhat+sherpa} & \textsc{hej}\\[-0.5ex]
 \multicolumn{1}{c}{$\eta^\mathrm{e}$}  & $\eta^\mathrm{e}$  & \multicolumn{3}{c}{\hspace{-6mm}$(\times 10^{-3})$} & & $(\times 10^{-3})$ & $(\times 10^{-3})$  \\[1ex]
\hline
$-1.1$~--~$-0.8$ & $ -0.9$ & $ 7.4$ & $0.1$ & ${}^{+1.3}_{-0.8}$ & $1.06\pm 0.03$ & $8.1{}^{+0.7}_{-1.0}$ & $7.6{}^{+3.4}_{-2.1}$ \\
$-0.8$~--~$-0.6$ & $ -0.7$ & $ 7.9$ & $0.1$ & ${}^{+1.3}_{-0.9}$ & $1.06\pm 0.03$ & $8.7{}^{+0.6}_{-1.0}$ & $8.4{}^{+3.7}_{-2.3}$ \\
$-0.6$~--~$-0.4$ & $ -0.5$ & $ 7.9$ & $0.1$ & ${}^{+1.3}_{-0.9}$ & $1.06\pm 0.03$ & $9.2{}^{+0.7}_{-1.1}$ & $8.9{}^{+3.9}_{-2.5}$ \\
$-0.4$~--~$-0.2$ & $ -0.3$ & $ 7.9$ & $0.1$ & ${}^{+1.3}_{-0.9}$ & $1.05\pm 0.02$ & $9.3{}^{+0.6}_{-1.0}$ & $9.2{}^{+4.0}_{-2.5}$ \\
$-0.2$~--~$0.0$ & $ -0.1$ & $ 7.8$ & $0.1$ & ${}^{+1.3}_{-0.9}$ & $1.05\pm 0.03$ & $9.6{}^{+0.7}_{-1.1}$ & $9.3{}^{+4.0}_{-2.5}$ \\
$0.0$~--~$0.2$ & $ 0.1$ & $ 7.9$ & $0.1$ & ${}^{+1.3}_{-0.9}$ & $1.05\pm 0.03$ & $9.6{}^{+0.7}_{-1.1}$ & $9.2{}^{+4.0}_{-2.5}$ \\
$0.2$~--~$0.4$ & $ 0.3$ & $ 8.0$ & $0.1$ & ${}^{+1.3}_{-0.9}$ & $1.05\pm 0.02$ & $9.3{}^{+0.6}_{-1.0}$ & $9.3{}^{+4.0}_{-2.5}$ \\
$0.4$~--~$0.6$ & $ 0.5$ & $ 8.0$ & $0.1$ & ${}^{+1.3}_{-0.9}$ & $1.06\pm 0.03$ & $9.2{}^{+0.7}_{-1.1}$ & $9.0{}^{+3.9}_{-2.5}$ \\
$0.6$~--~$0.8$ & $ 0.7$ & $ 8.0$ & $0.1$ & ${}^{+1.3}_{-0.9}$ & $1.06\pm 0.03$ & $8.7{}^{+0.6}_{-1.0}$ & $8.6{}^{+3.8}_{-2.4}$ \\
$0.8$~--~$1.1$ & $ 0.9$ & $ 7.5$ & $0.1$ & ${}^{+1.2}_{-0.8}$ & $1.06\pm 0.03$ & $8.1{}^{+0.7}_{-1.0}$ & $7.7{}^{+3.4}_{-2.1}$ \\
\end{tabular}
\end{ruledtabular}
\end{table}

\begin{table}[htbp]
\caption{\label{tab:results_lepton3eta}The measured differential cross section, normalized to the measured inclusive $W$ boson cross section, as a function of electron pseudorapidity for events with three or more jets produced in association with a $W$ boson. The first data uncertainty is statistical, the second is systematic. Also shown are predictions from NLO pQCD after hadronization corrections have been applied (also shown) and \textsc{hej} resummation predictions.}
\begin{ruledtabular}
\begin{tabular}{rc r@{$\pm$\hspace{-3mm}}l@{${}$\hspace{-3mm}}l rll}
\multicolumn{8}{c}{$1/\sigma_W\cdot  d\sigma/d\eta^\mathrm{electron}~\textrm{[three-jet bin]}~(1/\Delta y)$} \\
Bin interval  & Bin center  & \multicolumn{3}{c}{\hspace{-6mm}Measurement} & npQCD correction & NLO \textsc{blackhat+sherpa} & \textsc{hej}\\[-0.5ex]
 \multicolumn{1}{c}{$\eta^\mathrm{e}$}  & $\eta^\mathrm{e}$  & \multicolumn{3}{c}{\hspace{-6mm}$(\times 10^{-3})$} & & $(\times 10^{-3})$ & $(\times 10^{-3})$  \\[1ex]
\hline
$-1.1$~--~$-0.8$ & $ -0.9$ & $ 1.06$ & $0.04$ & ${}^{+0.24}_{-0.16}$ & $1.09\pm 0.04$ & $1.08{}^{+0.12}_{-0.19}$ & $1.06{}^{+0.74}_{-0.40}$ \\
$-0.8$~--~$-0.6$ & $ -0.7$ & $ 1.16$ & $0.04$ & ${}^{+0.26}_{-0.17}$ & $1.12\pm 0.05$ & $1.21{}^{+0.15}_{-0.21}$ & $1.19{}^{+0.81}_{-0.44}$ \\
$-0.6$~--~$-0.4$ & $ -0.5$ & $ 1.18$ & $0.04$ & ${}^{+0.26}_{-0.18}$ & $1.07\pm 0.04$ & $1.23{}^{+0.17}_{-0.22}$ & $1.21{}^{+0.84}_{-0.45}$ \\
$-0.4$~--~$-0.2$ & $ -0.3$ & $ 1.19$ & $0.04$ & ${}^{+0.27}_{-0.18}$ & $1.10\pm 0.04$ & $1.29{}^{+0.17}_{-0.23}$ & $1.25{}^{+0.86}_{-0.46}$ \\
$-0.2$~--~$0.0$ & $ -0.1$ & $ 1.19$ & $0.04$ & ${}^{+0.27}_{-0.18}$ & $1.09\pm 0.05$ & $1.30{}^{+0.18}_{-0.23}$ & $1.28{}^{+0.88}_{-0.47}$ \\
$0.0$~--~$0.2$ & $ 0.1$ & $ 1.20$ & $0.04$ & ${}^{+0.27}_{-0.18}$ & $1.09\pm 0.05$ & $1.30{}^{+0.18}_{-0.23}$ & $1.27{}^{+0.87}_{-0.47}$ \\
$0.2$~--~$0.4$ & $ 0.3$ & $ 1.22$ & $0.04$ & ${}^{+0.28}_{-0.18}$ & $1.10\pm 0.04$ & $1.29{}^{+0.17}_{-0.23}$ & $1.27{}^{+0.85}_{-0.46}$ \\
$0.4$~--~$0.6$ & $ 0.5$ & $ 1.20$ & $0.04$ & ${}^{+0.27}_{-0.18}$ & $1.07\pm 0.04$ & $1.23{}^{+0.17}_{-0.22}$ & $1.24{}^{+0.85}_{-0.46}$ \\
$0.6$~--~$0.8$ & $ 0.7$ & $ 1.16$ & $0.04$ & ${}^{+0.26}_{-0.17}$ & $1.12\pm 0.05$ & $1.21{}^{+0.15}_{-0.21}$ & $1.23{}^{+0.84}_{-0.46}$ \\
$0.8$~--~$1.1$ & $ 0.9$ & $ 1.07$ & $0.04$ & ${}^{+0.23}_{-0.16}$ & $1.09\pm 0.04$ & $1.08{}^{+0.12}_{-0.19}$ & $1.08{}^{+0.76}_{-0.41}$ \\
\end{tabular}
\end{ruledtabular}
\end{table}

\begin{table}[htbp]
\caption{\label{tab:results_lepton4eta}The measured differential cross section, normalized to the measured inclusive $W$ boson cross section, as a function of electron pseudorapidity for events with four or more jets produced in association with a $W$ boson. The first data uncertainty is statistical, the second is systematic. Also shown are predictions from NLO pQCD after hadronization corrections have been applied (also shown) and \textsc{hej} resummation predictions.}
\begin{ruledtabular}
\begin{tabular}{rc r@{$\pm$\hspace{-3mm}}l@{${}$\hspace{-3mm}}l rll}
\multicolumn{8}{c}{$1/\sigma_W\cdot  d\sigma/d\eta^\mathrm{electron}~\textrm{[four-jet bin]}~(1/\Delta y)$} \\
Bin interval  & Bin center  & \multicolumn{3}{c}{\hspace{-6mm}Measurement} & npQCD correction & NLO \textsc{blackhat+sherpa} & \textsc{hej}\\[-0.5ex]
 \multicolumn{1}{c}{$\eta^\mathrm{e}$}  & $\eta^\mathrm{e}$  & \multicolumn{3}{c}{\hspace{-6mm}$(\times 10^{-3})$} & & $(\times 10^{-3})$ & $(\times 10^{-3})$  \\[1ex]
\hline
$-1.1$~--~$-0.6$ & $ -0.8$ & $ 0.15$ & $0.01$ & ${}^{+0.05}_{-0.03}$ & $1.21\pm 0.14$ & $0.14{}^{+0.04}_{-0.04}$ & $0.12{}^{+0.11}_{-0.05}$ \\
$-0.6$~--~$-0.2$ & $ -0.4$ & $ 0.17$ & $0.01$ & ${}^{+0.05}_{-0.04}$ & $1.19\pm 0.08$ & $0.15{}^{+0.04}_{-0.04}$ & $0.13{}^{+0.12}_{-0.06}$ \\
$-0.2$~--~$0.2$ & $ 0.0$ & $ 0.17$ & $0.01$ & ${}^{+0.05}_{-0.04}$ & $1.16\pm 0.05$ & $0.15{}^{+0.04}_{-0.04}$ & $0.13{}^{+0.12}_{-0.06}$ \\
$0.2$~--~$0.6$ & $ 0.4$ & $ 0.17$ & $0.01$ & ${}^{+0.05}_{-0.04}$ & $1.19\pm 0.08$ & $0.15{}^{+0.04}_{-0.04}$ & $0.13{}^{+0.12}_{-0.06}$ \\
$0.6$~--~$1.1$ & $ 0.8$ & $ 0.14$ & $0.01$ & ${}^{+0.05}_{-0.03}$ & $1.21\pm 0.14$ & $0.14{}^{+0.04}_{-0.04}$ & $0.11{}^{+0.11}_{-0.05}$ \\
\end{tabular}
\end{ruledtabular}
\end{table}

\begin{table}[htbp]
\caption{\label{tab:results_dy12_2j}The measured differential cross section, normalized to the measured inclusive $W$ boson cross section, as a function of dijet rapidity separation of the leading two jets for events with two or more jets produced in association with a $W$ boson. The first data uncertainty is statistical, the second is systematic. Also shown are predictions from NLO pQCD after hadronization corrections have been applied (also shown) and \textsc{hej} resummation predictions.}
\begin{ruledtabular}
\begin{tabular}{rc r@{$\pm$\hspace{-3mm}}l@{${}$\hspace{-3mm}}l rll}
\multicolumn{8}{c}{$1/\sigma_W\cdot  d\sigma/d\Delta y_\mathrm{12}~\textrm{[two-jet bin]}~(1/\Delta y)$} \\
Bin interval  & Bin center  & \multicolumn{3}{c}{\hspace{-6mm}Measurement} & npQCD correction & NLO \textsc{blackhat+sherpa} & \textsc{hej}\\[-0.5ex]
\multicolumn{1}{c}{$\Delta y$}  & $\Delta y$ & \multicolumn{3}{c}{\hspace{-6mm}$(\times 10^{-3})$} & & $(\times 10^{-3})$ & $(\times 10^{-3})$  \\[1ex]
\hline
$0.0-0.4$ & $ 0.2$ & $ 7.47$ & $0.08$ & ${}^{+0.96}_{-0.70}$ & $1.05\pm 0.04$ & $8.13{}^{+0.63}_{-0.91}$ & $9.14{}^{+3.95}_{-2.51}$ \\
$0.4-0.8$ & $ 0.6$ & $ 7.94$ & $0.08$ & ${}^{+1.06}_{-0.76}$ & $1.06\pm 0.03$ & $8.45{}^{+0.68}_{-0.97}$ & $8.71{}^{+3.71}_{-2.37}$ \\
$0.8-1.2$ & $ 1.0$ & $ 7.46$ & $0.07$ & ${}^{+1.07}_{-0.75}$ & $1.03\pm 0.03$ & $8.25{}^{+0.69}_{-0.97}$ & $8.26{}^{+3.51}_{-2.24}$ \\
$1.2-1.6$ & $ 1.4$ & $ 6.08$ & $0.06$ & ${}^{+0.99}_{-0.67}$ & $1.02\pm 0.03$ & $6.69{}^{+0.49}_{-0.75}$ & $6.58{}^{+2.82}_{-1.79}$ \\
$1.6-2.1$ & $ 1.9$ & $ 4.67$ & $0.05$ & ${}^{+0.90}_{-0.57}$ & $1.04\pm 0.03$ & $5.21{}^{+0.34}_{-0.59}$ & $4.80{}^{+2.09}_{-1.32}$ \\
$2.1-2.6$ & $ 2.3$ & $ 3.10$ & $0.04$ & ${}^{+0.70}_{-0.43}$ & $1.07\pm 0.03$ & $3.79{}^{+0.32}_{-0.47}$ & $3.17{}^{+1.41}_{-0.88}$ \\
$2.6-3.2$ & $ 2.9$ & $ 1.67$ & $0.03$ & ${}^{+0.43}_{-0.25}$ & $1.10\pm 0.04$ & $2.23{}^{+0.10}_{-0.25}$ & $1.82{}^{+0.84}_{-0.52}$ \\
$3.2-4.0$ & $ 3.6$ & $ 0.60$ & $0.01$ & ${}^{+0.17}_{-0.09}$ & $1.15\pm 0.04$ & $1.01{}^{+0.05}_{-0.12}$ & $0.77{}^{+0.38}_{-0.23}$ \\
$4.0-5.0$ & $ 4.4$ & $ 0.11$ & $0.00$ & ${}^{+0.03}_{-0.02}$ & $1.31\pm 0.07$ & $0.26{}^{+0.01}_{-0.03}$ & $0.19{}^{+0.11}_{-0.06}$ \\
$5.0-6.0$ & $ 5.4$ & $ 0.007$ & $0.001$ & ${}^{+0.002}_{-0.001}$ & $1.69\pm 0.06$ & $0.028{}^{+0.001}_{-0.003}$ & $0.019{}^{+0.012}_{-0.007}$ \\
\end{tabular}
\end{ruledtabular}
\end{table}

\begin{table}[htbp]
\caption{\label{tab:results_dyFB_2j}The measured differential cross section, normalized to the measured inclusive $W$ boson cross section, as a function of dijet rapidity separation of the two most rapidity-separated jets for events with two or more jets produced in association with a $W$ boson. The first data uncertainty is statistical, the second is systematic. Also shown are predictions from NLO pQCD after hadronization corrections have been applied (also shown) and \textsc{hej} resummation predictions.}
\begin{ruledtabular}
\begin{tabular}{rc r@{$\pm$\hspace{-3mm}}l@{${}$\hspace{-3mm}}l rll}
\multicolumn{8}{c}{$1/\sigma_W\cdot  d\sigma/d\Delta y_\mathrm{FB}~\textrm{[two-jet bin]}~(1/\Delta y)$} \\
Bin interval  & Bin center  & \multicolumn{3}{c}{\hspace{-6mm}Measurement} & npQCD correction & NLO \textsc{blackhat+sherpa} & \textsc{hej}\\[-0.5ex]
\multicolumn{1}{c}{$\Delta y$}  & $\Delta y$ & \multicolumn{3}{c}{\hspace{-6mm}$(\times 10^{-3})$} & & $(\times 10^{-3})$ & $(\times 10^{-3})$  \\[1ex]
\hline
$0.0-0.4$ & $ 0.2$ & $ 6.45$ & $0.08$ & ${}^{+0.79}_{-0.59}$ & $1.04\pm 0.04$ & $7.45{}^{+0.27}_{-0.70}$ & $8.12{}^{+3.27}_{-2.14}$ \\
$0.4-0.8$ & $ 0.6$ & $ 7.34$ & $0.08$ & ${}^{+0.97}_{-0.67}$ & $1.06\pm 0.03$ & $7.97{}^{+0.42}_{-0.81}$ & $7.97{}^{+3.20}_{-2.09}$ \\
$0.8-1.2$ & $ 1.0$ & $ 7.14$ & $0.08$ & ${}^{+0.96}_{-0.68}$ & $1.03\pm 0.03$ & $8.03{}^{+0.57}_{-0.90}$ & $7.88{}^{+3.23}_{-2.09}$ \\
$1.2-1.6$ & $ 1.4$ & $ 6.04$ & $0.07$ & ${}^{+0.88}_{-0.62}$ & $1.02\pm 0.02$ & $6.70{}^{+0.48}_{-0.75}$ & $6.59{}^{+2.79}_{-1.78}$ \\
$1.6-2.1$ & $ 1.9$ & $ 4.91$ & $0.06$ & ${}^{+0.89}_{-0.58}$ & $1.05\pm 0.02$ & $5.38{}^{+0.43}_{-0.64}$ & $5.05{}^{+2.23}_{-1.40}$ \\
$2.1-2.6$ & $ 2.4$ & $ 3.51$ & $0.05$ & ${}^{+0.84}_{-0.52}$ & $1.06\pm 0.02$ & $4.03{}^{+0.46}_{-0.55}$ & $3.56{}^{+1.66}_{-1.02}$ \\
$2.6-3.2$ & $ 2.9$ & $ 2.18$ & $0.04$ & ${}^{+0.69}_{-0.39}$ & $1.10\pm 0.03$ & $2.46{}^{+0.24}_{-0.33}$ & $2.19{}^{+1.10}_{-0.66}$ \\
$3.2-4.0$ & $ 3.6$ & $ 0.92$ & $0.02$ & ${}^{+0.35}_{-0.18}$ & $1.14\pm 0.03$ & $1.16{}^{+0.15}_{-0.18}$ & $1.02{}^{+0.57}_{-0.33}$ \\
$4.0-5.0$ & $ 4.4$ & $ 0.19$ & $0.01$ & ${}^{+0.08}_{-0.04}$ & $1.28\pm 0.02$ & $0.31{}^{+0.04}_{-0.05}$ & $0.29{}^{+0.20}_{-0.10}$ \\
$5.0-6.0$ & $ 5.4$ & $ 0.013$ & $0.002$ & ${}^{+0.006}_{-0.003}$ & $1.69\pm 0.11$ & $0.036{}^{+0.007}_{-0.007}$ & $0.033{}^{+0.027}_{-0.013}$ \\
\end{tabular}
\end{ruledtabular}
\end{table}

\begin{table}[htbp]
\caption{\label{tab:results_dy12_3j}The measured differential cross section, normalized to the measured inclusive $W$ boson cross section, as a function of dijet rapidity separation of the leading two jets for events with three or more jets produced in association with a $W$ boson. The first data uncertainty is statistical, the second is systematic. Also shown are predictions from NLO pQCD after hadronization corrections have been applied (also shown) and \textsc{hej} resummation predictions.}
\begin{ruledtabular}
\begin{tabular}{rc r@{$\pm$\hspace{-3mm}}l@{${}$\hspace{-3mm}}l rll}
\multicolumn{8}{c}{$1/\sigma_W\cdot  d\sigma/d\Delta y_\mathrm{12}~\textrm{[three-jet bin]}~(1/\Delta y)$} \\
Bin interval  & Bin center  & \multicolumn{3}{c}{\hspace{-6mm}Measurement} & npQCD correction & NLO \textsc{blackhat+sherpa} & \textsc{hej}\\[-0.5ex]
\multicolumn{1}{c}{$\Delta y$}  & $\Delta y$ & \multicolumn{3}{c}{\hspace{-6mm}$(\times 10^{-3})$} & & $(\times 10^{-3})$ & $(\times 10^{-3})$  \\[1ex]
\hline
$0.0-0.4$ & $ 0.2$ & $ 1.13$ & $0.04$ & ${}^{+0.23}_{-0.16}$ & $1.09\pm 0.05$ & $1.13{}^{+0.17}_{-0.20}$ & $1.14{}^{+0.76}_{-0.42}$ \\
$0.4-0.8$ & $ 0.6$ & $ 1.14$ & $0.04$ & ${}^{+0.23}_{-0.16}$ & $1.09\pm 0.03$ & $1.16{}^{+0.17}_{-0.21}$ & $1.15{}^{+0.76}_{-0.42}$ \\
$0.8-1.2$ & $ 1.0$ & $ 1.03$ & $0.04$ & ${}^{+0.22}_{-0.15}$ & $1.09\pm 0.05$ & $1.11{}^{+0.15}_{-0.19}$ & $1.12{}^{+0.75}_{-0.41}$ \\
$1.2-1.6$ & $ 1.4$ & $ 0.94$ & $0.03$ & ${}^{+0.20}_{-0.14}$ & $1.04\pm 0.03$ & $0.90{}^{+0.13}_{-0.16}$ & $0.91{}^{+0.62}_{-0.34}$ \\
$1.6-2.1$ & $ 1.9$ & $ 0.73$ & $0.02$ & ${}^{+0.17}_{-0.11}$ & $1.07\pm 0.06$ & $0.66{}^{+0.03}_{-0.10}$ & $0.71{}^{+0.49}_{-0.26}$ \\
$2.1-2.6$ & $ 2.4$ & $ 0.48$ & $0.02$ & ${}^{+0.12}_{-0.07}$ & $1.08\pm 0.03$ & $0.52{}^{+0.08}_{-0.10}$ & $0.48{}^{+0.34}_{-0.18}$ \\
$2.6-3.2$ & $ 2.9$ & $ 0.26$ & $0.02$ & ${}^{+0.06}_{-0.04}$ & $1.15\pm 0.07$ & $0.33{}^{+0.04}_{-0.06}$ & $0.28{}^{+0.21}_{-0.11}$ \\
$3.2-4.0$ & $ 3.5$ & $ 0.10$ & $0.01$ & ${}^{+0.02}_{-0.01}$ & $1.17\pm 0.04$ & $0.14{}^{+0.02}_{-0.03}$ & $0.14{}^{+0.10}_{-0.05}$ \\
$4.0-5.0$ & $ 4.4$ & $ 0.018$ & $0.003$ & ${}^{+0.004}_{-0.003}$ & $1.30\pm 0.02$ & $0.040{}^{+0.003}_{-0.008}$ & $0.034{}^{+0.028}_{-0.014}$ \\
\end{tabular}
\end{ruledtabular}
\end{table}

\begin{table}[htbp]
\caption{\label{tab:results_dyFB_3j}The measured differential cross section, normalized to the measured inclusive $W$ boson cross section, as a function of dijet rapidity separation of the two most rapidity-separated jets for events with three or more jets produced in association with a $W$ boson. The first data uncertainty is statistical, the second is systematic. Also shown are predictions from NLO pQCD after hadronization corrections have been applied (also shown) and \textsc{hej} resummation predictions.}
\begin{ruledtabular}
\begin{tabular}{rc r@{$\pm$\hspace{-3mm}}l@{${}$\hspace{-3mm}}l rll}
\multicolumn{8}{c}{$1/\sigma_W\cdot  d\sigma/d\Delta y_\mathrm{FB}~\textrm{[three-jet bin]}~(1/\Delta y)$} \\
Bin interval  & Bin center  & \multicolumn{3}{c}{\hspace{-6mm}Measurement} & npQCD correction & NLO \textsc{blackhat+sherpa} & \textsc{hej}\\[-0.5ex]
\multicolumn{1}{c}{$\Delta y$}  & $\Delta y$ & \multicolumn{3}{c}{\hspace{-6mm}$(\times 10^{-3})$} & & $(\times 10^{-3})$ & $(\times 10^{-3})$  \\[1ex]
\hline
$0.0-0.4$ & $ 0.2$ & $ 0.13$ & $0.01$ & ${}^{+0.02}_{-0.02}$ & $1.04\pm 0.03$ & $0.11{}^{+0.01}_{-0.02}$ & $0.12{}^{+0.08}_{-0.04}$ \\
$0.4-0.8$ & $ 0.6$ & $ 0.40$ & $0.02$ & ${}^{+0.07}_{-0.05}$ & $1.09\pm 0.04$ & $0.43{}^{+0.07}_{-0.08}$ & $0.40{}^{+0.25}_{-0.14}$ \\
$0.8-1.2$ & $ 1.0$ & $ 0.71$ & $0.03$ & ${}^{+0.12}_{-0.09}$ & $1.10\pm 0.07$ & $0.76{}^{+0.09}_{-0.13}$ & $0.74{}^{+0.47}_{-0.26}$ \\
$1.2-1.6$ & $ 1.4$ & $ 0.96$ & $0.03$ & ${}^{+0.17}_{-0.12}$ & $1.05\pm 0.01$ & $0.92{}^{+0.13}_{-0.16}$ & $0.91{}^{+0.59}_{-0.33}$ \\
$1.6-2.1$ & $ 1.9$ & $ 1.01$ & $0.03$ & ${}^{+0.20}_{-0.14}$ & $1.08\pm 0.05$ & $0.94{}^{+0.11}_{-0.16}$ & $0.96{}^{+0.63}_{-0.35}$ \\
$2.1-2.6$ & $ 2.4$ & $ 0.87$ & $0.03$ & ${}^{+0.22}_{-0.14}$ & $1.07\pm 0.05$ & $0.87{}^{+0.12}_{-0.15}$ & $0.87{}^{+0.59}_{-0.32}$ \\
$2.6-3.2$ & $ 2.9$ & $ 0.67$ & $0.02$ & ${}^{+0.22}_{-0.13}$ & $1.11\pm 0.05$ & $0.70{}^{+0.11}_{-0.13}$ & $0.65{}^{+0.47}_{-0.25}$ \\
$3.2-4.0$ & $ 3.6$ & $ 0.36$ & $0.02$ & ${}^{+0.14}_{-0.08}$ & $1.12\pm 0.05$ & $0.37{}^{+0.03}_{-0.06}$ & $0.38{}^{+0.29}_{-0.15}$ \\
$4.0-5.0$ & $ 4.4$ & $ 0.09$ & $0.01$ & ${}^{+0.04}_{-0.02}$ & $1.20\pm 0.04$ & $0.13{}^{+0.02}_{-0.03}$ & $0.13{}^{+0.11}_{-0.05}$ \\
$5.0-6.0$ & $ 5.4$ & $ 0.006$ & $0.001$ & ${}^{+0.004}_{-0.002}$ & $1.64\pm 0.39$ & $0.025{}^{+0.005}_{-0.007}$ & $0.018{}^{+0.019}_{-0.008}$ \\
\end{tabular}
\end{ruledtabular}
\end{table}

\begin{table}[htbp]
\caption{\label{tab:results_dphi}
  The measured differential cross sections, normalized to the measured inclusive $W$ boson cross section, as a function of $\Delta\phi$ separation between the leading two jets and between the two most rapidity-separated jets for events with two or more jets produced in association with a $W$ boson. The first data uncertainty is statistical, the second is systematic. Also shown are predictions from NLO pQCD after hadronization corrections have been applied (also shown) and \textsc{hej} resummation predictions.}
\begin{ruledtabular}
\begin{tabular}{rc r@{$\pm$\hspace{-3mm}}l@{${}$\hspace{-3mm}}l rll}
\multicolumn{8}{c}{$1/\sigma_W\cdot  d\sigma/d\Delta\phi_\mathrm{12}~\textrm{[two-jet bin]}~(1/\Delta\phi)$} \\
Bin interval  & Bin center  & \multicolumn{3}{c}{\hspace{-6mm}Measurement} & npQCD correction & NLO \textsc{blackhat+sherpa} & \textsc{hej}\\[-0.5ex]
\multicolumn{1}{c}{$\Delta\phi$}  & $\Delta\phi$  & \multicolumn{3}{c}{\hspace{-6mm}$(\times 10^{-3})$} & & $(\times 10^{-3})$ & $(\times 10^{-3})$  \\[1ex]
\hline
$0.0-0.3$ & $ 0.2$ & $ 2.07$ & $0.04$ & ${}^{+0.47}_{-0.28}$ & $1.08\pm 0.03$ & $2.36{}^{+0.24}_{-0.30}$ & $2.24{}^{+1.01}_{-0.63}$ \\
$0.3-0.6$ & $ 0.5$ & $ 2.51$ & $0.05$ & ${}^{+0.56}_{-0.34}$ & $1.08\pm 0.03$ & $2.81{}^{+0.34}_{-0.38}$ & $2.79{}^{+1.24}_{-0.78}$ \\
$0.6-0.9$ & $ 0.7$ & $ 3.63$ & $0.06$ & ${}^{+0.78}_{-0.48}$ & $1.09\pm 0.02$ & $3.53{}^{+0.29}_{-0.41}$ & $3.51{}^{+1.55}_{-0.97}$ \\
$0.9-1.1$ & $ 1.0$ & $ 3.64$ & $0.07$ & ${}^{+0.76}_{-0.48}$ & $1.01\pm 0.03$ & $4.02{}^{+0.37}_{-0.48}$ & $4.46{}^{+1.95}_{-1.23}$ \\
$1.1-1.4$ & $ 1.2$ & $ 3.47$ & $0.06$ & ${}^{+0.69}_{-0.44}$ & $0.99\pm 0.03$ & $3.93{}^{+0.32}_{-0.46}$ & $4.38{}^{+1.94}_{-1.22}$ \\
$1.4-1.6$ & $ 1.5$ & $ 3.65$ & $0.06$ & ${}^{+0.69}_{-0.45}$ & $0.99\pm 0.03$ & $4.17{}^{+0.27}_{-0.46}$ & $4.62{}^{+2.05}_{-1.29}$ \\
$1.6-1.8$ & $ 1.7$ & $ 4.14$ & $0.07$ & ${}^{+0.74}_{-0.48}$ & $0.99\pm 0.02$ & $4.82{}^{+0.35}_{-0.56}$ & $5.11{}^{+2.29}_{-1.43}$ \\
$1.8-2.0$ & $ 1.9$ & $ 4.82$ & $0.08$ & ${}^{+0.80}_{-0.54}$ & $1.00\pm 0.03$ & $5.60{}^{+0.41}_{-0.63}$ & $5.92{}^{+2.59}_{-1.64}$ \\
$2.0-2.2$ & $ 2.1$ & $ 5.78$ & $0.09$ & ${}^{+0.89}_{-0.60}$ & $1.02\pm 0.04$ & $7.05{}^{+0.63}_{-0.86}$ & $6.74{}^{+2.94}_{-1.86}$ \\
$2.2-2.4$ & $ 2.3$ & $ 6.92$ & $0.09$ & ${}^{+0.98}_{-0.69}$ & $1.02\pm 0.01$ & $7.83{}^{+0.48}_{-0.85}$ & $7.92{}^{+3.45}_{-2.18}$ \\
$2.4-2.6$ & $ 2.5$ & $ 8.42$ & $0.10$ & ${}^{+1.13}_{-0.80}$ & $1.04\pm 0.02$ & $9.66{}^{+0.64}_{-1.07}$ & $9.27{}^{+4.00}_{-2.54}$ \\
$2.6-2.8$ & $ 2.7$ & $ 9.99$ & $0.12$ & ${}^{+1.29}_{-0.93}$ & $1.07\pm 0.02$ & $11.82{}^{+0.91}_{-1.39}$ & $10.61{}^{+4.57}_{-2.90}$ \\
$2.8-3.0$ & $ 2.9$ & $ 11.35$ & $0.14$ & ${}^{+1.44}_{-1.05}$ & $1.11\pm 0.02$ & $13.26{}^{+0.74}_{-1.44}$ & $11.80{}^{+5.08}_{-3.22}$ \\
$3.0-3.2$ & $ 3.1$ & $ 9.69$ & $0.12$ & ${}^{+1.23}_{-0.90}$ & $1.19\pm 0.05$ & $10.53{}^{+0.63}_{-1.17}$ & $8.78{}^{+3.77}_{-2.39}$ \\
%\end{tabular}
%\end{ruledtabular}
\hline\hline
%\begin{ruledtabular}
%\begin{tabular}{rc r@{$\pm$\hspace{-3mm}}l@{${}$\hspace{-3mm}}l rll}
\multicolumn{8}{c}{$1/\sigma_W\cdot  d\sigma/d\Delta\phi_\mathrm{FB}~\textrm{[two-jet bin]}~(1/\Delta\phi)$} \\
Bin interval  & Bin center  & \multicolumn{3}{c}{\hspace{-6mm}Measurement} & npQCD correction & NLO \textsc{blackhat+sherpa} & \textsc{hej}\\[-0.5ex]
\multicolumn{1}{c}{$\Delta\phi$}  & $\Delta\phi$  & \multicolumn{3}{c}{\hspace{-6mm}$(\times 10^{-3})$} & & $(\times 10^{-3})$ & $(\times 10^{-3})$  \\[1ex]
\hline
$0.0-0.3$ & $ 0.2$ & $ 2.35$ & $0.05$ & ${}^{+0.54}_{-0.32}$ & $1.07\pm 0.03$ & $2.54{}^{+0.35}_{-0.36}$ & $2.53{}^{+1.21}_{-0.74}$ \\
$0.3-0.6$ & $ 0.5$ & $ 2.73$ & $0.05$ & ${}^{+0.61}_{-0.37}$ & $1.07\pm 0.03$ & $2.97{}^{+0.43}_{-0.44}$ & $3.05{}^{+1.43}_{-0.88}$ \\
$0.6-0.9$ & $ 0.7$ & $ 3.67$ & $0.07$ & ${}^{+0.77}_{-0.48}$ & $1.08\pm 0.03$ & $3.64{}^{+0.37}_{-0.46}$ & $3.75{}^{+1.72}_{-1.06}$ \\
$0.9-1.1$ & $ 1.0$ & $ 3.64$ & $0.07$ & ${}^{+0.74}_{-0.47}$ & $1.00\pm 0.03$ & $4.09{}^{+0.41}_{-0.50}$ & $4.61{}^{+2.06}_{-1.29}$ \\
$1.1-1.4$ & $ 1.2$ & $ 3.53$ & $0.06$ & ${}^{+0.70}_{-0.45}$ & $0.99\pm 0.03$ & $4.03{}^{+0.38}_{-0.50}$ & $4.56{}^{+2.06}_{-1.28}$ \\
$1.4-1.6$ & $ 1.5$ & $ 3.77$ & $0.07$ & ${}^{+0.72}_{-0.46}$ & $1.00\pm 0.03$ & $4.28{}^{+0.31}_{-0.49}$ & $4.76{}^{+2.14}_{-1.34}$ \\
$1.6-1.8$ & $ 1.7$ & $ 4.26$ & $0.08$ & ${}^{+0.78}_{-0.51}$ & $1.00\pm 0.02$ & $4.90{}^{+0.39}_{-0.58}$ & $5.18{}^{+2.35}_{-1.46}$ \\
$1.8-2.0$ & $ 1.9$ & $ 4.89$ & $0.09$ & ${}^{+0.83}_{-0.55}$ & $1.00\pm 0.03$ & $5.65{}^{+0.43}_{-0.65}$ & $5.94{}^{+2.62}_{-1.65}$ \\
$2.0-2.2$ & $ 2.1$ & $ 5.79$ & $0.10$ & ${}^{+0.88}_{-0.61}$ & $1.02\pm 0.04$ & $7.03{}^{+0.62}_{-0.85}$ & $6.71{}^{+2.92}_{-1.85}$ \\
$2.2-2.4$ & $ 2.3$ & $ 6.83$ & $0.11$ & ${}^{+0.95}_{-0.67}$ & $1.01\pm 0.01$ & $7.73{}^{+0.43}_{-0.82}$ & $7.79{}^{+3.35}_{-2.13}$ \\
$2.4-2.6$ & $ 2.5$ & $ 8.24$ & $0.12$ & ${}^{+1.06}_{-0.77}$ & $1.04\pm 0.02$ & $9.45{}^{+0.53}_{-1.01}$ & $9.01{}^{+3.81}_{-2.44}$ \\
$2.6-2.8$ & $ 2.7$ & $ 9.80$ & $0.13$ & ${}^{+1.22}_{-0.88}$ & $1.07\pm 0.02$ & $11.55{}^{+0.76}_{-1.30}$ & $10.19{}^{+4.28}_{-2.74}$ \\
$2.8-3.0$ & $ 2.9$ & $ 11.03$ & $0.14$ & ${}^{+1.40}_{-1.01}$ & $1.12\pm 0.02$ & $13.01{}^{+0.54}_{-1.34}$ & $11.28{}^{+4.71}_{-3.02}$ \\
$3.0-3.2$ & $ 3.1$ & $ 9.24$ & $0.14$ & ${}^{+1.20}_{-0.89}$ & $1.20\pm 0.05$ & $10.32{}^{+0.45}_{-1.07}$ & $8.34{}^{+3.45}_{-2.22}$ \\
\end{tabular}
\end{ruledtabular}
\end{table}

\begin{table}[htbp]
\caption{\label{tab:results_dR_2j}The measured differential cross section, normalized to the measured inclusive $W$ boson cross section, as a function of $\Delta R$ separation between the two leading jets for events with two or more jets produced in association with a $W$ boson. The first data uncertainty is statistical, the second is systematic. Also shown are predictions from NLO pQCD after hadronization corrections have been applied (also shown) and \textsc{hej} resummation predictions.}
\begin{ruledtabular}
\begin{tabular}{rc r@{$\pm$\hspace{-3mm}}l@{${}$\hspace{-3mm}}l rll}
\multicolumn{8}{c}{$1/\sigma_W\cdot  d\sigma/d\Delta R_\mathrm{jj}~\textrm{[two-jet bin]}~(1/\Delta R)$} \\
Bin interval  & Bin center  & \multicolumn{3}{c}{\hspace{-6mm}Measurement} & npQCD correction & NLO \textsc{blackhat+sherpa} & \textsc{hej}\\[-0.5ex]
\multicolumn{1}{c}{$\Delta R$}  & $\Delta R$  & \multicolumn{3}{c}{\hspace{-6mm}$(\times 10^{-3})$} & & $(\times 10^{-3})$ & $(\times 10^{-3})$  \\[1ex]
\hline
$0.6-1.0$ & $ 0.8$ & $ 2.48$ & $0.04$ & ${}^{+0.48}_{-0.36}$ & $1.27\pm 0.04$ & $2.32{}^{+0.40}_{-0.36}$ & $1.79{}^{+0.75}_{-0.48}$ \\
$1.0-1.4$ & $ 1.2$ & $ 3.14$ & $0.05$ & ${}^{+0.55}_{-0.37}$ & $0.98\pm 0.03$ & $3.24{}^{+0.23}_{-0.36}$ & $3.88{}^{+1.67}_{-1.06}$ \\
$1.4-1.8$ & $ 1.6$ & $ 3.26$ & $0.05$ & ${}^{+0.56}_{-0.38}$ & $0.98\pm 0.03$ & $3.65{}^{+0.25}_{-0.39}$ & $4.30{}^{+1.87}_{-1.18}$ \\
$1.8-2.2$ & $ 2.0$ & $ 4.12$ & $0.06$ & ${}^{+0.69}_{-0.46}$ & $1.00\pm 0.02$ & $4.76{}^{+0.40}_{-0.55}$ & $5.20{}^{+2.26}_{-1.43}$ \\
$2.2-2.5$ & $ 2.3$ & $ 5.43$ & $0.08$ & ${}^{+0.88}_{-0.56}$ & $1.01\pm 0.03$ & $6.17{}^{+0.47}_{-0.70}$ & $6.56{}^{+2.86}_{-1.81}$ \\
$2.5-2.8$ & $ 2.6$ & $ 7.20$ & $0.09$ & ${}^{+1.05}_{-0.69}$ & $1.01\pm 0.01$ & $8.07{}^{+0.61}_{-0.90}$ & $8.31{}^{+3.60}_{-2.28}$ \\
$2.8-3.0$ & $ 2.9$ & $ 9.24$ & $0.11$ & ${}^{+1.20}_{-0.83}$ & $1.04\pm 0.03$ & $10.43{}^{+0.81}_{-1.21}$ & $10.31{}^{+4.43}_{-2.82}$ \\
$3.0-3.2$ & $ 3.1$ & $ 10.61$ & $0.13$ & ${}^{+1.39}_{-0.99}$ & $1.07\pm 0.04$ & $11.27{}^{+0.56}_{-1.17}$ & $10.95{}^{+4.73}_{-2.99}$ \\
$3.2-3.5$ & $ 3.4$ & $ 6.17$ & $0.08$ & ${}^{+1.02}_{-0.71}$ & $1.08\pm 0.02$ & $6.90{}^{+0.54}_{-0.82}$ & $5.79{}^{+2.51}_{-1.58}$ \\
$3.5-3.9$ & $ 3.7$ & $ 3.16$ & $0.05$ & ${}^{+0.70}_{-0.44}$ & $1.10\pm 0.02$ & $3.70{}^{+0.23}_{-0.44}$ & $2.90{}^{+1.29}_{-0.81}$ \\
$3.9-4.5$ & $ 4.2$ & $ 1.23$ & $0.03$ & ${}^{+0.33}_{-0.19}$ & $1.21\pm 0.03$ & $1.62{}^{+0.04}_{-0.16}$ & $1.21{}^{+0.57}_{-0.35}$ \\
$4.5-6.0$ & $ 5.0$ & $ 0.15$ & $0.00$ & ${}^{+0.04}_{-0.02}$ & $1.33\pm 0.00$ & $0.34{}^{+0.04}_{-0.05}$ & $0.19{}^{+0.11}_{-0.06}$ \\
\end{tabular}
\end{ruledtabular}
\end{table}

\begin{table}[htbp]
\caption{\label{tab:results_dy13_3j}The measured differential cross section, normalized to the measured inclusive $W$ boson cross section, as a function of the dijet rapidity separation of the leading and third jets for events with three or more jets produced in association with a $W$ boson. The first data uncertainty is statistical, the second is systematic. Also shown are predictions from NLO pQCD after hadronization corrections have been applied (also shown) and \textsc{hej} resummation predictions.}
\begin{ruledtabular}
\begin{tabular}{rc r@{$\pm$\hspace{-3mm}}l@{${}$\hspace{-3mm}}l rll}
\multicolumn{8}{c}{$1/\sigma_W\cdot  d\sigma/d\Delta y_\mathrm{13}~\textrm{[three-jet bin]}~(1/\Delta y)$} \\
Bin interval  & Bin center  & \multicolumn{3}{c}{\hspace{-6mm}Measurement} & npQCD correction & NLO \textsc{blackhat+sherpa} & \textsc{hej}\\[-0.5ex]
\multicolumn{1}{c}{$\Delta y$}  & $\Delta y$ & \multicolumn{3}{c}{\hspace{-6mm}$(\times 10^{-3})$} & & $(\times 10^{-3})$ & $(\times 10^{-3})$  \\[1ex]
\hline
$0.0-0.4$ & $ 0.1$ & $ 1.13$ & $0.04$ & ${}^{+0.21}_{-0.15}$ & $1.08\pm 0.06$ & $1.08{}^{+0.15}_{-0.19}$ & $1.09{}^{+0.72}_{-0.40}$ \\
$0.4-0.8$ & $ 0.6$ & $ 1.19$ & $0.04$ & ${}^{+0.23}_{-0.16}$ & $1.10\pm 0.06$ & $1.17{}^{+0.17}_{-0.21}$ & $1.13{}^{+0.75}_{-0.41}$ \\
$0.8-1.2$ & $ 1.0$ & $ 1.10$ & $0.04$ & ${}^{+0.21}_{-0.15}$ & $1.08\pm 0.04$ & $1.07{}^{+0.14}_{-0.18}$ & $1.09{}^{+0.73}_{-0.40}$ \\
$1.2-1.6$ & $ 1.4$ & $ 0.83$ & $0.03$ & ${}^{+0.18}_{-0.12}$ & $1.09\pm 0.04$ & $0.88{}^{+0.08}_{-0.14}$ & $0.89{}^{+0.60}_{-0.33}$ \\
$1.6-2.1$ & $ 1.8$ & $ 0.64$ & $0.03$ & ${}^{+0.16}_{-0.11}$ & $1.07\pm 0.04$ & $0.69{}^{+0.08}_{-0.12}$ & $0.70{}^{+0.48}_{-0.26}$ \\
$2.1-2.6$ & $ 2.3$ & $ 0.49$ & $0.02$ & ${}^{+0.15}_{-0.09}$ & $1.08\pm 0.01$ & $0.51{}^{+0.07}_{-0.09}$ & $0.51{}^{+0.36}_{-0.19}$ \\
$2.6-3.2$ & $ 2.9$ & $ 0.31$ & $0.02$ & ${}^{+0.12}_{-0.07}$ & $1.14\pm 0.06$ & $0.34{}^{+0.05}_{-0.06}$ & $0.31{}^{+0.23}_{-0.12}$ \\
$3.2-4.0$ & $ 3.6$ & $ 0.13$ & $0.01$ & ${}^{+0.05}_{-0.03}$ & $1.14\pm 0.06$ & $0.15{}^{+0.02}_{-0.03}$ & $0.14{}^{+0.11}_{-0.06}$ \\
$4.0-5.0$ & $ 4.4$ & $ 0.027$ & $0.003$ & ${}^{+0.013}_{-0.008}$ & $1.17\pm 0.09$ & $0.043{}^{+0.005}_{-0.008}$ & $0.046{}^{+0.039}_{-0.019}$ \\
$5.0-6.0$ & $ 5.2$ & $ 0.002$ & $0.001$ & ${}^{+0.001}_{-0.001}$ & $1.20\pm 0.10$ & $0.008{}^{+0.002}_{-0.003}$ & $0.005{}^{+0.005}_{-0.002}$ \\
\end{tabular}
\end{ruledtabular}
\end{table}

\begin{table}[htbp]
\caption{\label{tab:results_dy23_3j}The measured differential cross section, normalized to the measured inclusive $W$ boson cross section, as a function of the dijet rapidity separation of the second and third jets for events with three or more jets produced in association with a $W$ boson. The first data uncertainty is statistical, the second is systematic. Also shown are predictions from NLO pQCD after hadronization corrections have been applied (also shown) and \textsc{hej} resummation predictions.}
\begin{ruledtabular}
\begin{tabular}{rc r@{$\pm$\hspace{-3mm}}l@{${}$\hspace{-3mm}}l rll}
\multicolumn{8}{c}{$1/\sigma_W\cdot  d\sigma/d\Delta y_\mathrm{23}~\textrm{[three-jet bin]}~(1/\Delta y)$} \\
Bin interval  & Bin center  & \multicolumn{3}{c}{\hspace{-6mm}Measurement} & npQCD correction & NLO \textsc{blackhat+sherpa} & \textsc{hej}\\[-0.5ex]
\multicolumn{1}{c}{$\Delta y$}  & $\Delta y$ & \multicolumn{3}{c}{\hspace{-6mm}$(\times 10^{-3})$} & & $(\times 10^{-3})$ & $(\times 10^{-3})$  \\[1ex]
\hline
$0.0-0.4$ & $ 0.2$ & $ 1.08$ & $0.04$ & ${}^{+0.20}_{-0.15}$ & $1.08\pm 0.04$ & $0.94{}^{+0.12}_{-0.16}$ & $1.00{}^{+0.66}_{-0.36}$ \\
$0.4-0.8$ & $ 0.6$ & $ 1.13$ & $0.04$ & ${}^{+0.21}_{-0.15}$ & $1.10\pm 0.02$ & $1.12{}^{+0.19}_{-0.21}$ & $1.03{}^{+0.69}_{-0.38}$ \\
$0.8-1.2$ & $ 1.0$ & $ 1.09$ & $0.04$ & ${}^{+0.22}_{-0.15}$ & $1.08\pm 0.05$ & $1.12{}^{+0.15}_{-0.20}$ & $1.09{}^{+0.73}_{-0.40}$ \\
$1.2-1.6$ & $ 1.4$ & $ 0.88$ & $0.03$ & ${}^{+0.18}_{-0.13}$ & $1.05\pm 0.03$ & $0.90{}^{+0.12}_{-0.16}$ & $0.90{}^{+0.61}_{-0.33}$ \\
$1.6-2.1$ & $ 1.9$ & $ 0.66$ & $0.02$ & ${}^{+0.16}_{-0.10}$ & $1.10\pm 0.06$ & $0.74{}^{+0.10}_{-0.13}$ & $0.72{}^{+0.49}_{-0.27}$ \\
$2.1-2.6$ & $ 2.4$ & $ 0.47$ & $0.02$ & ${}^{+0.14}_{-0.09}$ & $1.11\pm 0.08$ & $0.55{}^{+0.08}_{-0.10}$ & $0.52{}^{+0.37}_{-0.20}$ \\
$2.6-3.2$ & $ 2.9$ & $ 0.31$ & $0.02$ & ${}^{+0.11}_{-0.06}$ & $1.09\pm 0.02$ & $0.33{}^{+0.04}_{-0.06}$ & $0.33{}^{+0.24}_{-0.13}$ \\
$3.2-4.0$ & $ 3.6$ & $ 0.14$ & $0.01$ & ${}^{+0.06}_{-0.03}$ & $1.15\pm 0.05$ & $0.15{}^{+0.00}_{-0.02}$ & $0.18{}^{+0.13}_{-0.07}$ \\
$4.0-5.0$ & $ 4.4$ & $ 0.028$ & $0.003$ & ${}^{+0.013}_{-0.007}$ & $1.22\pm 0.02$ & $0.057{}^{+0.004}_{-0.012}$ & $0.053{}^{+0.045}_{-0.022}$ \\
$5.0-6.0$ & $ 5.4$ & $ 0.002$ & $0.001$ & ${}^{+0.001}_{-0.001}$ & $1.46\pm 0.14$ & $0.009{}^{+0.001}_{-0.002}$ & $0.008{}^{+0.008}_{-0.004}$ \\
\end{tabular}
\end{ruledtabular}
\end{table}

\clearpage
\newpage

\begin{table}[htbp]
\caption{\label{tab:results_W1pt}The measured differential cross section, normalized to the measured inclusive $W$ boson cross section, as a function of the $W$ boson transverse momentum for events with one or more jets produced in association with a $W$ boson. The first data uncertainty is statistical, the second is systematic. Also shown are predictions from NLO pQCD after hadronization corrections have been applied (also shown).}
\begin{ruledtabular}
\begin{tabular}{rc r@{$\pm$\hspace{-9mm}}l@{${}$\hspace{-9mm}}l ll}
\multicolumn{7}{c}{$1/\sigma_W\cdot  d\sigma/dp_{T}^{W}~\textrm{[one-jet bin]}$~(GeV${}^{-1}$)} \\
Bin interval  & Bin center  & \multicolumn{3}{c}{\hspace{-10mm}Measurement} & npQCD correction & NLO \textsc{blackhat+sherpa} \\[-0.5ex]
\multicolumn{1}{c}{$\textrm{(GeV)}$}              &$\textrm{(GeV)}$             & \multicolumn{3}{c}{\hspace{-6mm}$(\times 10^{-6})$} & & $(\times 10^{-6})$ \\[1ex]
\hline
$0.0-10.0$ & $ 4.3$ & $ 625$ & $7$ & ${}^{+804}_{-236}$ & $2.21\pm 1.22$ & $450{}^{+167}_{-112}$ \\
$10.0-20.0$ & $ 15.6$ & $ 1823$ & $13$ & ${}^{+1331}_{-553}$ & $1.11\pm 0.13$ & $2256{}^{+803}_{-544}$ \\
$20.0-30.0$ & $ 22.3$ & $ 3798$ & $19$ & ${}^{+250}_{-266}$ & $0.93\pm 0.05$ & $3320{}^{+247}_{-565}$ \\
$30.0-40.0$ & $ 34.9$ & $ 2189$ & $16$ & ${}^{+156}_{-147}$ & $0.97\pm 0.03$ & $2787{}^{+175}_{-189}$ \\
$40.0-50.0$ & $ 44.7$ & $ 1164$ & $11$ & ${}^{+58}_{-48}$ & $0.99\pm 0.01$ & $1442{}^{+102}_{-109}$ \\
$50.0-65.0$ & $ 56.9$ & $ 584$ & $5$ & ${}^{+19}_{-20}$ & $1.00\pm 0.01$ & $679{}^{+51}_{-54}$ \\
$65.0-80.0$ & $ 71.9$ & $ 251.7$ & $2.8$ & ${}^{+5.7}_{-8.0}$ & $1.00\pm 0.01$ & $281.2{}^{+19.8}_{-21.9}$ \\
$80.0-100.0$ & $ 89.2$ & $ 106.2$ & $1.4$ & ${}^{+2.3}_{-3.7}$ & $1.00\pm 0.00$ & $119.7{}^{+9.3}_{-9.9}$ \\
$100.0-135.0$ & $ 115.4$ & $ 34.74$ & $0.47$ & ${}^{+0.73}_{-1.29}$ & $1.00\pm 0.01$ & $37.27{}^{+2.92}_{-3.07}$ \\
$135.0-200.0$ & $ 162.6$ & $ 6.43$ & $0.15$ & ${}^{+0.14}_{-0.26}$ & $1.01\pm 0.00$ & $6.75{}^{+0.54}_{-0.55}$ \\
$200.0-300.0$ & $ 239.0$ & $ 0.582$ & $0.032$ & ${}^{+0.019}_{-0.028}$ & $1.00\pm 0.01$ & $0.610{}^{+0.053}_{-0.060}$ \\
\end{tabular}
\end{ruledtabular}
\end{table}

\begin{table}[htbp]
\caption{\label{tab:results_W2pt}The measured differential cross section, normalized to the measured inclusive $W$ boson cross section, as a function of the $W$ boson transverse momentum for events with two or more jets produced in association with a $W$ boson. The first data uncertainty is statistical, the second is systematic. Also shown are predictions from NLO pQCD after hadronization corrections have been applied (also shown) and \textsc{hej} resummation predictions.}
\begin{ruledtabular}
\begin{tabular}{rc r@{$\pm$\hspace{-3mm}}l@{${}$\hspace{-3mm}}l rll}
\multicolumn{8}{c}{$1/\sigma_W\cdot  d\sigma/dp_{T}^{W}~\textrm{[two-jet bin]}$~(GeV${}^{-1}$)} \\
Bin interval  & Bin center  & \multicolumn{3}{c}{\hspace{-6mm}Measurement} & npQCD correction & NLO \textsc{blackhat+sherpa} & \textsc{hej}\\[-0.5ex]
\multicolumn{1}{c}{$\textrm{(GeV)}$}               &$\textrm{(GeV)}$             & \multicolumn{3}{c}{\hspace{-6mm}$(\times 10^{-6})$} & & $(\times 10^{-6})$ & $(\times 10^{-6})$  \\[1ex]
\hline
$0.0-15.0$ & $ 6.8$ & $ 214$ & $4$ & ${}^{+47}_{-30}$ & $1.34\pm 0.36$ & $230{}^{+11}_{-24}$ & $175{}^{+73}_{-47}$ \\
$15.0-25.0$ & $ 19.1$ & $ 279$ & $5$ & ${}^{+71}_{-40}$ & $1.08\pm 0.10$ & $325{}^{+28}_{-39}$ & $294{}^{+126}_{-80}$ \\
$25.0-35.0$ & $ 30.6$ & $ 254$ & $5$ & ${}^{+63}_{-37}$ & $1.04\pm 0.06$ & $307{}^{+25}_{-37}$ & $297{}^{+130}_{-82}$ \\
$35.0-45.0$ & $ 39.9$ & $ 240$ & $5$ & ${}^{+46}_{-31}$ & $1.00\pm 0.03$ & $277{}^{+20}_{-32}$ & $288{}^{+126}_{-79}$ \\
$45.0-55.0$ & $ 50.1$ & $ 201$ & $4$ & ${}^{+26}_{-20}$ & $0.98\pm 0.00$ & $227{}^{+16}_{-26}$ & $248{}^{+108}_{-68}$ \\
$55.0-65.0$ & $ 59.8$ & $ 141$ & $3$ & ${}^{+13}_{-11}$ & $1.00\pm 0.01$ & $163{}^{+12}_{-19}$ & $171{}^{+75}_{-47}$ \\
$65.0-80.0$ & $ 72.1$ & $ 86.8$ & $2.1$ & ${}^{+6.9}_{-6.0}$ & $1.00\pm 0.01$ & $98.8{}^{+8.2}_{-11.7}$ & $101.2{}^{+45.3}_{-28.4}$ \\
$80.0-105.0$ & $ 91.4$ & $ 39.3$ & $0.9$ & ${}^{+2.8}_{-2.5}$ & $1.00\pm 0.00$ & $43.9{}^{+3.4}_{-5.0}$ & $44.8{}^{+20.5}_{-12.8}$ \\
$105.0-140.0$ & $ 120.6$ & $ 13.09$ & $0.38$ & ${}^{+0.75}_{-0.78}$ & $1.00\pm 0.00$ & $14.01{}^{+0.97}_{-1.52}$ & $14.50{}^{+6.76}_{-4.20}$ \\
$140.0-190.0$ & $ 161.5$ & $ 3.41$ & $0.14$ & ${}^{+0.18}_{-0.20}$ & $0.98\pm 0.01$ & $3.41{}^{+0.26}_{-0.38}$ & $3.68{}^{+1.75}_{-1.08}$ \\
$190.0-250.0$ & $ 215.9$ & $ 0.628$ & $0.049$ & ${}^{+0.035}_{-0.038}$ & $0.94\pm 0.04$ & $0.617{}^{+0.044}_{-0.070}$ & $0.640{}^{+0.317}_{-0.194}$ \\
\end{tabular}
\end{ruledtabular}
\end{table}

\begin{table}[htbp]
\caption{\label{tab:results_W3pt}The measured differential cross section, normalized to the measured inclusive $W$ boson cross section, as a function of the $W$ boson transverse momentum for events with three or more jets produced in association with a $W$ boson. The first data uncertainty is statistical, the second is systematic. Also shown are predictions from NLO pQCD after hadronization corrections have been applied (also shown) and \textsc{hej} resummation predictions.}
\begin{ruledtabular}
\begin{tabular}{rc r@{$\pm$\hspace{-3mm}}l@{${}$\hspace{-3mm}}l rll}
\multicolumn{8}{c}{$1/\sigma_W\cdot  d\sigma/dp_{T}^{W}~\textrm{[three-jet bin]}$~(GeV${}^{-1}$)} \\
Bin interval  & Bin center  & \multicolumn{3}{c}{\hspace{-6mm}Measurement} & npQCD correction & NLO \textsc{blackhat+sherpa} & \textsc{hej}\\[-0.5ex]
\multicolumn{1}{c}{$\textrm{(GeV)}$}               &$\textrm{(GeV)}$             & \multicolumn{3}{c}{\hspace{-6mm}$(\times 10^{-6})$} & & $(\times 10^{-6})$ & $(\times 10^{-6})$  \\[1ex]
\hline
$0.0-15.0$ & $ 7.2$ & $ 17.8$ & $0.7$ & ${}^{+5.6}_{-3.4}$ & $1.41\pm 0.09$ & $21.5{}^{+3.2}_{-4.1}$ & $15.6{}^{+10.7}_{-5.8}$ \\
$15.0-30.0$ & $ 21.2$ & $ 39.3$ & $1.2$ & ${}^{+11.5}_{-7.0}$ & $1.13\pm 0.06$ & $38.3{}^{+3.3}_{-6.4}$ & $39.6{}^{+27.5}_{-14.8}$ \\
$30.0-45.0$ & $ 38.3$ & $ 36.8$ & $1.0$ & ${}^{+9.3}_{-6.1}$ & $1.06\pm 0.02$ & $37.9{}^{+4.7}_{-6.6}$ & $40.3{}^{+28.0}_{-15.0}$ \\
$45.0-60.0$ & $ 52.4$ & $ 27.0$ & $0.9$ & ${}^{+5.9}_{-4.1}$ & $1.06\pm 0.05$ & $29.6{}^{+4.3}_{-5.3}$ & $29.9{}^{+20.5}_{-11.1}$ \\
$60.0-75.0$ & $ 67.1$ & $ 18.7$ & $0.7$ & ${}^{+3.6}_{-2.5}$ & $1.04\pm 0.05$ & $19.5{}^{+2.9}_{-3.5}$ & $19.9{}^{+13.5}_{-7.3}$ \\
$75.0-90.0$ & $ 82.4$ & $ 12.2$ & $0.6$ & ${}^{+2.0}_{-1.5}$ & $1.01\pm 0.03$ & $12.2{}^{+1.8}_{-2.2}$ & $12.1{}^{+8.2}_{-4.5}$ \\
$90.0-115.0$ & $ 101.7$ & $ 6.38$ & $0.29$ & ${}^{+0.89}_{-0.70}$ & $1.02\pm 0.02$ & $6.35{}^{+0.96}_{-1.11}$ & $6.19{}^{+4.14}_{-2.27}$ \\
$115.0-150.0$ & $ 130.3$ & $ 2.55$ & $0.16$ & ${}^{+0.31}_{-0.27}$ & $0.99\pm 0.02$ & $2.28{}^{+0.30}_{-0.39}$ & $2.33{}^{+1.55}_{-0.85}$ \\
$150.0-250.0$ & $ 190.7$ & $ 0.446$ & $0.036$ & ${}^{+0.051}_{-0.048}$ & $1.00\pm 0.06$ & $0.408{}^{+0.059}_{-0.070}$ & $0.363{}^{+0.242}_{-0.134}$ \\
\end{tabular}
\end{ruledtabular}
\end{table}

\begin{table}[htbp]
\caption{\label{tab:results_W4pt}The measured differential cross section, normalized to the measured inclusive $W$ boson cross section, as a function of the $W$ boson transverse momentum for events with four or more jets produced in association with a $W$ boson. The first data uncertainty is statistical, the second is systematic. Also shown are predictions from NLO pQCD after hadronization corrections have been applied (also shown) and \textsc{hej} resummation predictions.}
\begin{ruledtabular}
\begin{tabular}{rc r@{$\pm$\hspace{-3mm}}l@{${}$\hspace{-3mm}}l rll}
\multicolumn{8}{c}{$1/\sigma_W\cdot  d\sigma/dp_{T}^{W}~\textrm{[four-jet bin]}$~(GeV${}^{-1}$)} \\
Bin interval  & Bin center  & \multicolumn{3}{c}{\hspace{-6mm}Measurement} & npQCD correction & NLO \textsc{blackhat+sherpa} & \textsc{hej}\\[-0.5ex]
\multicolumn{1}{c}{$\textrm{(GeV)}$}               &$\textrm{(GeV)}$             & \multicolumn{3}{c}{\hspace{-6mm}$(\times 10^{-6})$} & & $(\times 10^{-6})$ & $(\times 10^{-6})$  \\[1ex]
\hline
$0.0-20.0$ & $ 9.5$ & $ 2.38$ & $0.30$ & ${}^{+0.97}_{-0.55}$ & $1.42\pm 0.20$ & $2.52{}^{+0.65}_{-0.68}$ & $1.81{}^{+1.70}_{-0.81}$ \\
$20.0-35.0$ & $ 25.9$ & $ 4.7$ & $0.5$ & ${}^{+1.8}_{-1.0}$ & $1.22\pm 0.14$ & $4.4{}^{+1.1}_{-1.2}$ & $4.0{}^{+3.6}_{-1.7}$ \\
$35.0-50.0$ & $ 43.8$ & $ 4.8$ & $0.4$ & ${}^{+1.6}_{-1.0}$ & $1.16\pm 0.10$ & $4.2{}^{+1.1}_{-1.1}$ & $3.5{}^{+3.3}_{-1.6}$ \\
$50.0-70.0$ & $ 59.9$ & $ 3.5$ & $0.3$ & ${}^{+1.1}_{-0.8}$ & $1.17\pm 0.04$ & $3.2{}^{+0.9}_{-0.9}$ & $2.6{}^{+2.4}_{-1.2}$ \\
$70.0-100.0$ & $ 84.4$ & $ 1.78$ & $0.16$ & ${}^{+0.49}_{-0.39}$ & $1.12\pm 0.05$ & $1.54{}^{+0.37}_{-0.41}$ & $1.42{}^{+1.31}_{-0.63}$ \\
$100.0-135.0$ & $ 116.8$ & $ 0.67$ & $0.10$ & ${}^{+0.17}_{-0.15}$ & $1.14\pm 0.18$ & $0.62{}^{+0.15}_{-0.16}$ & $0.48{}^{+0.43}_{-0.21}$ \\
$135.0-200.0$ & $ 162.2$ & $ 0.169$ & $0.039$ & ${}^{+0.043}_{-0.042}$ & $0.99\pm 0.03$ & $0.139{}^{+0.038}_{-0.038}$ & $0.117{}^{+0.103}_{-0.051}$ \\
\end{tabular}
\end{ruledtabular}
\end{table}

\begin{table}[htbp]
\caption{\label{tab:results_dijetpt2}The measured differential cross section, normalized to the measured inclusive $W$ boson cross section, as a function of dijet transverse momentum for events with two or more jets produced in association with a $W$ boson. The first data uncertainty is statistical, the second is systematic. Also shown are predictions from NLO pQCD after hadronization corrections have been applied (also shown) and \textsc{hej} resummation predictions.}
\begin{ruledtabular}
\begin{tabular}{rc r@{$\pm$\hspace{-3mm}}l@{${}$\hspace{-3mm}}l rll}
\multicolumn{8}{c}{$1/\sigma_W\cdot  d\sigma/dp_{T}^\mathrm{jj}~\textrm{[two-jet bin]}$~(GeV${}^{-1}$)} \\
Bin interval  & Bin center  & \multicolumn{3}{c}{\hspace{-6mm}Measurement} & npQCD correction & NLO \textsc{blackhat+sherpa} & \textsc{hej}\\[-0.5ex]
\multicolumn{1}{c}{$\textrm{(GeV)}$}               &$\textrm{(GeV)}$             & \multicolumn{3}{c}{\hspace{-6mm}$(\times 10^{-6})$} & & $(\times 10^{-6})$ & $(\times 10^{-6})$  \\[1ex]
\hline
$0.0-10.0$ & $ 4.6$ & $ 169$ & $2$ & ${}^{+33}_{-20}$ & $1.40\pm 0.09$ & $200{}^{+17}_{-25}$ & $143{}^{+61}_{-39}$ \\
$10.0-20.0$ & $ 14.4$ & $ 265$ & $3$ & ${}^{+50}_{-31}$ & $1.13\pm 0.02$ & $311{}^{+22}_{-36}$ & $270{}^{+115}_{-73}$ \\
$20.0-30.0$ & $ 20.7$ & $ 272$ & $3$ & ${}^{+50}_{-32}$ & $1.04\pm 0.03$ & $310{}^{+23}_{-36}$ & $295{}^{+126}_{-80}$ \\
$30.0-40.0$ & $ 31.4$ & $ 251$ & $3$ & ${}^{+47}_{-29}$ & $1.02\pm 0.03$ & $280{}^{+17}_{-31}$ & $284{}^{+122}_{-77}$ \\
$40.0-50.0$ & $ 46.2$ & $ 244$ & $3$ & ${}^{+44}_{-28}$ & $1.03\pm 0.04$ & $283{}^{+27}_{-35}$ & $274{}^{+119}_{-75}$ \\
\end{tabular}
\end{ruledtabular}
\end{table}

\begin{table}[htbp]
\caption{\label{tab:results_dijetpt3}The measured differential cross section, normalized to the measured inclusive $W$ boson cross section, as a function of dijet transverse momentum for events with three or more jets produced in association with a $W$ boson. The first data uncertainty is statistical, the second is systematic. Also shown are predictions from NLO pQCD after hadronization corrections have been applied (also shown) and \textsc{hej} resummation predictions.}
\begin{ruledtabular}
\begin{tabular}{rc r@{$\pm$\hspace{-3mm}}l@{${}$\hspace{-3mm}}l rll}
\multicolumn{8}{c}{$1/\sigma_W\cdot  d\sigma/dp_{T}^\mathrm{jj}~\textrm{[three-jet bin]}$~(GeV${}^{-1}$)} \\
Bin interval  & Bin center  & \multicolumn{3}{c}{\hspace{-6mm}Measurement} & npQCD correction & NLO \textsc{blackhat+sherpa} & \textsc{hej}\\[-0.5ex]
\multicolumn{1}{c}{$\textrm{(GeV)}$}               &$\textrm{(GeV)}$             & \multicolumn{3}{c}{\hspace{-6mm}$(\times 10^{-6})$} & & $(\times 10^{-6})$ & $(\times 10^{-6})$  \\[1ex]
\hline
$0.0-15.0$ & $ 6.8$ & $ 17.6$ & $0.6$ & ${}^{+4.3}_{-2.7}$ & $1.23\pm 0.06$ & $18.3{}^{+1.1}_{-2.9}$ & $19.6{}^{+14.2}_{-7.5}$ \\
$15.0-30.0$ & $ 21.1$ & $ 31.8$ & $0.8$ & ${}^{+7.6}_{-4.9}$ & $1.13\pm 0.06$ & $35.2{}^{+4.9}_{-6.4}$ & $33.3{}^{+23.0}_{-12.4}$ \\
$30.0-45.0$ & $ 30.4$ & $ 34.0$ & $0.9$ & ${}^{+7.8}_{-5.1}$ & $1.12\pm 0.07$ & $37.0{}^{+4.7}_{-6.6}$ & $36.0{}^{+24.4}_{-13.3}$ \\
$45.0-65.0$ & $ 55.8$ & $ 29.8$ & $0.7$ & ${}^{+6.6}_{-4.4}$ & $1.07\pm 0.06$ & $30.2{}^{+4.3}_{-5.5}$ & $30.1{}^{+20.7}_{-11.2}$ \\
$65.0-85.0$ & $ 74.5$ & $ 17.0$ & $0.5$ & ${}^{+3.6}_{-2.5}$ & $1.05\pm 0.06$ & $18.1{}^{+2.8}_{-3.3}$ & $17.7{}^{+12.0}_{-6.5}$ \\
$85.0-110.0$ & $ 96.6$ & $ 8.0$ & $0.3$ & ${}^{+1.6}_{-1.2}$ & $0.98\pm 0.06$ & $7.6{}^{+0.9}_{-1.2}$ & $8.0{}^{+5.4}_{-3.0}$ \\
$110.0-140.0$ & $ 123.5$ & $ 3.11$ & $0.17$ & ${}^{+0.61}_{-0.48}$ & $0.96\pm 0.03$ & $2.96{}^{+0.39}_{-0.50}$ & $3.09{}^{+2.07}_{-1.14}$ \\
$140.0-175.0$ & $ 155.2$ & $ 1.09$ & $0.09$ & ${}^{+0.21}_{-0.17}$ & $0.98\pm 0.09$ & $1.04{}^{+0.12}_{-0.17}$ & $1.00{}^{+0.67}_{-0.37}$ \\
$175.0-210.0$ & $ 191.4$ & $ 0.37$ & $0.05$ & ${}^{+0.07}_{-0.06}$ & $0.92\pm 0.09$ & $0.34{}^{+0.05}_{-0.06}$ & $0.33{}^{+0.22}_{-0.12}$ \\
$210.0-280.0$ & $ 238.8$ & $ 0.093$ & $0.019$ & ${}^{+0.019}_{-0.017}$ & $0.84\pm 0.11$ & $0.071{}^{+0.009}_{-0.013}$ & $0.072{}^{+0.048}_{-0.027}$ \\
\end{tabular}
\end{ruledtabular}
\end{table}

\begin{table}[htbp]
\caption{\label{tab:results_dijetmass2}The measured differential cross section, normalized to the measured inclusive $W$ boson cross section, as a function of dijet invariant mass for events with two or more jets produced in association with a $W$ boson. The first data uncertainty is statistical, the second is systematic. Also shown are predictions from NLO pQCD after hadronization corrections have been applied (also shown) and \textsc{hej} resummation predictions.}
\begin{ruledtabular}
\begin{tabular}{rc r@{$\pm$\hspace{-3mm}}l@{${}$\hspace{-3mm}}l rll}
\multicolumn{8}{c}{$1/\sigma_W\cdot  d\sigma/dm_\mathrm{jj}~\textrm{[two-jet bin]}$~(GeV${}^{-1}$)} \\
Bin interval  & Bin center  & \multicolumn{3}{c}{\hspace{-6mm}Measurement} & npQCD correction & NLO \textsc{blackhat+sherpa} & \textsc{hej}\\[-0.5ex]
\multicolumn{1}{c}{$\textrm{(GeV)}$}               &$\textrm{(GeV)}$             & \multicolumn{3}{c}{\hspace{-6mm}$(\times 10^{-6})$} & & $(\times 10^{-6})$ & $(\times 10^{-6})$  \\[1ex]
\hline
$20.0-30.0$ & $ 25.1$ & $ 39.9$ & $1.3$ & ${}^{+9.8}_{-6.0}$ & $0.96\pm 0.13$ & $52.6{}^{+5.6}_{-6.8}$ & $63.7{}^{+25.8}_{-16.7}$ \\

$30.0-40.0$ & $ 34.7$ & $ 98$ & $2$ & ${}^{+23}_{-14}$ & $1.04\pm 0.04$ & $110{}^{+10}_{-13}$ & $120{}^{+50}_{-32}$ \\
$40.0-50.0$ & $ 45.1$ & $ 144$ & $2$ & ${}^{+31}_{-19}$ & $1.05\pm 0.06$ & $171{}^{+13}_{-19}$ & $183{}^{+76}_{-49}$ \\
$50.0-60.0$ & $ 54.0$ & $ 195$ & $3$ & ${}^{+37}_{-23}$ & $1.08\pm 0.04$ & $226{}^{+19}_{-26}$ & $222{}^{+93}_{-60}$ \\
$60.0-80.0$ & $ 70.7$ & $ 186$ & $2$ & ${}^{+30}_{-20}$ & $1.06\pm 0.02$ & $204{}^{+15}_{-23}$ & $199{}^{+84}_{-54}$ \\
\end{tabular}
\end{ruledtabular}
\end{table}

\begin{table}[htbp]
\caption{\label{tab:results_dijetmass3}The measured differential cross section, normalized to the measured inclusive $W$ boson cross section, as a function of dijet invariant mass for events with three or more jets produced in association with a $W$ boson. The first data uncertainty is statistical, the second is systematic. Also shown are predictions from NLO pQCD after hadronization corrections have been applied (also shown) and \textsc{hej} resummation predictions.}
\begin{ruledtabular}
\begin{tabular}{rc r@{$\pm$\hspace{-3mm}}l@{${}$\hspace{-3mm}}l rll}
\multicolumn{8}{c}{$1/\sigma_W\cdot  d\sigma/dm_\mathrm{jj}~\textrm{[three-jet bin]}$~(GeV${}^{-1}$)} \\
Bin interval  & Bin center  & \multicolumn{3}{c}{\hspace{-6mm}Measurement} & npQCD correction & NLO \textsc{blackhat+sherpa} & \textsc{hej}\\[-0.5ex]
\multicolumn{1}{c}{$\textrm{(GeV)}$}               &$\textrm{(GeV)}$             & \multicolumn{3}{c}{\hspace{-6mm}$(\times 10^{-6})$} & & $(\times 10^{-6})$ & $(\times 10^{-6})$  \\[1ex]
\hline
$30.0-45.0$ & $ 36.8$ & $ 9.1$ & $0.5$ & ${}^{+3.2}_{-1.8}$ & $1.18\pm 0.11$ & $9.9{}^{+1.9}_{-2.0}$ & $9.9{}^{+6.8}_{-3.7}$ \\
$45.0-65.0$ & $ 54.6$ & $ 17.1$ & $0.6$ & ${}^{+5.1}_{-3.1}$ & $1.18\pm 0.04$ & $18.5{}^{+3.2}_{-3.5}$ & $17.9{}^{+12.4}_{-6.7}$ \\
$65.0-85.0$ & $ 68.3$ & $ 20.9$ & $0.6$ & ${}^{+5.3}_{-3.3}$ & $1.10\pm 0.06$ & $21.3{}^{+3.1}_{-3.9}$ & $21.8{}^{+14.9}_{-8.1}$ \\
$85.0-110.0$ & $ 98.5$ & $ 18.6$ & $0.5$ & ${}^{+3.9}_{-2.7}$ & $1.08\pm 0.04$ & $18.0{}^{+1.8}_{-3.0}$ & $19.1{}^{+13.1}_{-7.1}$ \\
$110.0-140.0$ & $ 124.3$ & $ 13.4$ & $0.4$ & ${}^{+2.4}_{-1.8}$ & $1.07\pm 0.05$ & $13.6{}^{+1.8}_{-2.4}$ & $13.6{}^{+9.3}_{-5.0}$ \\
$140.0-175.0$ & $ 156.0$ & $ 8.5$ & $0.3$ & ${}^{+1.4}_{-1.1}$ & $1.06\pm 0.04$ & $9.0{}^{+1.1}_{-1.6}$ & $8.5{}^{+5.8}_{-3.1}$ \\
$175.0-210.0$ & $ 191.0$ & $ 5.13$ & $0.23$ & ${}^{+0.79}_{-0.64}$ & $1.07\pm 0.03$ & $5.62{}^{+0.70}_{-0.96}$ & $5.38{}^{+3.67}_{-2.00}$ \\
$210.0-260.0$ & $ 234.4$ & $ 2.78$ & $0.14$ & ${}^{+0.42}_{-0.35}$ & $1.01\pm 0.00$ & $2.89{}^{+0.25}_{-0.46}$ & $2.80{}^{+1.97}_{-1.05}$ \\
$260.0-330.0$ & $ 291.6$ & $ 1.24$ & $0.08$ & ${}^{+0.18}_{-0.15}$ & $1.05\pm 0.02$ & $1.47{}^{+0.14}_{-0.25}$ & $1.26{}^{+0.90}_{-0.48}$ \\
\end{tabular}
\end{ruledtabular}
\end{table}

\begin{table}[htbp]
\caption{\label{tab:results_HT1}The measured differential cross section, normalized to the measured inclusive $W$ boson cross section, as a function of 
$H_T$ (the scalar sum of the transverse energies of the $W$ boson and all jets) for events with one or more jets produced in association with a $W$ boson. The first data uncertainty is statistical, the second is systematic. Also shown are predictions from NLO pQCD after hadronization corrections have been applied (also shown).}
\begin{ruledtabular}
\begin{tabular}{rc r@{$\pm$\hspace{-9mm}}l@{${}$\hspace{-9mm}}l ll}
\multicolumn{7}{c}{$1/\sigma_W\cdot  d\sigma/dH_{T}~\textrm{[one-jet bin]}$~(GeV${}^{-1}$)} \\
Bin interval  & Bin center  & \multicolumn{3}{c}{\hspace{-10mm}Measurement} & npQCD correction & NLO \textsc{blackhat+sherpa} \\[-0.5ex]
\multicolumn{1}{c}{$\textrm{(GeV)}$}              &$\textrm{(GeV)}$             & \multicolumn{3}{c}{\hspace{-6mm}$(\times 10^{-6})$} & & $(\times 10^{-6})$ \\[1ex]
\hline
$100.0-120.0$ & $ 109.8$ & $ 3294$ & $13$ & ${}^{+515}_{-322}$ & $1.04\pm 0.02$ & $3487{}^{+90}_{-152}$ \\
$120.0-140.0$ & $ 129.4$ & $ 1050$ & $9$ & ${}^{+31}_{-33}$ & $0.99\pm 0.01$ & $1284{}^{+47}_{-69}$ \\
$140.0-170.0$ & $ 153.3$ & $ 461$ & $3$ & ${}^{+26}_{-22}$ & $0.99\pm 0.01$ & $515{}^{+54}_{-50}$ \\
$170.0-200.0$ & $ 183.4$ & $ 184$ & $2$ & ${}^{+10}_{-9}$ & $0.98\pm 0.01$ & $192{}^{+29}_{-24}$ \\
$200.0-245.0$ & $ 219.9$ & $ 72.6$ & $0.8$ & ${}^{+3.6}_{-3.7}$ & $0.98\pm 0.01$ & $70.3{}^{+12.1}_{-9.4}$ \\
$245.0-300.0$ & $ 269.6$ & $ 24.6$ & $0.4$ & ${}^{+1.4}_{-1.4}$ & $0.98\pm 0.01$ & $22.7{}^{+4.5}_{-3.3}$ \\
$300.0-360.0$ & $ 327.0$ & $ 8.62$ & $0.18$ & ${}^{+0.59}_{-0.59}$ & $0.99\pm 0.00$ & $7.28{}^{+1.49}_{-1.07}$ \\
$360.0-470.0$ & $ 406.8$ & $ 2.35$ & $0.07$ & ${}^{+0.19}_{-0.19}$ & $0.98\pm 0.02$ & $1.89{}^{+0.42}_{-0.30}$ \\
\end{tabular}
\end{ruledtabular}
\end{table}

\begin{table}[htbp]
\caption{\label{tab:results_HT2}The measured differential cross section, normalized to the measured inclusive $W$ boson cross section, as a function of $H_T$ (the scalar sum of the transverse energies of the $W$ boson and all jets) for events with two or more jets produced in association with a $W$ boson. The first data uncertainty is statistical, the second is systematic. Also shown are predictions from NLO pQCD after hadronization corrections have been applied (also shown) and \textsc{hej} resummation predictions.}
\begin{ruledtabular}
\begin{tabular}{rc r@{$\pm$\hspace{-3mm}}l@{${}$\hspace{-3mm}}l rll}
\multicolumn{8}{c}{$1/\sigma_W\cdot  d\sigma/dH_{T}~\textrm{[two-jet bin]}$~(GeV${}^{-1}$)} \\
Bin interval  & Bin center  & \multicolumn{3}{c}{\hspace{-6mm}Measurement} & npQCD correction & NLO \textsc{blackhat+sherpa} & \textsc{hej}\\[-0.5ex]
\multicolumn{1}{c}{$\textrm{(GeV)}$}               &$\textrm{(GeV)}$             & \multicolumn{3}{c}{\hspace{-6mm}$(\times 10^{-6})$} & & $(\times 10^{-6})$ & $(\times 10^{-6})$  \\[1ex]
\hline
$120.0-135.0$ & $ 126.7$ & $ 160$ & $4$ & ${}^{+77}_{-36}$ & $1.35\pm 0.06$ & $199{}^{+14}_{-22}$ & $201{}^{+81}_{-53}$ \\
$135.0-155.0$ & $ 144.2$ & $ 242$ & $3$ & ${}^{+58}_{-36}$ & $1.06\pm 0.03$ & $286{}^{+19}_{-32}$ & $222{}^{+91}_{-59}$ \\
$155.0-185.0$ & $ 168.4$ & $ 154$ & $1$ & ${}^{+18}_{-14}$ & $1.02\pm 0.02$ & $176{}^{+11}_{-20}$ & $146{}^{+64}_{-40}$ \\
$185.0-230.0$ & $ 204.9$ & $ 68.0$ & $0.7$ & ${}^{+5.7}_{-4.9}$ & $0.99\pm 0.02$ & $74.9{}^{+5.8}_{-8.7}$ & $69.5{}^{+32.7}_{-20.0}$ \\
$230.0-280.0$ & $ 252.5$ & $ 25.6$ & $0.4$ & ${}^{+2.0}_{-1.9}$ & $0.98\pm 0.02$ & $27.6{}^{+3.0}_{-3.5}$ & $27.4{}^{+13.6}_{-8.2}$ \\
$280.0-340.0$ & $ 306.9$ & $ 9.51$ & $0.20$ & ${}^{+0.77}_{-0.75}$ & $0.99\pm 0.02$ & $9.67{}^{+1.12}_{-1.23}$ & $10.27{}^{+5.27}_{-3.15}$ \\
$340.0-390.0$ & $ 362.6$ & $ 3.80$ & $0.14$ & ${}^{+0.34}_{-0.33}$ & $0.97\pm 0.02$ & $3.64{}^{+0.45}_{-0.48}$ & $4.12{}^{+2.17}_{-1.29}$ \\
$390.0-460.0$ & $ 420.4$ & $ 1.62$ & $0.07$ & ${}^{+0.15}_{-0.15}$ & $0.97\pm 0.01$ & $1.46{}^{+0.18}_{-0.20}$ & $1.64{}^{+0.88}_{-0.52}$ \\
\end{tabular}
\end{ruledtabular}
\end{table}

\begin{table}[htbp]
\caption{\label{tab:results_HT3}The measured differential cross section, normalized to the measured inclusive $W$ boson cross section, as a function of $H_T$ (the scalar sum of the transverse energies of the $W$ boson and all jets) for events with three or more jets produced in association with a $W$ boson. The first data uncertainty is statistical, the second is systematic. Also shown are predictions from NLO pQCD after hadronization corrections have been applied (also shown) and \textsc{hej} resummation predictions.}
\begin{ruledtabular}
\begin{tabular}{rc r@{$\pm$\hspace{-3mm}}l@{${}$\hspace{-3mm}}l rll}
\multicolumn{8}{c}{$1/\sigma_W\cdot  d\sigma/dH_{T}~\textrm{[three-jet bin]}$~(GeV${}^{-1}$)} \\
Bin interval  & Bin center  & \multicolumn{3}{c}{\hspace{-6mm}Measurement} & npQCD correction & NLO \textsc{blackhat+sherpa} & \textsc{hej}\\[-0.5ex]
\multicolumn{1}{c}{$\textrm{(GeV)}$}               &$\textrm{(GeV)}$             & \multicolumn{3}{c}{\hspace{-6mm}$(\times 10^{-6})$} & & $(\times 10^{-6})$ & $(\times 10^{-6})$  \\[1ex]
\hline
$140.0-165.0$ & $ 151.2$ & $ 9.4$ & $0.5$ & ${}^{+4.0}_{-2.1}$ & $1.53\pm 0.04$ & $11.7{}^{+2.6}_{-2.6}$ & $14.9{}^{+10.6}_{-5.6}$ \\
$165.0-200.0$ & $ 180.5$ & $ 22.9$ & $0.7$ & ${}^{+6.9}_{-4.2}$ & $1.13\pm 0.07$ & $24.0{}^{+3.8}_{-4.8}$ & $22.6{}^{+15.4}_{-8.3}$ \\
$200.0-240.0$ & $ 218.0$ & $ 17.2$ & $0.4$ & ${}^{+3.4}_{-2.5}$ & $1.04\pm 0.03$ & $17.2{}^{+2.2}_{-3.3}$ & $15.9{}^{+10.8}_{-5.9}$ \\
$240.0-290.0$ & $ 262.4$ & $ 9.1$ & $0.3$ & ${}^{+1.3}_{-1.1}$ & $1.01\pm 0.03$ & $9.1{}^{+1.4}_{-1.9}$ & $8.2{}^{+5.6}_{-3.0}$ \\
$290.0-340.0$ & $ 312.2$ & $ 4.17$ & $0.17$ & ${}^{+0.53}_{-0.50}$ & $1.01\pm 0.04$ & $4.05{}^{+0.54}_{-0.79}$ & $3.73{}^{+2.53}_{-1.38}$ \\
$340.0-390.0$ & $ 362.5$ & $ 1.91$ & $0.11$ & ${}^{+0.25}_{-0.26}$ & $0.97\pm 0.02$ & $1.82{}^{+0.28}_{-0.37}$ & $1.73{}^{+1.18}_{-0.64}$ \\
$390.0-450.0$ & $ 416.6$ & $ 0.86$ & $0.07$ & ${}^{+0.12}_{-0.13}$ & $0.97\pm 0.00$ & $0.82{}^{+0.13}_{-0.17}$ & $0.77{}^{+0.52}_{-0.29}$ \\
\end{tabular}
\end{ruledtabular}
\end{table}

\begin{table}[htbp]
\caption{\label{tab:results_HT4}The measured differential cross section, normalized to the measured inclusive $W$ boson cross section, as a function of $H_T$ (the scalar sum of the transverse energies of the $W$ boson and all jets) for events with four or more jets produced in association with a $W$ boson. The first data uncertainty is statistical, the second is systematic. Also shown are predictions from NLO pQCD after hadronization corrections have been applied (also shown) and \textsc{hej} resummation predictions.}
\begin{ruledtabular}
\begin{tabular}{rc r@{$\pm$\hspace{-3mm}}l@{${}$\hspace{-3mm}}l rll}
\multicolumn{8}{c}{$1/\sigma_W\cdot  d\sigma/dH_{T}~\textrm{[four-jet bin]}$~(GeV${}^{-1}$)} \\
Bin interval  & Bin center  & \multicolumn{3}{c}{\hspace{-6mm}Measurement} & npQCD correction & NLO \textsc{blackhat+sherpa} & \textsc{hej}\\[-0.5ex]
\multicolumn{1}{c}{$\textrm{(GeV)}$}               &$\textrm{(GeV)}$             & \multicolumn{3}{c}{\hspace{-6mm}$(\times 10^{-6})$} & & $(\times 10^{-6})$ & $(\times 10^{-6})$  \\[1ex]
\hline
$160.0-200.0$ & $ 177.0$ & $ 0.65$ & $0.15$ & ${}^{+0.52}_{-0.19}$ & $1.78\pm 0.26$ & $0.84{}^{+0.30}_{-0.25}$ & $0.90{}^{+0.91}_{-0.41}$ \\
$200.0-250.0$ & $ 221.9$ & $ 2.27$ & $0.21$ & ${}^{+0.95}_{-0.51}$ & $1.27\pm 0.16$ & $2.10{}^{+0.53}_{-0.56}$ & $1.79{}^{+1.70}_{-0.80}$ \\
$250.0-310.0$ & $ 277.1$ & $ 1.94$ & $0.14$ & ${}^{+0.54}_{-0.38}$ & $1.13\pm 0.07$ & $1.61{}^{+0.42}_{-0.43}$ & $1.28{}^{+1.14}_{-0.56}$ \\
$310.0-370.0$ & $ 338.2$ & $ 0.92$ & $0.09$ & ${}^{+0.27}_{-0.24}$ & $1.06\pm 0.04$ & $0.78{}^{+0.19}_{-0.21}$ & $0.60{}^{+0.53}_{-0.26}$ \\
$370.0-450.0$ & $ 406.1$ & $ 0.232$ & $0.078$ & ${}^{+0.083}_{-0.086}$ & $0.96\pm 0.03$ & $0.294{}^{+0.064}_{-0.075}$ & $0.242{}^{+0.210}_{-0.104}$ \\
\end{tabular}
\end{ruledtabular}
\end{table}

\begin{table}[htbp]
\caption{\label{tab:results_njet_ht_1}The measured mean number of jets with $p_T>20$~GeV and $|y|<3.2$
as a function of the scalar sum of transverse energies of the $W$ boson and jets in events with one or more jets produced in association with a $W$ boson.
The first data uncertainty is statistical, the second is systematic.
Also shown are predictions from NLO pQCD predictions.} 
\begin{ruledtabular}
\begin{tabular}{rc r@{$\pm$\hspace{-9mm}}l@{${}$\hspace{-9mm}}l ll}
\multicolumn{7}{c}{$\langle N_\mathrm{jet}\rangle~\textrm{versus}~H_{T}~\textrm{[one-jet bin]}$} \\
Bin interval  & Bin center  & \multicolumn{3}{c}{\hspace{-10mm}Measurement} & npQCD correction & NLO \textsc{blackhat+sherpa} \\[-0.5ex]
\multicolumn{1}{c}{$\textrm{(GeV)}$}              &$\textrm{(GeV)}$             & \multicolumn{3}{c}{\hspace{-6mm}} & &  \\[1ex]
\hline
$100.0-120.0$ & $ 110.0$ & $ 0.996$ & $0.000$ & ${}^{+0.003}_{-0.006}$ & $1.00\pm 0.00$ & $1.000{}^{+0.000}_{-0.000}$ \\
$120.0-140.0$ & $ 130.0$ & $ 1.168$ & $0.000$ & ${}^{+0.044}_{-0.022}$ & $1.03\pm 0.00$ & $1.177{}^{+0.003}_{-0.003}$ \\
$140.0-170.0$ & $ 155.0$ & $ 1.465$ & $0.001$ & ${}^{+0.043}_{-0.027}$ & $1.02\pm 0.00$ & $1.485{}^{+0.003}_{-0.003}$ \\
$170.0-200.0$ & $ 185.0$ & $ 1.714$ & $0.001$ & ${}^{+0.038}_{-0.027}$ & $1.02\pm 0.01$ & $1.763{}^{+0.006}_{-0.006}$ \\
$200.0-245.0$ & $ 222.5$ & $ 1.928$ & $0.002$ & ${}^{+0.037}_{-0.029}$ & $1.01\pm 0.01$ & $2.001{}^{+0.005}_{-0.005}$ \\
$245.0-300.0$ & $ 272.5$ & $ 2.140$ & $0.003$ & ${}^{+0.040}_{-0.032}$ & $1.01\pm 0.01$ & $2.205{}^{+0.007}_{-0.007}$ \\
$300.0-360.0$ & $ 330.0$ & $ 2.314$ & $0.001$ & ${}^{+0.047}_{-0.044}$ & $1.01\pm 0.01$ & $2.374{}^{+0.011}_{-0.011}$ \\
$360.0-470.0$ & $ 415.0$ & $ 2.478$ & $0.012$ & ${}^{+0.059}_{-0.061}$ & $1.00\pm 0.01$ & $2.469{}^{+0.014}_{-0.014}$ \\
\end{tabular}
\end{ruledtabular}
\end{table}

\begin{table}[htbp]
\caption{\label{tab:results_njet_ht_2}The measured mean number of jets with $p_T>20$~GeV and $|y|<3.2$
as a function of the scalar sum of transverse energies of the $W$ boson and jets in events with two or more jets produced in association with a $W$ boson.
The first data uncertainty is statistical, the second is systematic.
Also shown are predictions from NLO pQCD and \textsc{hej} resummation predictions. }
\begin{ruledtabular}
\begin{tabular}{rc r@{$\pm$\hspace{-3mm}}l@{${}$\hspace{-3mm}}l rll}
\multicolumn{8}{c}{$\langle N_\mathrm{jet}\rangle~\textrm{versus}~H_{T}~\textrm{[two-jet bin]}$} \\
Bin interval  & Bin center  & \multicolumn{3}{c}{\hspace{-6mm}Measurement} & npQCD correction & NLO \textsc{blackhat+sherpa} & \textsc{hej}\\[-0.5ex]
\multicolumn{1}{c}{$\textrm{(GeV)}$}               &$\textrm{(GeV)}$             & \multicolumn{3}{c}{\hspace{-6mm}} & & &  \\[1ex]
\hline
$120.0-135.0$ & $ 127.5$ & $ 1.992$ & $0.003$ & ${}^{+0.002}_{-0.002}$ & $1.00\pm 0.01$ & $2.000{}^{+0.000}_{-0.000}$ & $2.011{}^{+0.003}_{-0.002}$ \\
$135.0-155.0$ & $ 145.0$ & $ 2.012$ & $0.002$ & ${}^{+0.006}_{-0.002}$ & $1.00\pm 0.01$ & $2.039{}^{+0.003}_{-0.003}$ & $2.049{}^{+0.011}_{-0.008}$ \\
$155.0-185.0$ & $ 170.0$ & $ 2.132$ & $0.000$ & ${}^{+0.018}_{-0.009}$ & $1.00\pm 0.01$ & $2.184{}^{+0.006}_{-0.006}$ & $2.153{}^{+0.027}_{-0.021}$ \\
$185.0-230.0$ & $ 207.5$ & $ 2.306$ & $0.001$ & ${}^{+0.026}_{-0.017}$ & $1.01\pm 0.01$ & $2.326{}^{+0.008}_{-0.008}$ & $2.288{}^{+0.046}_{-0.035}$ \\
$230.0-280.0$ & $ 255.0$ & $ 2.489$ & $0.002$ & ${}^{+0.028}_{-0.019}$ & $1.01\pm 0.01$ & $2.481{}^{+0.006}_{-0.006}$ & $2.412{}^{+0.060}_{-0.047}$ \\
$280.0-340.0$ & $ 310.0$ & $ 2.635$ & $0.003$ & ${}^{+0.035}_{-0.030}$ & $1.01\pm 0.01$ & $2.579{}^{+0.021}_{-0.021}$ & $2.482{}^{+0.065}_{-0.052}$ \\
$340.0-390.0$ & $ 365.0$ & $ 2.702$ & $0.000$ & ${}^{+0.046}_{-0.048}$ & $1.01\pm 0.01$ & $2.696{}^{+0.012}_{-0.012}$ & $2.531{}^{+0.069}_{-0.055}$ \\
$390.0-460.0$ & $ 425.0$ & $ 2.739$ & $0.029$ & ${}^{+0.054}_{-0.058}$ & $1.00\pm 0.01$ & $2.743{}^{+0.020}_{-0.020}$ & $2.549{}^{+0.066}_{-0.054}$ \\
\end{tabular}
\end{ruledtabular}
\end{table}

\clearpage
\newpage

\begin{table}[htbp]
\caption{\label{tab:results_njet_dy}
The measured mean number of jets with $p_T>20$~GeV and $|y|<3.2$
as a function of both the 
dijet rapidity separation of the two leading $p_T$ jets ($\Delta y_{12}$) and the most rapidity-separated jets ($\Delta y_{FB}$)
in events with two or more jets produced in association with a $W$ boson.
The first data uncertainty is statistical, the second is systematic.
Also shown are predictions from NLO pQCD and \textsc{hej} resummation predictions.}
\begin{ruledtabular}
\begin{tabular}{rc r@{$\pm$\hspace{-3mm}}l@{${}$\hspace{-3mm}}l rll}
\multicolumn{8}{c}{$\langle N_\mathrm{jet}\rangle~\textrm{versus}~\Delta y_{12}$} \\
Bin interval  & Bin center  & \multicolumn{3}{c}{\hspace{-6mm}Measurement} & npQCD correction & NLO \textsc{blackhat+sherpa} & \textsc{hej}\\[-0.5ex]
\multicolumn{1}{c}{$\Delta y$}  & $\Delta y$ & \multicolumn{3}{c}{} & & & \\[1ex]
\hline
$0.0-0.4$ & $ 0.2$ & $ 2.174$ & $0.000$ & ${}^{+0.010}_{-0.008}$ & $1.00\pm 0.00$ & $2.156{}^{+0.002}_{-0.002}$ & $2.138{}^{+0.025}_{-0.019}$ \\
$0.4-0.8$ & $ 0.6$ & $ 2.160$ & $0.000$ & ${}^{+0.009}_{-0.008}$ & $1.00\pm 0.00$ & $2.154{}^{+0.002}_{-0.002}$ & $2.145{}^{+0.027}_{-0.020}$ \\
$0.8-1.2$ & $ 1.0$ & $ 2.156$ & $0.002$ & ${}^{+0.009}_{-0.007}$ & $1.01\pm 0.00$ & $2.152{}^{+0.002}_{-0.002}$ & $2.149{}^{+0.028}_{-0.021}$ \\
$1.2-1.6$ & $ 1.4$ & $ 2.178$ & $0.001$ & ${}^{+0.009}_{-0.007}$ & $1.00\pm 0.00$ & $2.152{}^{+0.002}_{-0.002}$ & $2.153{}^{+0.030}_{-0.022}$ \\
$1.6-2.1$ & $ 1.8$ & $ 2.182$ & $0.001$ & ${}^{+0.007}_{-0.007}$ & $1.00\pm 0.00$ & $2.144{}^{+0.008}_{-0.008}$ & $2.162{}^{+0.032}_{-0.024}$ \\
$2.1-2.6$ & $ 2.3$ & $ 2.173$ & $0.001$ & ${}^{+0.006}_{-0.006}$ & $1.00\pm 0.00$ & $2.153{}^{+0.005}_{-0.005}$ & $2.166{}^{+0.032}_{-0.024}$ \\
$2.6-3.2$ & $ 2.9$ & $ 2.181$ & $0.000$ & ${}^{+0.007}_{-0.007}$ & $1.00\pm 0.00$ & $2.163{}^{+0.005}_{-0.005}$ & $2.173{}^{+0.035}_{-0.026}$ \\
$3.2-4.0$ & $ 3.6$ & $ 2.187$ & $0.003$ & ${}^{+0.010}_{-0.010}$ & $1.00\pm 0.00$ & $2.158{}^{+0.006}_{-0.006}$ & $2.193{}^{+0.035}_{-0.027}$ \\
$4.0-5.0$ & $ 4.5$ & $ 2.198$ & $0.004$ & ${}^{+0.025}_{-0.024}$ & $1.00\pm 0.00$ & $2.167{}^{+0.021}_{-0.021}$ & $2.192{}^{+0.034}_{-0.025}$ \\
$5.0-6.0$ & $ 5.5$ & $ 2.117$ & $0.011$ & ${}^{+0.069}_{-0.060}$ & $1.00\pm 0.00$ & $2.140{}^{+0.043}_{-0.043}$ & $2.209{}^{+0.047}_{-0.034}$ \\
%\end{tabular}
%\end{ruledtabular}
%\end{table}
\hline\hline
%\begin{table}[htbp]
%\caption{\label{tab:results_njet_dy_FB}.}
%\begin{ruledtabular}
%\begin{tabular}{rc r@{$\pm$\hspace{-3mm}}l@{${}$\hspace{-3mm}}l rll}
\multicolumn{8}{c}{$\langle N_\mathrm{jet}\rangle~\textrm{versus}~\Delta y_{FB}$} \\
Bin interval  & Bin center  & \multicolumn{3}{c}{\hspace{-6mm}Measurement} & npQCD correction & NLO \textsc{blackhat+sherpa} & \textsc{hej}\\[-0.5ex]
\multicolumn{1}{c}{$\Delta y$}  & $\Delta y$ & \multicolumn{3}{c}{} & & & \\[1ex]
\hline
$0.0-0.4$ & $ 0.2$ & $ 2.023$ & $0.001$ & ${}^{+0.002}_{-0.002}$ & $1.00\pm 0.00$ & $2.015{}^{+0.000}_{-0.000}$ & $2.015{}^{+0.002}_{-0.002}$ \\
$0.4-0.8$ & $ 0.6$ & $ 2.056$ & $0.001$ & ${}^{+0.003}_{-0.003}$ & $1.00\pm 0.00$ & $2.056{}^{+0.001}_{-0.001}$ & $2.052{}^{+0.008}_{-0.006}$ \\
$0.8-1.2$ & $ 1.0$ & $ 2.115$ & $0.001$ & ${}^{+0.005}_{-0.005}$ & $1.00\pm 0.00$ & $2.101{}^{+0.001}_{-0.001}$ & $2.099{}^{+0.016}_{-0.012}$ \\
$1.2-1.6$ & $ 1.4$ & $ 2.179$ & $0.000$ & ${}^{+0.007}_{-0.007}$ & $1.00\pm 0.00$ & $2.148{}^{+0.002}_{-0.002}$ & $2.149{}^{+0.025}_{-0.019}$ \\
$1.6-2.1$ & $ 1.8$ & $ 2.240$ & $0.000$ & ${}^{+0.008}_{-0.008}$ & $1.00\pm 0.00$ & $2.195{}^{+0.003}_{-0.003}$ & $2.206{}^{+0.034}_{-0.026}$ \\
$2.1-2.6$ & $ 2.3$ & $ 2.296$ & $0.003$ & ${}^{+0.010}_{-0.010}$ & $1.00\pm 0.00$ & $2.243{}^{+0.006}_{-0.006}$ & $2.273{}^{+0.044}_{-0.034}$ \\
$2.6-3.2$ & $ 2.9$ & $ 2.371$ & $0.001$ & ${}^{+0.012}_{-0.012}$ & $1.00\pm 0.00$ & $2.295{}^{+0.007}_{-0.007}$ & $2.342{}^{+0.055}_{-0.043}$ \\
$3.2-4.0$ & $ 3.6$ & $ 2.462$ & $0.001$ & ${}^{+0.015}_{-0.015}$ & $1.00\pm 0.00$ & $2.326{}^{+0.026}_{-0.026}$ & $2.431{}^{+0.063}_{-0.051}$ \\
$4.0-5.0$ & $ 4.5$ & $ 2.548$ & $0.005$ & ${}^{+0.033}_{-0.032}$ & $1.00\pm 0.00$ & $2.422{}^{+0.040}_{-0.040}$ & $2.542{}^{+0.067}_{-0.057}$ \\
$5.0-6.0$ & $ 5.5$ & $ 2.579$ & $0.008$ & ${}^{+0.237}_{-0.221}$ & $1.00\pm 0.01$ & $2.714{}^{+0.187}_{-0.187}$ & $2.661{}^{+0.083}_{-0.071}$ \\
\end{tabular}
\end{ruledtabular}
\end{table}

\clearpage
\newpage

\begin{table}[htbp]
\caption{\label{tab:results_prob_dy}
The third jet emission probability for jets with $p_T>20$~GeV and $|y|<3.2$ measured as a function of both the
dijet rapidity separation of the two leading $p_T$ jets ($\Delta y_{12}$) with and without a requirement that the third jet be emitted into the 
rapidity interval defined by the leading two jets, and as a function of the most rapidity-separated jets ($\Delta y_{FB}$),
in events with two or more jets produced in association with a $W$ boson.
The first data uncertainty is statistical, the second is systematic.
Also shown are predictions from NLO pQCD and \textsc{hej} resummation predictions for each of the configurations.}
\begin{ruledtabular}
\begin{tabular}{rc r@{$\pm$\hspace{-3mm}}l@{${}$\hspace{-3mm}}l rll}
\multicolumn{8}{c}{$\textrm{Probability of third jet emission versus}~\Delta y_{12}$} \\
Bin interval  & Bin center  & \multicolumn{3}{c}{\hspace{-6mm}Measurement} & npQCD correction & NLO \textsc{blackhat+sherpa} & \textsc{hej}\\[-0.5ex]
\multicolumn{1}{c}{$\Delta y$}  & $\Delta y$ & \multicolumn{3}{c}{} & & & \\[1ex]
\hline
$0.0-0.4$ & $ 0.2$ & $ 0.151$ & $0.005$ & ${}^{+0.011}_{-0.009}$ & $1.04\pm 0.01$ & $0.139{}^{+0.002}_{-0.002}$ & $0.124{}^{+0.020}_{-0.016}$ \\
$0.4-0.8$ & $ 0.6$ & $ 0.143$ & $0.005$ & ${}^{+0.010}_{-0.008}$ & $1.03\pm 0.00$ & $0.137{}^{+0.002}_{-0.002}$ & $0.132{}^{+0.022}_{-0.017}$ \\
$0.8-1.2$ & $ 1.0$ & $ 0.138$ & $0.005$ & ${}^{+0.009}_{-0.007}$ & $1.06\pm 0.02$ & $0.135{}^{+0.002}_{-0.002}$ & $0.136{}^{+0.023}_{-0.018}$ \\
$1.2-1.6$ & $ 1.4$ & $ 0.155$ & $0.005$ & ${}^{+0.008}_{-0.007}$ & $1.02\pm 0.01$ & $0.134{}^{+0.002}_{-0.002}$ & $0.138{}^{+0.024}_{-0.018}$ \\
$1.6-2.1$ & $ 1.8$ & $ 0.157$ & $0.005$ & ${}^{+0.006}_{-0.006}$ & $1.03\pm 0.03$ & $0.127{}^{+0.008}_{-0.008}$ & $0.147{}^{+0.027}_{-0.020}$ \\
$2.1-2.6$ & $ 2.3$ & $ 0.154$ & $0.007$ & ${}^{+0.004}_{-0.004}$ & $1.02\pm 0.01$ & $0.137{}^{+0.005}_{-0.005}$ & $0.151{}^{+0.027}_{-0.020}$ \\
$2.6-3.2$ & $ 2.9$ & $ 0.157$ & $0.012$ & ${}^{+0.005}_{-0.004}$ & $1.05\pm 0.03$ & $0.147{}^{+0.005}_{-0.005}$ & $0.156{}^{+0.029}_{-0.022}$ \\
$3.2-4.0$ & $ 3.6$ & $ 0.173$ & $0.022$ & ${}^{+0.013}_{-0.011}$ & $1.03\pm 0.01$ & $0.140{}^{+0.006}_{-0.006}$ & $0.176{}^{+0.030}_{-0.023}$ \\
$4.0-5.0$ & $ 4.5$ & $ 0.163$ & $0.025$ & ${}^{+0.023}_{-0.023}$ & $0.98\pm 0.03$ & $0.150{}^{+0.020}_{-0.020}$ & $0.175{}^{+0.028}_{-0.021}$ \\
%\end{tabular}
%\end{ruledtabular}
%\end{table}
\hline\hline
%\begin{table}[htbp]
%\caption{\label{tab:results_prob_dy_12restrict}.}
%\begin{ruledtabular}
%\begin{tabular}{rc r@{$\pm$\hspace{-3mm}}l@{${}$\hspace{-3mm}}l rll}
\multicolumn{8}{c}{$\textrm{Probability of third jet emission into the rapidity gap versus}~\Delta y_{12}$} \\
Bin interval  & Bin center  & \multicolumn{3}{c}{\hspace{-6mm}Measurement} & npQCD correction & NLO \textsc{blackhat+sherpa} & \textsc{hej}\\[0.5ex]
\multicolumn{1}{c}{$\Delta y$}  & $\Delta y$ & \multicolumn{3}{c}{} & & & \\[1ex]
\hline
$0.0-0.4$ & $ 0.2$ & $ 0.007$ & $0.001$ & ${}^{+0.001}_{-0.000}$ & $0.99\pm 0.00$ & $0.005{}^{+0.000}_{-0.000}$ & $0.006{}^{+0.001}_{-0.001}$ \\
$0.4-0.8$ & $ 0.6$ & $ 0.021$ & $0.002$ & ${}^{+0.001}_{-0.001}$ & $1.08\pm 0.04$ & $0.016{}^{+0.001}_{-0.001}$ & $0.018{}^{+0.003}_{-0.002}$ \\
$0.8-1.2$ & $ 1.0$ & $ 0.041$ & $0.003$ & ${}^{+0.002}_{-0.002}$ & $1.09\pm 0.05$ & $0.033{}^{+0.001}_{-0.001}$ & $0.033{}^{+0.005}_{-0.004}$ \\
$1.2-1.6$ & $ 1.4$ & $ 0.067$ & $0.003$ & ${}^{+0.002}_{-0.002}$ & $1.03\pm 0.01$ & $0.047{}^{+0.002}_{-0.002}$ & $0.049{}^{+0.008}_{-0.006}$ \\
$1.6-2.1$ & $ 1.8$ & $ 0.086$ & $0.004$ & ${}^{+0.002}_{-0.003}$ & $1.03\pm 0.03$ & $0.058{}^{+0.003}_{-0.003}$ & $0.066{}^{+0.011}_{-0.008}$ \\
$2.1-2.6$ & $ 2.3$ & $ 0.096$ & $0.005$ & ${}^{+0.002}_{-0.002}$ & $1.00\pm 0.01$ & $0.068{}^{+0.010}_{-0.010}$ & $0.083{}^{+0.014}_{-0.010}$ \\
$2.6-3.2$ & $ 2.9$ & $ 0.108$ & $0.009$ & ${}^{+0.004}_{-0.005}$ & $1.02\pm 0.04$ & $0.094{}^{+0.007}_{-0.007}$ & $0.102{}^{+0.017}_{-0.013}$ \\
$3.2-4.0$ & $ 3.6$ & $ 0.126$ & $0.025$ & ${}^{+0.011}_{-0.013}$ & $1.00\pm 0.01$ & $0.083{}^{+0.017}_{-0.017}$ & $0.122{}^{+0.021}_{-0.016}$ \\
$4.0-5.0$ & $ 4.5$ & $ 0.111$ & $0.017$ & ${}^{+0.019}_{-0.020}$ & $0.95\pm 0.06$ & $0.131{}^{+0.029}_{-0.029}$ & $0.142{}^{+0.024}_{-0.019}$ \\
%\end{tabular}
%\end{ruledtabular}
%\end{table}
\hline\hline

%\begin{table}[htbp]
%\caption{\label{tab:results_prob_dy_FB}.}
%\begin{ruledtabular}
%\begin{tabular}{rc r@{$\pm$\hspace{-3mm}}l@{${}$\hspace{-3mm}}l rll}
\multicolumn{8}{c}{$\textrm{Probability of third jet emission versus}~\Delta y_{FB}$} \\
Bin interval  & Bin center  & \multicolumn{3}{c}{\hspace{-6mm}Measurement} & npQCD correction & NLO \textsc{blackhat+sherpa} & \textsc{hej}\\[-0.5ex]
\multicolumn{1}{c}{$\Delta y$}  & $\Delta y$ & \multicolumn{3}{c}{} & & & \\[1ex]
\hline
$0.0-0.4$ & $ 0.2$ & $ 0.020$ & $0.002$ & ${}^{+0.001}_{-0.001}$ & $1.00\pm 0.00$ & $0.015{}^{+0.000}_{-0.000}$ & $0.015{}^{+0.002}_{-0.002}$ \\
$0.4-0.8$ & $ 0.6$ & $ 0.054$ & $0.003$ & ${}^{+0.002}_{-0.002}$ & $1.03\pm 0.01$ & $0.054{}^{+0.001}_{-0.001}$ & $0.050{}^{+0.008}_{-0.006}$ \\
$0.8-1.2$ & $ 1.0$ & $ 0.100$ & $0.004$ & ${}^{+0.003}_{-0.003}$ & $1.07\pm 0.04$ & $0.095{}^{+0.002}_{-0.002}$ & $0.095{}^{+0.015}_{-0.011}$ \\
$1.2-1.6$ & $ 1.4$ & $ 0.159$ & $0.005$ & ${}^{+0.004}_{-0.004}$ & $1.02\pm 0.01$ & $0.137{}^{+0.002}_{-0.002}$ & $0.139{}^{+0.021}_{-0.017}$ \\
$1.6-2.1$ & $ 1.8$ & $ 0.205$ & $0.006$ & ${}^{+0.005}_{-0.006}$ & $1.03\pm 0.02$ & $0.175{}^{+0.003}_{-0.003}$ & $0.190{}^{+0.029}_{-0.023}$ \\
$2.1-2.6$ & $ 2.3$ & $ 0.247$ & $0.007$ & ${}^{+0.009}_{-0.009}$ & $1.00\pm 0.02$ & $0.215{}^{+0.006}_{-0.006}$ & $0.244{}^{+0.036}_{-0.028}$ \\
$2.6-3.2$ & $ 2.9$ & $ 0.307$ & $0.010$ & ${}^{+0.019}_{-0.016}$ & $1.01\pm 0.02$ & $0.285{}^{+0.007}_{-0.007}$ & $0.298{}^{+0.042}_{-0.034}$ \\
$3.2-4.0$ & $ 3.6$ & $ 0.387$ & $0.015$ & ${}^{+0.041}_{-0.027}$ & $0.99\pm 0.02$ & $0.315{}^{+0.025}_{-0.025}$ & $0.375{}^{+0.047}_{-0.039}$ \\
$4.0-5.0$ & $ 4.5$ & $ 0.460$ & $0.025$ & ${}^{+0.065}_{-0.039}$ & $0.94\pm 0.03$ & $0.409{}^{+0.038}_{-0.038}$ & $0.448{}^{+0.047}_{-0.042}$ \\
$5.0-6.0$ & $ 5.5$ & $ 0.485$ & $0.081$ & ${}^{+0.122}_{-0.118}$ & $0.97\pm 0.06$ & $0.687{}^{+0.182}_{-0.182}$ & $0.546{}^{+0.060}_{-0.051}$ \\
\end{tabular}
\end{ruledtabular}
\end{table}

\clearpage
\newpage

\begin{table}[htbp]
\caption{\label{tab:results_gapfraction_dy}
The gap fraction for events with two or more jets produced in association with a $W$ boson, measured as a function of the
dijet rapidity separation of the two leading $p_T$ jets ($\Delta y_{12}$) and as a function of the two most rapidity-separated jets ($\Delta y_{FB}$).
The first data uncertainty is statistical, the second is systematic.
Also shown are predictions from NLO pQCD and \textsc{hej} resummation predictions for each of the configurations.}
\begin{ruledtabular}
\begin{tabular}{rc r@{$\pm$\hspace{-3mm}}l@{${}$\hspace{-3mm}}l rll}
\multicolumn{8}{c}{$\textrm{Gap fraction versus}~\Delta y_{12}$} \\
Bin interval  & Bin center  & \multicolumn{3}{c}{\hspace{-6mm}Measurement} & npQCD correction & NLO \textsc{blackhat+sherpa} & \textsc{hej}\\[-0.5ex]
\multicolumn{1}{c}{$\Delta y$}  & $\Delta y$ & \multicolumn{3}{c}{} & & & \\[1ex]
\hline
$0.0-0.4$ & $ 0.2$ & $ 0.993$ & $0.001$ & ${}^{+0.000}_{-0.001}$ & $0.99\pm 0.00$ & $0.995{}^{+0.000}_{-0.000}$ & $0.994{}^{+0.001}_{-0.001}$ \\
$0.4-0.8$ & $ 0.6$ & $ 0.979$ & $0.002$ & ${}^{+0.001}_{-0.001}$ & $1.08\pm 0.04$ & $0.984{}^{+0.001}_{-0.001}$ & $0.982{}^{+0.002}_{-0.003}$ \\
$0.8-1.2$ & $ 1.0$ & $ 0.959$ & $0.003$ & ${}^{+0.002}_{-0.002}$ & $1.09\pm 0.05$ & $0.967{}^{+0.001}_{-0.001}$ & $0.967{}^{+0.004}_{-0.005}$ \\
$1.2-1.6$ & $ 1.4$ & $ 0.933$ & $0.003$ & ${}^{+0.002}_{-0.002}$ & $1.03\pm 0.01$ & $0.953{}^{+0.002}_{-0.002}$ & $0.951{}^{+0.006}_{-0.008}$ \\
$1.6-2.1$ & $ 1.8$ & $ 0.914$ & $0.004$ & ${}^{+0.003}_{-0.002}$ & $1.03\pm 0.03$ & $0.942{}^{+0.003}_{-0.003}$ & $0.934{}^{+0.008}_{-0.011}$ \\
$2.1-2.6$ & $ 2.3$ & $ 0.904$ & $0.005$ & ${}^{+0.002}_{-0.002}$ & $1.00\pm 0.01$ & $0.932{}^{+0.010}_{-0.010}$ & $0.917{}^{+0.010}_{-0.014}$ \\
$2.6-3.2$ & $ 2.9$ & $ 0.892$ & $0.009$ & ${}^{+0.005}_{-0.004}$ & $1.02\pm 0.04$ & $0.906{}^{+0.007}_{-0.007}$ & $0.898{}^{+0.013}_{-0.017}$ \\
$3.2-4.0$ & $ 3.6$ & $ 0.874$ & $0.025$ & ${}^{+0.013}_{-0.011}$ & $1.00\pm 0.01$ & $0.917{}^{+0.017}_{-0.017}$ & $0.878{}^{+0.016}_{-0.021}$ \\
$4.0-5.0$ & $ 4.5$ & $ 0.889$ & $0.017$ & ${}^{+0.020}_{-0.019}$ & $0.95\pm 0.06$ & $0.869{}^{+0.029}_{-0.029}$ & $0.858{}^{+0.019}_{-0.024}$ \\
%\end{tabular}
%\end{ruledtabular}
%\end{table}
\hline\hline

%\begin{table}[htbp]
%\caption{\label{tab:results_gapfraction_dy_FB}.}
%\begin{ruledtabular}
%\begin{tabular}{rc r@{$\pm$\hspace{-3mm}}l@{${}$\hspace{-3mm}}l rll}
\multicolumn{8}{c}{$\textrm{Gap fraction versus}~\Delta y_{FB}$} \\
Bin interval  & Bin center  & \multicolumn{3}{c}{\hspace{-6mm}Measurement} & npQCD correction & NLO \textsc{blackhat+sherpa} & \textsc{hej}\\[-0.5ex]
\multicolumn{1}{c}{$\Delta y$}  & $\Delta y$ & \multicolumn{3}{c}{} & & & \\[1ex]
\hline
$0.0-0.4$ & $ 0.2$ & $ 0.980$ & $0.002$ & ${}^{+0.001}_{-0.001}$ & $1.00\pm 0.00$ & $0.985{}^{+0.000}_{-0.000}$ & $0.985{}^{+0.002}_{-0.002}$ \\
$0.4-0.8$ & $ 0.6$ & $ 0.946$ & $0.003$ & ${}^{+0.002}_{-0.002}$ & $1.03\pm 0.01$ & $0.946{}^{+0.001}_{-0.001}$ & $0.950{}^{+0.006}_{-0.008}$ \\
$0.8-1.2$ & $ 1.0$ & $ 0.900$ & $0.004$ & ${}^{+0.003}_{-0.003}$ & $1.07\pm 0.04$ & $0.905{}^{+0.002}_{-0.002}$ & $0.905{}^{+0.011}_{-0.015}$ \\
$1.2-1.6$ & $ 1.4$ & $ 0.841$ & $0.005$ & ${}^{+0.004}_{-0.004}$ & $1.02\pm 0.01$ & $0.863{}^{+0.002}_{-0.002}$ & $0.861{}^{+0.017}_{-0.021}$ \\
$1.6-2.1$ & $ 1.8$ & $ 0.795$ & $0.006$ & ${}^{+0.006}_{-0.005}$ & $1.03\pm 0.02$ & $0.825{}^{+0.003}_{-0.003}$ & $0.810{}^{+0.023}_{-0.029}$ \\
$2.1-2.6$ & $ 2.3$ & $ 0.753$ & $0.007$ & ${}^{+0.009}_{-0.009}$ & $1.00\pm 0.02$ & $0.785{}^{+0.006}_{-0.006}$ & $0.756{}^{+0.028}_{-0.036}$ \\
$2.6-3.2$ & $ 2.9$ & $ 0.693$ & $0.010$ & ${}^{+0.016}_{-0.019}$ & $1.01\pm 0.02$ & $0.715{}^{+0.007}_{-0.007}$ & $0.702{}^{+0.034}_{-0.042}$ \\
$3.2-4.0$ & $ 3.6$ & $ 0.613$ & $0.015$ & ${}^{+0.027}_{-0.041}$ & $0.99\pm 0.02$ & $0.685{}^{+0.025}_{-0.025}$ & $0.625{}^{+0.039}_{-0.047}$ \\
$4.0-5.0$ & $ 4.5$ & $ 0.540$ & $0.025$ & ${}^{+0.039}_{-0.065}$ & $0.94\pm 0.03$ & $0.591{}^{+0.038}_{-0.038}$ & $0.552{}^{+0.042}_{-0.047}$ \\
$5.0-6.0$ & $ 5.5$ & $ 0.515$ & $0.081$ & ${}^{+0.118}_{-0.122}$ & $0.97\pm 0.06$ & $0.313{}^{+0.182}_{-0.182}$ & $0.454{}^{+0.051}_{-0.060}$ \\
\end{tabular}
\end{ruledtabular}
\end{table}

\clearpage
\newpage

\begin{table}[htbp]
\caption{\label{tab:NPC_jet1eta}Non-perturbative correction (applied to NLO pQCD predictions) as a function of leading jet rapidity for events with one or more jets produced in association with a $W$ boson. The first number is the \textsc{siscone} to midpoint cone jet algorithm correction, the second is the correction for underlying event and hadronization effects, the third is the total correction factor.}
\begin{ruledtabular}
\begin{tabular}{rccc} 
Bin interval  & $\sigma^\mathrm{midpoint}_\mathrm{particle} / \sigma^\textsc{siscone}_\mathrm{particle}$  &  $\sigma^\textsc{siscone}_\mathrm{particle}/\sigma^\textsc{siscone}_\mathrm{parton}$ & Total correction  \\[1ex]
\hline
$-3.2$~--~$-2.8$ & $ 0.98$ & $1.01$ & $0.99\pm 0.02$ \\
$-2.8$~--~$-2.4$ & $ 0.99$ & $1.02$ & $1.01\pm 0.02$ \\
$-2.4$~--~$-2.0$ & $ 1.00$ & $1.01$ & $1.01\pm 0.01$ \\
$-2.0$~--~$-1.6$ & $ 1.00$ & $1.01$ & $1.01\pm 0.01$ \\
$-1.6$~--~$-1.2$ & $ 1.01$ & $1.01$ & $1.01\pm 0.01$ \\
$-1.2$~--~$-0.8$ & $ 1.01$ & $1.01$ & $1.01\pm 0.01$ \\
$-0.8$~--~$-0.4$ & $ 1.01$ & $1.01$ & $1.02\pm 0.02$ \\
$-0.4$~--~$0.0$ & $ 1.01$ & $1.01$ & $1.02\pm 0.02$ \\
$0.0$~--~$0.4$ & $ 1.01$ & $1.01$ & $1.02\pm 0.02$ \\
$0.4$~--~$0.8$ & $ 1.01$ & $1.01$ & $1.02\pm 0.02$ \\
$0.8$~--~$1.2$ & $ 1.01$ & $1.01$ & $1.01\pm 0.01$ \\
$1.2$~--~$1.6$ & $ 1.01$ & $1.01$ & $1.01\pm 0.01$ \\
$1.6$~--~$2.0$ & $ 1.00$ & $1.01$ & $1.01\pm 0.01$ \\
$2.0$~--~$2.4$ & $ 1.00$ & $1.01$ & $1.01\pm 0.01$ \\
$2.4$~--~$2.8$ & $ 0.99$ & $1.02$ & $1.01\pm 0.02$ \\
$2.8$~--~$3.2$ & $ 0.98$ & $1.01$ & $0.99\pm 0.02$ \\
\end{tabular}
\end{ruledtabular}
\end{table}

\begin{table}[htbp]
\caption{\label{tab:NPC_jet2eta}Non-perturbative correction (applied to NLO pQCD predictions) as a function of second jet rapidity for events with two or more jets produced in association with a $W$ boson. The first number is the \textsc{siscone} to midpoint cone jet algorithm correction, the second is the correction for underlying event and hadronization effects, the third is the total correction factor.}
\begin{ruledtabular}
\begin{tabular}{rccc} 
Bin interval  & $\sigma^\mathrm{midpoint}_\mathrm{particle} / \sigma^\textsc{siscone}_\mathrm{particle}$  &  $\sigma^\textsc{siscone}_\mathrm{particle}/\sigma^\textsc{siscone}_\mathrm{parton}$ & Total correction  \\[1ex]
\hline
$-3.2$~--~$-2.8$ & $ 1.01$ & $1.09$ & $1.10\pm 0.04$ \\
$-2.8$~--~$-2.4$ & $ 1.01$ & $1.07$ & $1.09\pm 0.03$ \\
$-2.4$~--~$-2.0$ & $ 1.01$ & $1.07$ & $1.08\pm 0.03$ \\
$-2.0$~--~$-1.6$ & $ 1.02$ & $1.04$ & $1.05\pm 0.02$ \\
$-1.6$~--~$-1.2$ & $ 1.01$ & $1.04$ & $1.05\pm 0.02$ \\
$-1.2$~--~$-0.8$ & $ 1.01$ & $1.03$ & $1.05\pm 0.02$ \\
$-0.8$~--~$-0.4$ & $ 1.01$ & $1.04$ & $1.05\pm 0.03$ \\
$-0.4$~--~$0.0$ & $ 1.01$ & $1.04$ & $1.06\pm 0.03$ \\
$0.0$~--~$0.4$ & $ 1.01$ & $1.04$ & $1.06\pm 0.03$ \\
$0.4$~--~$0.8$ & $ 1.01$ & $1.04$ & $1.05\pm 0.03$ \\
$0.8$~--~$1.2$ & $ 1.01$ & $1.03$ & $1.05\pm 0.02$ \\
$1.2$~--~$1.6$ & $ 1.01$ & $1.04$ & $1.05\pm 0.02$ \\
$1.6$~--~$2.0$ & $ 1.02$ & $1.04$ & $1.05\pm 0.02$ \\
$2.0$~--~$2.4$ & $ 1.01$ & $1.07$ & $1.08\pm 0.03$ \\
$2.4$~--~$2.8$ & $ 1.01$ & $1.07$ & $1.09\pm 0.03$ \\
$2.8$~--~$3.2$ & $ 1.01$ & $1.09$ & $1.10\pm 0.04$ \\
\end{tabular}
\end{ruledtabular}
\end{table}

\begin{table}[htbp]
\caption{\label{tab:NPC_jet3eta}Non-perturbative correction (applied to NLO pQCD predictions) as a function of third jet rapidity for events with three or more jets produced in association with a $W$ boson. The first number is the \textsc{siscone} to midpoint cone jet algorithm correction, the second is the correction for underlying event and hadronization effects, the third is the total correction factor.}
\begin{ruledtabular}
\begin{tabular}{rccc} 
Bin interval  & $\sigma^\mathrm{midpoint}_\mathrm{particle} / \sigma^\textsc{siscone}_\mathrm{particle}$  &  $\sigma^\textsc{siscone}_\mathrm{particle}/\sigma^\textsc{siscone}_\mathrm{parton}$ & Total correction  \\[1ex]
\hline
$-3.2$~--~$-2.4$ & $ 0.98$ & $1.09$ & $1.07\pm 0.05$ \\
$-2.4$~--~$-1.6$ & $ 1.01$ & $1.08$ & $1.09\pm 0.04$ \\
$-1.6$~--~$-0.8$ & $ 1.01$ & $1.09$ & $1.10\pm 0.04$ \\
$-0.8$~--~$0.0$ & $ 1.00$ & $1.09$ & $1.09\pm 0.05$ \\
$0.0$~--~$0.8$ & $ 1.00$ & $1.09$ & $1.09\pm 0.05$ \\
$0.8$~--~$1.6$ & $ 1.01$ & $1.09$ & $1.10\pm 0.04$ \\
$1.6$~--~$2.4$ & $ 1.01$ & $1.08$ & $1.09\pm 0.04$ \\
$2.4$~--~$3.2$ & $ 0.98$ & $1.09$ & $1.07\pm 0.05$ \\
\end{tabular}
\end{ruledtabular}
\end{table}

\begin{table}[htbp]
\caption{\label{tab:NPC_jet4eta}Non-perturbative correction (applied to NLO pQCD predictions) as a function of fourth jet rapidity for events with four or more jets produced in association with a $W$ boson. The first number is the \textsc{siscone} to midpoint cone jet algorithm correction, the second is the correction for underlying event and hadronization effects, the third is the total correction factor.}
\begin{ruledtabular}
\begin{tabular}{rccc} 
Bin interval  & $\sigma^\mathrm{midpoint}_\mathrm{particle} / \sigma^\textsc{siscone}_\mathrm{particle}$  &  $\sigma^\textsc{siscone}_\mathrm{particle}/\sigma^\textsc{siscone}_\mathrm{parton}$ & Total correction  \\[1ex]
\hline
$-3.2$~--~$-2.4$ & $ 1.01$ & $1.21$ & $1.23\pm 0.15$ \\
$-2.4$~--~$-1.6$ & $ 1.03$ & $1.18$ & $1.22\pm 0.09$ \\
$-1.6$~--~$-0.8$ & $ 1.00$ & $1.14$ & $1.13\pm 0.07$ \\
$-0.8$~--~$0.0$ & $ 1.00$ & $1.23$ & $1.23\pm 0.13$ \\
$0.0$~--~$0.8$ & $ 1.00$ & $1.23$ & $1.23\pm 0.13$ \\
$0.8$~--~$1.6$ & $ 1.00$ & $1.14$ & $1.13\pm 0.07$ \\
$1.6$~--~$2.4$ & $ 1.03$ & $1.18$ & $1.22\pm 0.09$ \\
$2.4$~--~$3.2$ & $ 1.01$ & $1.21$ & $1.23\pm 0.15$ \\
\end{tabular}
\end{ruledtabular}
\end{table}

\begin{table}[htbp]
\caption{\label{tab:NPC_lepton1pt}Non-perturbative correction (applied to NLO pQCD predictions) as a function of electron transverse momentum for events with one or more jets produced in association with a $W$ boson. The first number is the \textsc{siscone} to midpoint cone jet algorithm correction, the second is the correction for underlying event and hadronization effects, the third is the total correction factor.}
\begin{ruledtabular}
\begin{tabular}{rccc} 
Bin interval  & $\sigma^\mathrm{midpoint}_\mathrm{particle} / \sigma^\textsc{siscone}_\mathrm{particle}$  &  $\sigma^\textsc{siscone}_\mathrm{particle}/\sigma^\textsc{siscone}_\mathrm{parton}$ & Total correction  \\[1ex]
\hline
$15.0$~--~$25.0$ & $ 1.01$ & $1.03$ & $1.04\pm 0.01$ \\
$25.0$~--~$35.0$ & $ 1.01$ & $1.01$ & $1.02\pm 0.02$ \\
$35.0$~--~$45.0$ & $ 1.00$ & $1.04$ & $1.05\pm 0.03$ \\
$45.0$~--~$55.0$ & $ 1.01$ & $0.95$ & $0.96\pm 0.01$ \\
$55.0$~--~$70.0$ & $ 1.00$ & $0.98$ & $0.98\pm 0.00$ \\
$70.0$~--~$90.0$ & $ 1.00$ & $0.99$ & $0.99\pm 0.00$ \\
$90.0$~--~$130.0$ & $ 1.00$ & $0.98$ & $0.98\pm 0.00$ \\
$130.0$~--~$200.0$ & $ 1.00$ & $0.96$ & $0.96\pm 0.01$ \\
$200.0$~--~$300.0$ & $ 1.01$ & $1.01$ & $1.01\pm 0.03$ \\
\end{tabular}
\end{ruledtabular}
\end{table}

\begin{table}[htbp]
\caption{\label{tab:NPC_lepton2pt}Non-perturbative correction (applied to NLO pQCD predictions) as a function of electron transverse momentum for events with two or more jets produced in association with a $W$ boson. The first number is the \textsc{siscone} to midpoint cone jet algorithm correction, the second is the correction for underlying event and hadronization effects, the third is the total correction factor.}
\begin{ruledtabular}
\begin{tabular}{rccc} 
Bin interval  & $\sigma^\mathrm{midpoint}_\mathrm{particle} / \sigma^\textsc{siscone}_\mathrm{particle}$  &  $\sigma^\textsc{siscone}_\mathrm{particle}/\sigma^\textsc{siscone}_\mathrm{parton}$ & Total correction  \\[1ex]
\hline
$15.0$~--~$25.0$ & $ 1.01$ & $1.06$ & $1.07\pm 0.03$ \\
$25.0$~--~$35.0$ & $ 1.01$ & $1.07$ & $1.08\pm 0.03$ \\
$35.0$~--~$45.0$ & $ 1.02$ & $1.08$ & $1.10\pm 0.03$ \\
$45.0$~--~$55.0$ & $ 1.01$ & $1.01$ & $1.02\pm 0.02$ \\
$55.0$~--~$70.0$ & $ 1.01$ & $0.98$ & $0.99\pm 0.02$ \\
$70.0$~--~$100.0$ & $ 1.00$ & $0.99$ & $0.99\pm 0.02$ \\
$100.0$~--~$170.0$ & $ 0.99$ & $0.99$ & $0.97\pm 0.02$ \\
$170.0$~--~$250.0$ & $ 0.95$ & $1.01$ & $0.96\pm 0.02$ \\
\end{tabular}
\end{ruledtabular}
\end{table}

\begin{table}[htbp]
\caption{\label{tab:NPC_lepton3pt}Non-perturbative correction (applied to NLO pQCD predictions) as a function of electron transverse momentum for events with three or more jets produced in association with a $W$ boson. The first number is the \textsc{siscone} to midpoint cone jet algorithm correction, the second is the correction for underlying event and hadronization effects, the third is the total correction factor.}
\begin{ruledtabular}
\begin{tabular}{rccc} 
Bin interval  & $\sigma^\mathrm{midpoint}_\mathrm{particle} / \sigma^\textsc{siscone}_\mathrm{particle}$  &  $\sigma^\textsc{siscone}_\mathrm{particle}/\sigma^\textsc{siscone}_\mathrm{parton}$ & Total correction  \\[1ex]
\hline
$15.0$~--~$25.0$ & $ 1.00$ & $1.11$ & $1.12\pm 0.04$ \\
$25.0$~--~$35.0$ & $ 1.01$ & $1.12$ & $1.13\pm 0.05$ \\
$35.0$~--~$50.0$ & $ 1.01$ & $1.10$ & $1.11\pm 0.05$ \\
$50.0$~--~$70.0$ & $ 1.01$ & $1.02$ & $1.03\pm 0.03$ \\
$70.0$~--~$110.0$ & $ 0.99$ & $1.03$ & $1.01\pm 0.02$ \\
$110.0$~--~$150.0$ & $ 0.95$ & $1.01$ & $0.97\pm 0.04$ \\
$150.0$~--~$250.0$ & $ 0.94$ & $1.04$ & $0.98\pm 0.04$ \\
\end{tabular}
\end{ruledtabular}
\end{table}

\begin{table}[htbp]
\caption{\label{tab:NPC_lepton4pt}Non-perturbative correction (applied to NLO pQCD predictions) as a function of electron transverse momentum for events with four or more jets produced in association with a $W$ boson. The first number is the \textsc{siscone} to midpoint cone jet algorithm correction, the second is the correction for underlying event and hadronization effects, the third is the total correction factor.}
\begin{ruledtabular}
\begin{tabular}{rccc} 
Bin interval  & $\sigma^\mathrm{midpoint}_\mathrm{particle} / \sigma^\textsc{siscone}_\mathrm{particle}$  &  $\sigma^\textsc{siscone}_\mathrm{particle}/\sigma^\textsc{siscone}_\mathrm{parton}$ & Total correction  \\[1ex]
\hline
$15.0$~--~$25.0$ & $ 1.01$ & $1.30$ & $1.31\pm 0.08$ \\
$25.0$~--~$35.0$ & $ 1.03$ & $1.15$ & $1.18\pm 0.12$ \\
$35.0$~--~$50.0$ & $ 1.02$ & $1.22$ & $1.24\pm 0.17$ \\
$50.0$~--~$70.0$ & $ 0.97$ & $1.15$ & $1.12\pm 0.04$ \\
$70.0$~--~$130.0$ & $ 0.98$ & $1.09$ & $1.06\pm 0.03$ \\
$130.0$~--~$200.0$ & $ 0.90$ & $1.18$ & $1.06\pm 0.07$ \\
\end{tabular}
\end{ruledtabular}
\end{table}

\begin{table}[htbp]
\caption{\label{tab:NPC_lepton1eta}Non-perturbative correction (applied to NLO pQCD predictions) as a function of electron pseudorapidity for events with one or more jets produced in association with a $W$ boson. The first number is the \textsc{siscone} to midpoint cone jet algorithm correction, the second is the correction for underlying event and hadronization effects, the third is the total correction factor.}
\begin{ruledtabular}
\begin{tabular}{rccc} 
Bin interval  & $\sigma^\mathrm{midpoint}_\mathrm{particle} / \sigma^\textsc{siscone}_\mathrm{particle}$  &  $\sigma^\textsc{siscone}_\mathrm{particle}/\sigma^\textsc{siscone}_\mathrm{parton}$ & Total correction  \\[1ex]
\hline
$-1.1$~--~$-0.8$ & $ 1.01$ & $1.01$ & $1.02\pm 0.02$ \\
$-0.8$~--~$-0.6$ & $ 1.01$ & $1.01$ & $1.01\pm 0.02$ \\
$-0.6$~--~$-0.4$ & $ 1.01$ & $1.01$ & $1.01\pm 0.02$ \\
$-0.4$~--~$-0.2$ & $ 1.01$ & $1.01$ & $1.01\pm 0.02$ \\
$-0.2$~--~$0.0$ & $ 1.01$ & $1.01$ & $1.01\pm 0.02$ \\
$0.0$~--~$0.2$ & $ 1.01$ & $1.01$ & $1.01\pm 0.02$ \\
$0.2$~--~$0.4$ & $ 1.01$ & $1.01$ & $1.01\pm 0.02$ \\
$0.4$~--~$0.6$ & $ 1.01$ & $1.01$ & $1.01\pm 0.02$ \\
$0.6$~--~$0.8$ & $ 1.01$ & $1.01$ & $1.01\pm 0.02$ \\
$0.8$~--~$1.1$ & $ 1.01$ & $1.01$ & $1.02\pm 0.02$ \\
\end{tabular}
\end{ruledtabular}
\end{table}

\begin{table}[htbp]
\caption{\label{tab:NPC_lepton2eta}Non-perturbative correction (applied to NLO pQCD predictions) as a function of electron pseudorapidity for events with two or more jets produced in association with a $W$ boson. The first number is the \textsc{siscone} to midpoint cone jet algorithm correction, the second is the correction for underlying event and hadronization effects, the third is the total correction factor.}
\begin{ruledtabular}
\begin{tabular}{rccc} 
Bin interval  & $\sigma^\mathrm{midpoint}_\mathrm{particle} / \sigma^\textsc{siscone}_\mathrm{particle}$  &  $\sigma^\textsc{siscone}_\mathrm{particle}/\sigma^\textsc{siscone}_\mathrm{parton}$ & Total correction  \\[1ex]
\hline
$-1.1$~--~$-0.8$ & $ 1.01$ & $1.04$ & $1.06\pm 0.03$ \\
$-0.8$~--~$-0.6$ & $ 1.01$ & $1.05$ & $1.06\pm 0.03$ \\
$-0.6$~--~$-0.4$ & $ 1.01$ & $1.05$ & $1.06\pm 0.03$ \\
$-0.4$~--~$-0.2$ & $ 1.01$ & $1.04$ & $1.05\pm 0.02$ \\
$-0.2$~--~$0.0$ & $ 1.01$ & $1.04$ & $1.05\pm 0.03$ \\
$0.0$~--~$0.2$ & $ 1.01$ & $1.04$ & $1.05\pm 0.03$ \\
$0.2$~--~$0.4$ & $ 1.01$ & $1.04$ & $1.05\pm 0.02$ \\
$0.4$~--~$0.6$ & $ 1.01$ & $1.05$ & $1.06\pm 0.03$ \\
$0.6$~--~$0.8$ & $ 1.01$ & $1.05$ & $1.06\pm 0.03$ \\
$0.8$~--~$1.1$ & $ 1.01$ & $1.04$ & $1.06\pm 0.03$ \\
\end{tabular}
\end{ruledtabular}
\end{table}

\begin{table}[htbp]
\caption{\label{tab:NPC_lepton3eta}Non-perturbative correction (applied to NLO pQCD predictions) as a function of electron pseudorapidity for events with three or more jets produced in association with a $W$ boson. The first number is the \textsc{siscone} to midpoint cone jet algorithm correction, the second is the correction for underlying event and hadronization effects, the third is the total correction factor.}
\begin{ruledtabular}
\begin{tabular}{rccc} 
Bin interval  & $\sigma^\mathrm{midpoint}_\mathrm{particle} / \sigma^\textsc{siscone}_\mathrm{particle}$  &  $\sigma^\textsc{siscone}_\mathrm{particle}/\sigma^\textsc{siscone}_\mathrm{parton}$ & Total correction  \\[1ex]
\hline
$-1.1$~--~$-0.8$ & $ 1.01$ & $1.08$ & $1.09\pm 0.04$ \\
$-0.8$~--~$-0.6$ & $ 1.00$ & $1.11$ & $1.12\pm 0.05$ \\
$-0.6$~--~$-0.4$ & $ 1.00$ & $1.06$ & $1.07\pm 0.04$ \\
$-0.4$~--~$-0.2$ & $ 1.00$ & $1.10$ & $1.10\pm 0.04$ \\
$-0.2$~--~$0.0$ & $ 1.01$ & $1.08$ & $1.09\pm 0.05$ \\
$0.0$~--~$0.2$ & $ 1.01$ & $1.08$ & $1.09\pm 0.05$ \\
$0.2$~--~$0.4$ & $ 1.00$ & $1.10$ & $1.10\pm 0.04$ \\
$0.4$~--~$0.6$ & $ 1.00$ & $1.06$ & $1.07\pm 0.04$ \\
$0.6$~--~$0.8$ & $ 1.00$ & $1.11$ & $1.12\pm 0.05$ \\
$0.8$~--~$1.1$ & $ 1.01$ & $1.08$ & $1.09\pm 0.04$ \\
\end{tabular}
\end{ruledtabular}
\end{table}

\begin{table}[htbp]
\caption{\label{tab:NPC_lepton4eta}Non-perturbative correction (applied to NLO pQCD predictions) as a function of electron pseudorapidity for events with four or more jets produced in association with a $W$ boson. The first number is the \textsc{siscone} to midpoint cone jet algorithm correction, the second is the correction for underlying event and hadronization effects, the third is the total correction factor.}
\begin{ruledtabular}
\begin{tabular}{rccc} 
Bin interval  & $\sigma^\mathrm{midpoint}_\mathrm{particle} / \sigma^\textsc{siscone}_\mathrm{particle}$  &  $\sigma^\textsc{siscone}_\mathrm{particle}/\sigma^\textsc{siscone}_\mathrm{parton}$ & Total correction  \\[1ex]
\hline
$-1.1$~--~$-0.6$ & $ 1.01$ & $1.20$ & $1.21\pm 0.14$ \\
$-0.6$~--~$-0.2$ & $ 1.01$ & $1.17$ & $1.19\pm 0.08$ \\
$-0.2$~--~$0.2$ & $ 0.97$ & $1.20$ & $1.16\pm 0.05$ \\
$0.2$~--~$0.6$ & $ 1.01$ & $1.17$ & $1.19\pm 0.08$ \\
$0.6$~--~$1.1$ & $ 1.01$ & $1.20$ & $1.21\pm 0.14$ \\
\end{tabular}
\end{ruledtabular}
\end{table}

\begin{table}[htbp]
\caption{\label{tab:NPC_dy12_2j}Non-perturbative correction (applied to NLO pQCD predictions) as a function of dijet rapidity separation of the leading two jets for events with two or more jets produced in association with a $W$ boson. The first number is the \textsc{siscone} to midpoint cone jet algorithm correction, the second is the correction for underlying event and hadronization effects, the third is the total correction factor.}
\begin{ruledtabular}
\begin{tabular}{rccc} 
Bin interval  & $\sigma^\mathrm{midpoint}_\mathrm{particle} / \sigma^\textsc{siscone}_\mathrm{particle}$  &  $\sigma^\textsc{siscone}_\mathrm{particle}/\sigma^\textsc{siscone}_\mathrm{parton}$ & Total correction  \\[1ex]
\hline
$0.0$~--~$0.4$ & $ 1.01$ & $1.04$ & $1.05\pm 0.04$ \\
$0.4$~--~$0.8$ & $ 1.00$ & $1.05$ & $1.06\pm 0.03$ \\
$0.8$~--~$1.2$ & $ 1.01$ & $1.02$ & $1.03\pm 0.03$ \\
$1.2$~--~$1.6$ & $ 1.02$ & $1.01$ & $1.02\pm 0.03$ \\
$1.6$~--~$2.1$ & $ 1.02$ & $1.02$ & $1.04\pm 0.03$ \\
$2.1$~--~$2.6$ & $ 1.02$ & $1.04$ & $1.07\pm 0.03$ \\
$2.6$~--~$3.2$ & $ 1.02$ & $1.08$ & $1.10\pm 0.04$ \\
$3.2$~--~$4.0$ & $ 1.01$ & $1.14$ & $1.15\pm 0.04$ \\
$4.0$~--~$5.0$ & $ 1.02$ & $1.29$ & $1.31\pm 0.07$ \\
$5.0$~--~$6.0$ & $ 1.03$ & $1.64$ & $1.69\pm 0.06$ \\
\end{tabular}
\end{ruledtabular}
\end{table}

\begin{table}[htbp]
\caption{\label{tab:NPC_dyFB_2j}Non-perturbative correction (applied to NLO pQCD predictions) as a function of dijet rapidity separation of the two most rapidity-separated jets for events with two or more jets produced in association with a $W$ boson. The first number is the \textsc{siscone} to midpoint cone jet algorithm correction, the second is the correction for underlying event and hadronization effects, the third is the total correction factor.}
\begin{ruledtabular}
%
\end{ruledtabular}
\end{table}

\begin{table}[htbp]
\caption{\label{tab:NPC_dy12_3j}Non-perturbative correction (applied to NLO pQCD predictions) as a function of dijet rapidity separation of the leading two jets for events with three or more jets produced in association with a $W$ boson. The first number is the \textsc{siscone} to midpoint cone jet algorithm correction, the second is the correction for underlying event and hadronization effects, the third is the total correction factor.}
\begin{ruledtabular}
%
\end{ruledtabular}
\end{table}

\begin{table}[htbp]
\caption{\label{tab:NPC_dyFB_3j}Non-perturbative correction (applied to NLO pQCD predictions) as a function of dijet rapidity separation of the two most rapidity-separated jets for events with three or more jets produced in association with a $W$ boson. The first number is the \textsc{siscone} to midpoint cone jet algorithm correction, the second is the correction for underlying event and hadronization effects, the third is the total correction factor.}
\begin{ruledtabular}
%
\end{ruledtabular}
\end{table}

\begin{table}[htbp]
\caption{\label{tab:NPC_dphi12}Non-perturbative correction (applied to NLO pQCD predictions) as a function of $\Delta\phi$ separation between the leading two jets for events with two or more jets produced in association with a $W$ boson. The first number is the \textsc{siscone} to midpoint cone jet algorithm correction, the second is the correction for underlying event and hadronization effects, the third is the total correction factor.}
\begin{ruledtabular}
%
\end{ruledtabular}
\end{table}

\clearpage
\newpage

\begin{table}[htbp]
\caption{\label{tab:NPC_dphiFB}Non-perturbative correction (applied to NLO pQCD predictions) as a function of $\Delta\phi$ separation between the two most rapidity-separated jets for events with two or more jets produced in association with a $W$ boson. The first number is the \textsc{siscone} to midpoint cone jet algorithm correction, the second is the correction for underlying event and hadronization effects, the third is the total correction factor.}
\begin{ruledtabular}
%
\end{ruledtabular}
\end{table}

\begin{table}[htbp]
\caption{\label{tab:NPC_dR}Non-perturbative correction (applied to NLO pQCD predictions) as a function of $\Delta R$ separation between the two leading jets for events with two or more jets produced in association with a $W$ boson. The first number is the \textsc{siscone} to midpoint cone jet algorithm correction, the second is the correction for underlying event and hadronization effects, the third is the total correction factor.}
\begin{ruledtabular}
%
\end{ruledtabular}
\end{table}

\begin{table}[htbp]
\caption{\label{tab:NPC_dy13_3j}Non-perturbative correction (applied to NLO pQCD predictions) as a function of the dijet rapidity separation of the leading and third jets for events with three or more jets produced in association with a $W$ boson. The first number is the \textsc{siscone} to midpoint cone jet algorithm correction, the second is the correction for underlying event and hadronization effects, the third is the total correction factor.}
\begin{ruledtabular}
%
\end{ruledtabular}
\end{table}

\begin{table}[htbp]
\caption{\label{tab:NPC_dy23_3j}Non-perturbative correction (applied to NLO pQCD predictions) as a function of the dijet rapidity separation of the second and third jets for events with three or more jets produced in association with a $W$ boson. The first number is the \textsc{siscone} to midpoint cone jet algorithm correction, the second is the correction for underlying event and hadronization effects, the third is the total correction factor.}
\begin{ruledtabular}
%
\end{ruledtabular}
\end{table}

\begin{table}[htbp]
\caption{\label{tab:NPC_W1pt}Non-perturbative correction (applied to NLO pQCD predictions) as a function of $W$ boson transverse momentum for events with one or more jets produced in association with a $W$ boson. The first number is the \textsc{siscone} to midpoint cone jet algorithm correction, the second is the correction for underlying event and hadronization effects, the third is the total correction factor.}
\begin{ruledtabular}
%
\end{ruledtabular}
\end{table}

\begin{table}[htbp]
\caption{\label{tab:NPC_W2pt}Non-perturbative correction (applied to NLO pQCD predictions) as a function of $W$ boson transverse momentum for events with two or more jets produced in association with a $W$ boson. The first number is the \textsc{siscone} to midpoint cone jet algorithm correction, the second is the correction for underlying event and hadronization effects, the third is the total correction factor.}
\begin{ruledtabular}
%
\end{ruledtabular}
\end{table}

\begin{table}[htbp]
\caption{\label{tab:NPC_W3pt}Non-perturbative correction (applied to NLO pQCD predictions) as a function of $W$ boson transverse momentum for events with three or more jets produced in association with a $W$ boson. The first number is the \textsc{siscone} to midpoint cone jet algorithm correction, the second is the correction for underlying event and hadronization effects, the third is the total correction factor.}
\begin{ruledtabular}
%
\end{ruledtabular}
\end{table}

\begin{table}[htbp]
\caption{\label{tab:NPC_W4pt}Non-perturbative correction (applied to NLO pQCD predictions) as a function of $W$ boson transverse momentum for events with four or more jets produced in association with a $W$ boson. The first number is the \textsc{siscone} to midpoint cone jet algorithm correction, the second is the correction for underlying event and hadronization effects, the third is the total correction factor.}
\begin{ruledtabular}
%
\end{ruledtabular}
\end{table}

\begin{table}[htbp]
\caption{\label{tab:NPC_dijetpt2}Non-perturbative correction (applied to NLO pQCD predictions) as a function of dijet transverse momentum for events with two or more jets produced in association with a $W$ boson. The first number is the \textsc{siscone} to midpoint cone jet algorithm correction, the second is the correction for underlying event and hadronization effects, the third is the total correction factor.}
\begin{ruledtabular}
%
\end{ruledtabular}
\end{table}

\begin{table}[htbp]
\caption{\label{tab:NPC_dijetpt3}Non-perturbative correction (applied to NLO pQCD predictions) as a function of dijet transverse momentum for events with three or more jets produced in association with a $W$ boson. The first number is the \textsc{siscone} to midpoint cone jet algorithm correction, the second is the correction for underlying event and hadronization effects, the third is the total correction factor.}
\begin{ruledtabular}
%
\end{ruledtabular}
\end{table}

\begin{table}[htbp]
\caption{\label{tab:NPC_dijetmass2}Non-perturbative correction (applied to NLO pQCD predictions) as a function of dijet invariant mass for events with two or more jets produced in association with a $W$ boson. The first number is the \textsc{siscone} to midpoint cone jet algorithm correction, the second is the correction for underlying event and hadronization effects, the third is the total correction factor.}
\begin{ruledtabular}
%
\end{ruledtabular}
\end{table}

\begin{table}[htbp]
\caption{\label{tab:NPC_dijetmass3}Non-perturbative correction (applied to NLO pQCD predictions) as a function of dijet invariant mass for events with three or more jets produced in association with a $W$ boson. The first number is the \textsc{siscone} to midpoint cone jet algorithm correction, the second is the correction for underlying event and hadronization effects, the third is the total correction factor.}
\begin{ruledtabular}
%
\end{ruledtabular}
\end{table}

\newpage
\clearpage

\begin{table}[htbp]
\caption{\label{tab:NPC_HT1}Non-perturbative correction (applied to NLO pQCD predictions) as a function of the scalar sum of the transverse energies of the W boson and all jets for events with one or more jets produced in association with a $W$ boson. The first number is the \textsc{siscone} to midpoint cone jet algorithm correction, the second is the correction for underlying event and hadronization effects, the third is the total correction factor.}
\begin{ruledtabular}
%
\end{ruledtabular}
\end{table}

\begin{table}[htbp]
\caption{\label{tab:NPC_HT2}Non-perturbative correction (applied to NLO pQCD predictions) as a function of the scalar sum of the transverse energies of the W boson and all jets for events with two or more jets produced in association with a $W$ boson. The first number is the \textsc{siscone} to midpoint cone jet algorithm correction, the second is the correction for underlying event and hadronization effects, the third is the total correction factor.}
\begin{ruledtabular}
%
\end{ruledtabular}
\end{table}

\begin{table}[htbp]
\caption{\label{tab:NPC_HT3}Non-perturbative correction (applied to NLO pQCD predictions) as a function of the scalar sum of the transverse energies of the W boson and all jets for events with three or more jets produced in association with a $W$ boson. The first number is the \textsc{siscone} to midpoint cone jet algorithm correction, the second is the correction for underlying event and hadronization effects, the third is the total correction factor.}
\begin{ruledtabular}
%
\end{ruledtabular}
\end{table}

\begin{table}[htbp]
\caption{\label{tab:NPC_HT4}Non-perturbative correction (applied to NLO pQCD predictions) as a function of the scalar sum of the transverse energies of the W boson and all jets for events with four or more jets produced in association with a $W$ boson. The first number is the \textsc{siscone} to midpoint cone jet algorithm correction, the second is the correction for underlying event and hadronization effects, the third is the total correction factor.}
\begin{ruledtabular}
%
\end{ruledtabular}
\end{table}

\begin{table}[htbp]
\caption{\label{tab:NPC_njet_HT1}Non-perturbative correction (applied to NLO pQCD predictions) as a function of the mean number of jets with $p_T>20$~GeV and $|y|<3.2$ as a function of the scalar sum of the transverse energies of the W boson and all jets for events with one or more jets produced in association with a $W$ boson. The first number is the \textsc{siscone} to midpoint cone jet algorithm correction, the second is the correction for underlying event and hadronization effects, the third is the total correction factor.}
\begin{ruledtabular}
\begin{tabular}{rccc} 
Bin interval  & $\sigma^\mathrm{midpoint}_\mathrm{particle} / \sigma^\textsc{siscone}_\mathrm{particle}$  &  $\sigma^\textsc{siscone}_\mathrm{particle}/\sigma^\textsc{siscone}_\mathrm{parton}$ & Total correction  \\[1ex]
\hline
$100.0$~--~$120.0$ & $ 1.00$ & $1.00$ & $1.00\pm 0.00$ \\
$120.0$~--~$140.0$ & $ 1.00$ & $1.03$ & $1.03\pm 0.00$ \\
$140.0$~--~$170.0$ & $ 1.00$ & $1.02$ & $1.02\pm 0.00$ \\
$170.0$~--~$200.0$ & $ 1.00$ & $1.01$ & $1.02\pm 0.01$ \\
$200.0$~--~$245.0$ & $ 1.00$ & $1.01$ & $1.01\pm 0.01$ \\
$245.0$~--~$300.0$ & $ 1.00$ & $1.01$ & $1.01\pm 0.01$ \\
$300.0$~--~$360.0$ & $ 0.99$ & $1.01$ & $1.01\pm 0.01$ \\
$360.0$~--~$470.0$ & $ 0.99$ & $1.01$ & $1.00\pm 0.01$ \\
\end{tabular}
\end{ruledtabular}
\end{table}

\begin{table}[htbp]
\caption{\label{tab:NPC_njet_HT2}Non-perturbative correction (applied to NLO pQCD predictions) as a function of the mean number of jets with $p_T>20$~GeV and $|y|<3.2$ as a function of the scalar sum of the transverse energies of the W boson and all jets for events with two or more jets produced in association with a $W$ boson. The first number is the \textsc{siscone} to midpoint cone jet algorithm correction, the second is the correction for underlying event and hadronization effects, the third is the total correction factor.}
\begin{ruledtabular}
\begin{tabular}{rccc} 
Bin interval  & $\sigma^\mathrm{midpoint}_\mathrm{particle} / \sigma^\textsc{siscone}_\mathrm{particle}$  &  $\sigma^\textsc{siscone}_\mathrm{particle}/\sigma^\textsc{siscone}_\mathrm{parton}$ & Total correction  \\[1ex]
\hline
$120.0$~--~$135.0$ & $ 1.00$ & $1.00$ & $1.00\pm 0.00$ \\
$135.0$~--~$155.0$ & $ 1.00$ & $1.01$ & $1.01\pm 0.00$ \\
$155.0$~--~$185.0$ & $ 1.00$ & $1.01$ & $1.00\pm 0.00$ \\
$185.0$~--~$230.0$ & $ 1.00$ & $1.01$ & $1.01\pm 0.01$ \\
$230.0$~--~$280.0$ & $ 1.00$ & $1.01$ & $1.01\pm 0.01$ \\
$280.0$~--~$340.0$ & $ 1.00$ & $1.01$ & $1.01\pm 0.01$ \\
$340.0$~--~$390.0$ & $ 0.99$ & $1.01$ & $1.01\pm 0.01$ \\
$390.0$~--~$460.0$ & $ 0.99$ & $1.01$ & $1.00\pm 0.01$ \\
\end{tabular}
\end{ruledtabular}
\end{table}

\begin{table}[htbp]
  \caption{\label{tab:NPC_njet_dy12}Non-perturbative correction (applied to NLO pQCD predictions) as a function of the mean number of jets with $p_T>20$~GeV and $|y|<3.2$ as a function of the dijet rapidity separation of the two leading $p_T$ jets for events with two or more jets produced in association with a $W$ boson. The first number is the \textsc{siscone} to midpoint cone jet algorithm correction, the second is the correction for underlying event and hadronization effects, the third is the total correction factor.}
\begin{ruledtabular}
\begin{tabular}{rccc} 
Bin interval  & $\sigma^\mathrm{midpoint}_\mathrm{particle} / \sigma^\textsc{siscone}_\mathrm{particle}$  &  $\sigma^\textsc{siscone}_\mathrm{particle}/\sigma^\textsc{siscone}_\mathrm{parton}$ & Total correction  \\[1ex]
\hline
$0.0$~--~$0.4$ & $ 1.00$ & $1.01$ & $1.00\pm 0.00$ \\
$0.4$~--~$0.8$ & $ 1.00$ & $1.00$ & $1.00\pm 0.00$ \\
$0.8$~--~$1.2$ & $ 1.00$ & $1.01$ & $1.01\pm 0.00$ \\
$1.2$~--~$1.6$ & $ 1.00$ & $1.00$ & $1.00\pm 0.00$ \\
$1.6$~--~$2.1$ & $ 1.00$ & $1.00$ & $1.00\pm 0.00$ \\
$2.1$~--~$2.6$ & $ 1.00$ & $1.00$ & $1.00\pm 0.00$ \\
$2.6$~--~$3.2$ & $ 1.00$ & $1.00$ & $1.00\pm 0.00$ \\
$3.2$~--~$4.0$ & $ 1.00$ & $1.00$ & $1.00\pm 0.00$ \\
$4.0$~--~$5.0$ & $ 1.00$ & $1.00$ & $1.00\pm 0.00$ \\
$5.0$~--~$6.0$ & $ 1.01$ & $0.99$ & $1.00\pm 0.00$ \\
\end{tabular}
\end{ruledtabular}
\end{table}

\begin{table}[htbp]
\caption{\label{tab:NPC_njet_dyFB}Non-perturbative correction (applied to NLO pQCD predictions) as a function of the mean number of jets with $p_T>20$~GeV and $|y|<3.2$ as a function of the dijet rapidity separation of the most rapidity-separated jets for events with two or more jets produced in association with a $W$ boson. The first number is the \textsc{siscone} to midpoint cone jet algorithm correction, the second is the correction for underlying event and hadronization effects, the third is the total correction factor.}
\begin{ruledtabular}
\begin{tabular}{rccc} 
Bin interval  & $\sigma^\mathrm{midpoint}_\mathrm{particle} / \sigma^\textsc{siscone}_\mathrm{particle}$  &  $\sigma^\textsc{siscone}_\mathrm{particle}/\sigma^\textsc{siscone}_\mathrm{parton}$ & Total correction  \\[1ex]
\hline
$0.0$~--~$0.4$ & $ 1.00$ & $1.00$ & $1.00\pm 0.00$ \\
$0.4$~--~$0.8$ & $ 1.00$ & $1.00$ & $1.00\pm 0.00$ \\
$0.8$~--~$1.2$ & $ 1.00$ & $1.00$ & $1.00\pm 0.00$ \\
$1.2$~--~$1.6$ & $ 1.00$ & $1.01$ & $1.00\pm 0.00$ \\
$1.6$~--~$2.1$ & $ 1.00$ & $1.00$ & $1.00\pm 0.00$ \\
$2.1$~--~$2.6$ & $ 1.00$ & $1.00$ & $1.00\pm 0.00$ \\
$2.6$~--~$3.2$ & $ 1.00$ & $1.00$ & $1.00\pm 0.00$ \\
$3.2$~--~$4.0$ & $ 1.00$ & $1.00$ & $1.00\pm 0.00$ \\
$4.0$~--~$5.0$ & $ 1.00$ & $0.99$ & $0.99\pm 0.00$ \\
$5.0$~--~$6.0$ & $ 1.00$ & $0.99$ & $1.00\pm 0.01$ \\
\end{tabular}
\end{ruledtabular}
\end{table}

\begin{table}[htbp]
\caption{\label{tab:NPC_prob_dy12}Non-perturbative correction (applied to NLO pQCD predictions) as a function of third jet emission probability as a function of dijet rapidity separation of the two leading $p_T$ jets for events with two or more jets produced in association with a $W$ boson. The first number is the \textsc{siscone} to midpoint cone jet algorithm correction, the second is the correction for underlying event and hadronization effects, the third is the total correction factor.}
\begin{ruledtabular}
\begin{tabular}{rccc} 
Bin interval  & $\sigma^\mathrm{midpoint}_\mathrm{particle} / \sigma^\textsc{siscone}_\mathrm{particle}$  &  $\sigma^\textsc{siscone}_\mathrm{particle}/\sigma^\textsc{siscone}_\mathrm{parton}$ & Total correction  \\[1ex]
\hline
$0.0$~--~$0.4$ & $ 0.98$ & $1.06$ & $1.04\pm 0.01$ \\
$0.4$~--~$0.8$ & $ 1.00$ & $1.04$ & $1.03\pm 0.00$ \\
$0.8$~--~$1.2$ & $ 0.99$ & $1.08$ & $1.06\pm 0.02$ \\
$1.2$~--~$1.6$ & $ 0.99$ & $1.04$ & $1.02\pm 0.01$ \\
$1.6$~--~$2.1$ & $ 0.98$ & $1.04$ & $1.03\pm 0.03$ \\
$2.1$~--~$2.6$ & $ 1.02$ & $1.00$ & $1.02\pm 0.01$ \\
$2.6$~--~$3.2$ & $ 1.00$ & $1.05$ & $1.05\pm 0.03$ \\
$3.2$~--~$4.0$ & $ 0.99$ & $1.04$ & $1.03\pm 0.01$ \\
$4.0$~--~$5.0$ & $ 1.04$ & $0.94$ & $0.98\pm 0.03$ \\
\end{tabular}
\end{ruledtabular}
\end{table}

\begin{table}[htbp]
\caption{\label{tab:NPC_prob_dyFB}Non-perturbative correction (applied to NLO pQCD predictions) as a function of third jet emission probability as a function of dijet rapidity separation of the two most rapidity-separated jets for events with two or more jets produced in association with a $W$ boson. The first number is the \textsc{siscone} to midpoint cone jet algorithm correction, the second is the correction for underlying event and hadronization effects, the third is the total correction factor.}
\begin{ruledtabular}
\begin{tabular}{rccc} 
Bin interval  & $\sigma^\mathrm{midpoint}_\mathrm{particle} / \sigma^\textsc{siscone}_\mathrm{particle}$  &  $\sigma^\textsc{siscone}_\mathrm{particle}/\sigma^\textsc{siscone}_\mathrm{parton}$ & Total correction  \\[1ex]
\hline
$0.0$~--~$0.4$ & $ 0.96$ & $1.04$ & $1.00\pm 0.00$ \\
$0.4$~--~$0.8$ & $ 0.98$ & $1.05$ & $1.03\pm 0.01$ \\
$0.8$~--~$1.2$ & $ 0.98$ & $1.09$ & $1.07\pm 0.04$ \\
$1.2$~--~$1.6$ & $ 0.97$ & $1.06$ & $1.02\pm 0.01$ \\
$1.6$~--~$2.1$ & $ 0.99$ & $1.04$ & $1.03\pm 0.02$ \\
$2.1$~--~$2.6$ & $ 0.99$ & $1.01$ & $1.00\pm 0.02$ \\
$2.6$~--~$3.2$ & $ 0.99$ & $1.02$ & $1.01\pm 0.02$ \\
$3.2$~--~$4.0$ & $ 1.01$ & $0.98$ & $0.99\pm 0.02$ \\
$4.0$~--~$5.0$ & $ 1.01$ & $0.93$ & $0.94\pm 0.03$ \\
$5.0$~--~$6.0$ & $ 1.02$ & $0.95$ & $0.97\pm 0.06$ \\
\end{tabular}
\end{ruledtabular}
\end{table}

\begin{table}[htbp]
\caption{\label{tab:NPC_prob_dy12restrict}Non-perturbative correction (applied to NLO pQCD predictions) as a function of third jet emission probability as a function of dijet rapidity separation of the two leading $p_T$ jets, with a requirement that the third jet be emitted between the leading two jets in rapidity, for events with two or more jets produced in association with a $W$ boson. The first number is the \textsc{siscone} to midpoint cone jet algorithm correction, the second is the correction for underlying event and hadronization effects, the third is the total correction factor.}
\begin{ruledtabular}
\begin{tabular}{rccc} 
Bin interval  & $\sigma^\mathrm{midpoint}_\mathrm{particle} / \sigma^\textsc{siscone}_\mathrm{particle}$  &  $\sigma^\textsc{siscone}_\mathrm{particle}/\sigma^\textsc{siscone}_\mathrm{parton}$ & Total correction  \\[1ex]
\hline
$0.0$~--~$0.4$ & $ 0.95$ & $1.04$ & $0.99\pm 0.00$ \\
$0.4$~--~$0.8$ & $ 0.99$ & $1.08$ & $1.08\pm 0.04$ \\
$0.8$~--~$1.2$ & $ 0.98$ & $1.11$ & $1.09\pm 0.05$ \\
$1.2$~--~$1.6$ & $ 0.99$ & $1.05$ & $1.03\pm 0.01$ \\
$1.6$~--~$2.1$ & $ 1.00$ & $1.03$ & $1.03\pm 0.03$ \\
$2.1$~--~$2.6$ & $ 1.01$ & $0.99$ & $1.00\pm 0.01$ \\
$2.6$~--~$3.2$ & $ 0.99$ & $1.03$ & $1.02\pm 0.04$ \\
$3.2$~--~$4.0$ & $ 0.98$ & $1.02$ & $1.00\pm 0.01$ \\
$4.0$~--~$5.0$ & $ 1.05$ & $0.90$ & $0.95\pm 0.06$ \\
\end{tabular}
\end{ruledtabular}
\end{table}

\begin{table}[htbp]
\caption{\label{tab:NPC_gap_dy12}Non-perturbative correction (applied to NLO pQCD predictions) to the measured fraction of $W+2\textrm{-jet}$ events having no additional jet emission with $p_T>20$~GeV and $|y|<3.2$ into the rapidity interval defined by the two highest $p_T$ jets as a function of the rapidity interval.
The first number is the \textsc{siscone} to midpoint cone jet algorithm correction, the second is the correction for underlying event and hadronization effects, the third is the total correction factor.}
\begin{ruledtabular}
\begin{tabular}{rccc} 
Bin interval  & $\sigma^\mathrm{midpoint}_\mathrm{particle} / \sigma^\textsc{siscone}_\mathrm{particle}$  &  $\sigma^\textsc{siscone}_\mathrm{particle}/\sigma^\textsc{siscone}_\mathrm{parton}$ & Total correction  \\[1ex]
\hline
$0.0$~--~$0.4$ & $ 0.95$ & $1.04$ & $0.99\pm 0.00$ \\
$0.4$~--~$0.8$ & $ 0.99$ & $1.08$ & $1.08\pm 0.04$ \\
$0.8$~--~$1.2$ & $ 0.98$ & $1.11$ & $1.09\pm 0.05$ \\
$1.2$~--~$1.6$ & $ 0.99$ & $1.05$ & $1.03\pm 0.01$ \\
$1.6$~--~$2.1$ & $ 1.00$ & $1.03$ & $1.03\pm 0.03$ \\
$2.1$~--~$2.6$ & $ 1.01$ & $0.99$ & $1.00\pm 0.01$ \\
$2.6$~--~$3.2$ & $ 0.99$ & $1.03$ & $1.02\pm 0.04$ \\
$3.2$~--~$4.0$ & $ 0.98$ & $1.02$ & $1.00\pm 0.01$ \\
$4.0$~--~$5.0$ & $ 1.05$ & $0.90$ & $0.95\pm 0.06$ \\
$5.0$~--~$6.0$ & $ 1.05$ & $0.90$ & $0.95\pm 0.00$ \\
\end{tabular}
\end{ruledtabular}
\end{table}

\begin{table}[htbp]
\caption{\label{tab:NPC_gap_dyFB}Non-perturbative correction (applied to NLO pQCD predictions) to the measured fraction of $W+2\textrm{-jet}$ events having no additional jet emission with $p_T>20$~GeV and $|y|<3.2$ into the rapidity interval defined by the two most rapidity-separated jets as a function of the rapidity interval.
The first number is the \textsc{siscone} to midpoint cone jet algorithm correction, the second is the correction for underlying event and hadronization effects, the third is the total correction factor.}
\begin{ruledtabular}
\begin{tabular}{rccc} 
Bin interval  & $\sigma^\mathrm{midpoint}_\mathrm{particle} / \sigma^\textsc{siscone}_\mathrm{particle}$  &  $\sigma^\textsc{siscone}_\mathrm{particle}/\sigma^\textsc{siscone}_\mathrm{parton}$ & Total correction  \\[1ex]
\hline
$0.0$~--~$0.4$ & $ 0.96$ & $1.04$ & $1.00\pm 0.00$ \\
$0.4$~--~$0.8$ & $ 0.98$ & $1.05$ & $1.03\pm 0.01$ \\
$0.8$~--~$1.2$ & $ 0.98$ & $1.09$ & $1.07\pm 0.04$ \\
$1.2$~--~$1.6$ & $ 0.97$ & $1.06$ & $1.02\pm 0.01$ \\
$1.6$~--~$2.1$ & $ 0.99$ & $1.04$ & $1.03\pm 0.02$ \\
$2.1$~--~$2.6$ & $ 0.99$ & $1.01$ & $1.00\pm 0.02$ \\
$2.6$~--~$3.2$ & $ 0.99$ & $1.02$ & $1.01\pm 0.02$ \\
$3.2$~--~$4.0$ & $ 1.01$ & $0.98$ & $0.99\pm 0.02$ \\
$4.0$~--~$5.0$ & $ 1.01$ & $0.93$ & $0.94\pm 0.03$ \\
$5.0$~--~$6.0$ & $ 1.02$ & $0.95$ & $0.97\pm 0.06$ \\
\end{tabular}
\end{ruledtabular}
\end{table}

\clearpage
\newpage
\section{Statistical Uncertainty Correlations On Unfolded Distributions}

\begin{center}
{\bf To appear as an Electronic Physics Auxiliary Publication (EPAPS)}
\end{center}

In this appendix, we provide the bin-to-bin correlations on the statistical uncertainties for each measurement made in the 
main text of the paper, in both graphical and tabular format.    These correlations are constructed from the covariance matrix 
of the unfolded data, where the correlation $corr_{i,j} = \frac{cov_{i,j}}{\sqrt{cov_{i,i}}\sqrt{cov_{j,j}}}$ and $cov_{x,y}$ 
represents the covariance between each bin $x$ and $y$.   
Each correlation matrix is symmetric and has value one along the diagonal by construction.   
Numerical values of the statistical correlations are given in white font on the plots to assist the reader in resolving the colors.

\begin{figure}[htbp]
  \begin{center}
    \includegraphics[width=0.35\textwidth]{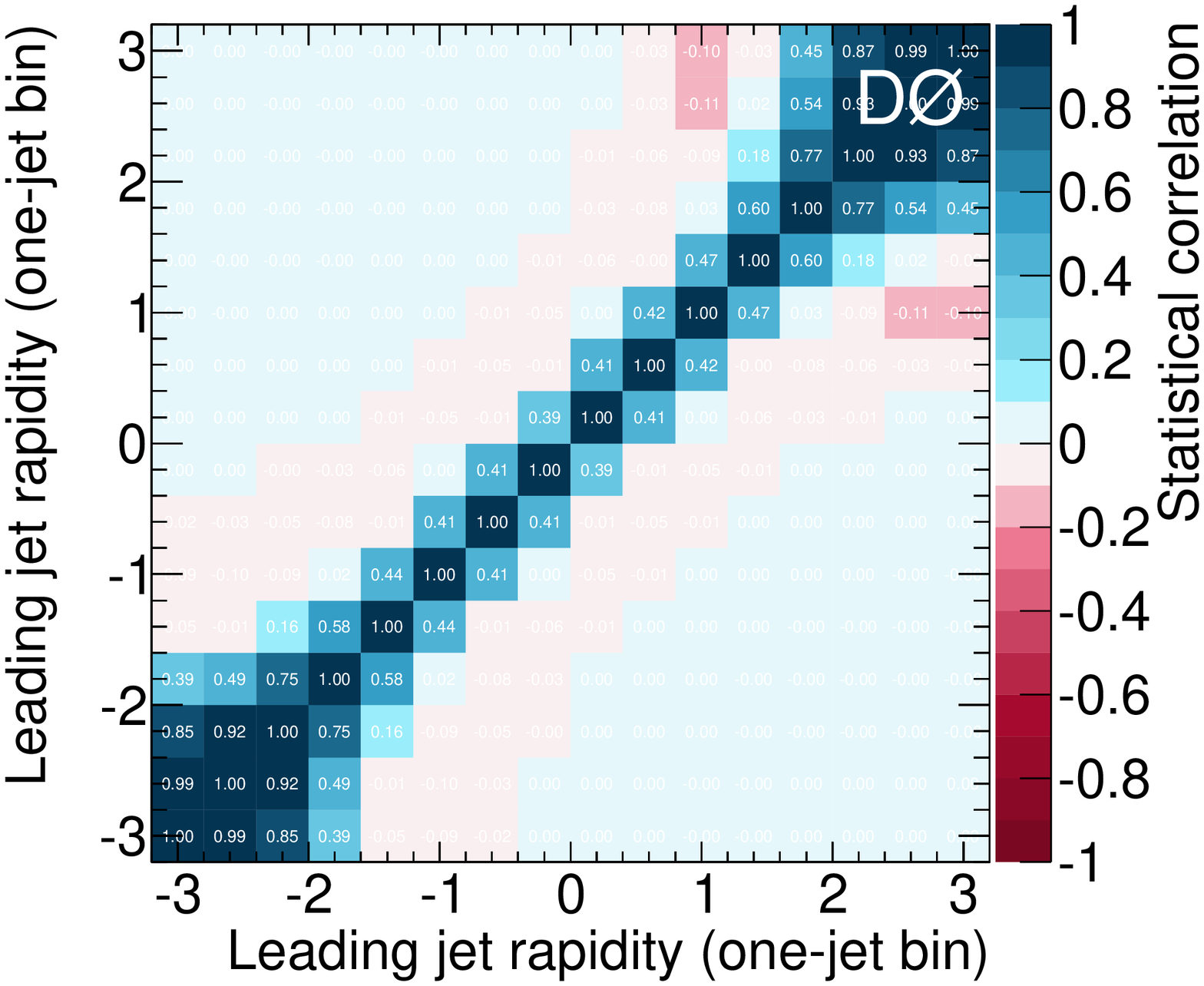}
    \includegraphics[width=0.35\textwidth]{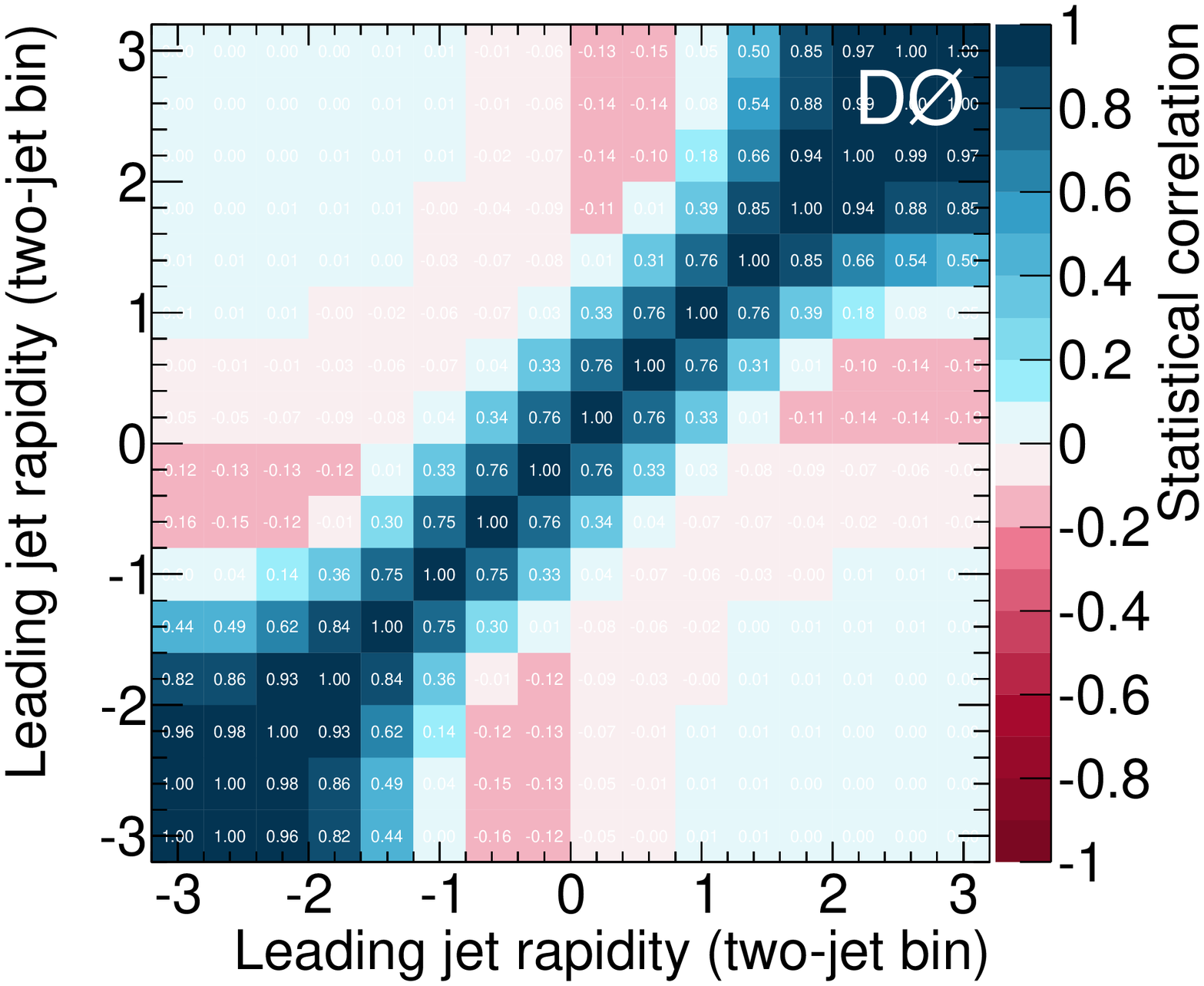}
    \includegraphics[width=0.35\textwidth]{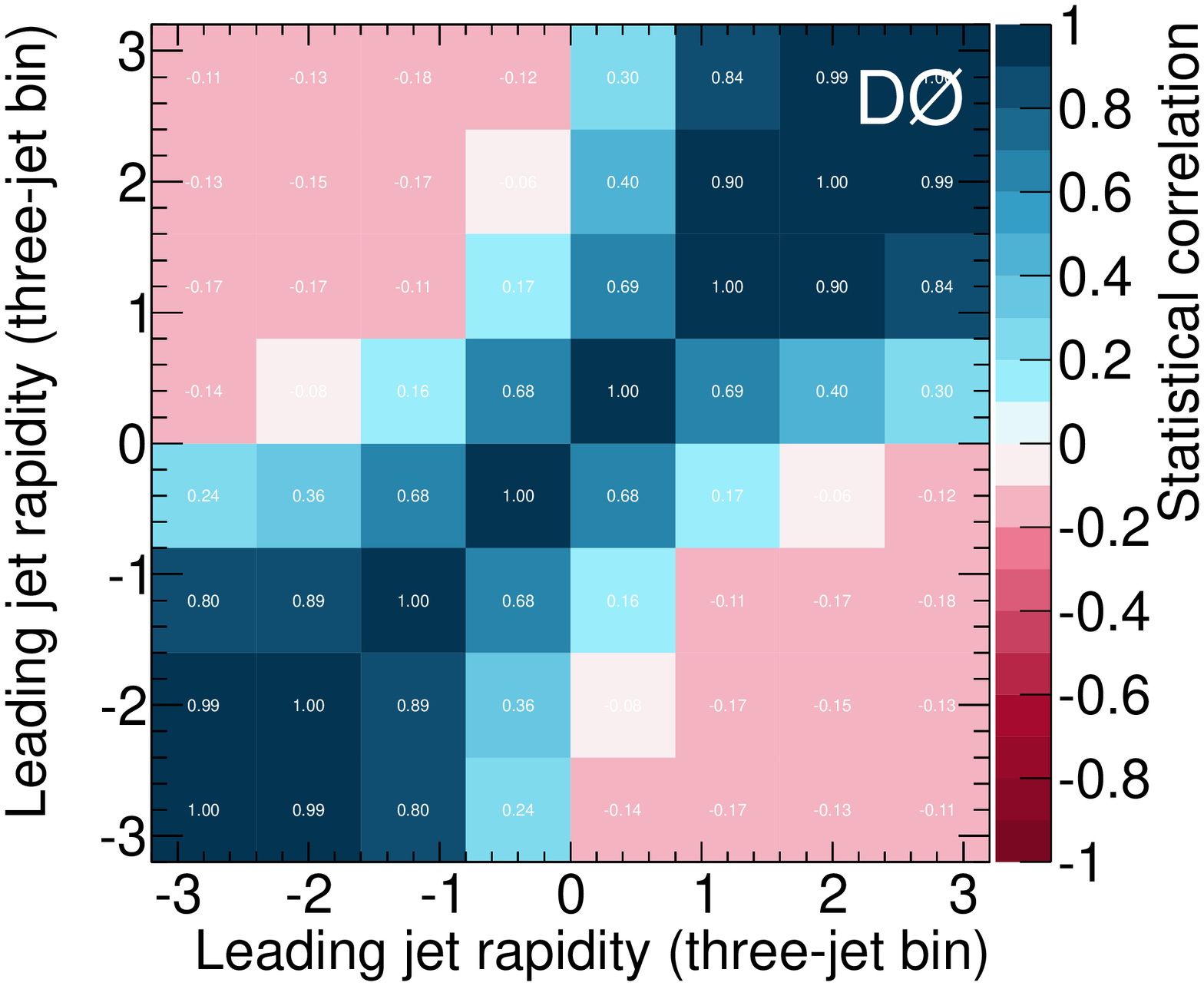}
    \includegraphics[width=0.35\textwidth]{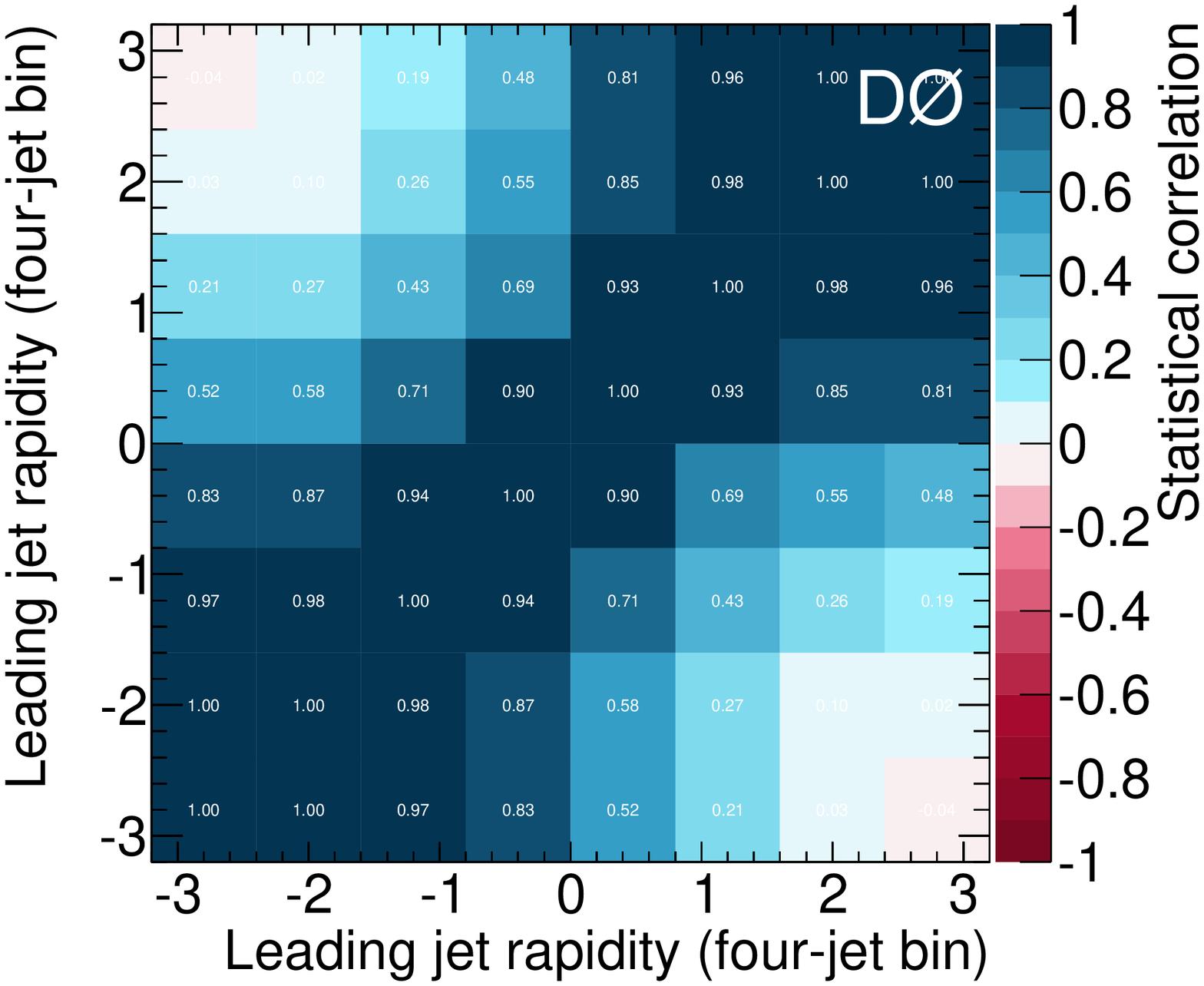}
    \caption{Statistical uncertainty correlations between bins for unfolded jet rapidity distributions.
     \label{fig:correlations_jeteta}
    }
  \end{center}
\end{figure}

\begin{figure}[htbp]
  \begin{center}
    \includegraphics[width=0.35\textwidth]{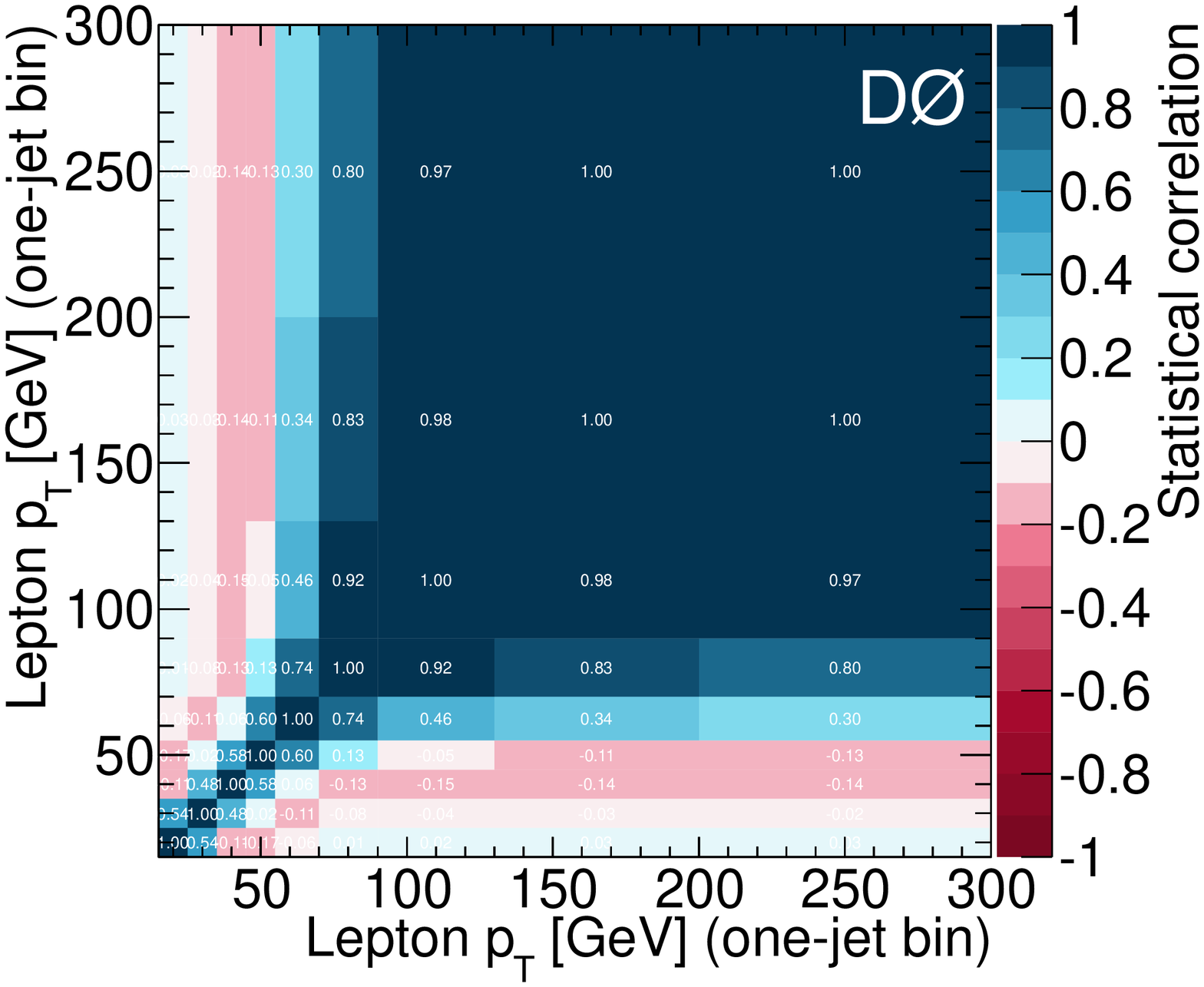}
    \includegraphics[width=0.35\textwidth]{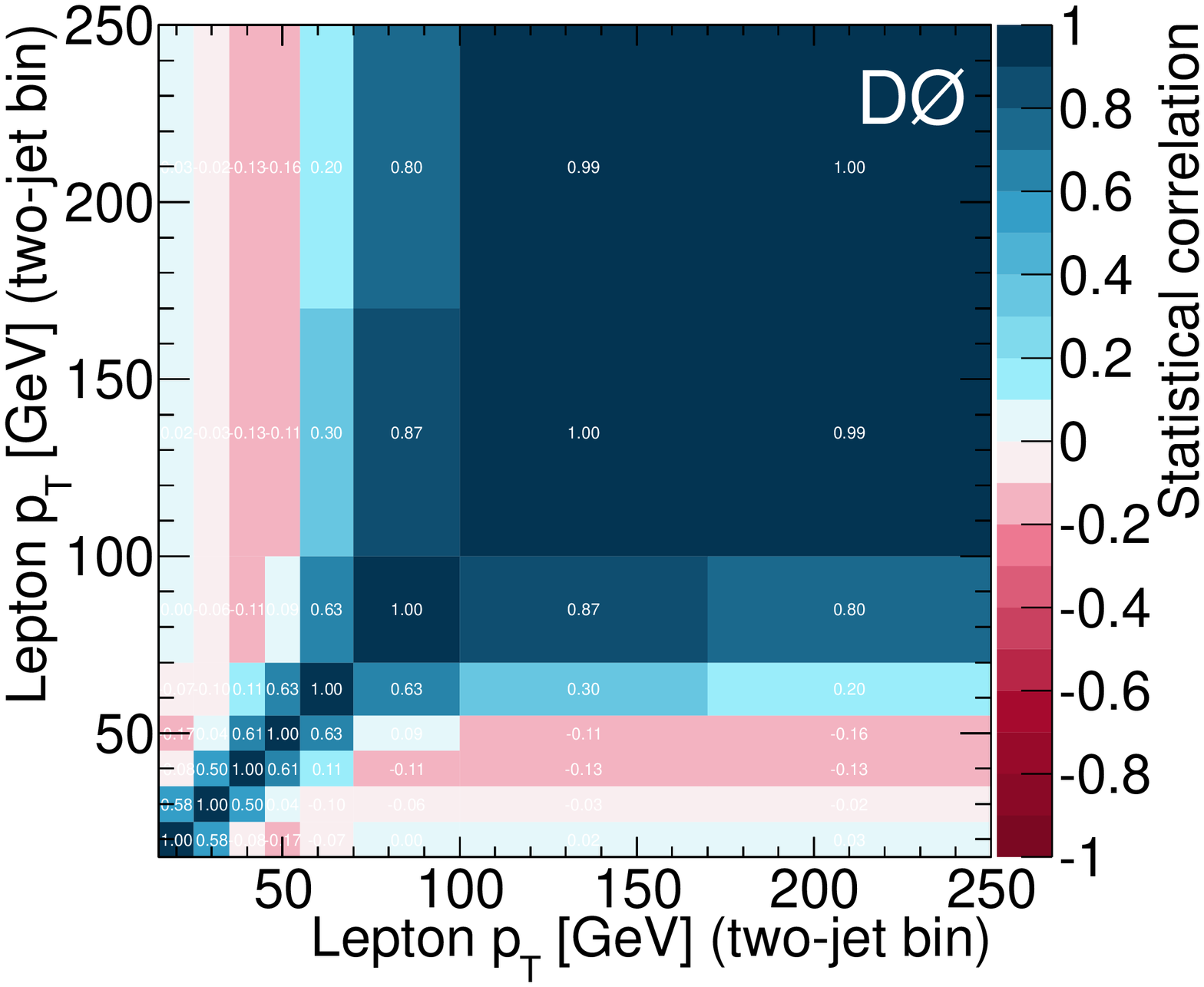}
    \includegraphics[width=0.35\textwidth]{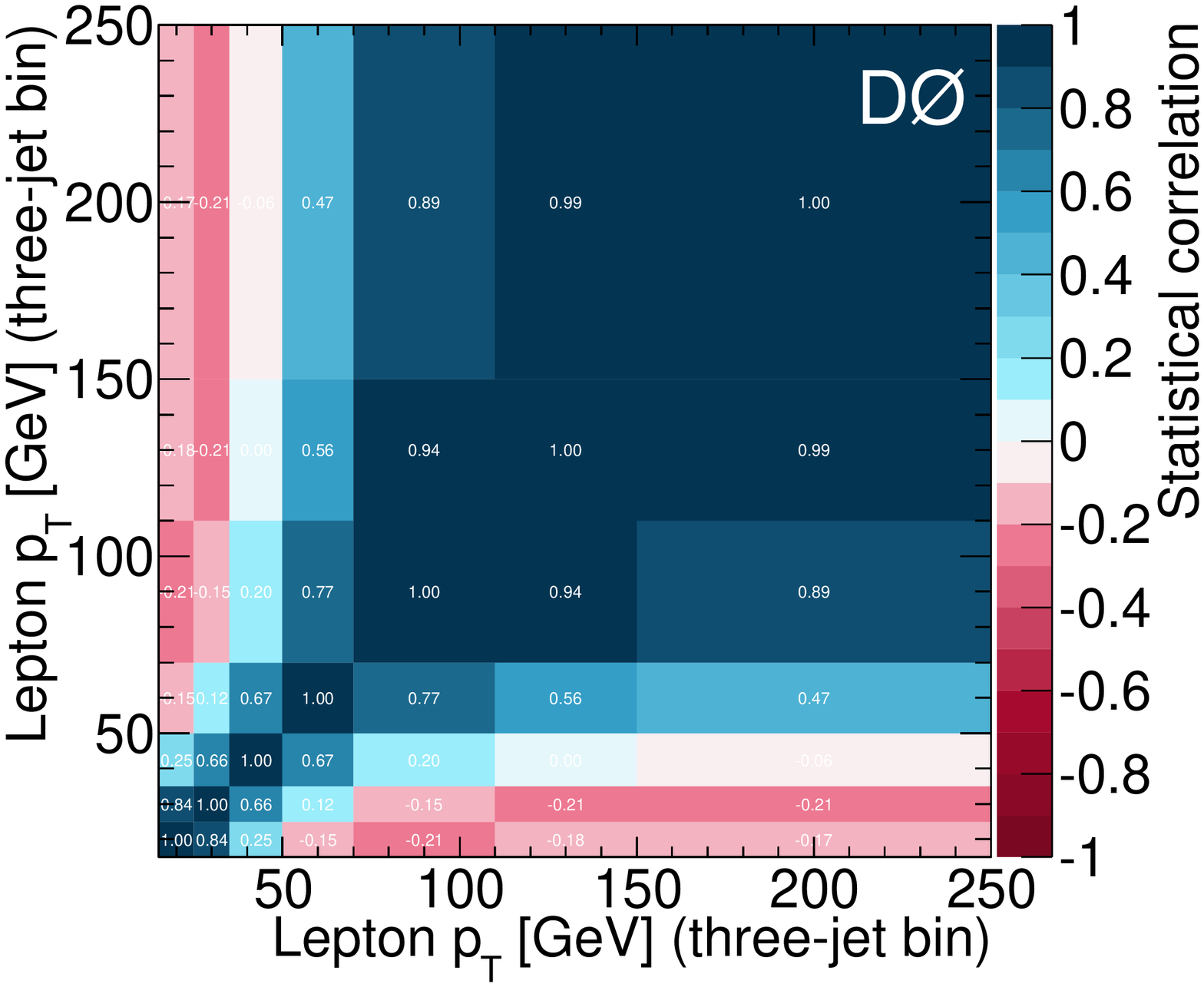}
    \includegraphics[width=0.35\textwidth]{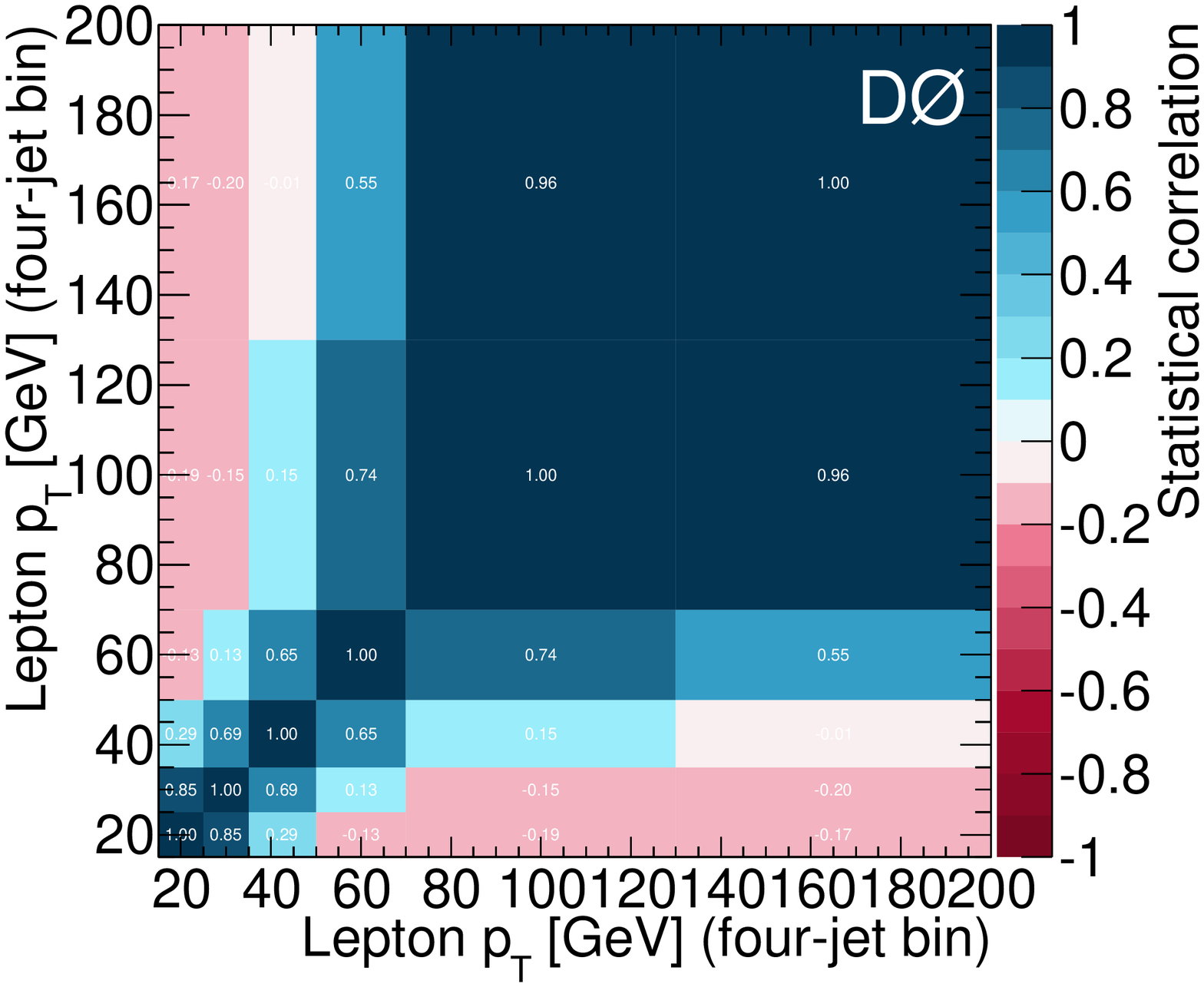}
    \caption{Statistical uncertainty correlations between bins for unfolded lepton transverse momentum distributions.
     \label{fig:correlations_leptonpt}
    }
  \end{center}
\end{figure}

\begin{figure}[htbp]
  \begin{center}
    \includegraphics[width=0.35\textwidth]{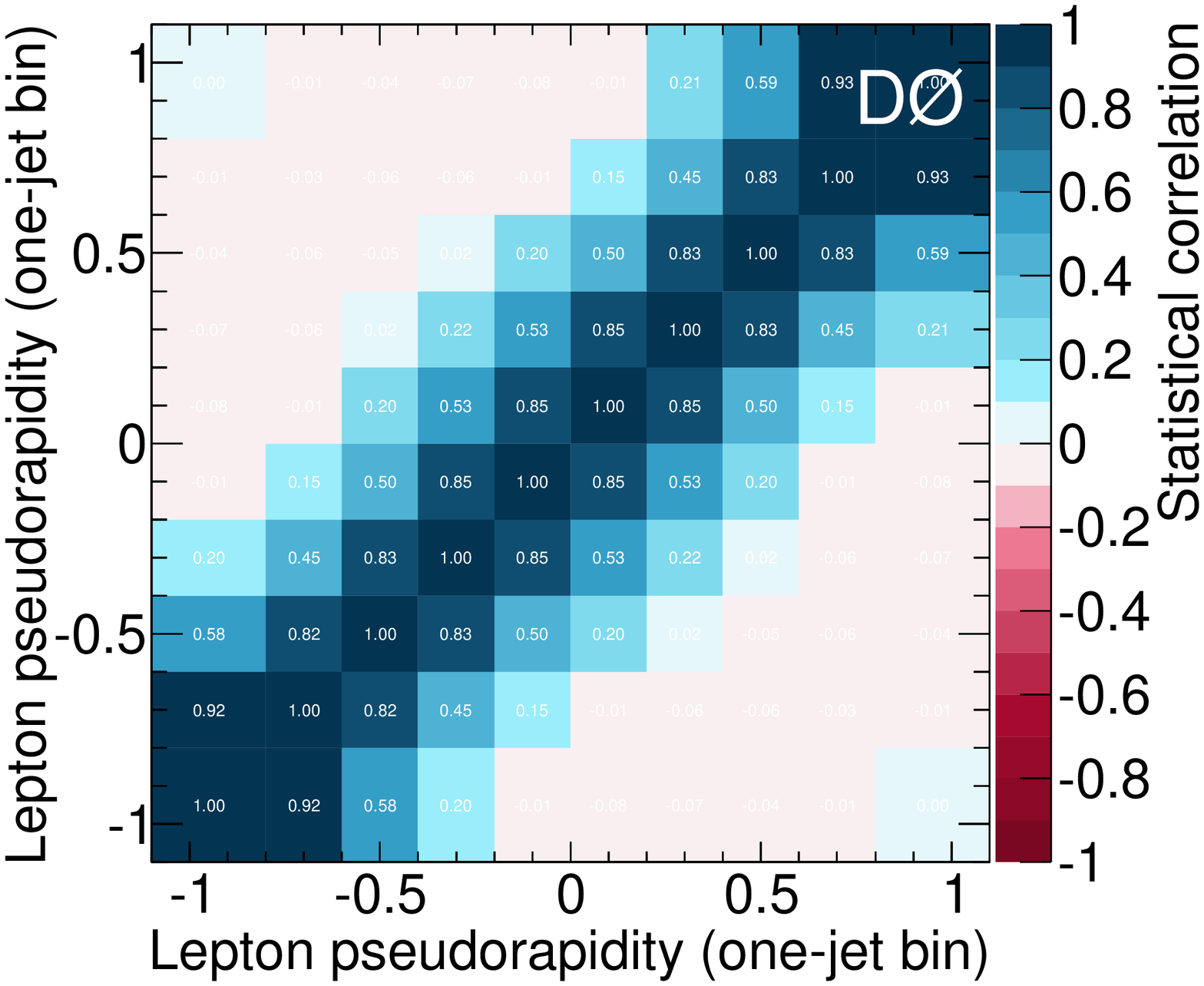}
    \includegraphics[width=0.35\textwidth]{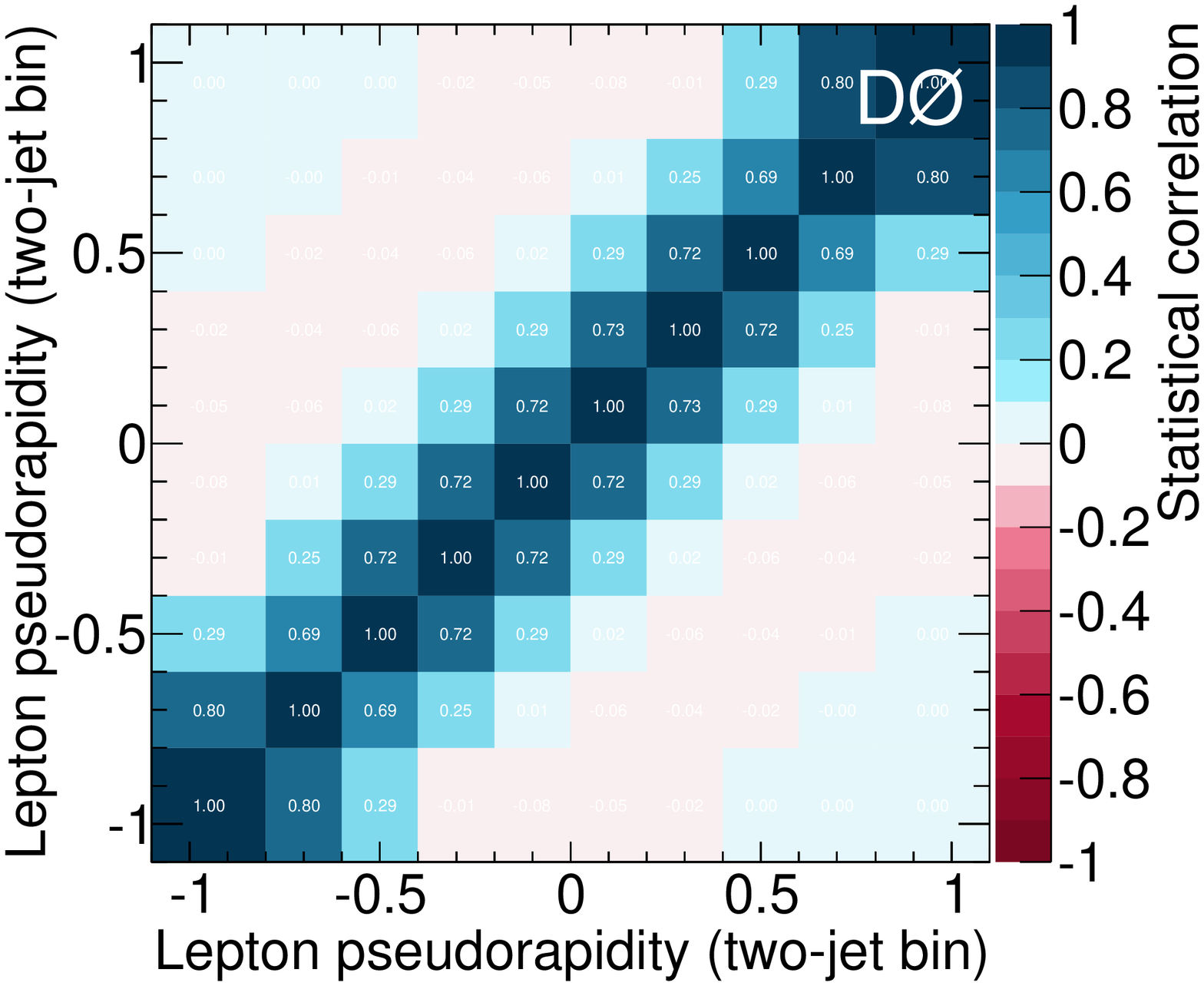}
    \includegraphics[width=0.35\textwidth]{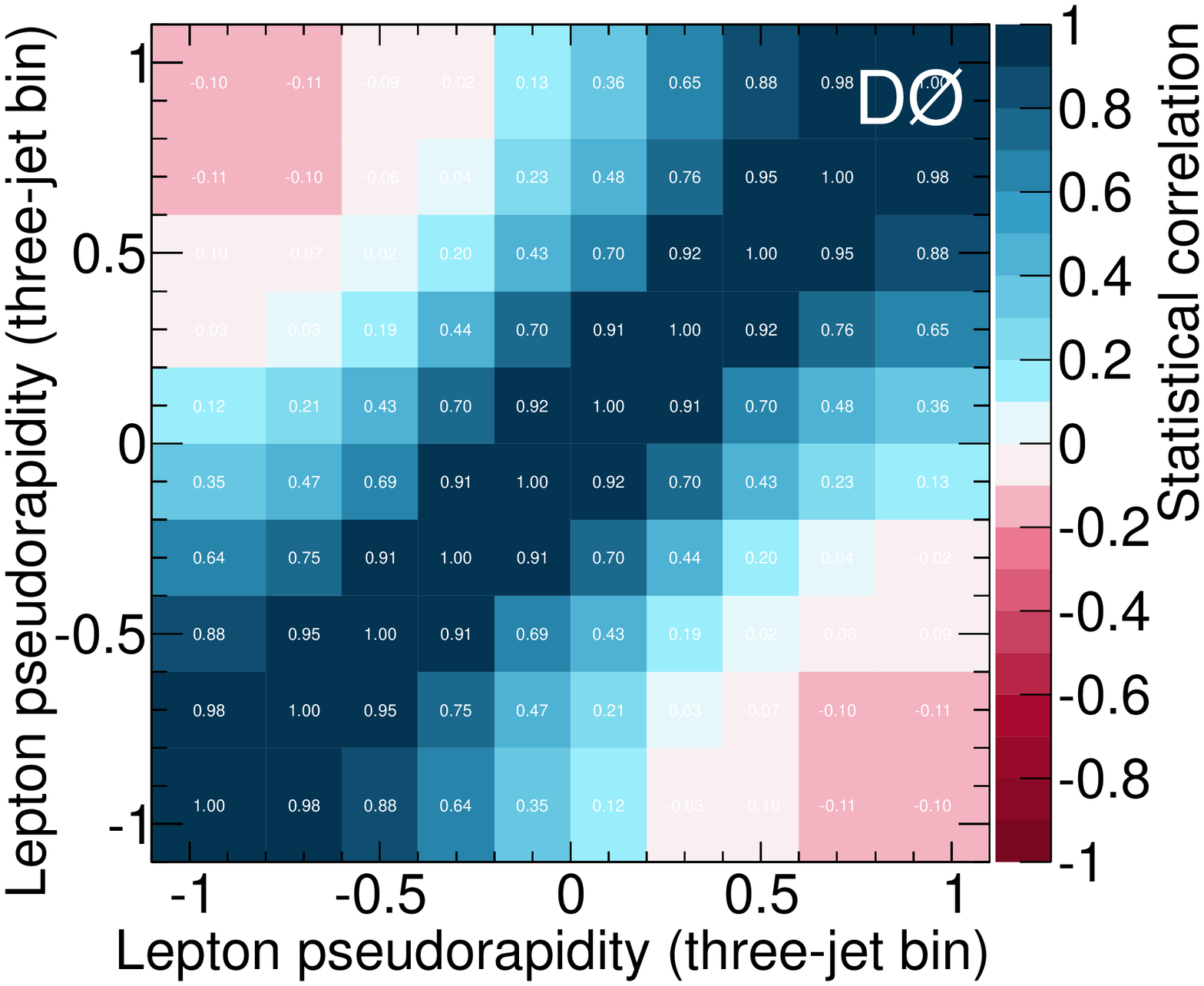}
    \includegraphics[width=0.35\textwidth]{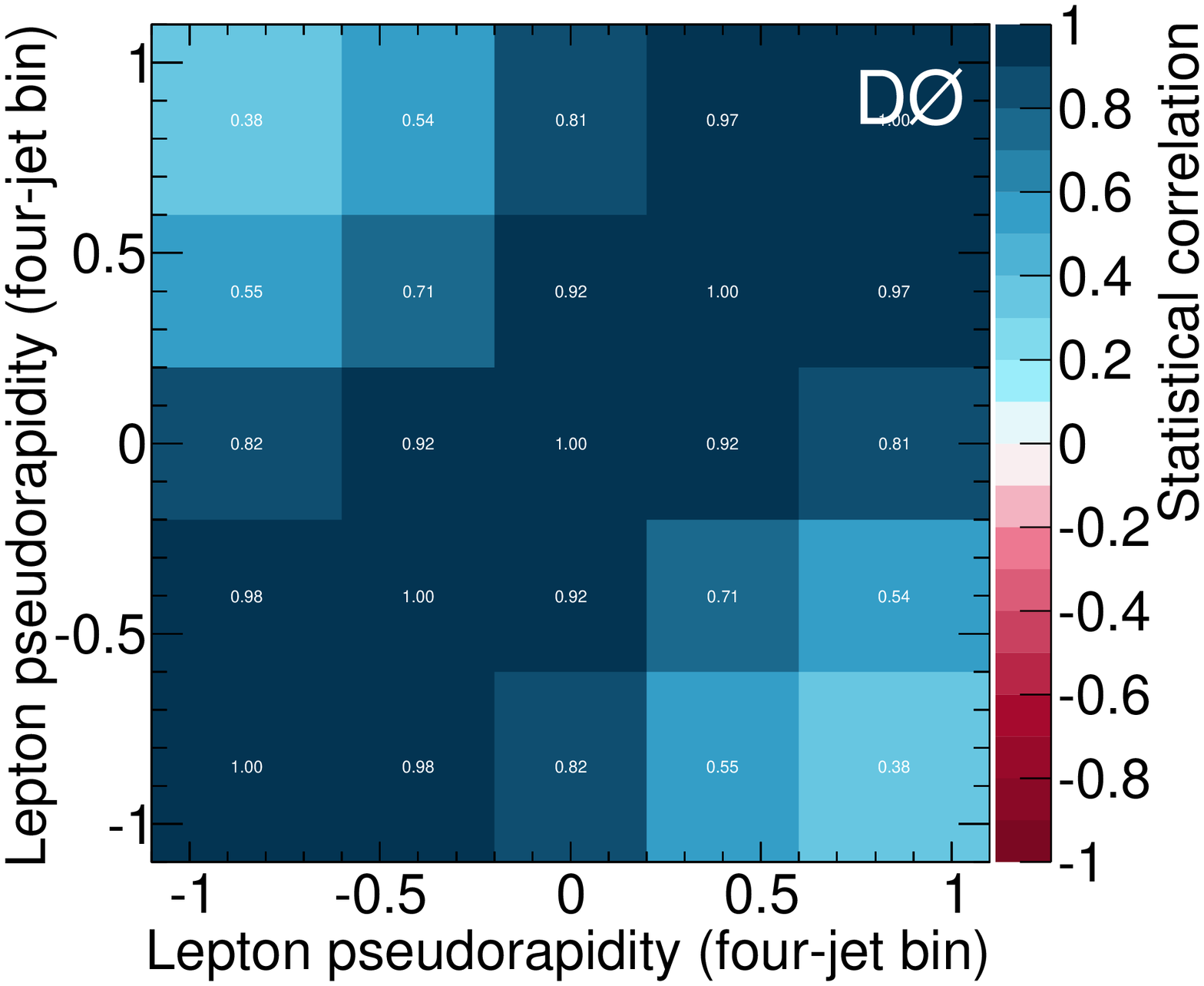}
    \caption{Statistical uncertainty correlations between bins for unfolded lepton pseudorapidity distributions.
      \label{fig:correlations_leptoneta}
    }
  \end{center}
\end{figure}

\begin{figure}[htbp]
  \begin{center}
    \includegraphics[width=0.35\textwidth]{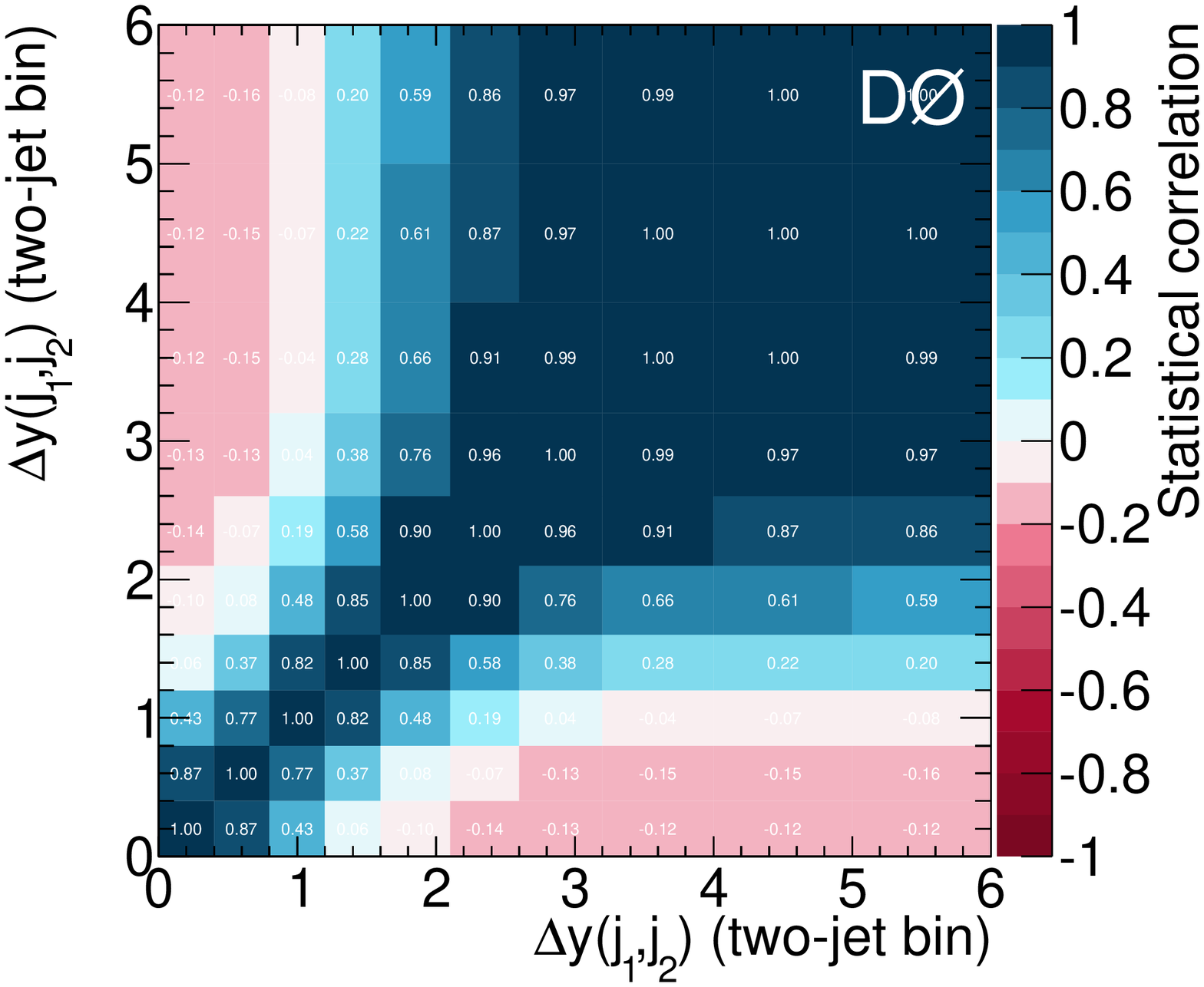}
    \includegraphics[width=0.35\textwidth]{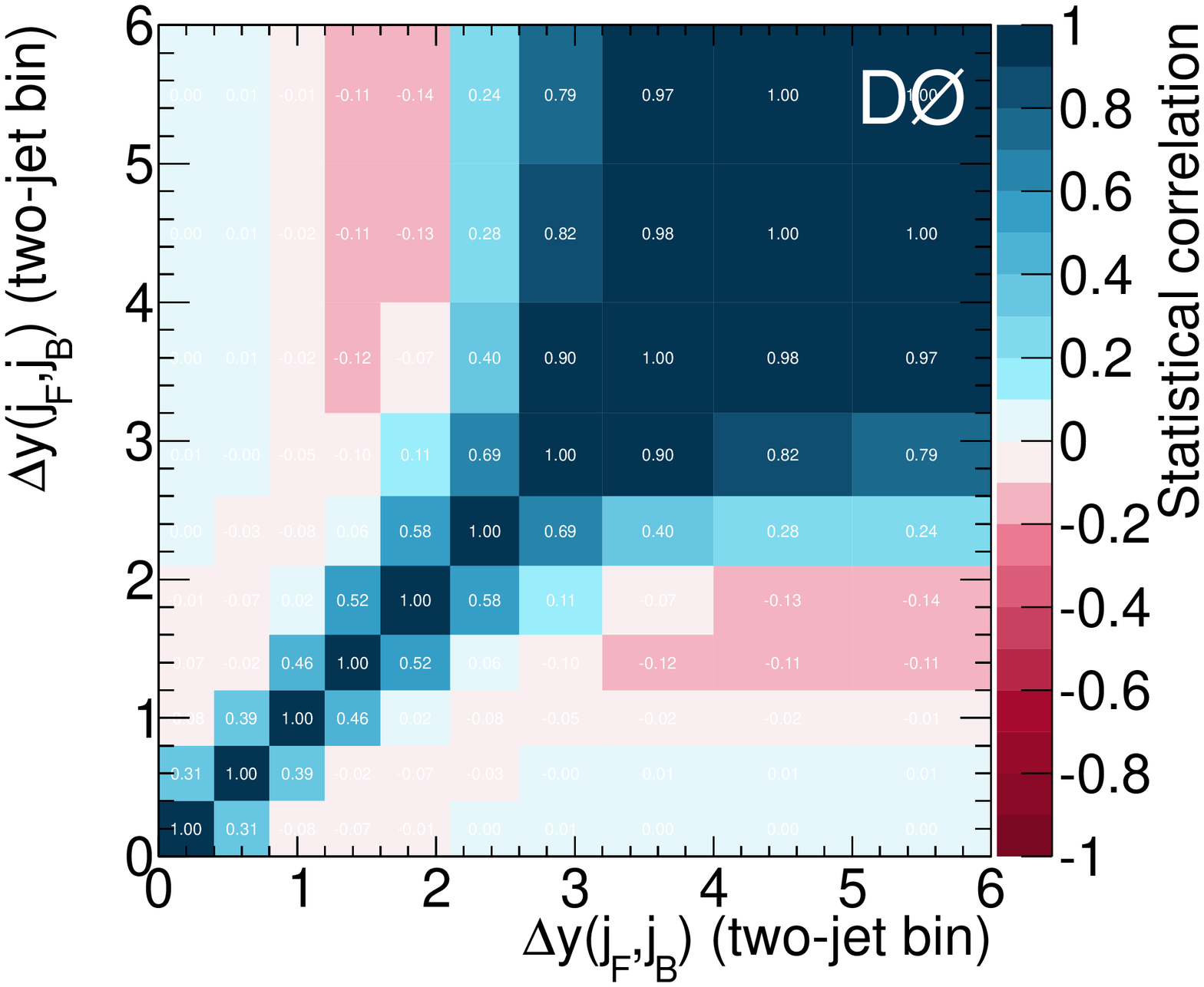}
    \includegraphics[width=0.35\textwidth]{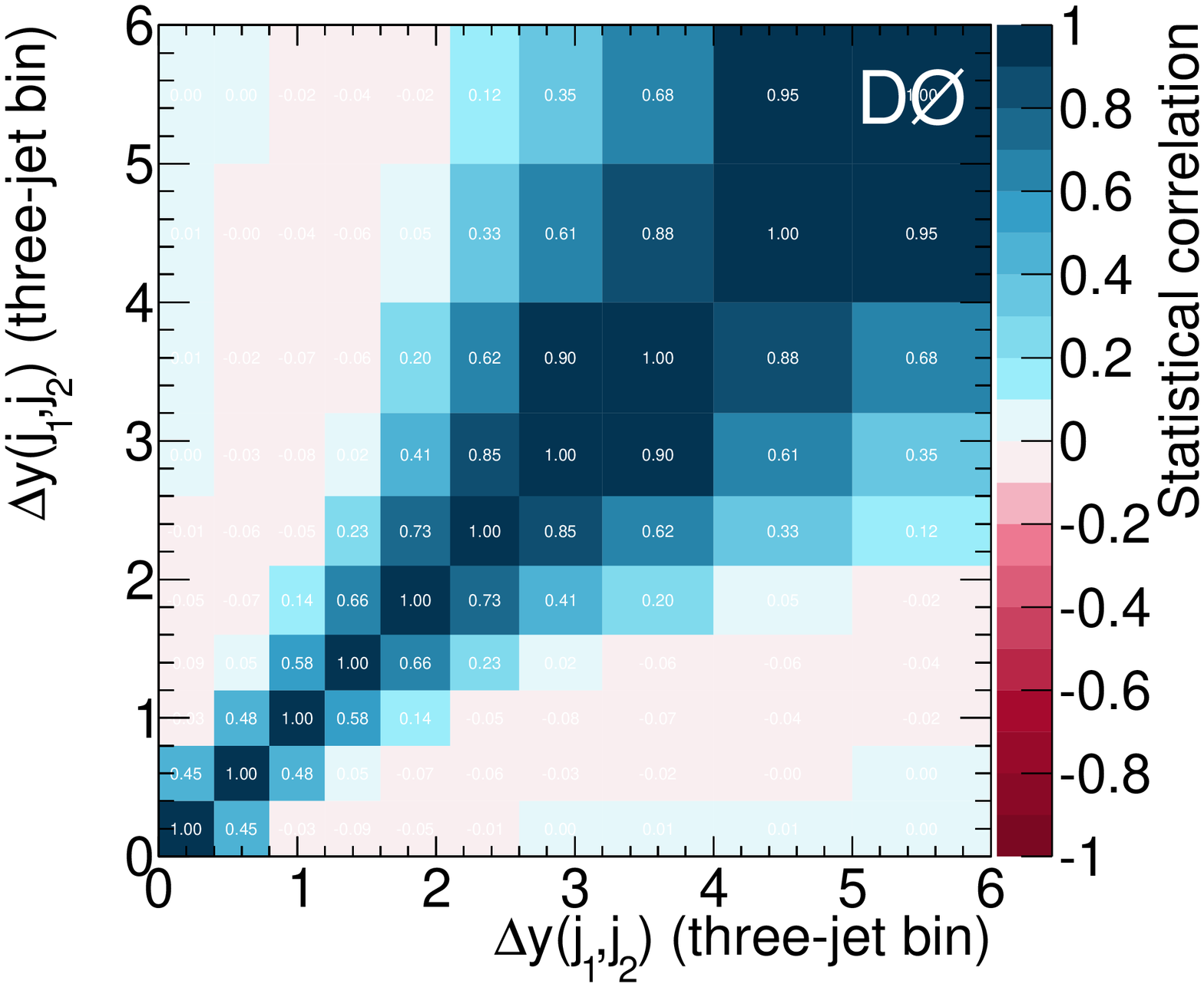}
    \includegraphics[width=0.35\textwidth]{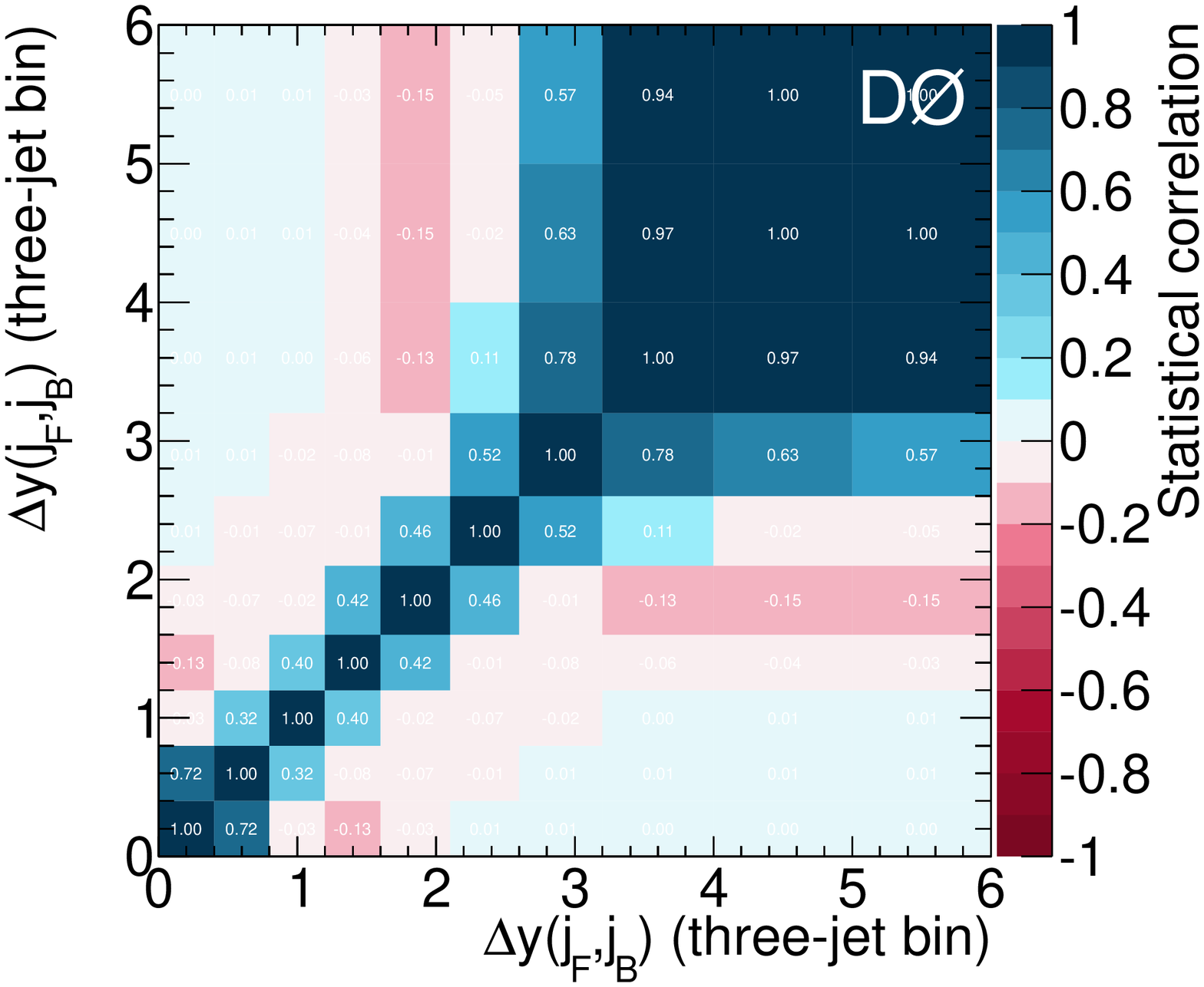}
    \caption{Statistical uncertainty correlations between bins for unfolded dijet rapidity separations for the two highest $p_T$ or two most rapidity-separated jets.
      \label{fig:correlations_jetdeltarap}
    }
  \end{center}
\end{figure}

\begin{figure}[htbp]
  \begin{center}
    \includegraphics[width=0.35\textwidth]{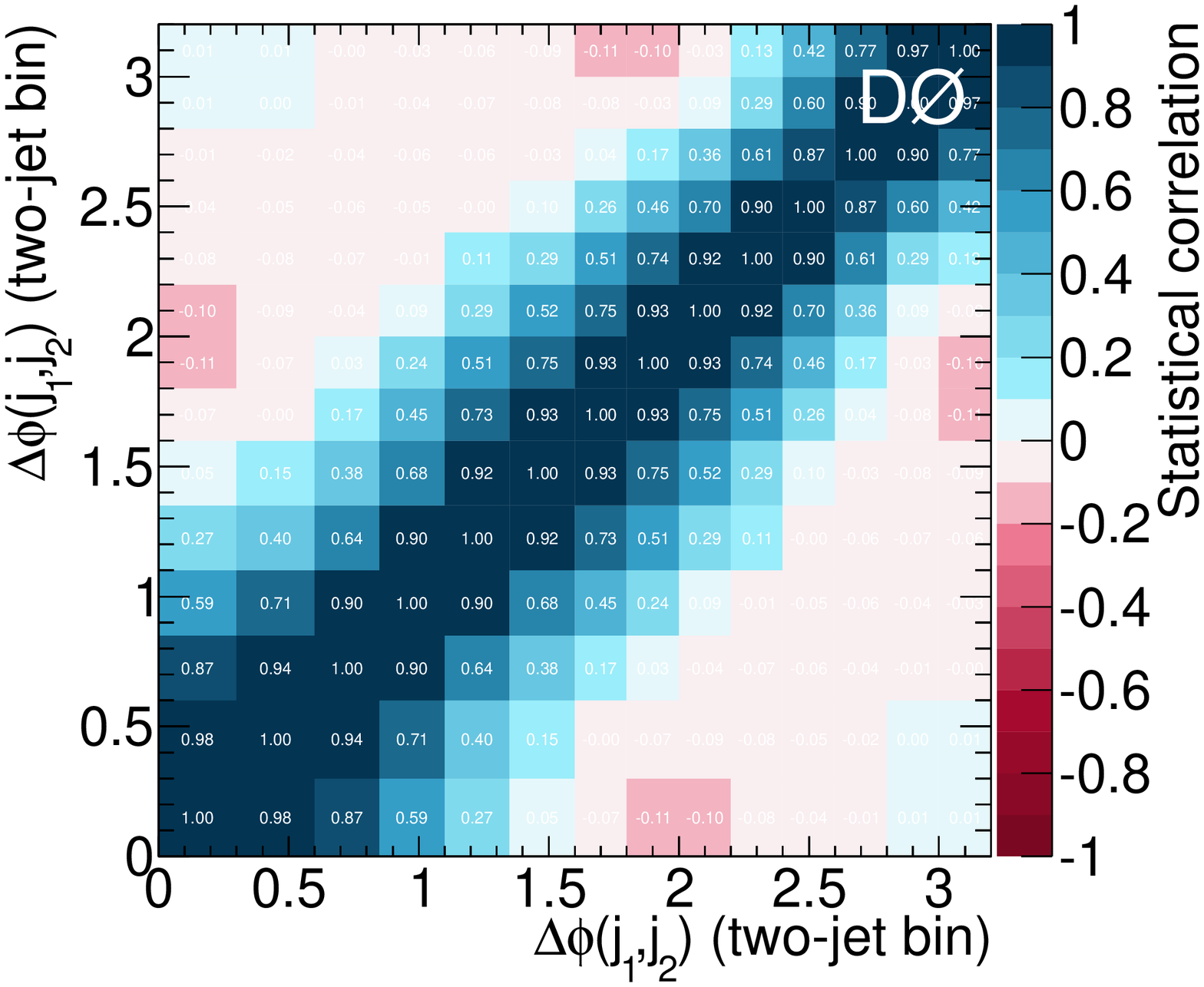}
    \includegraphics[width=0.35\textwidth]{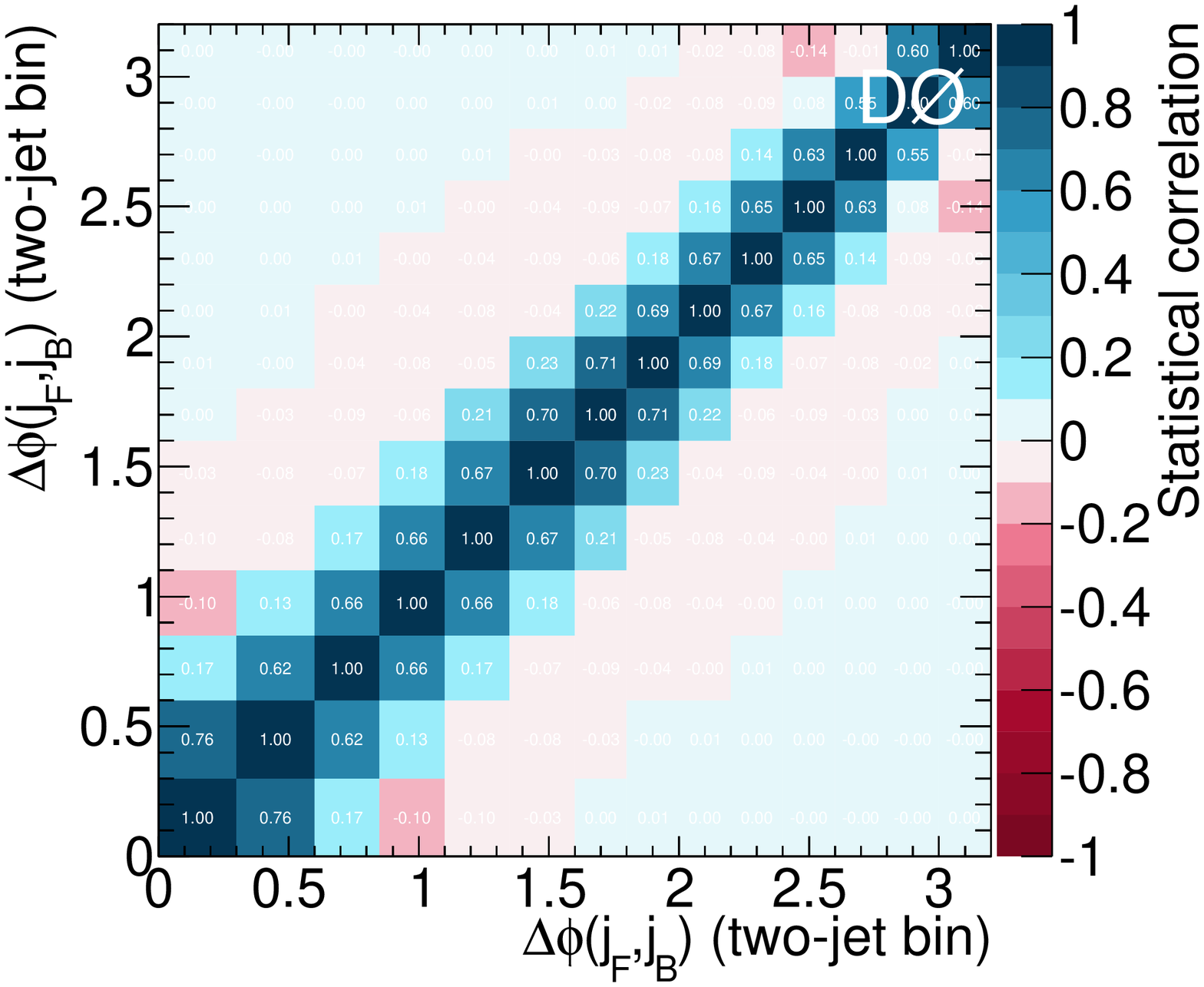}
    \includegraphics[width=0.35\textwidth]{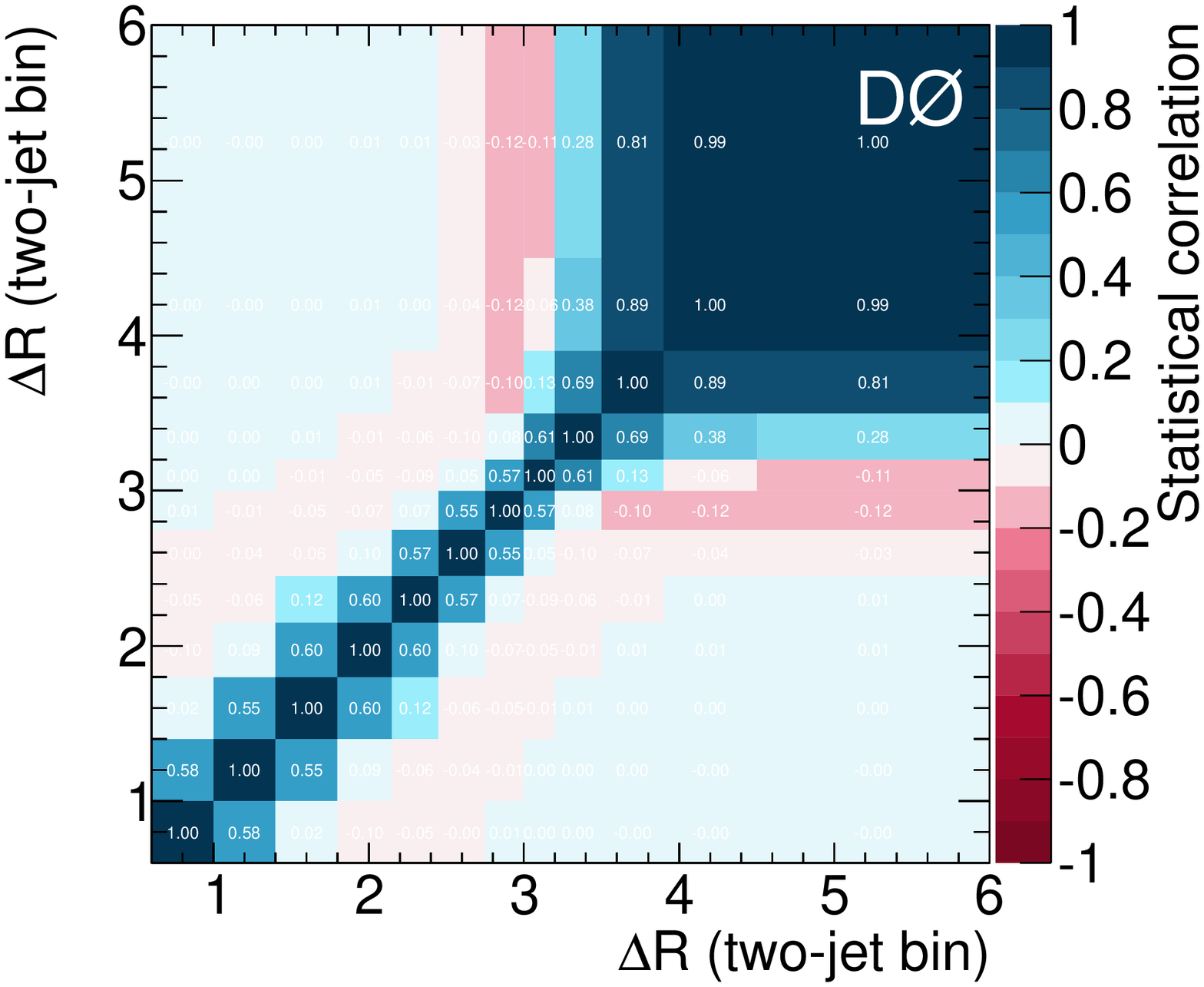}
    \caption{Statistical uncertainty correlations between bins for unfolded azimuthal angle separation between the two highest $p_T$ or two most rapidity-separated jets, 
      and the dijet angular separation in $\eta-\phi$ space.
      \label{fig:correlations_jetdphi_dR}
    }
  \end{center}
\end{figure}

\begin{figure}[htbp]
  \begin{center}
    \includegraphics[width=0.35\textwidth]{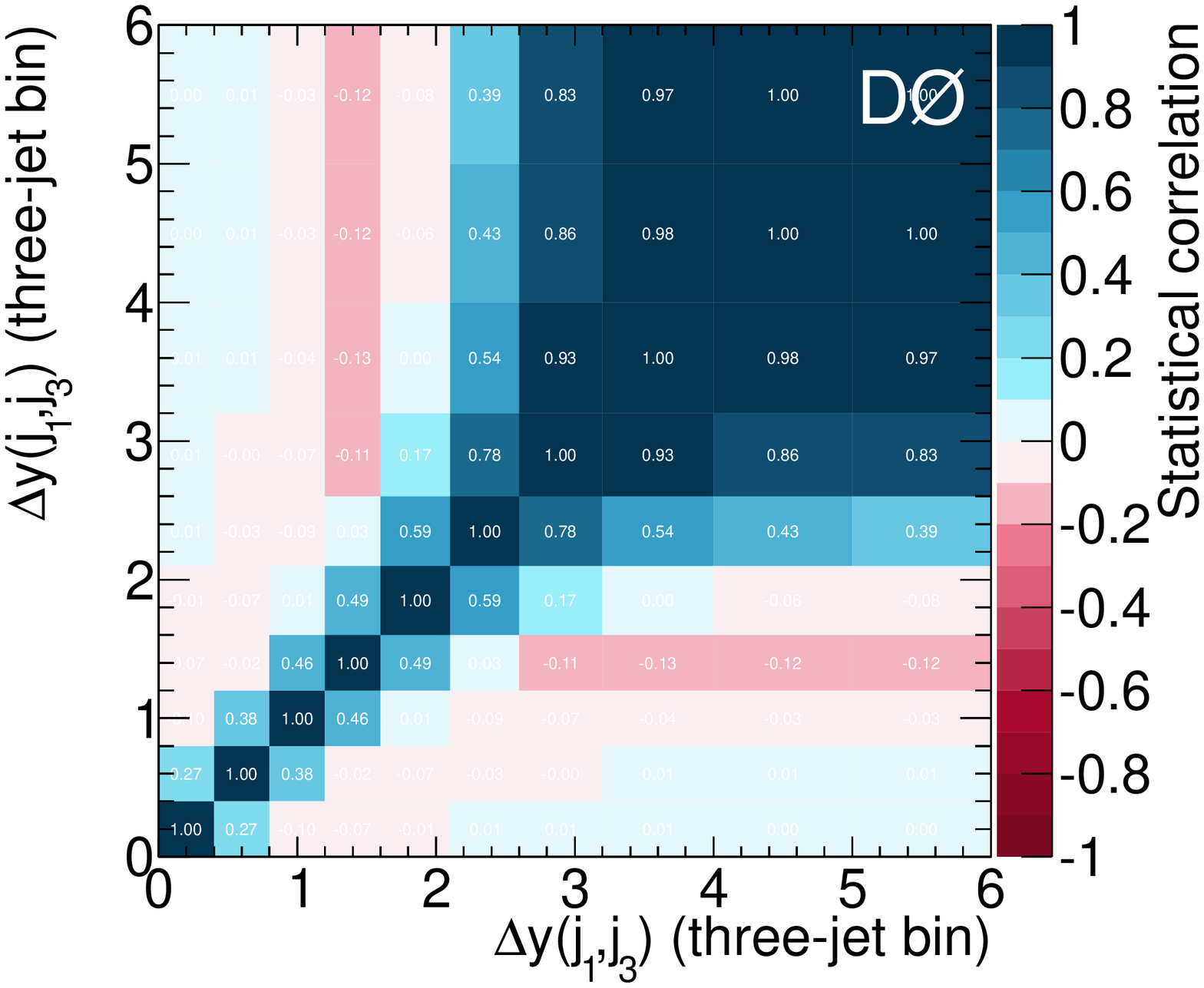}
    \includegraphics[width=0.35\textwidth]{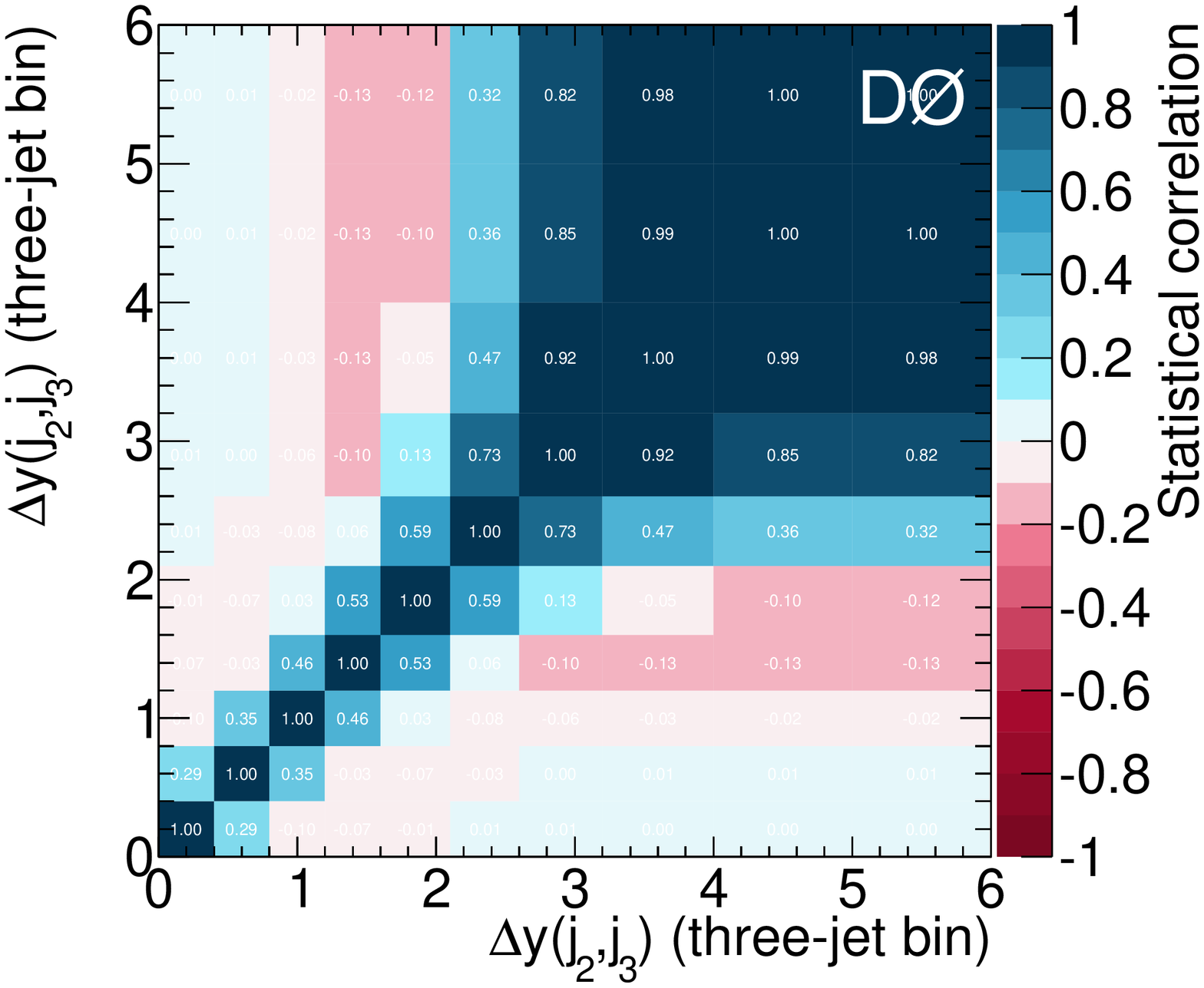}
    \caption{Statistical uncertainty correlations between bins for unfolded dijet rapidity separation between the highest and second highest $p_T$ jets and the second and third highest $p_T$ 
      jets in $W+3\textrm{-jet}$ events.
      \label{fig:correlations_jetDeltaRap3j}
    }
  \end{center}
\end{figure}

\begin{figure}[htbp]
  \begin{center}
    \includegraphics[width=0.35\textwidth]{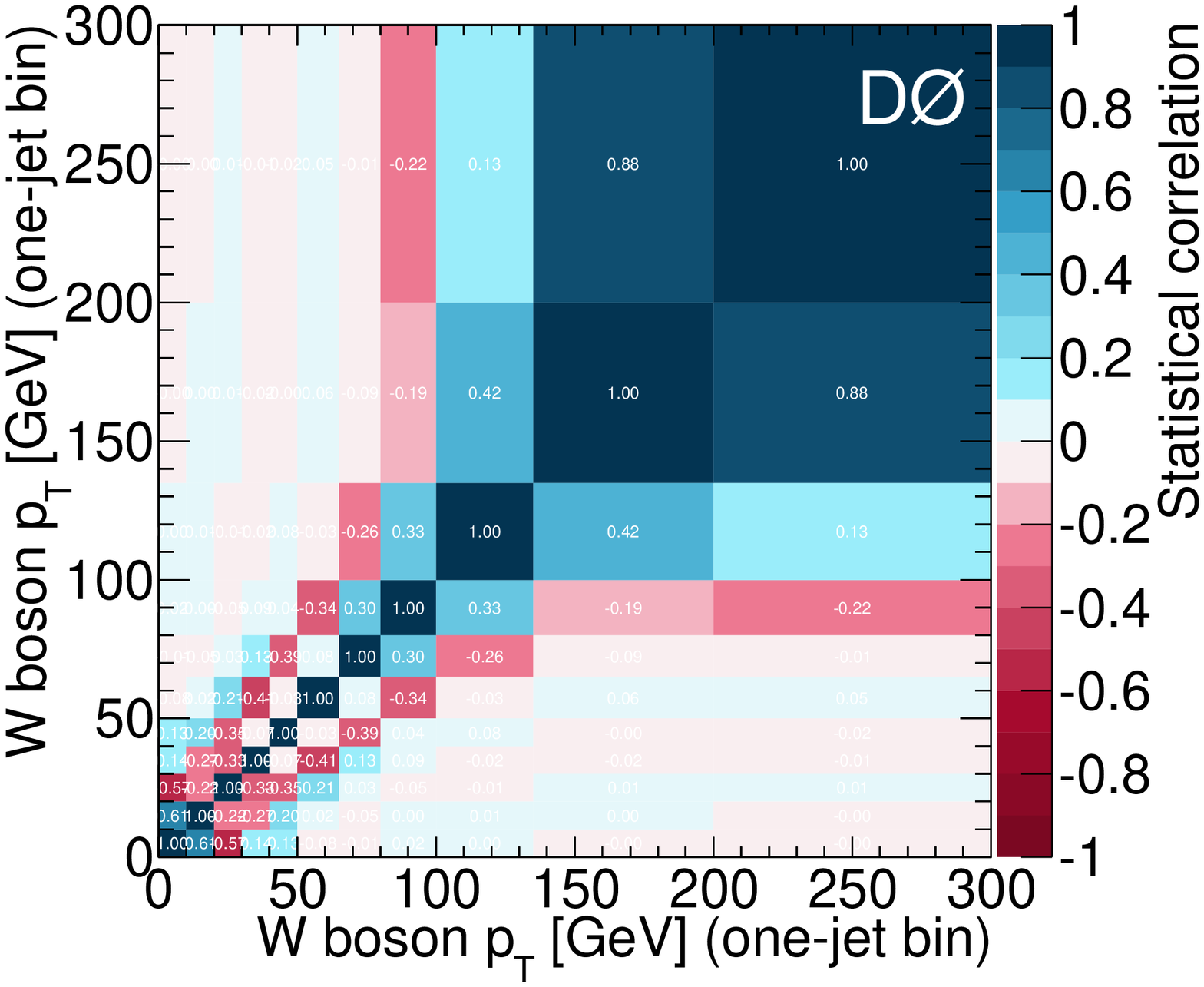}
    \includegraphics[width=0.35\textwidth]{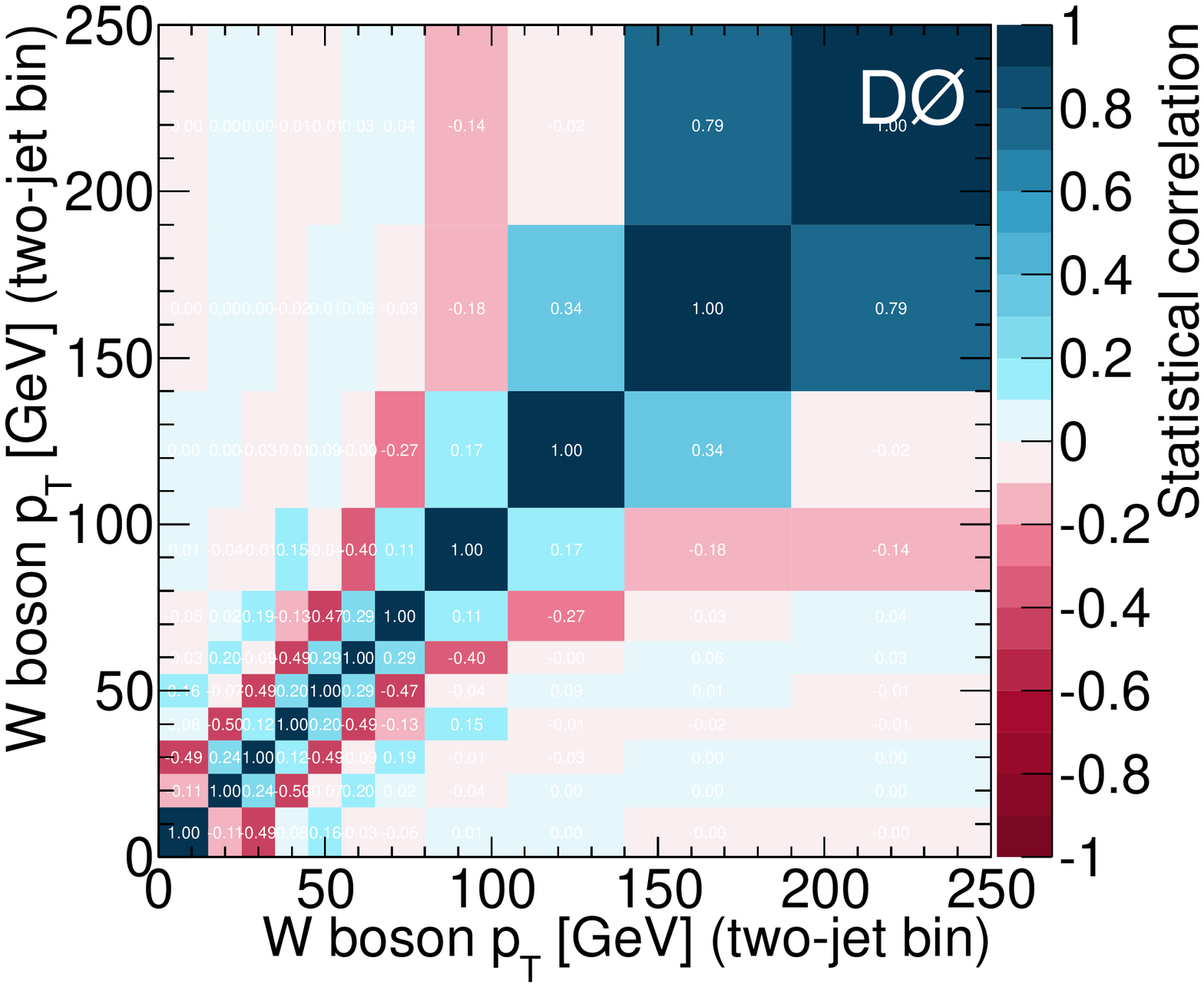}
    \includegraphics[width=0.35\textwidth]{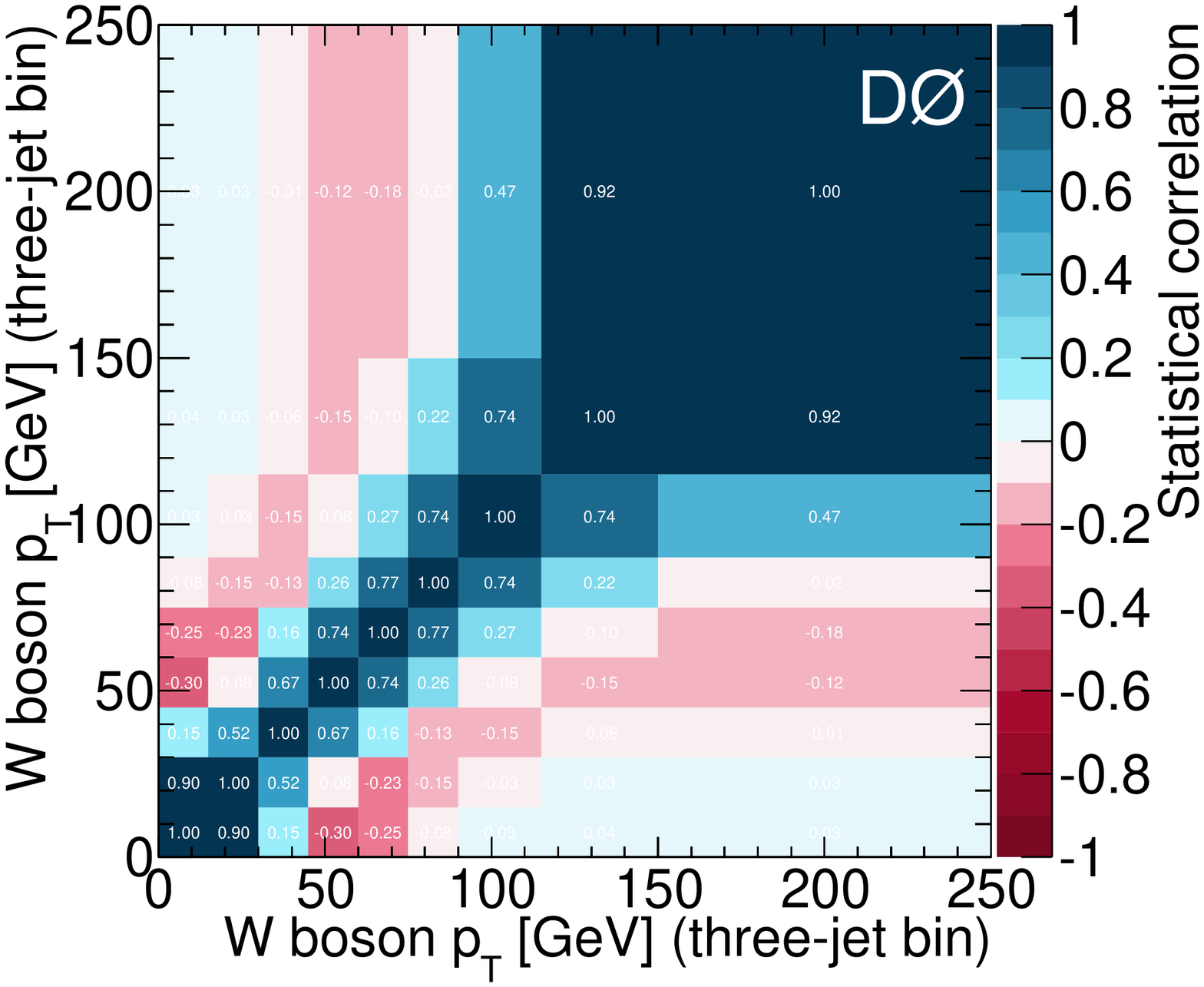}
    \includegraphics[width=0.35\textwidth]{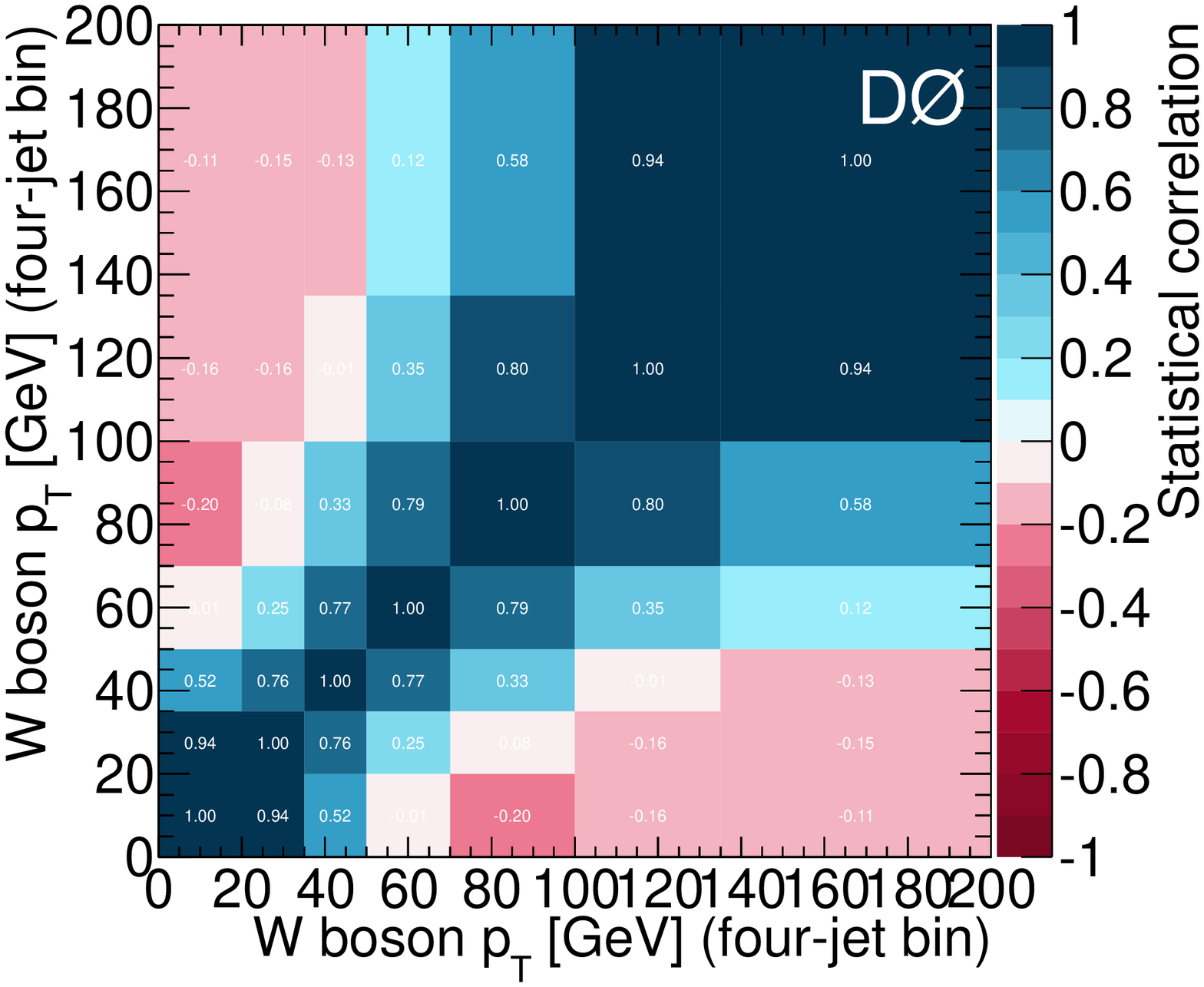}
    \caption{Statistical uncertainty correlations between bins for unfolded $W$ boson transverse momentum distributions.
      \label{fig:correlations_Wpt}
    }
  \end{center}
\end{figure}

\begin{figure}[htbp]
  \begin{center}
    \includegraphics[width=0.35\textwidth]{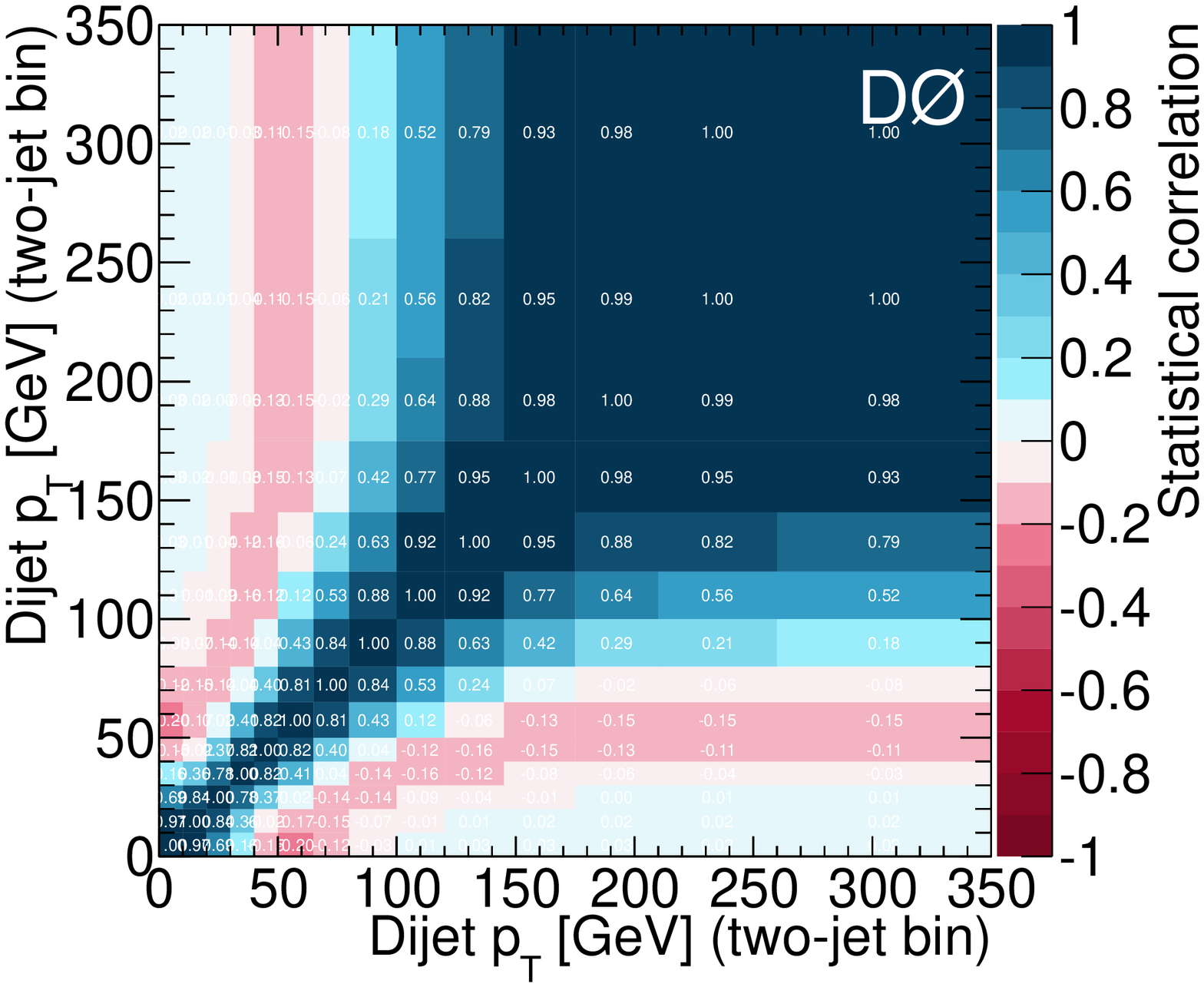}
    \includegraphics[width=0.35\textwidth]{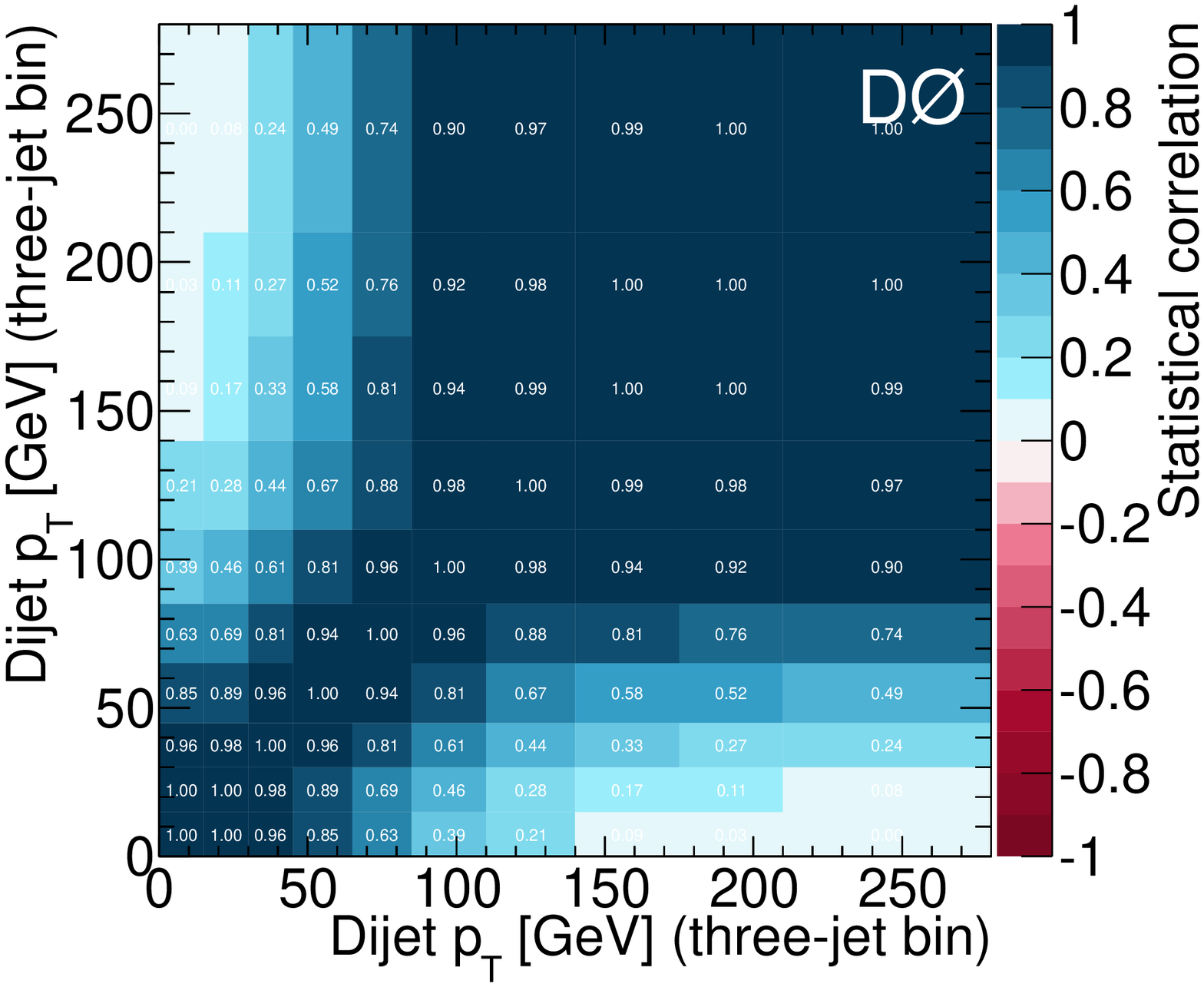}
    \includegraphics[width=0.35\textwidth]{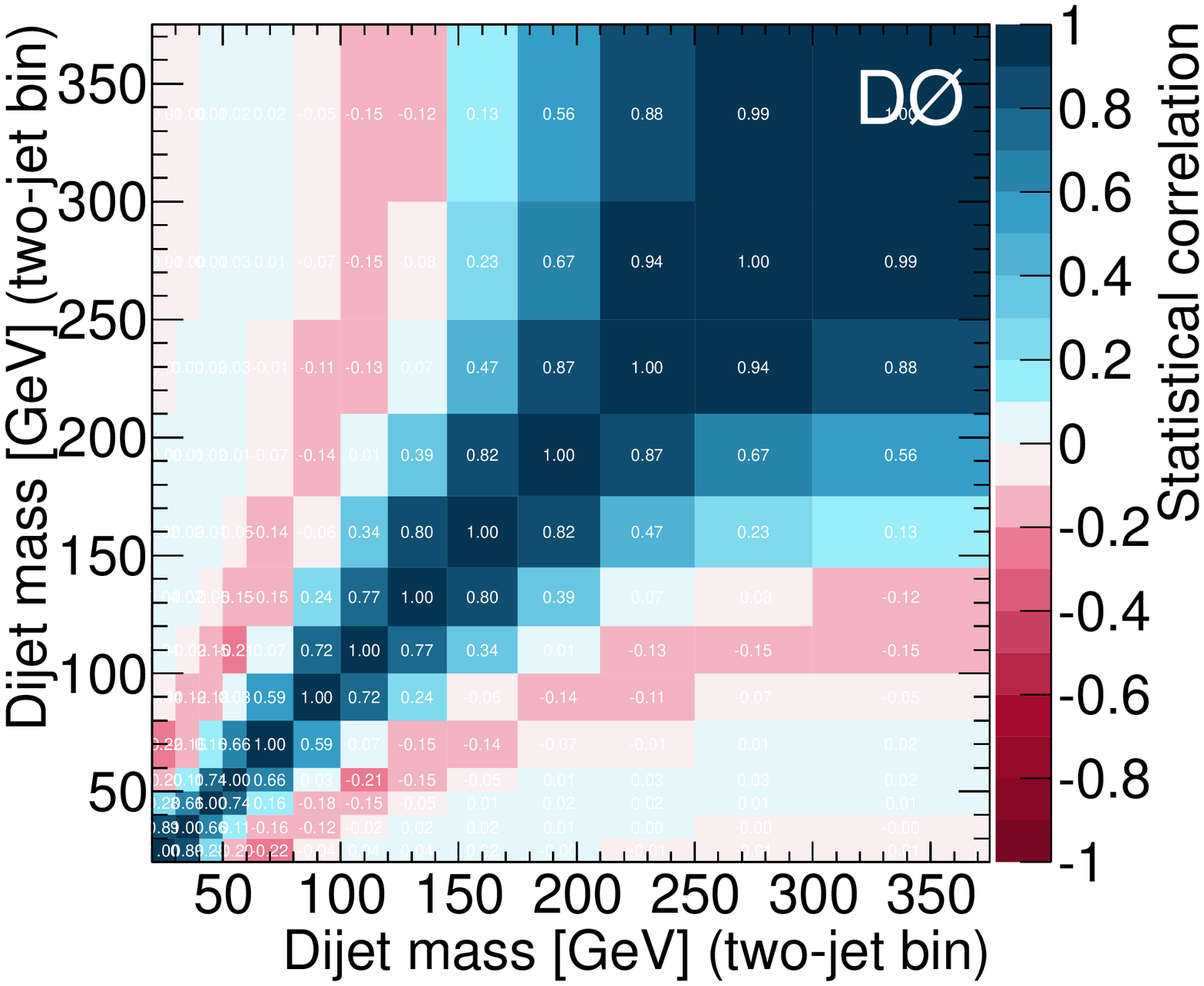}
    \includegraphics[width=0.35\textwidth]{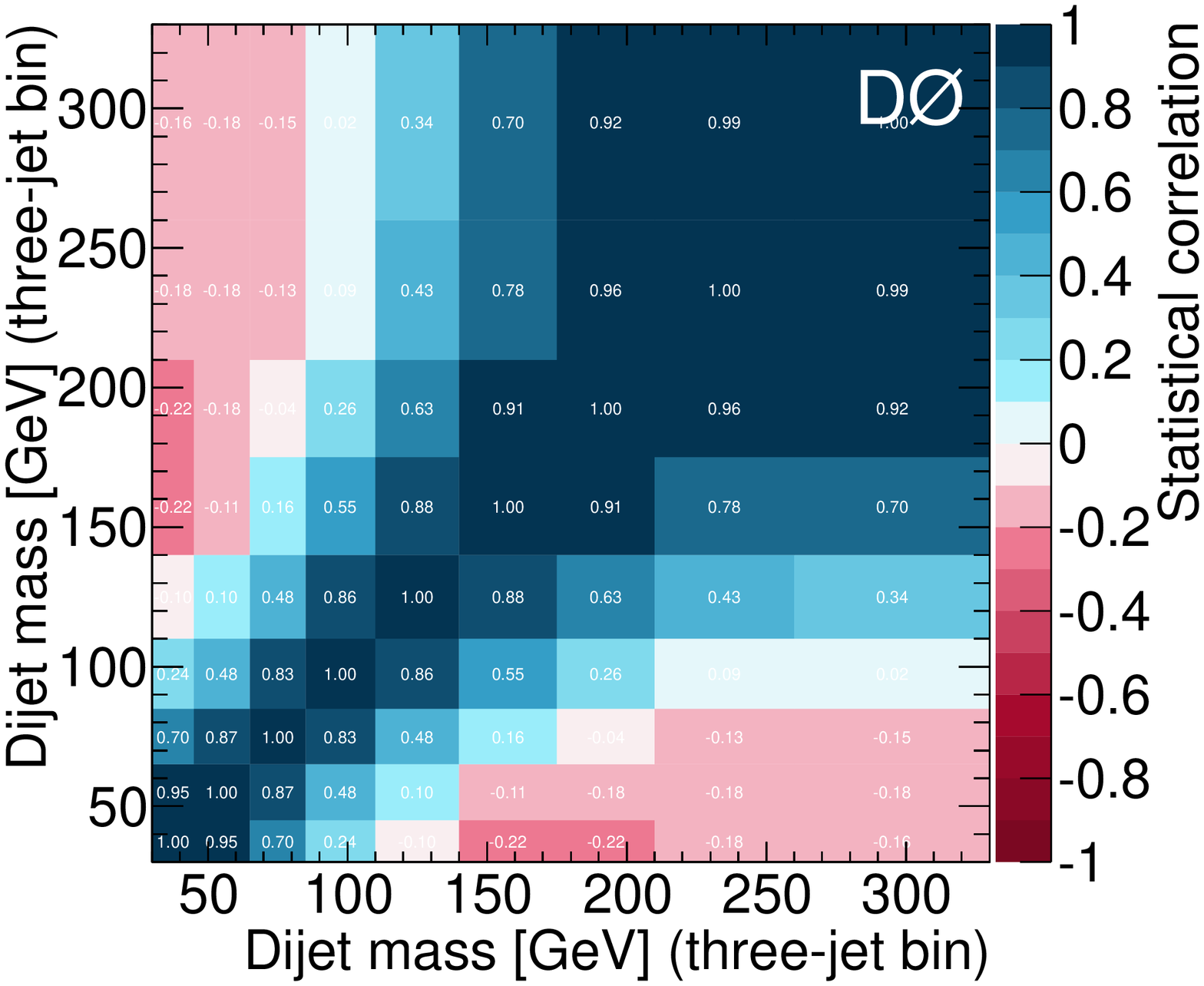}
    \caption{Statistical uncertainty correlations between bins for unfolded dijet $p_T$ and invariant mass distributions.
      \label{fig:correlations_dijetptmass}
    }
  \end{center}
\end{figure}

\begin{figure}[htbp]
  \begin{center}
    \includegraphics[width=0.35\textwidth]{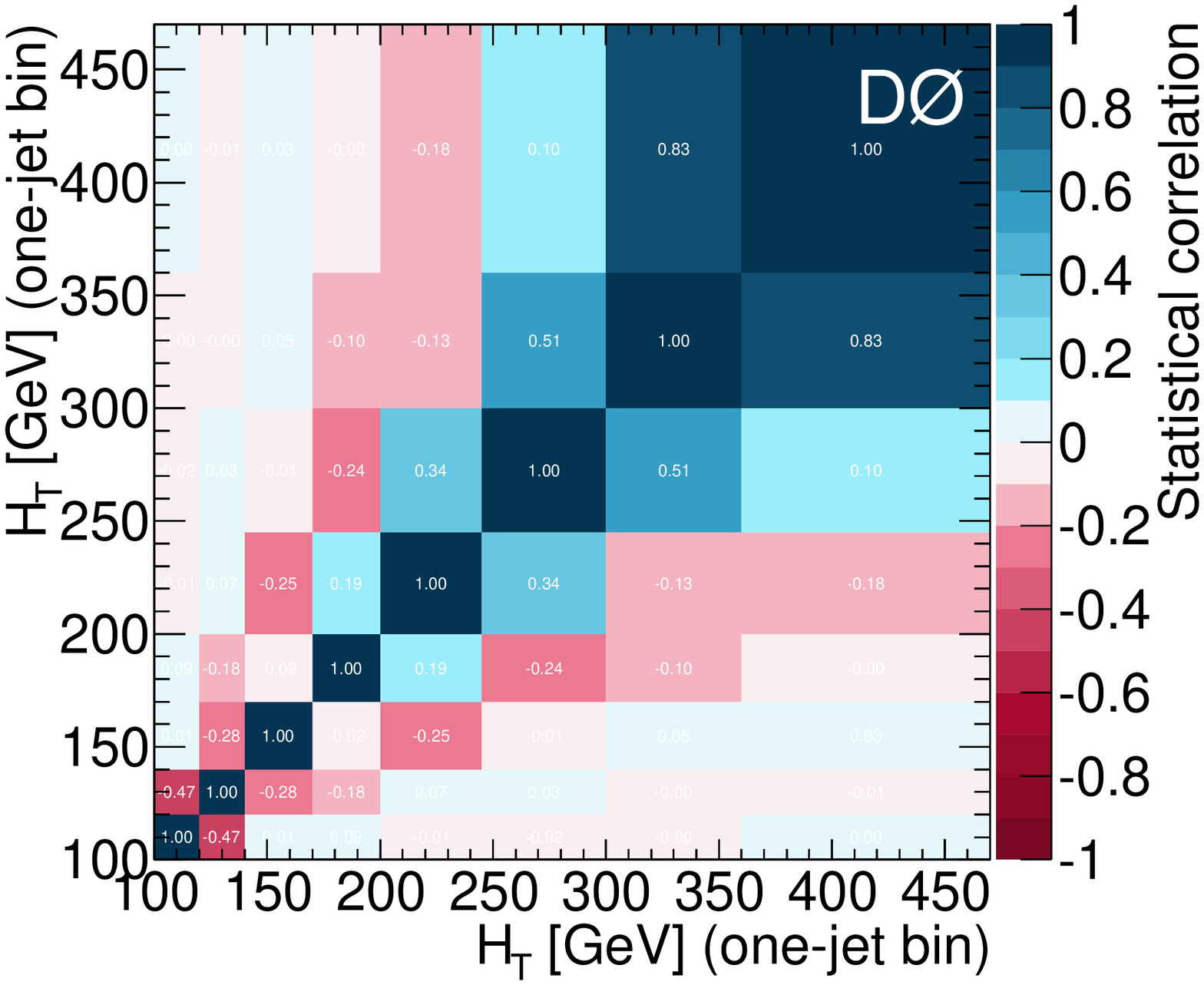}
    \includegraphics[width=0.35\textwidth]{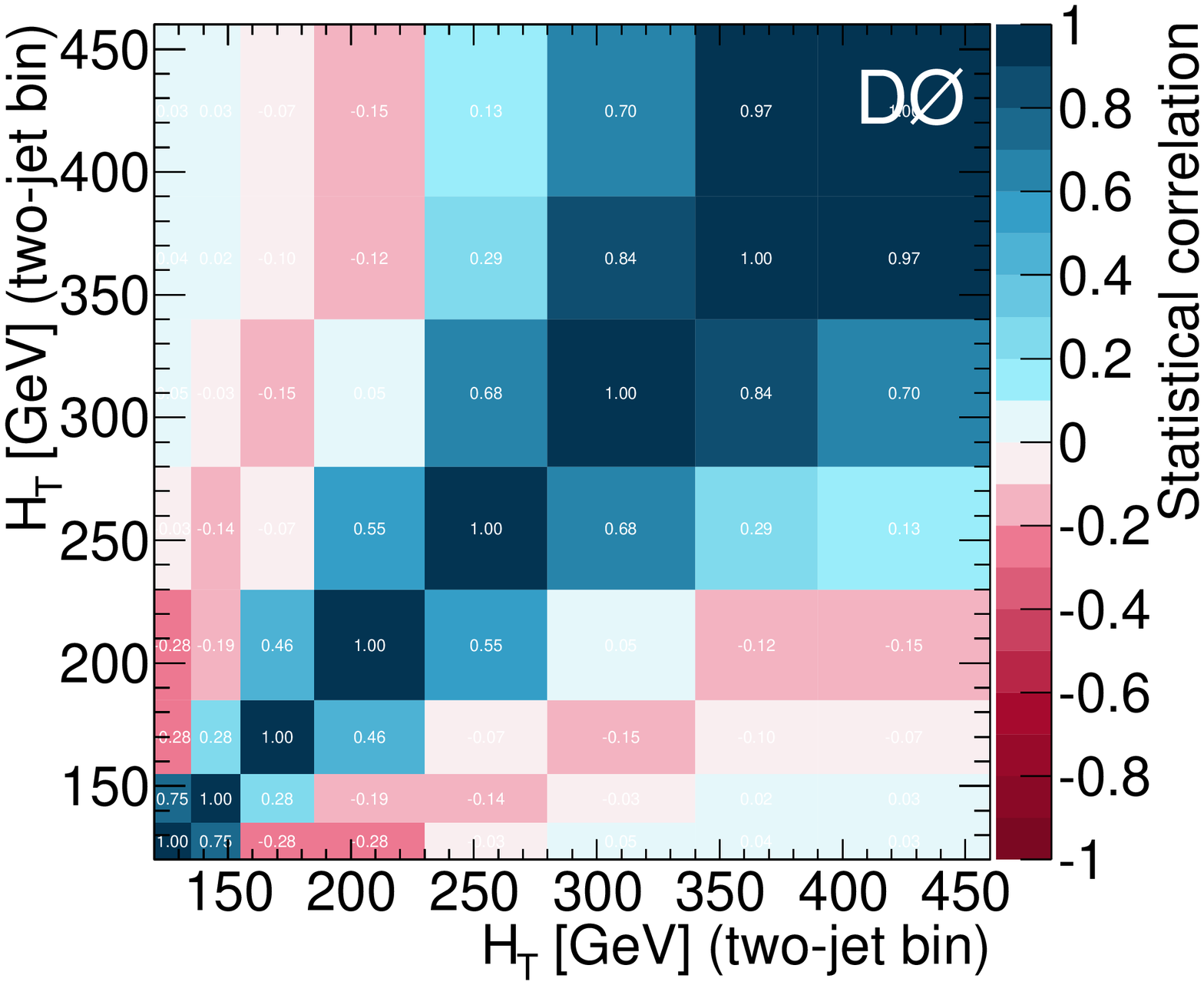}
    \includegraphics[width=0.35\textwidth]{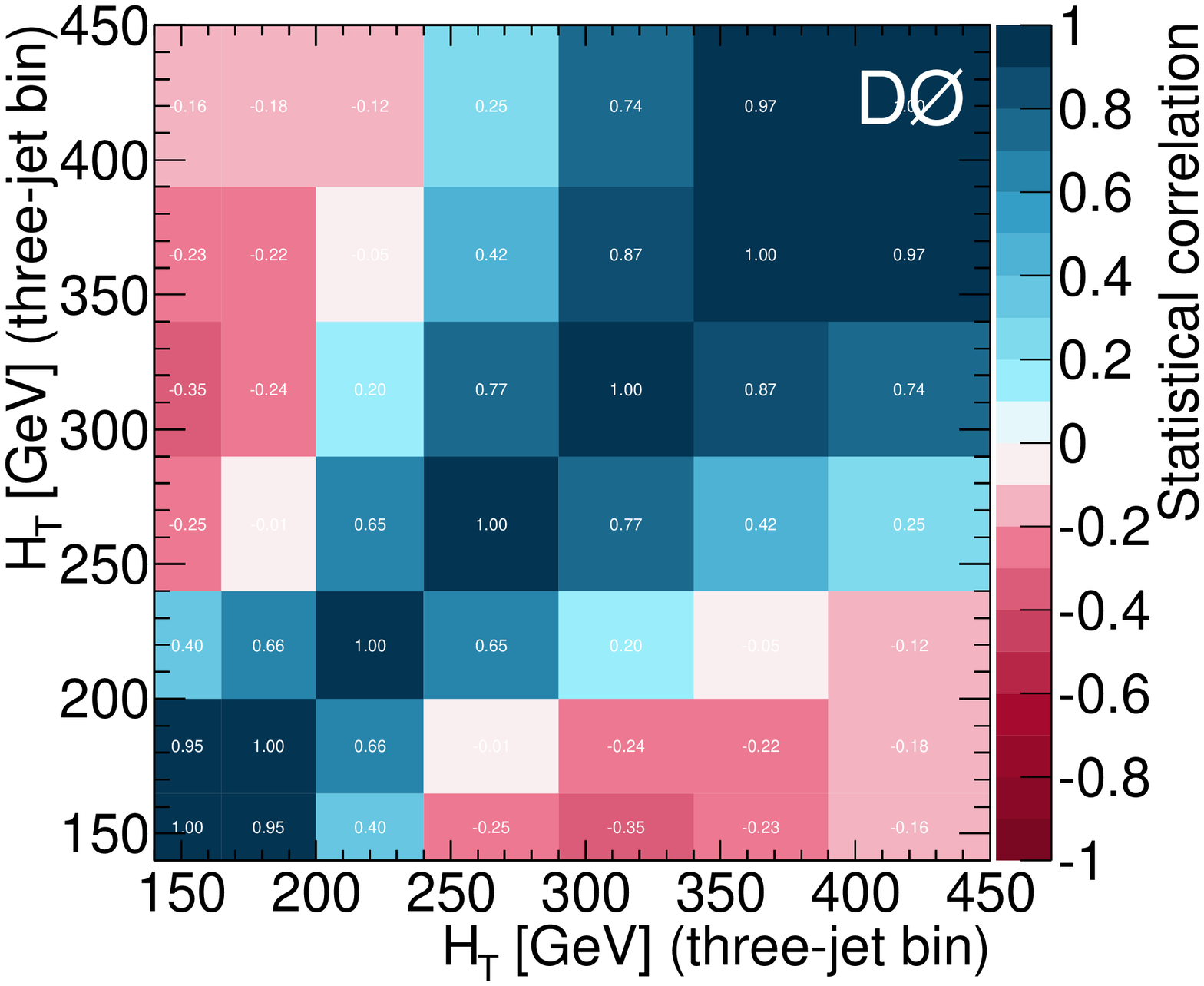}
    \includegraphics[width=0.35\textwidth]{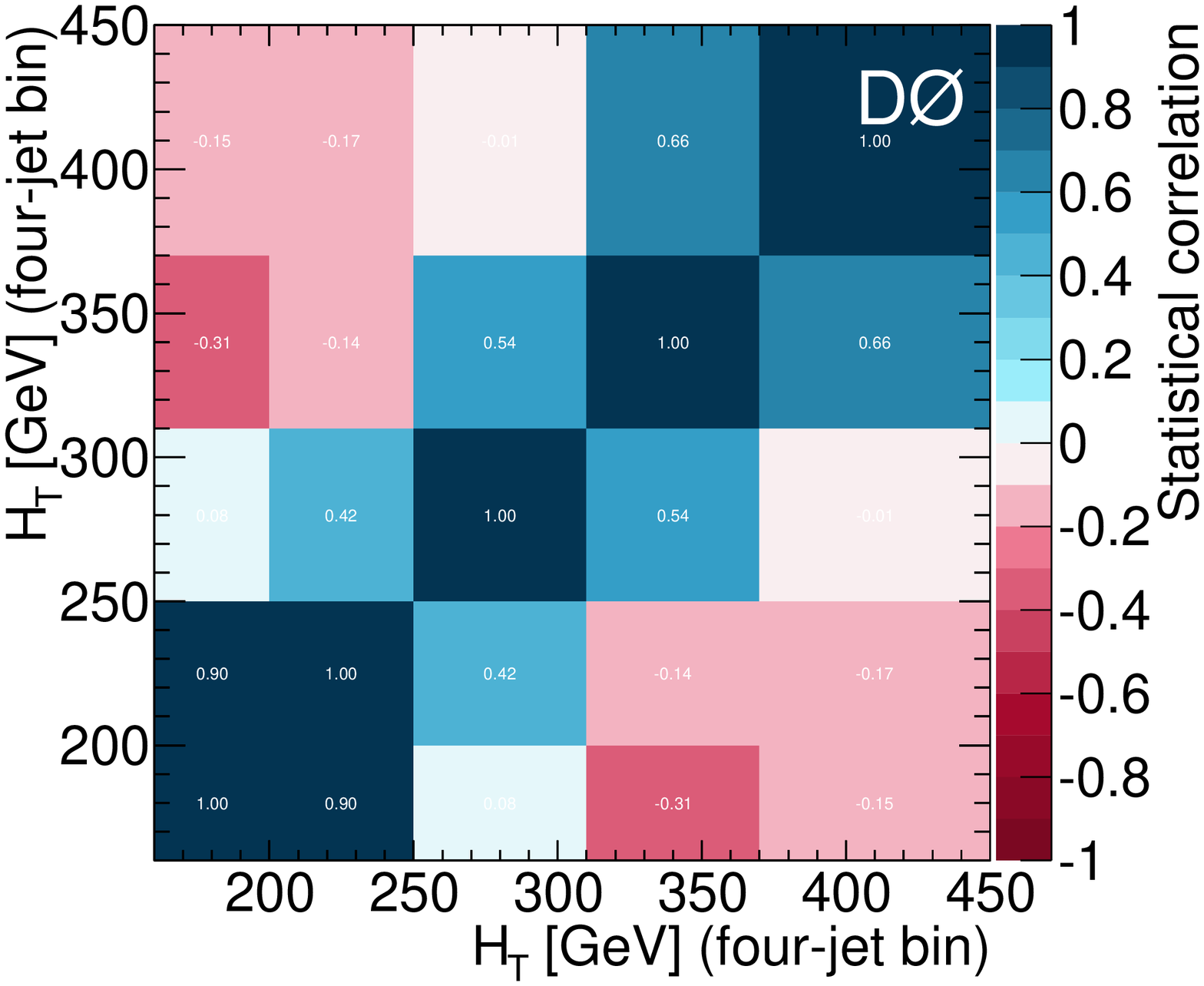}
    \caption{Statistical uncertainty correlations between bins for unfolded $H_T$ (scalar sum of the transverse energies of the $W$ boson and all jets) distributions.
      \label{fig:correlations_jetHT}
    }
  \end{center}
\end{figure}

\begin{figure}[htbp]
  \begin{center}
    \includegraphics[width=0.35\textwidth]{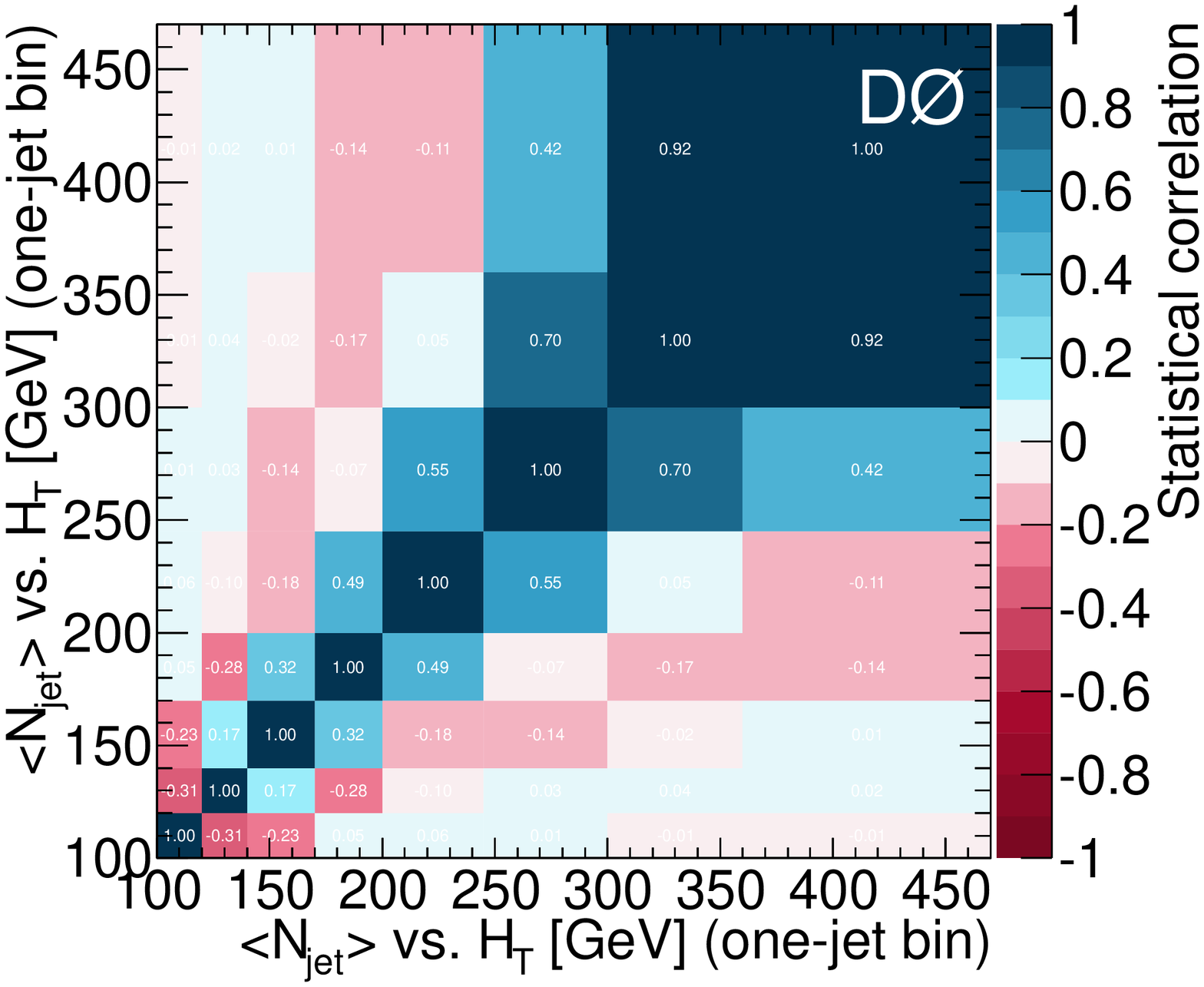}
    \includegraphics[width=0.35\textwidth]{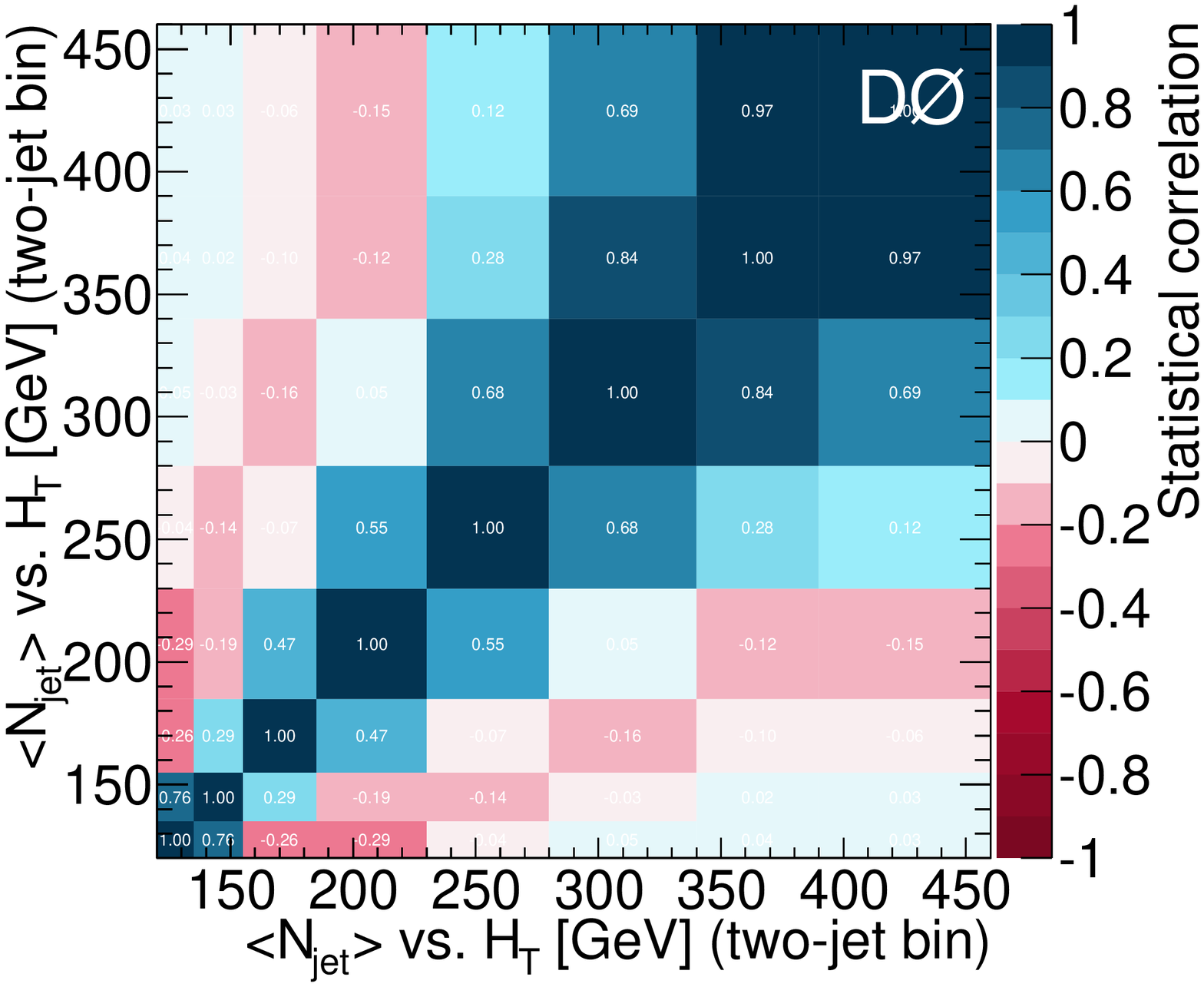}
    \includegraphics[width=0.35\textwidth]{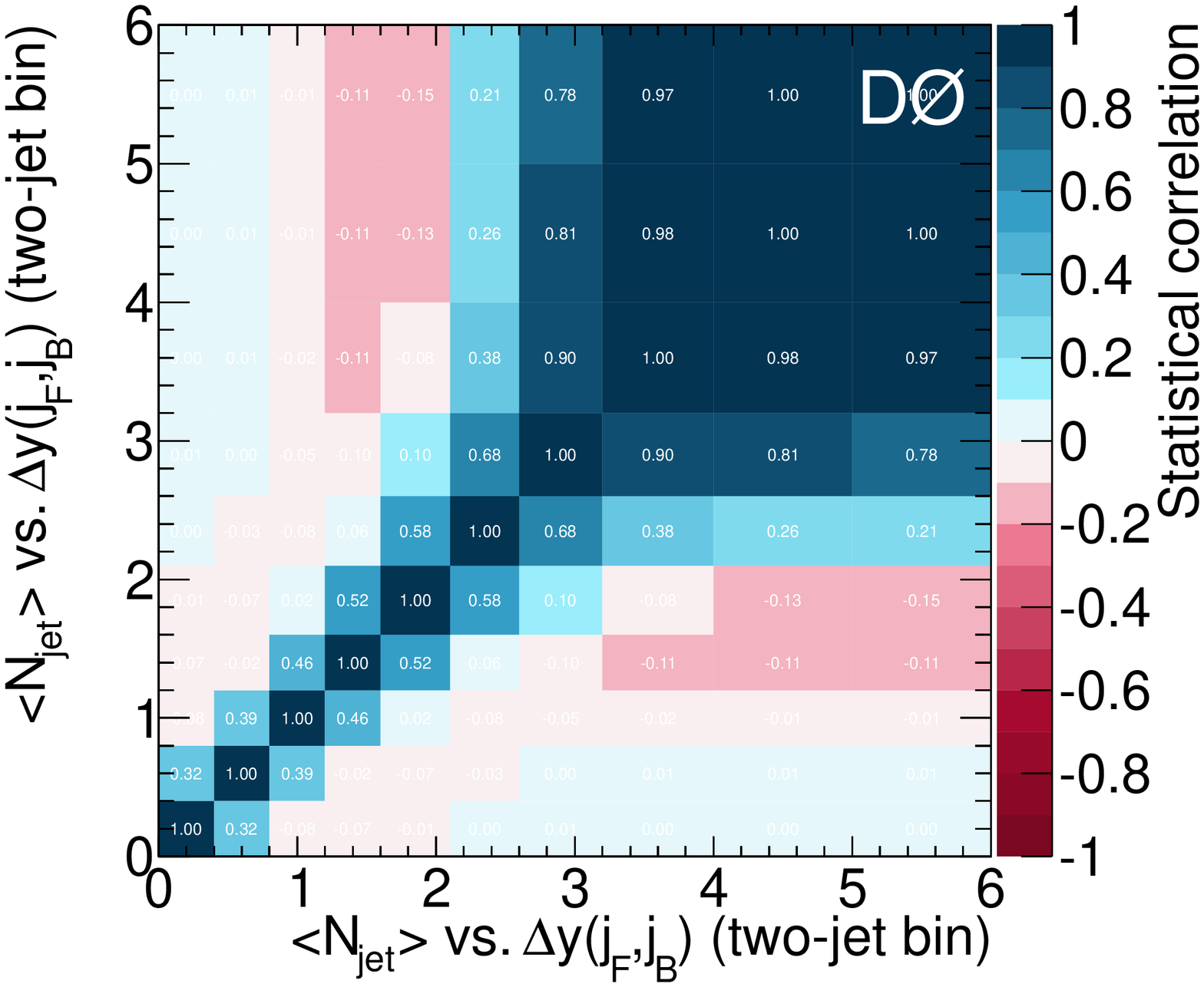}
    \includegraphics[width=0.35\textwidth]{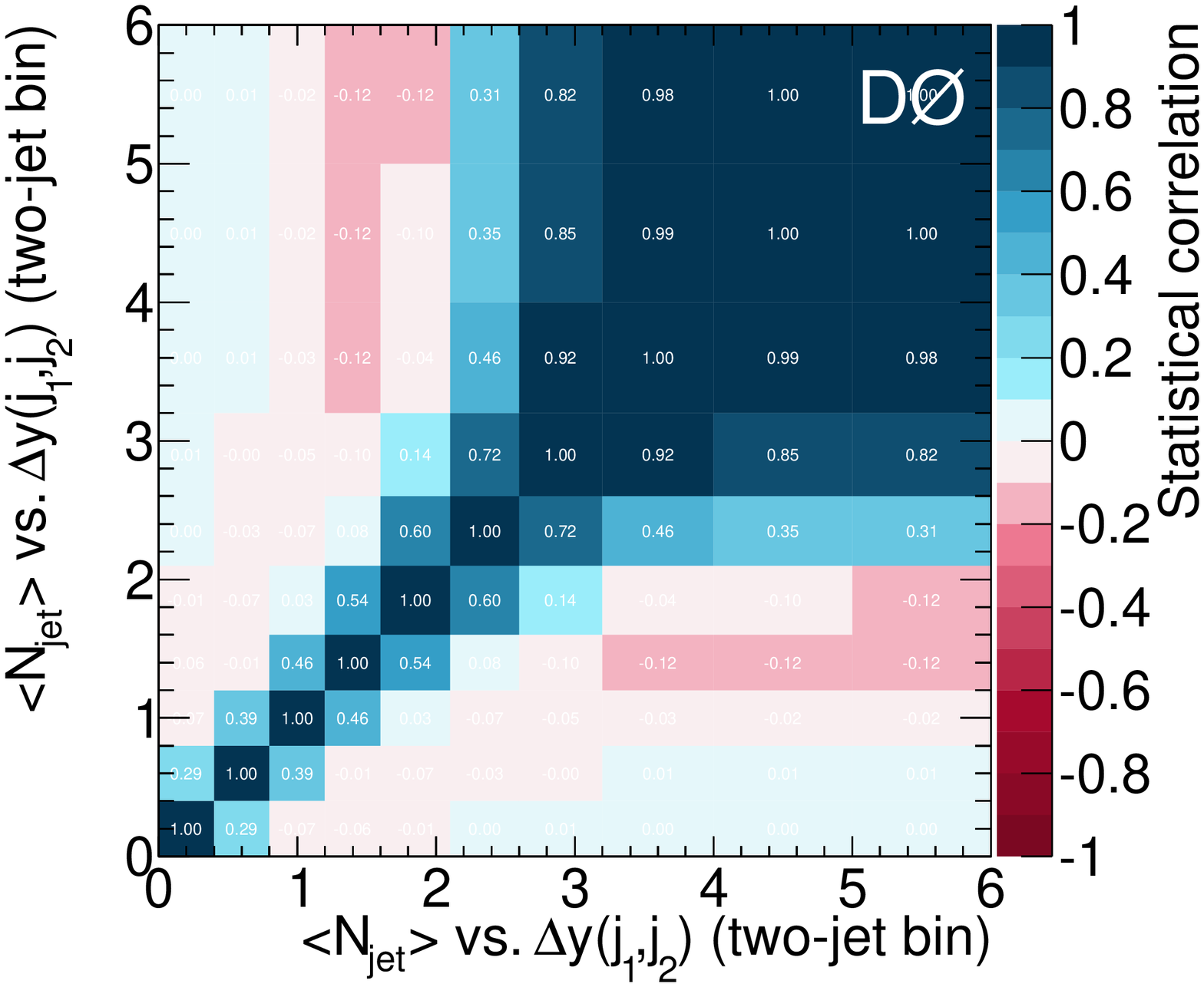}
    \caption{Statistical uncertainty correlations between bins for the unfolded mean number of jets as a function of $H_T$ in the inclusive one and two jet multiplicity bins
      and as a function of the dijet rapidity separations for the two highest $p_T$ or two most rapidity-separated jets.
      \label{fig:correlations_njet}
    }
  \end{center}
\end{figure}

\newpage
\clearpage

\begin{sidewaystable}[htbp]
\caption{\label{tab:statCorr-jet1eta}Statistical correlation matrix for the leading jet rapidity distribution.}
\tiny
\begin{ruledtabular}

\end{ruledtabular}
%\end{sidewaystable}

   \vspace{2\baselineskip}

%\begin{sidewaystable}[htbp]
\caption{\label{tab:statCorr-jet2eta}Statistical correlation matrix for the second jet rapidity distribution.}
\tiny
\begin{ruledtabular}
%
\end{ruledtabular}
\end{sidewaystable}

\begin{table}[htbp]
\caption{\label{tab:statCorr-jet3eta}Statistical correlation matrix for the third jet rapidity distribution.}
\tiny
\begin{ruledtabular}
%
\end{ruledtabular}
\end{table}

\begin{table}[htbp]
\caption{\label{tab:statCorr-jet4eta}Statistical correlation matrix for the fourth jet rapidity distribution.}
\tiny
\begin{ruledtabular}
%
\end{ruledtabular}
\end{table}

\begin{table}[htbp]
\caption{\label{tab:statCorr-lepton1pt}Statistical correlation matrix for the lepton $p_T$ [one-jet] distribution.}
\tiny
\begin{ruledtabular}
%
\end{ruledtabular}
\end{table}

\begin{table}[htbp]
\caption{\label{tab:statCorr-lepton2pt}Statistical correlation matrix for the lepton $p_T$ [two-jet] distribution.}
\tiny
\begin{ruledtabular}
%
\end{ruledtabular}
\end{table}

\begin{table}[htbp]
\caption{\label{tab:statCorr-lepton3pt}Statistical correlation matrix for the lepton $p_T$ [three-jet] distribution.}
\tiny
\begin{ruledtabular}
%
\end{ruledtabular}
\end{table}

\begin{table}[htbp]
\caption{\label{tab:statCorr-lepton4pt}Statistical correlation matrix for the lepton $p_T$ [four-jet] distribution.}
\tiny
\begin{ruledtabular}
%
\end{ruledtabular}
\end{table}

\begin{table}[htbp]
\caption{\label{tab:statCorr-lepton1eta}Statistical correlation matrix for the lepton pseudorapidity [one-jet] distribution.}
\tiny
\begin{ruledtabular}
%
\end{ruledtabular}
\end{table}

\begin{table}[htbp]
\caption{\label{tab:statCorr-lepton2eta}Statistical correlation matrix for the lepton pseudorapidity [two-jet] distribution.}
\tiny
\begin{ruledtabular}
%
\end{ruledtabular}
\end{table}

\begin{table}[htbp]
\caption{\label{tab:statCorr-lepton3eta}Statistical correlation matrix for the lepton pseudorapidity [three-jet] distribution.}
\tiny
\begin{ruledtabular}
%
\end{ruledtabular}
\end{table}

\begin{table}[htbp]
\caption{\label{tab:statCorr-lepton4eta}Statistical correlation matrix for the lepton pseudorapidity [four-jet] distribution.}
\tiny
\begin{ruledtabular}
%
\end{ruledtabular}
\end{table}

\begin{table}[htbp]
\caption{\label{tab:statCorr-jet2DeltaRap}Statistical correlation matrix for the $\Delta y_{12}$ [two-jet] distribution.}
\tiny
\begin{ruledtabular}
%
\end{ruledtabular}
\end{table}

\begin{table}[htbp]
\caption{\label{tab:statCorr-jet2DeltaRapFB}Statistical correlation matrix for the $\Delta y_{FB}$ [two-jet] distribution.}
\tiny
\begin{ruledtabular}
%
\end{ruledtabular}
\end{table}

\begin{table}[htbp]
\caption{\label{tab:statCorr-jet3DeltaRap_3jet}Statistical correlation matrix for the $\Delta y_{12}$ [three-jet] distribution.}
\tiny
\begin{ruledtabular}
%
\end{ruledtabular}
\end{table}

\begin{table}[htbp]
\caption{\label{tab:statCorr-jet3DeltaRapFB}Statistical correlation matrix for the $\Delta y_{FB}$ [three-jet] distribution.}
\tiny
\begin{ruledtabular}
%
\end{ruledtabular}
\end{table}

\begin{table}[htbp]
\caption{\label{tab:statCorr-jet2dPhi}Statistical correlation matrix for the $\Delta\phi_{12}$ distribution.}
\tiny
\begin{ruledtabular}
%
\end{ruledtabular}
\end{table}

\begin{table}[htbp]
\caption{\label{tab:statCorr-jet2dPhiFB}Statistical correlation matrix for the $\Delta\phi_{FB}$ distribution.}
\tiny
\begin{ruledtabular}
%
\end{ruledtabular}
\end{table}

\clearpage
\newpage

\begin{table}[htbp]
\caption{\label{tab:statCorr-jet2dR}Statistical correlation matrix for the $\Delta R_{jj}$ distribution.}
\tiny
\begin{ruledtabular}
%
\end{ruledtabular}
\end{table}

\begin{table}[htbp]
\caption{\label{tab:statCorr-jet3DeltaRap13}Statistical correlation matrix for the $\Delta y_{13}$ distribution.}
\tiny
\begin{ruledtabular}
%
\end{ruledtabular}
\end{table}

\begin{table}[htbp]
\caption{\label{tab:statCorr-jet3DeltaRap23}Statistical correlation matrix for the $\Delta y_{23}$ distribution.}
\tiny
\begin{ruledtabular}
%
\end{ruledtabular}
\end{table}

\clearpage
\newpage

\begin{table}[htbp]
\caption{\label{tab:statCorr-W1pt}Statistical correlation matrix for the $W$ boson $p_T$ [one-jet] distribution.}
\tiny
\begin{ruledtabular}
%
\end{ruledtabular}
\end{table}

\begin{table}[htbp]
\caption{\label{tab:statCorr-W2pt}Statistical correlation matrix for the $W$ boson $p_T$ [two-jet] distribution.}
\tiny
\begin{ruledtabular}
%
\end{ruledtabular}
\end{table}

\begin{table}[htbp]
\caption{\label{tab:statCorr-W3pt}Statistical correlation matrix for the $W$ boson $p_T$ [three-jet] distribution.}
\tiny
\begin{ruledtabular}
%
\end{ruledtabular}
\end{table}

\begin{table}[htbp]
\caption{\label{tab:statCorr-W4pt}Statistical correlation matrix for the $W$ boson $p_T$ [four-jet] distribution.}
\tiny
\begin{ruledtabular}
%
\end{ruledtabular}
\end{table}

\begin{sidewaystable}[htbp]
\caption{\label{tab:statCorr-dijet2pt}Statistical correlation matrix for the dijet $p_T$ [two-jet] distribution.}
\tiny
\begin{ruledtabular}
%
\end{ruledtabular}
%\end{sidewaystable}

   \vspace{2\baselineskip}

%\begin{sidewaystable}[htbp]
\caption{\label{tab:statCorr-dijet3pt}Statistical correlation matrix for the dijet $p_T$ [three-jet] distribution.}
\tiny
\begin{ruledtabular}
%
\end{ruledtabular}
\end{sidewaystable}

\begin{sidewaystable}[htbp]
\caption{\label{tab:statCorr-dijet2mass}Statistical correlation matrix for the dijet invariant mass [two-jet] distribution.}
\tiny
\begin{ruledtabular}
%
\end{ruledtabular}
%\end{sidewaystable}

   \vspace{2\baselineskip}

%\begin{sidewaystable}[htbp]
\caption{\label{tab:statCorr-dijet3mass}Statistical correlation matrix for the dijet invariant mass [three-jet] distribution.}
\tiny
\begin{ruledtabular}
%
\end{ruledtabular}
\end{sidewaystable}

\begin{table}[htbp]
\caption{\label{tab:statCorr-jet1HT}Statistical correlation matrix for the $H_T$ [one-jet] distribution.}
\tiny
\begin{ruledtabular}
%
\end{ruledtabular}
\end{table}

\begin{table}[htbp]
\caption{\label{tab:statCorr-jet2HT}Statistical correlation matrix for the $H_T$ [two-jet] distribution.}
\tiny
\begin{ruledtabular}
%
\end{ruledtabular}
\end{table}

\begin{table}[htbp]
\caption{\label{tab:statCorr-jet3HT}Statistical correlation matrix for the $H_T$ [three-jet] distribution.}
\tiny
\begin{ruledtabular}
%
\end{ruledtabular}
\end{table}

\begin{table}[htbp]
\caption{\label{tab:statCorr-jet4HT}Statistical correlation matrix for the $H_T$ [four-jet] distribution.}
\tiny
\begin{ruledtabular}
%
\end{ruledtabular}
\end{table}

\clearpage
\newpage
\section{Normalized Inverse Covariance Matrices From Unfolding}

\begin{center}
{\bf To appear as an Electronic Physics Auxiliary Publication (EPAPS)}
\end{center}

The inverse of the statistical covariance matrix can be a more directly useful quantity for fitting models to the unfolded data, 
and so we provide them in this appendix.   The values are normalized to $1.0$ along the diagonal.

\begin{figure}[htbp]
  \begin{center}
    \includegraphics[width=0.35\textwidth]{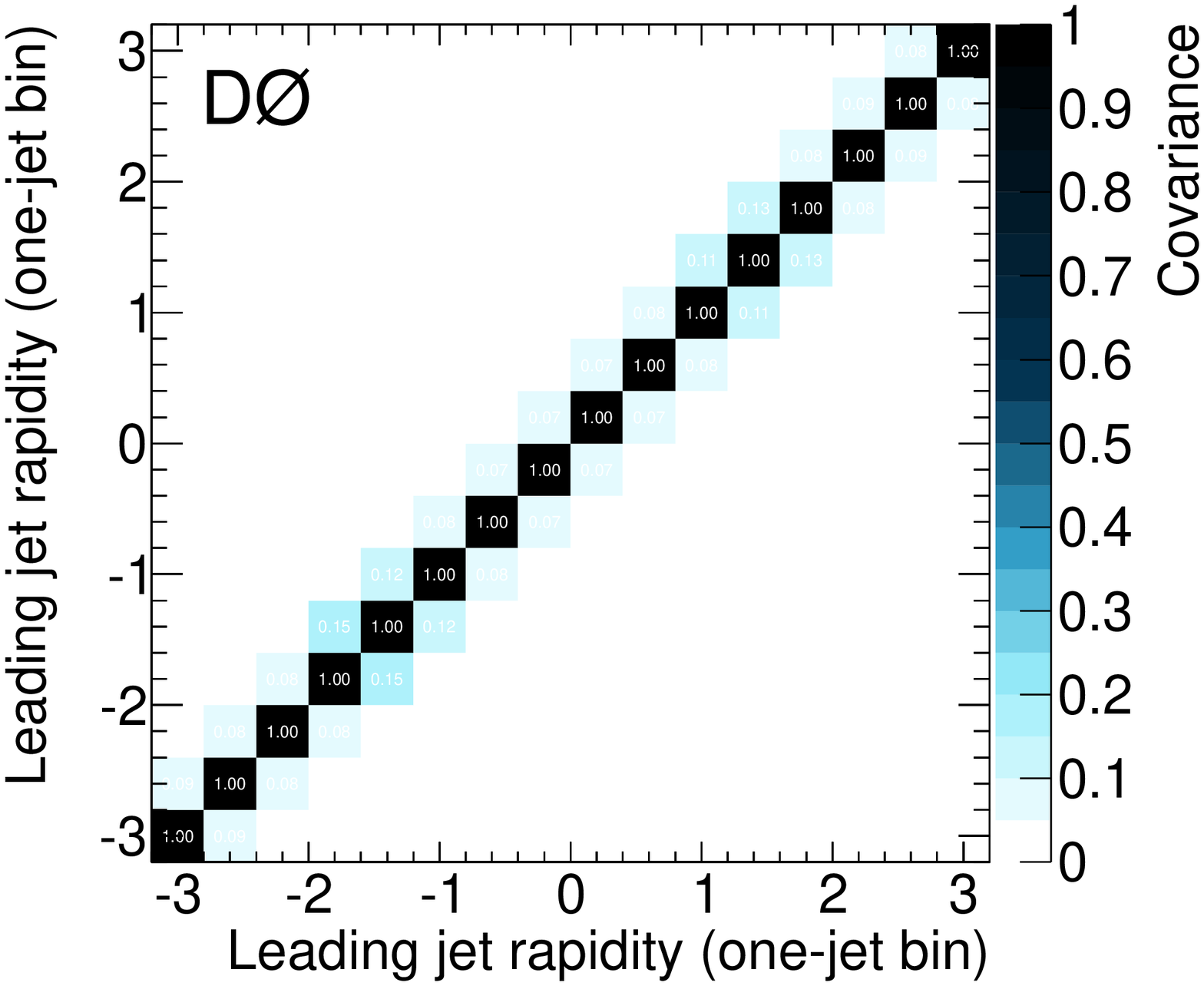}
    \includegraphics[width=0.35\textwidth]{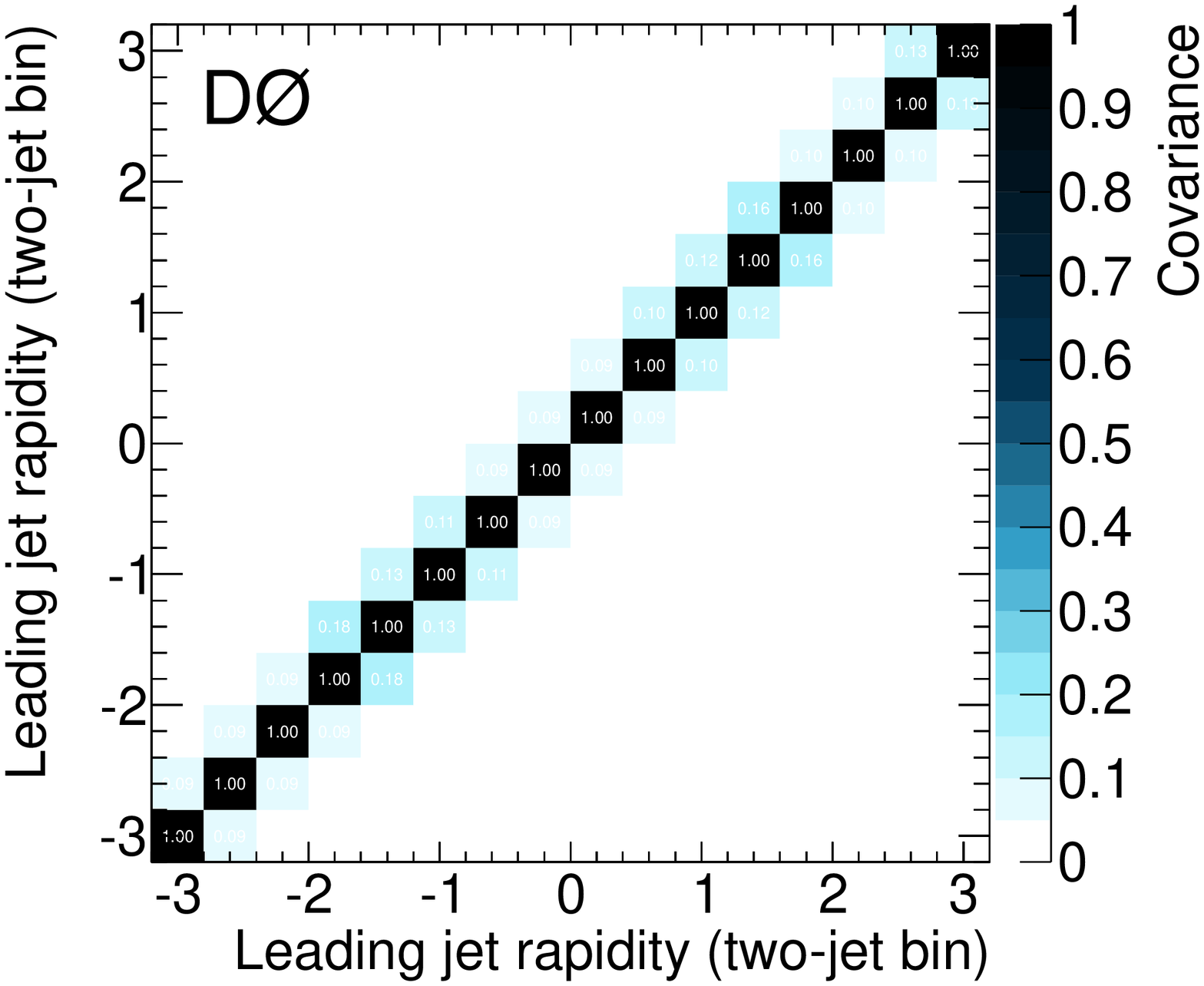}
    \includegraphics[width=0.35\textwidth]{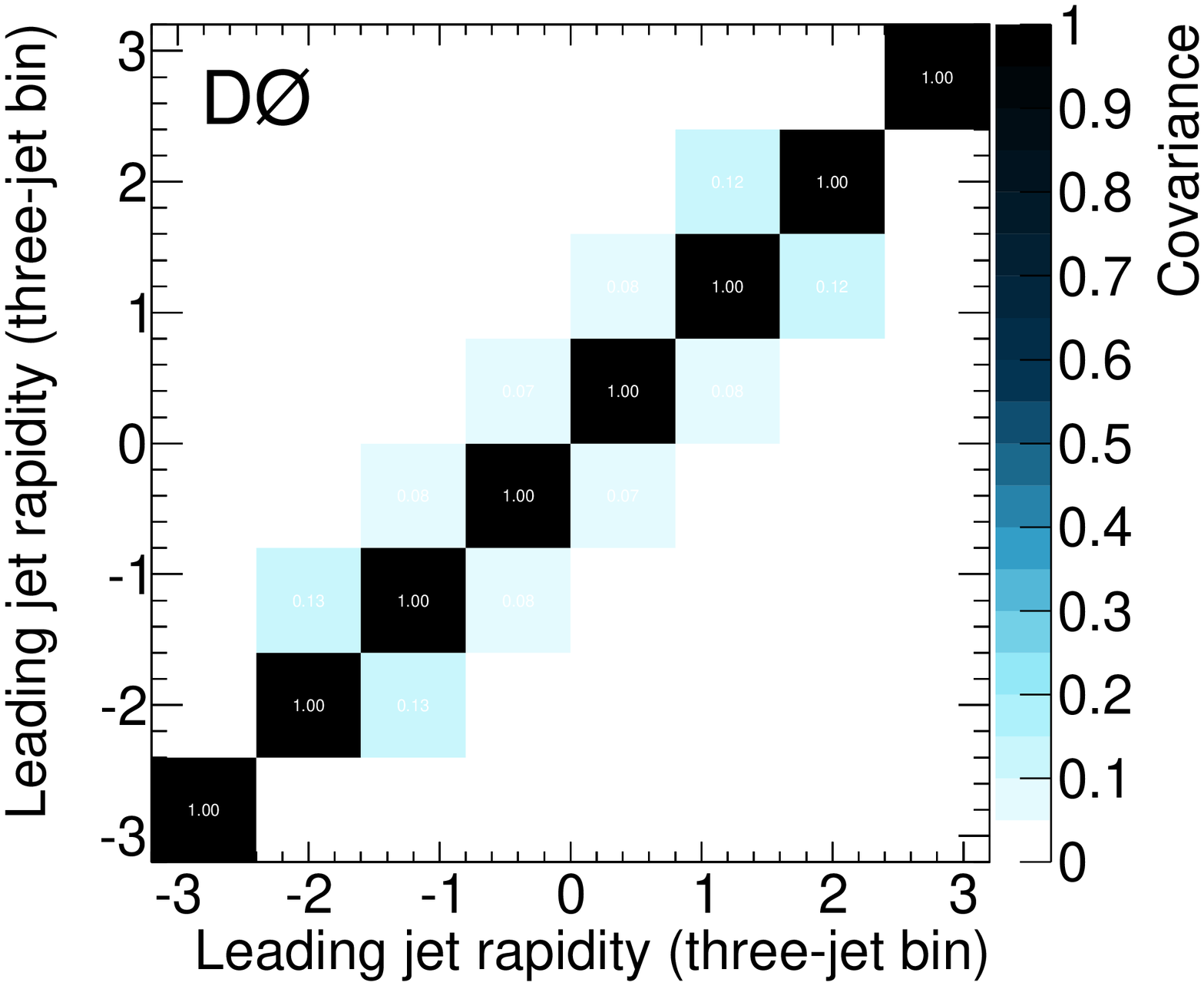}
    \includegraphics[width=0.35\textwidth]{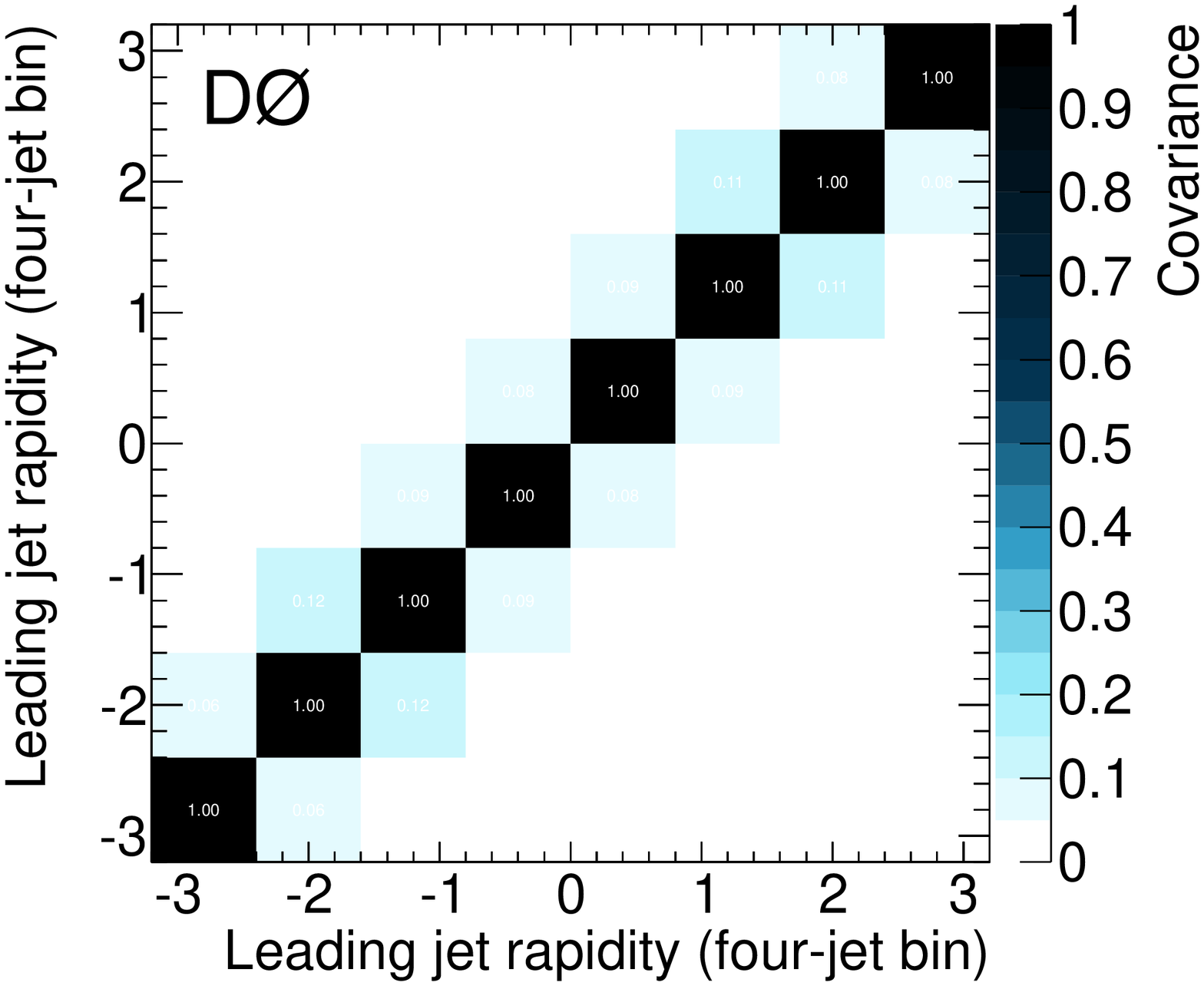}
    \caption{Normalized inverse covariance matrices for unfolded jet rapidity distributions.
     \label{fig:normInvCov_jeteta}
    }
  \end{center}
\end{figure}

\begin{figure}[htbp]
  \begin{center}
    \includegraphics[width=0.35\textwidth]{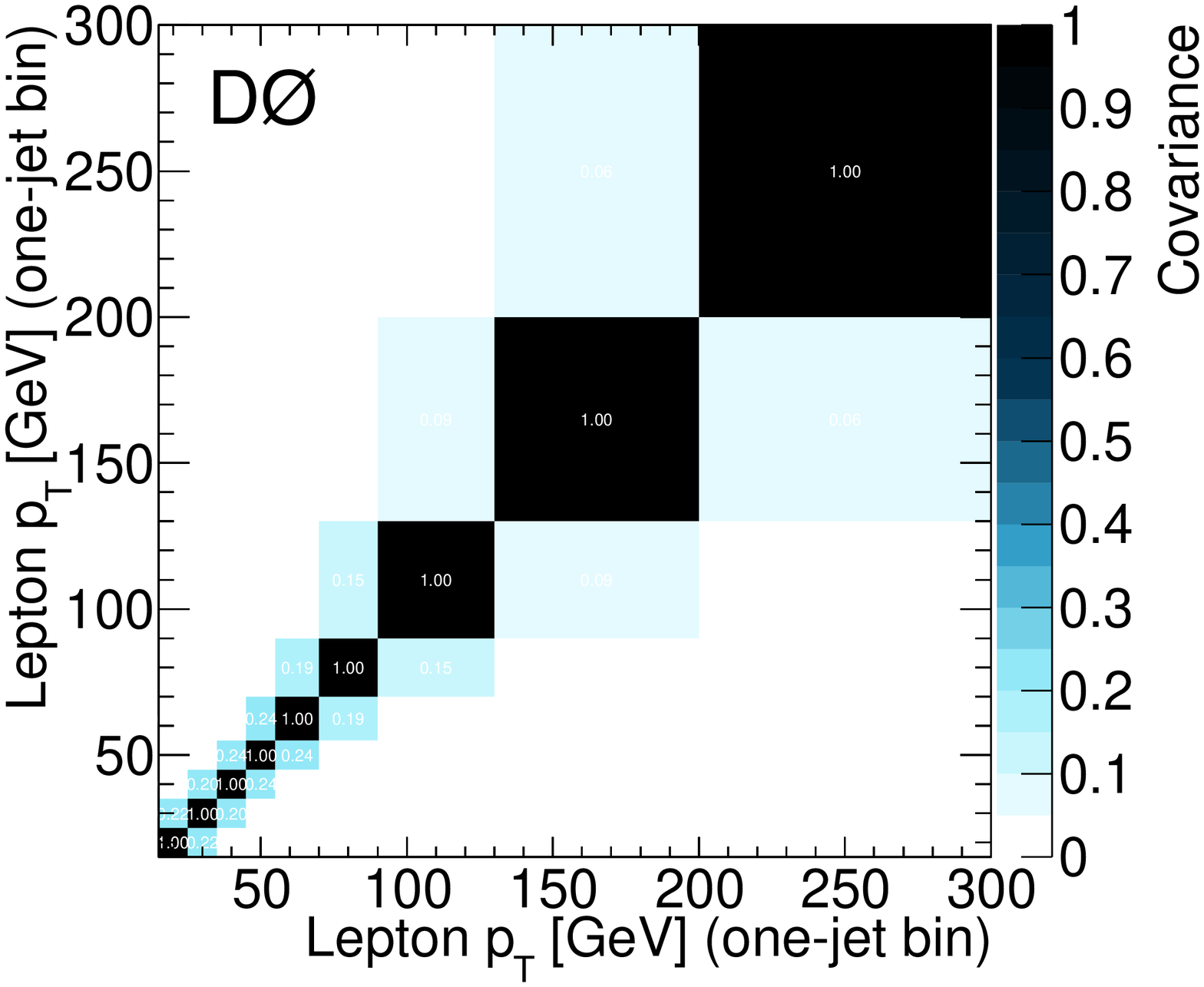}
    \includegraphics[width=0.35\textwidth]{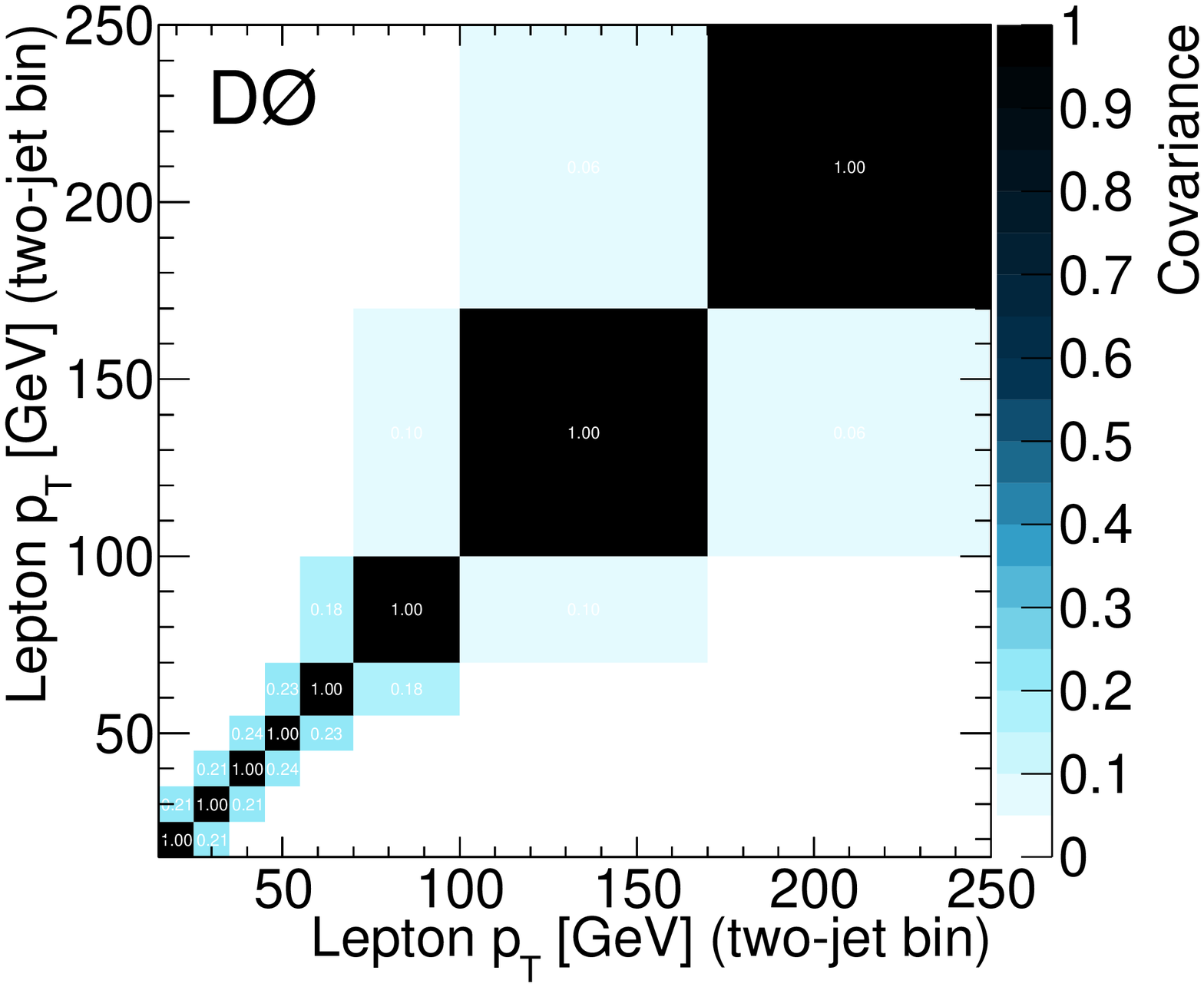}
    \includegraphics[width=0.35\textwidth]{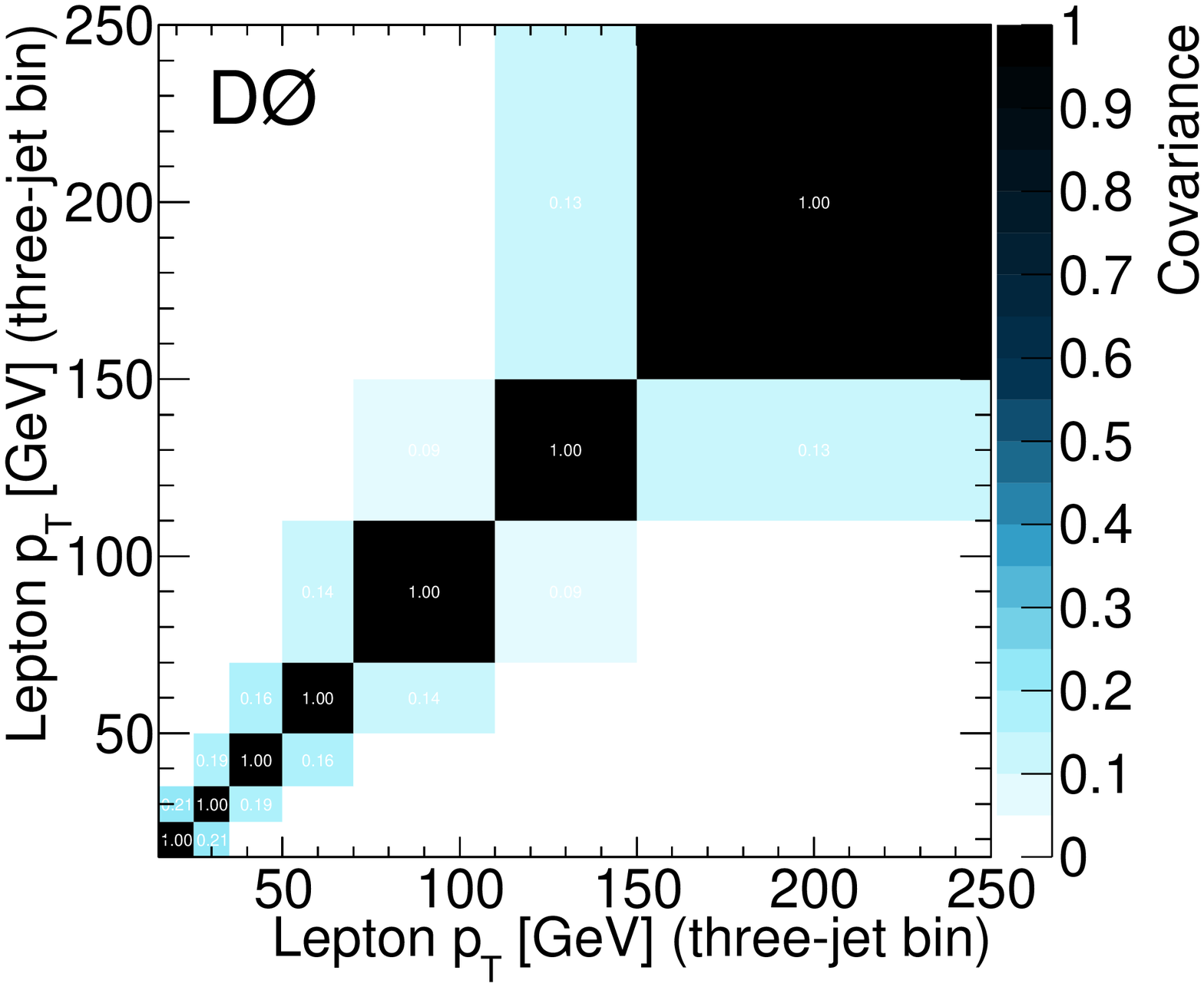}
    \includegraphics[width=0.35\textwidth]{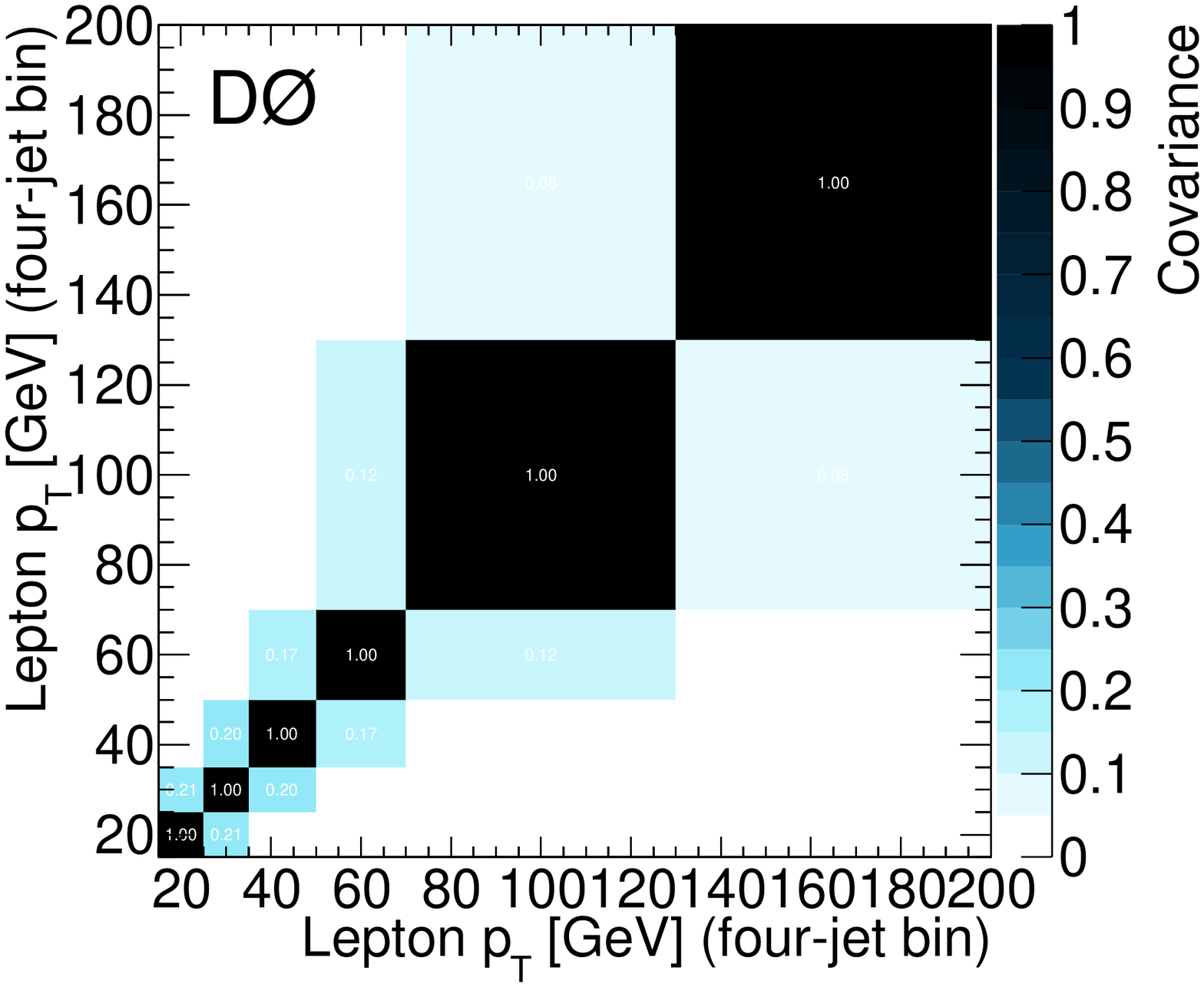}
    \caption{Normalized inverse covariance matrices for unfolded lepton transverse momentum distributions.
    \label{fig:normInvCov_leptonpt}
    }
  \end{center}
\end{figure}

\begin{figure}[htbp]
  \begin{center}
    \includegraphics[width=0.35\textwidth]{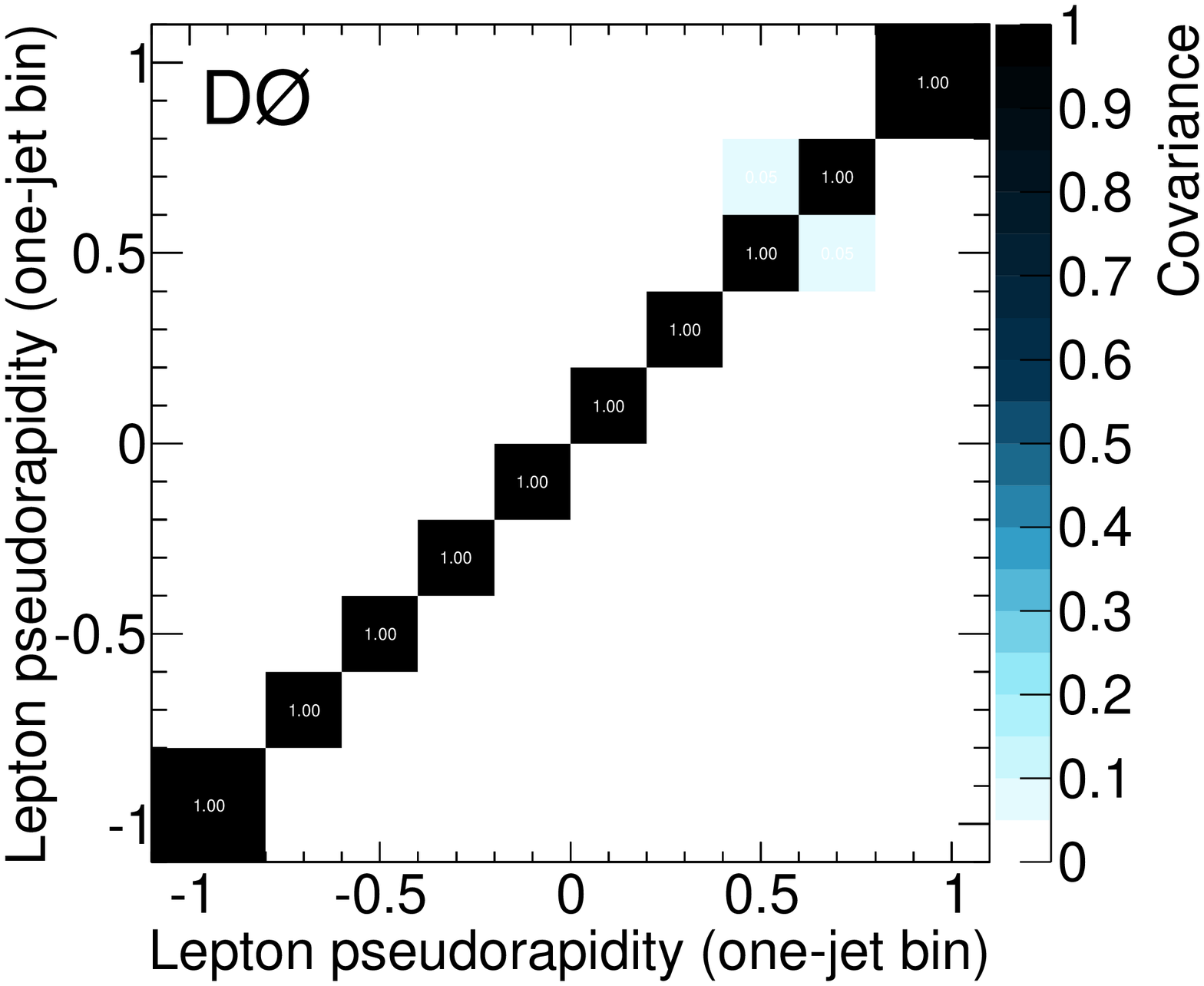}
    \includegraphics[width=0.35\textwidth]{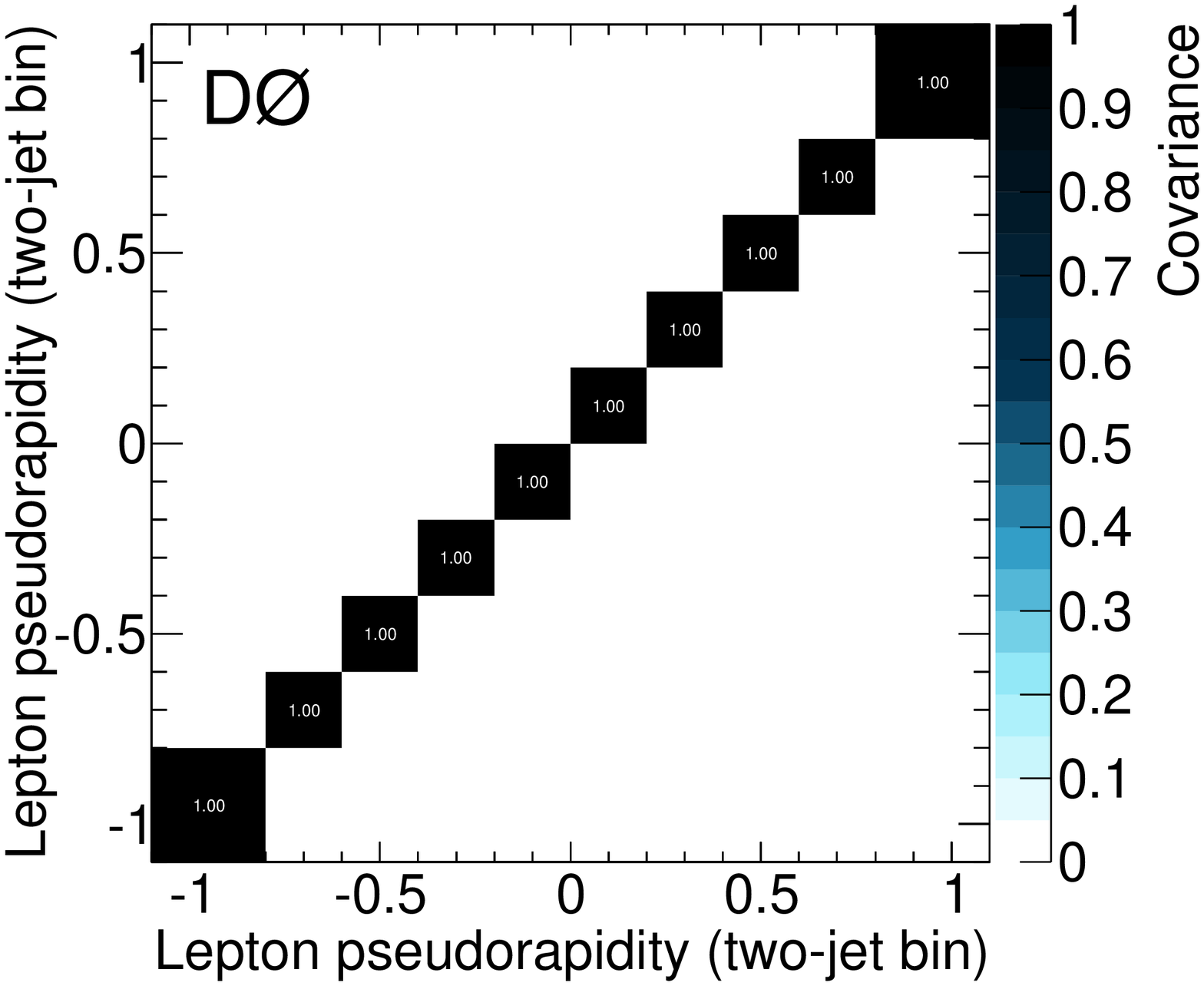}
    \includegraphics[width=0.35\textwidth]{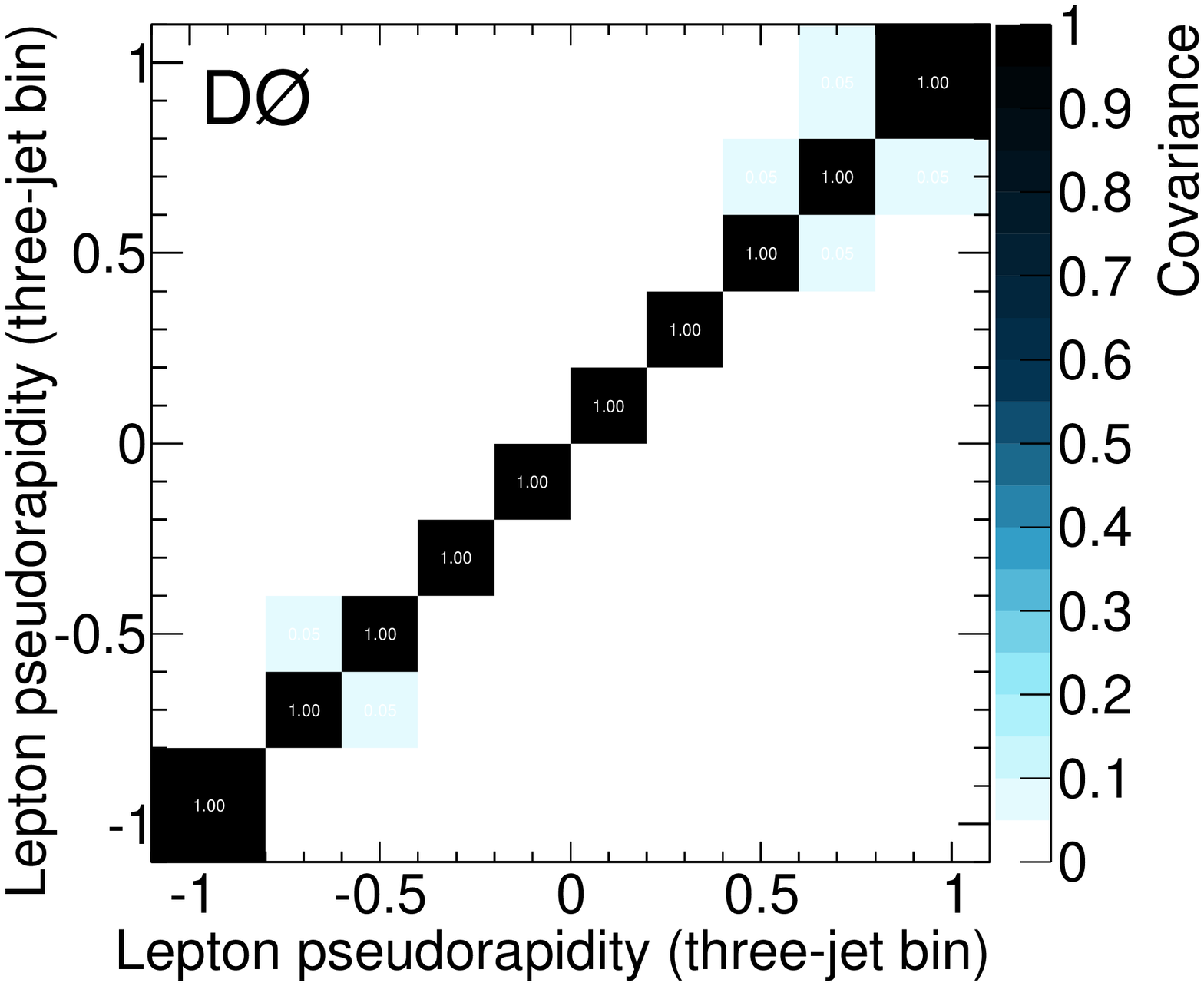}
    \includegraphics[width=0.35\textwidth]{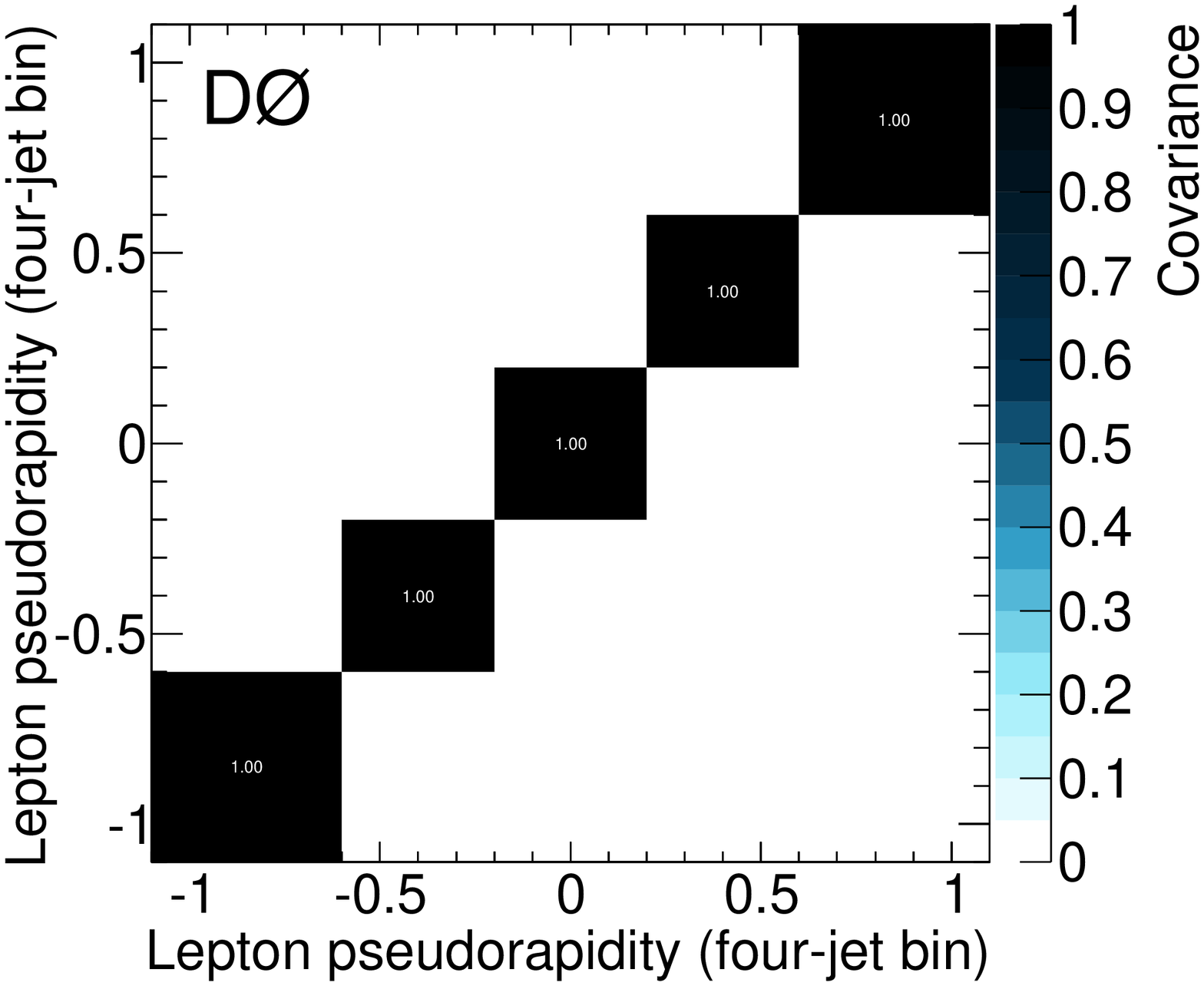}
    \caption{Normalized inverse covariance matrices for unfolded lepton pseudorapidity distributions.
      \label{fig:normInvCov_leptoneta}
    }
  \end{center}
\end{figure}

\begin{figure}[htbp]
  \begin{center}
    \includegraphics[width=0.35\textwidth]{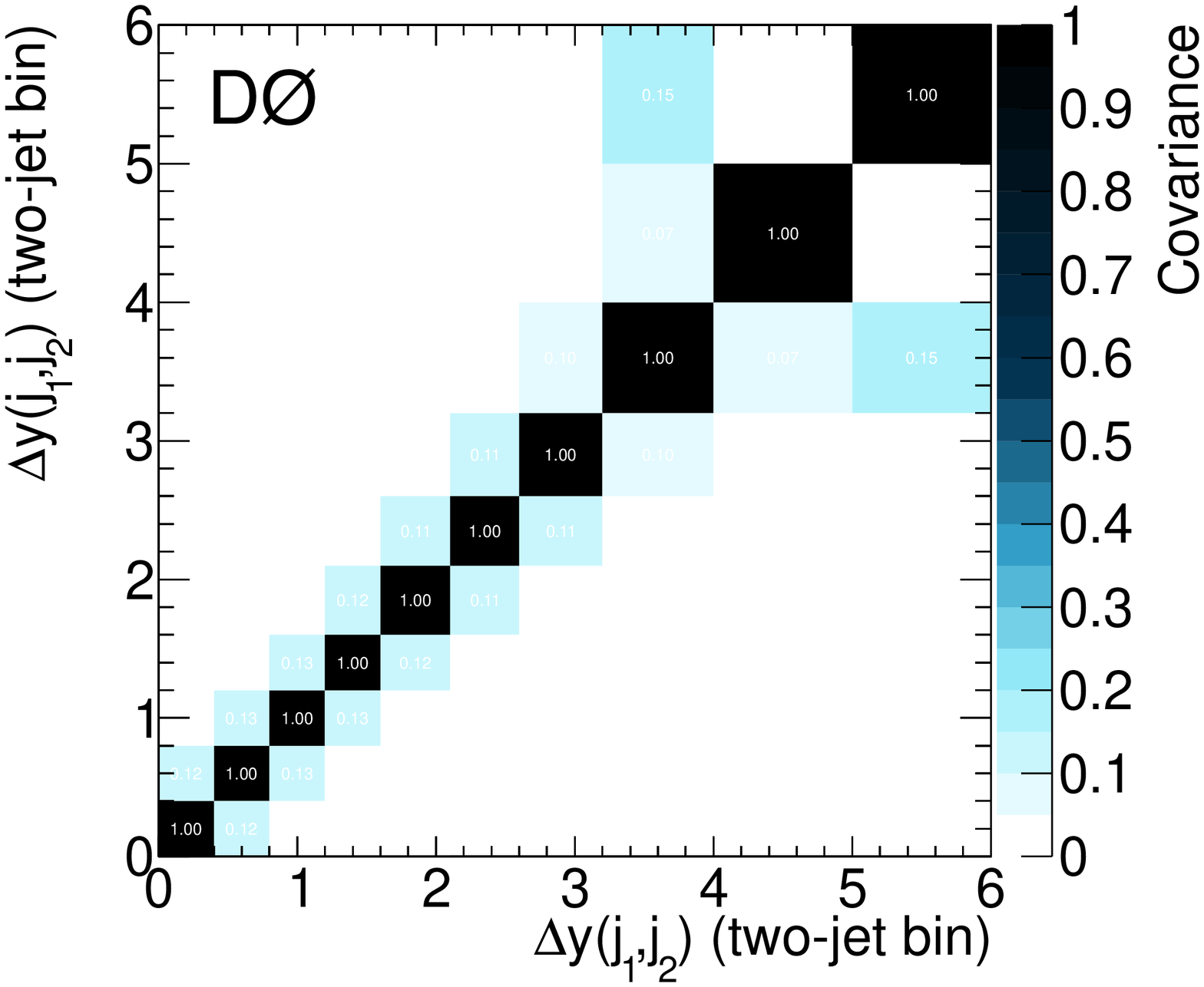}
    \includegraphics[width=0.35\textwidth]{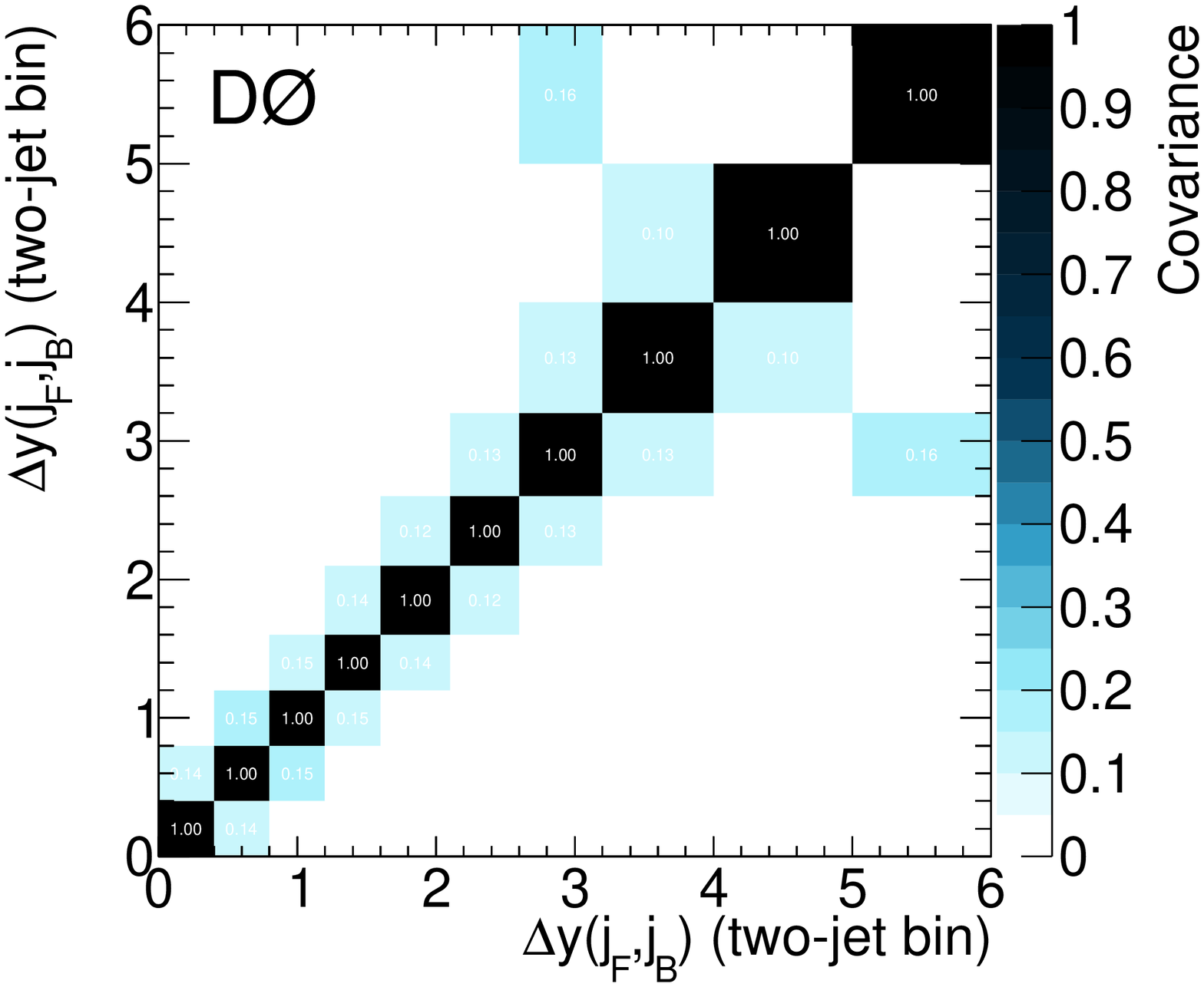}
    \includegraphics[width=0.35\textwidth]{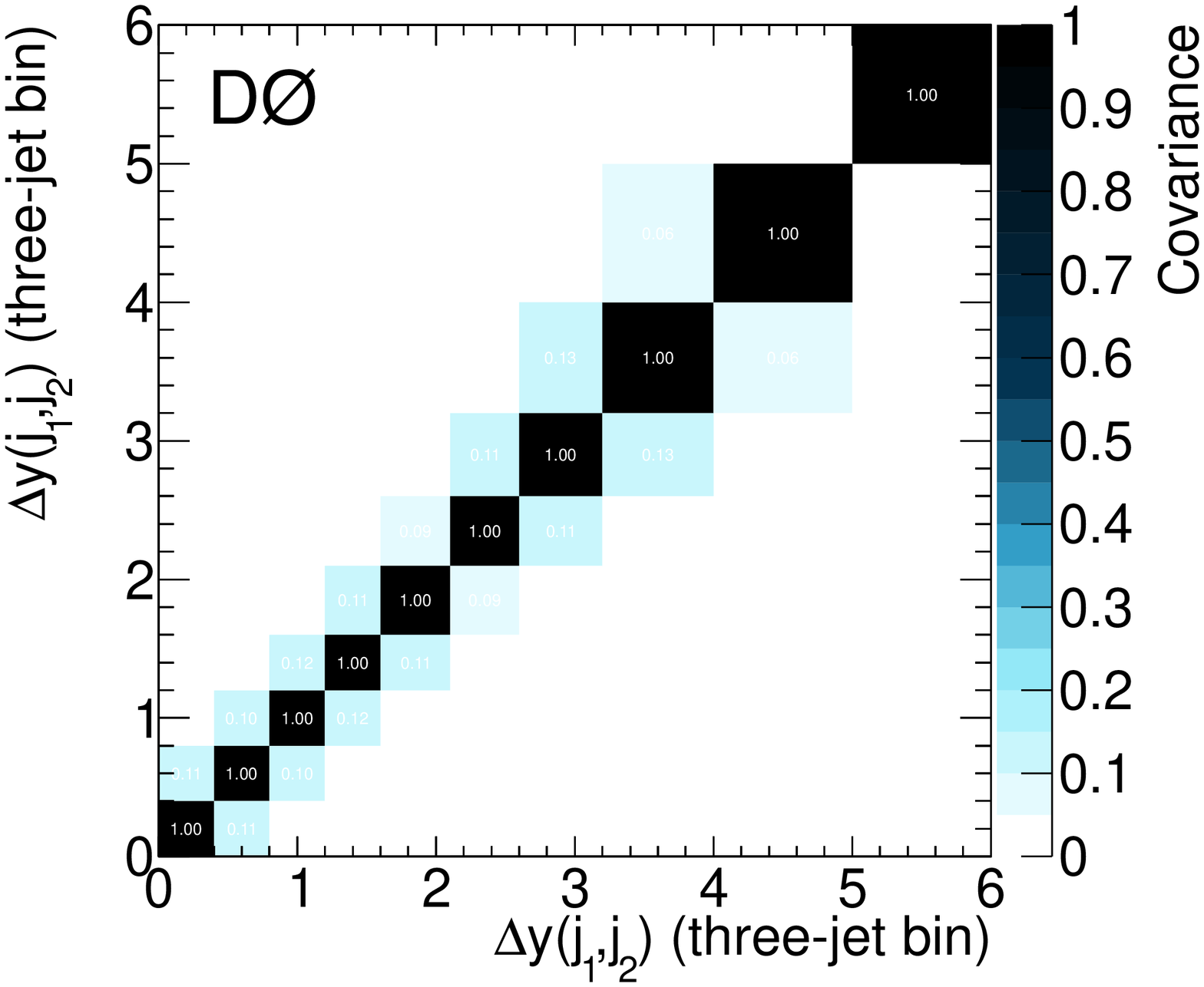}
    \includegraphics[width=0.35\textwidth]{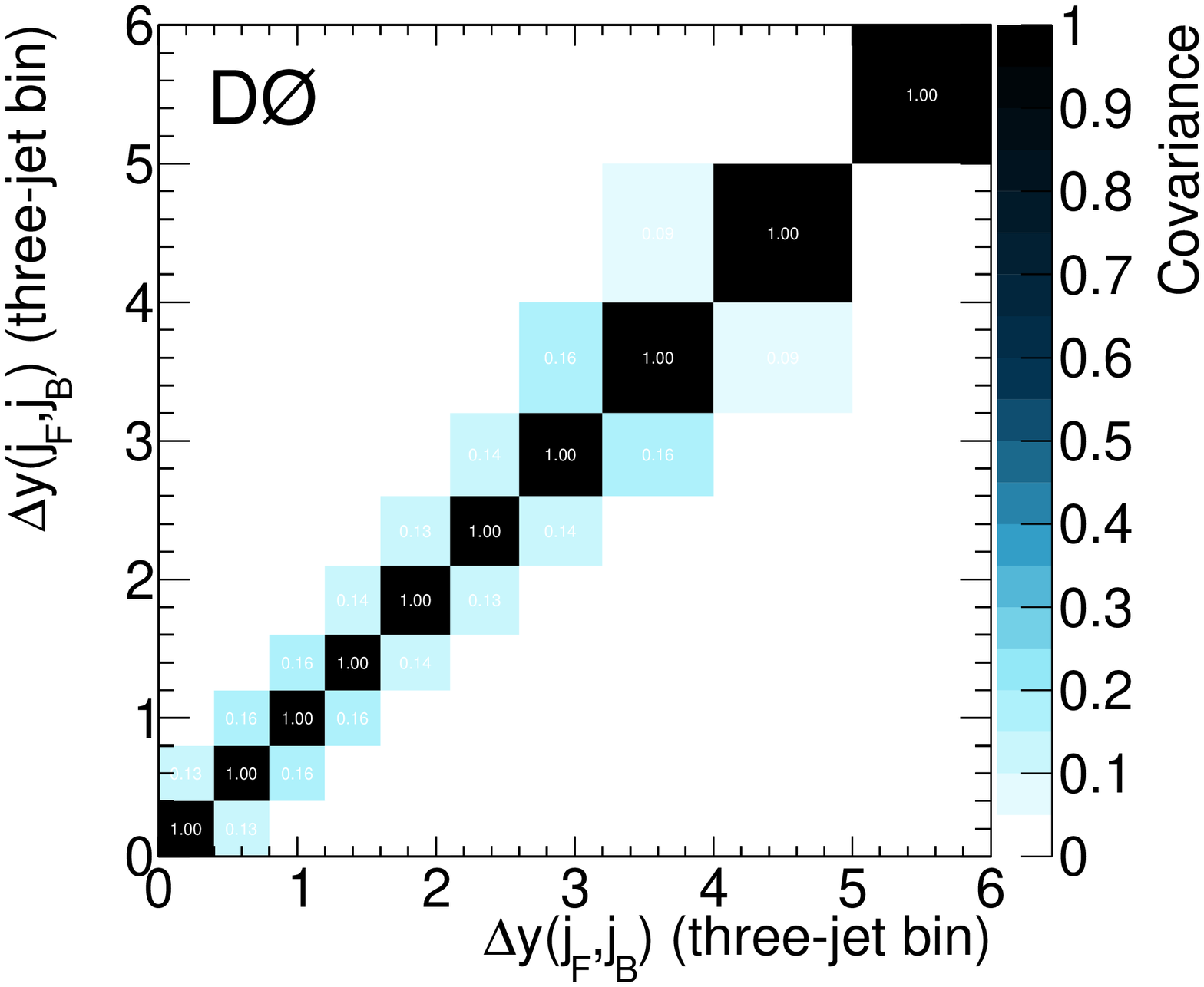}
    \caption{Normalized inverse covariance matrices for unfolded dijet rapidity separations for the two highest $p_T$ or two most rapidity-separated jets.    
      \label{fig:normInvCov_jetdeltarap}
    }
  \end{center}
\end{figure}

\begin{figure}[htbp]
  \begin{center}
    \includegraphics[width=0.35\textwidth]{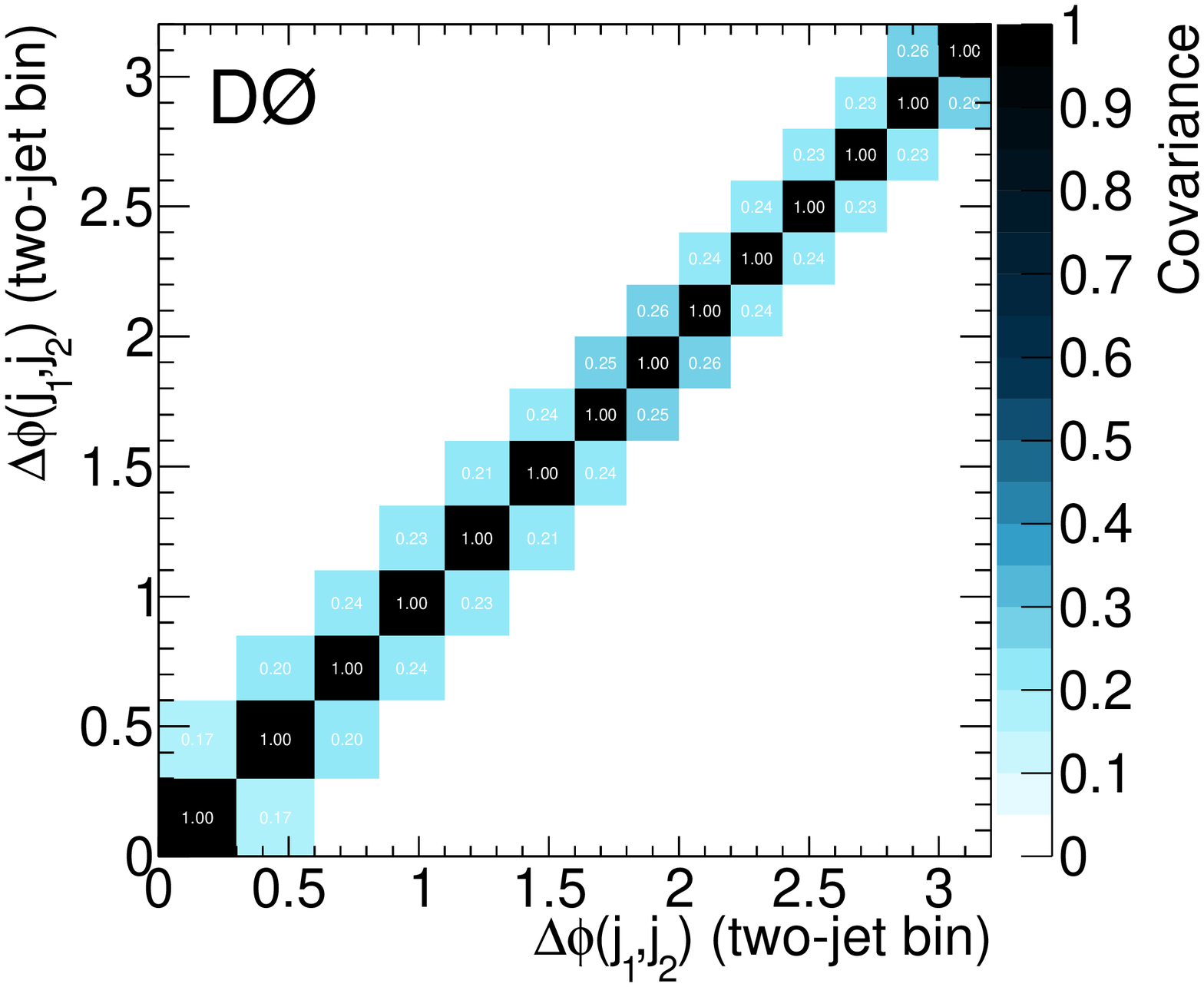}
    \includegraphics[width=0.35\textwidth]{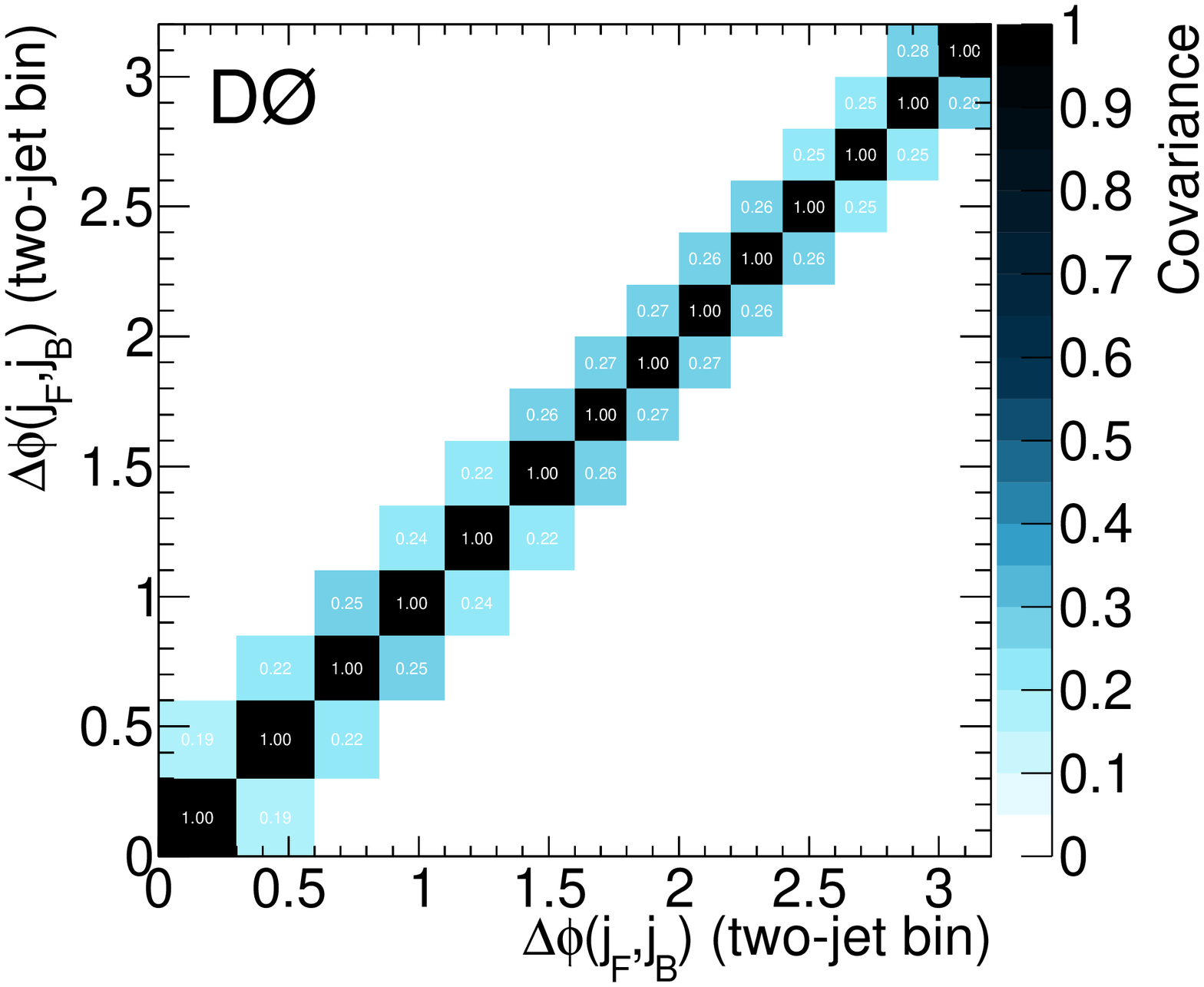}
    \includegraphics[width=0.35\textwidth]{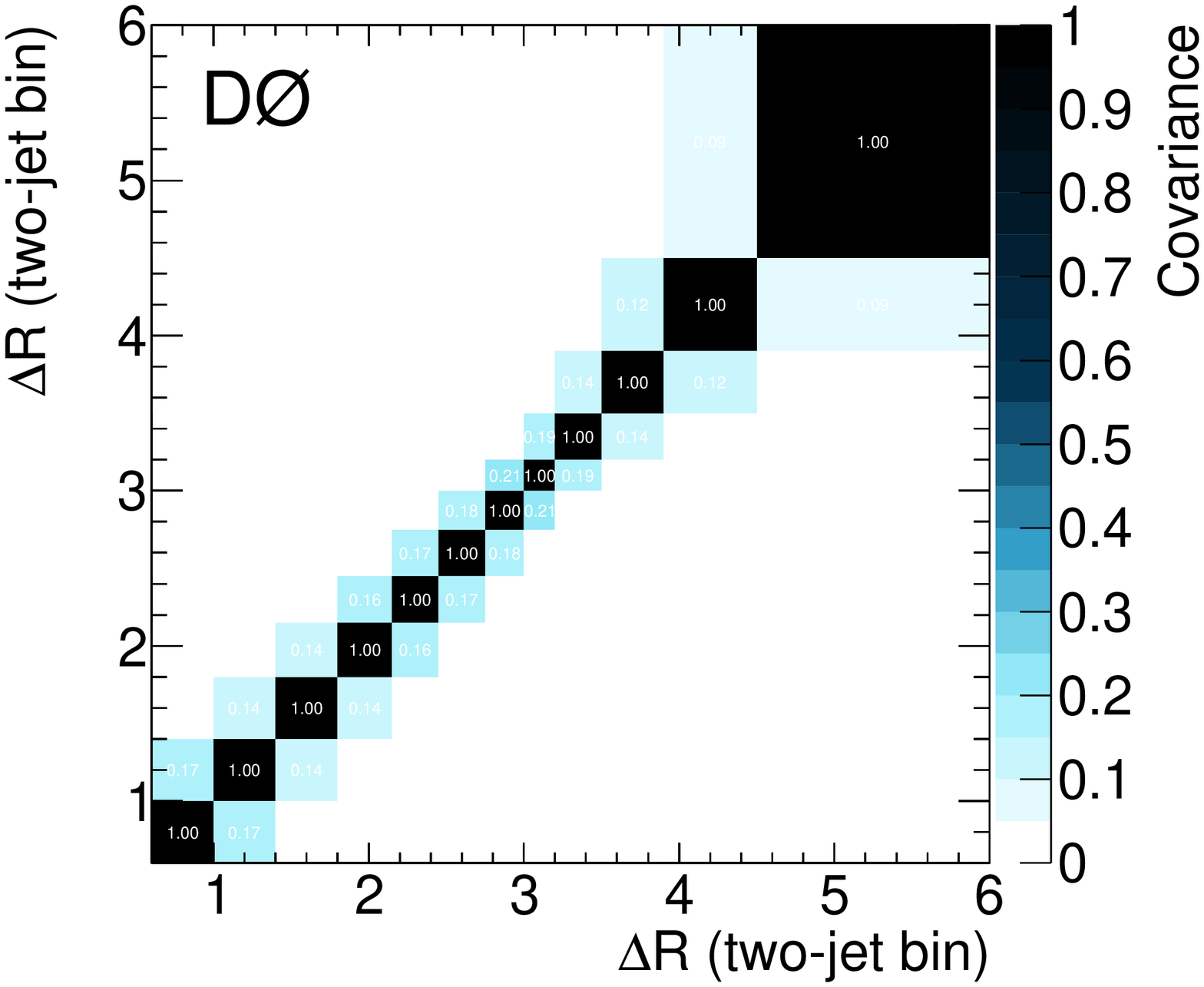}
    \caption{Normalized inverse covariance matrices for unfolded azimuthal angle separation between the two highest $p_T$ or two most rapidity-separated jets, 
      and the dijet angular separation in $\eta-\phi$ space.
      \label{fig:normInvCov_jetdphi_dR}
    }
  \end{center}
\end{figure}

\begin{figure}[htbp]
  \begin{center}
    \includegraphics[width=0.35\textwidth]{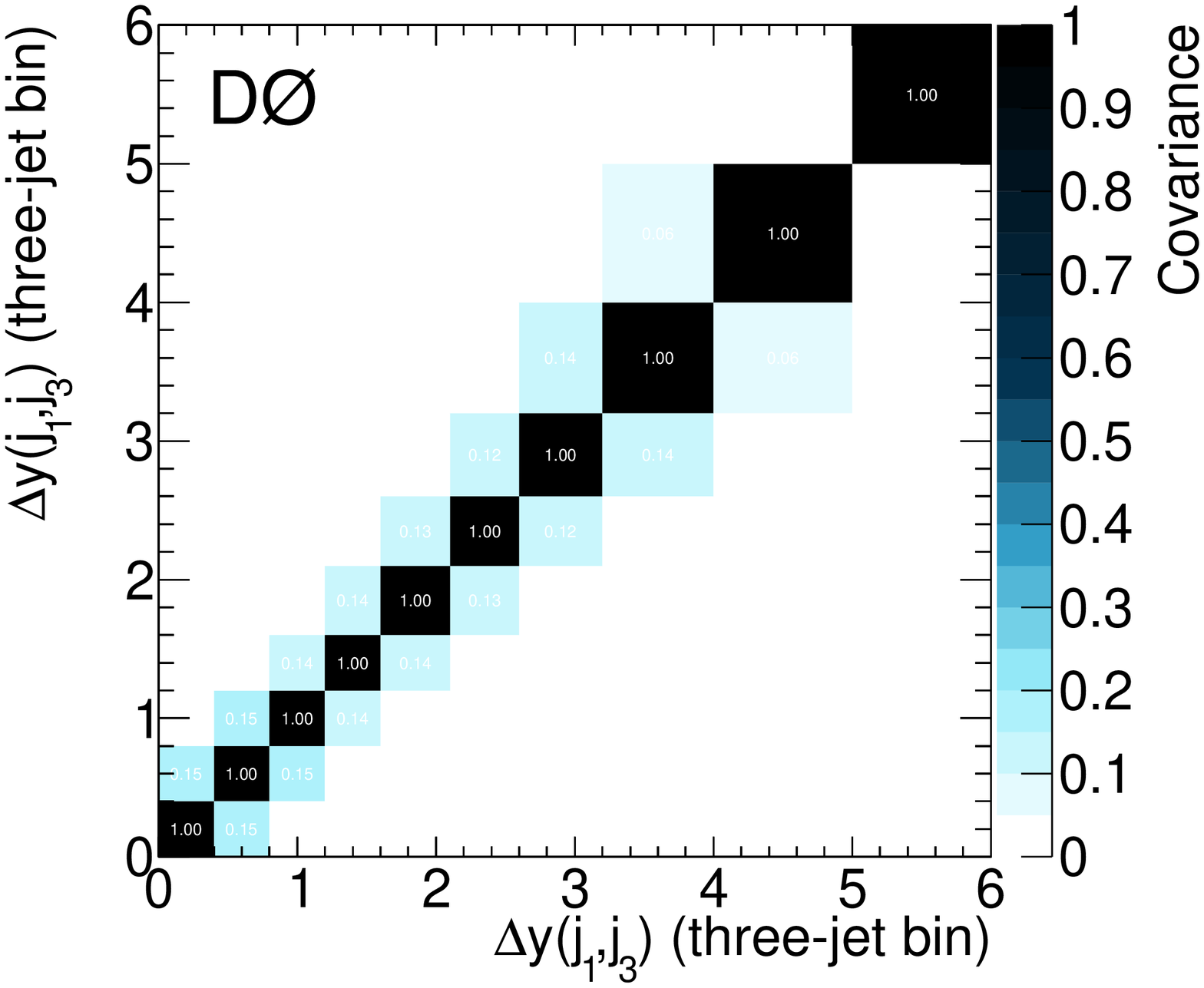}
    \includegraphics[width=0.35\textwidth]{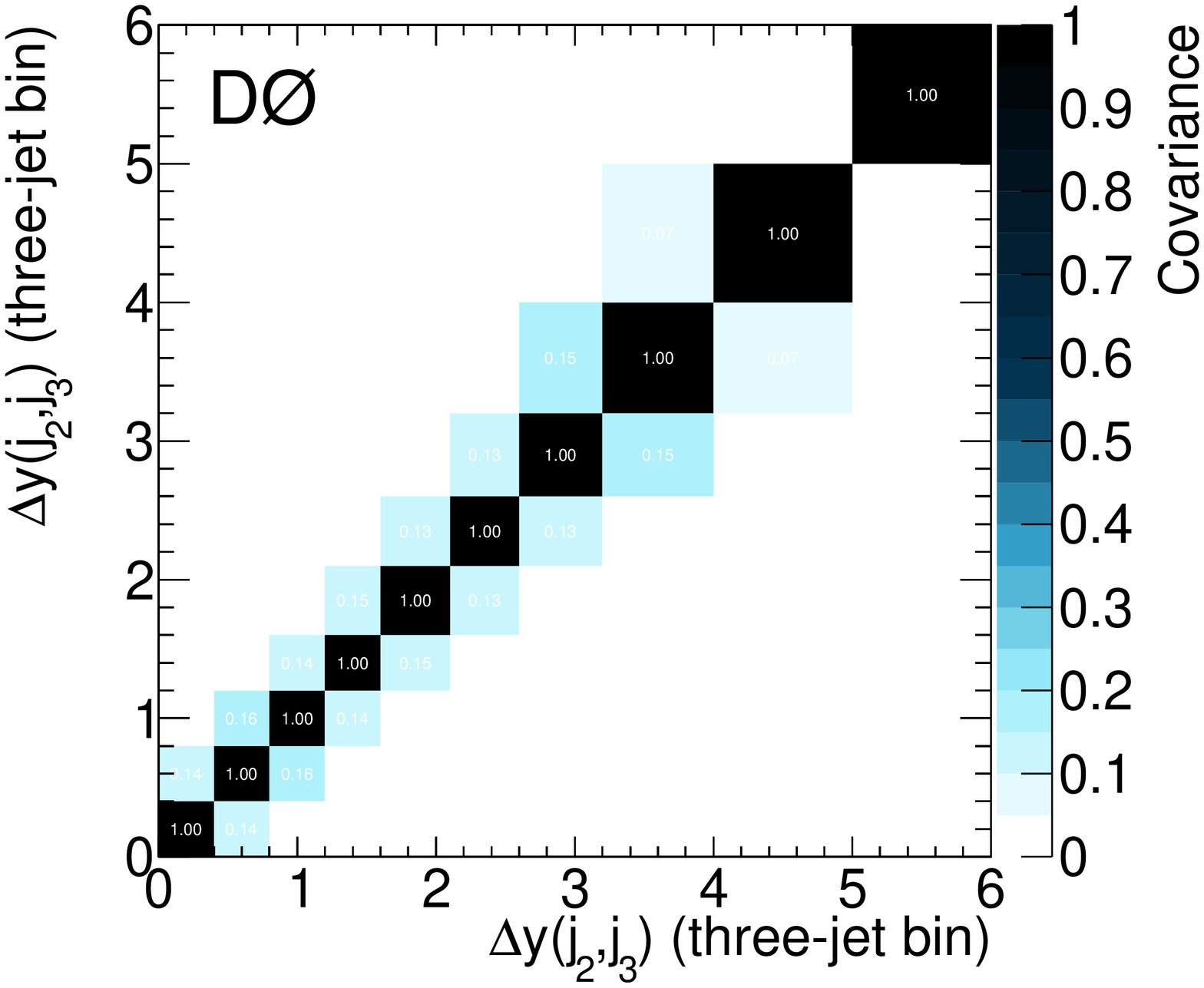}
    \caption{Normalized inverse covariance matrices for unfolded dijet rapidity separation between the highest and second highest $p_T$ jets and the second and third highest $p_T$ 
      jets in $W+3\textrm{-jet}$ events.
      \label{fig:normInvCov_jetDeltaRap3j}
    }
  \end{center}
\end{figure}

\begin{figure}[htbp]
  \begin{center}
    \includegraphics[width=0.35\textwidth]{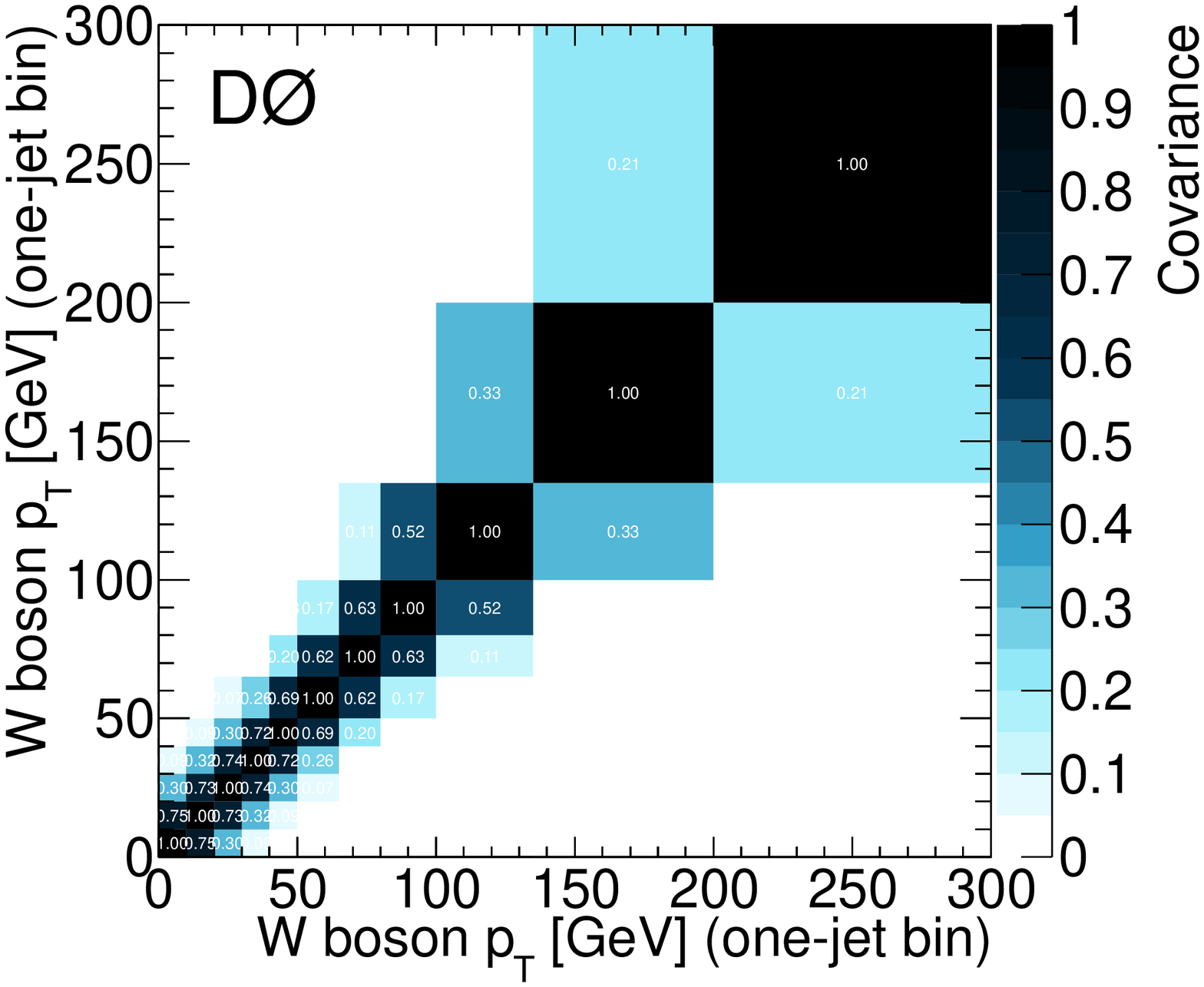}
    \includegraphics[width=0.35\textwidth]{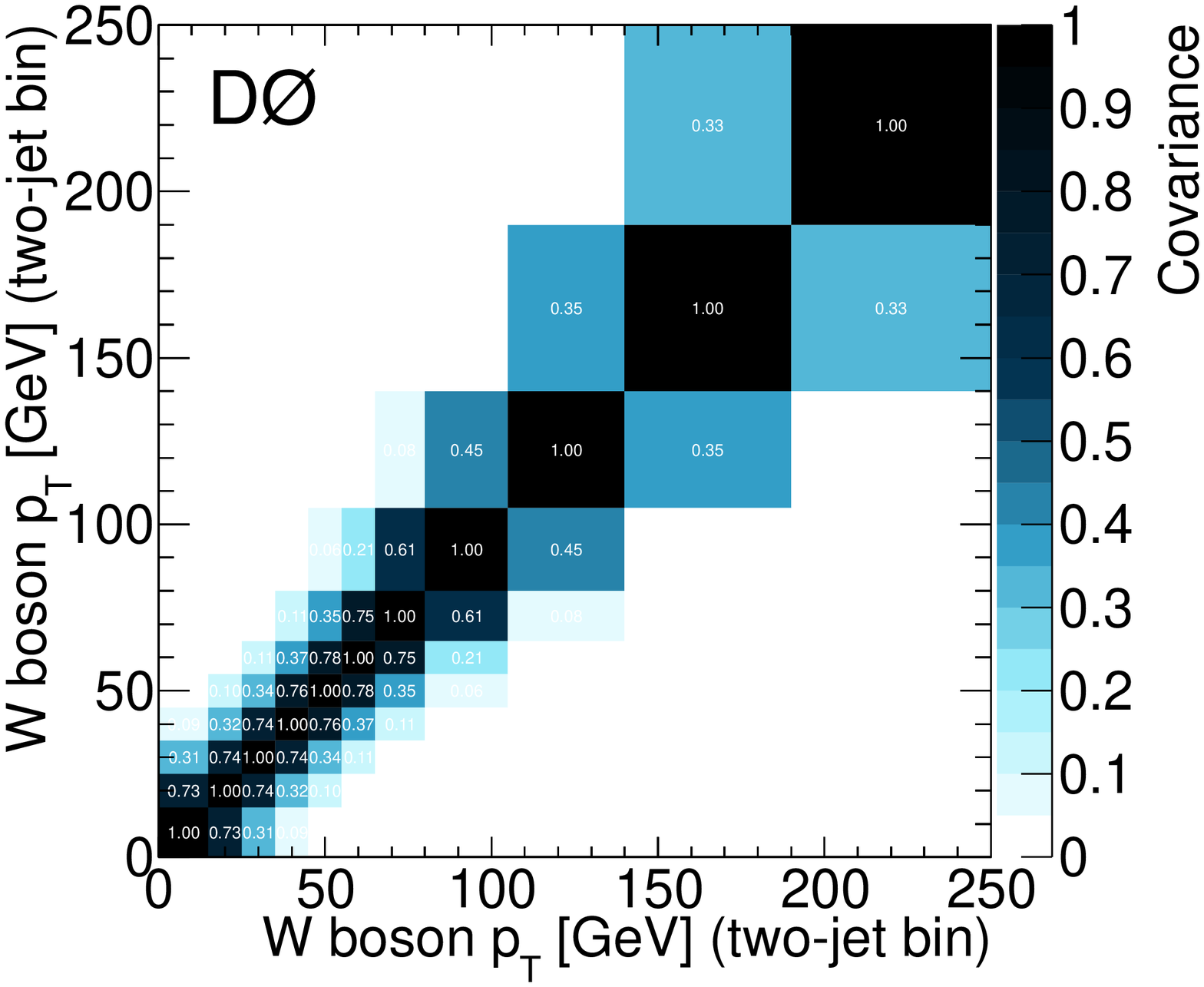}
    \includegraphics[width=0.35\textwidth]{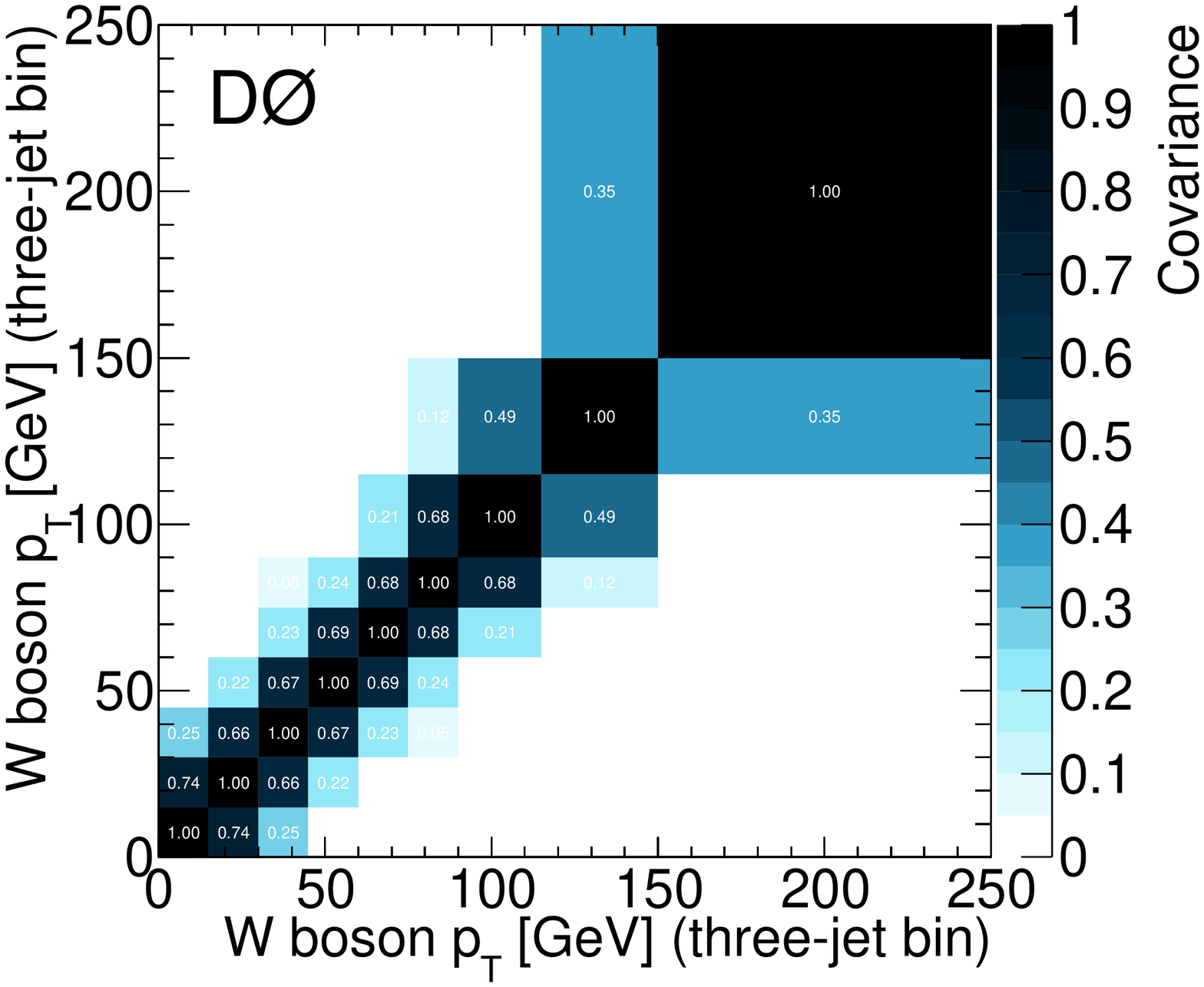}
    \includegraphics[width=0.35\textwidth]{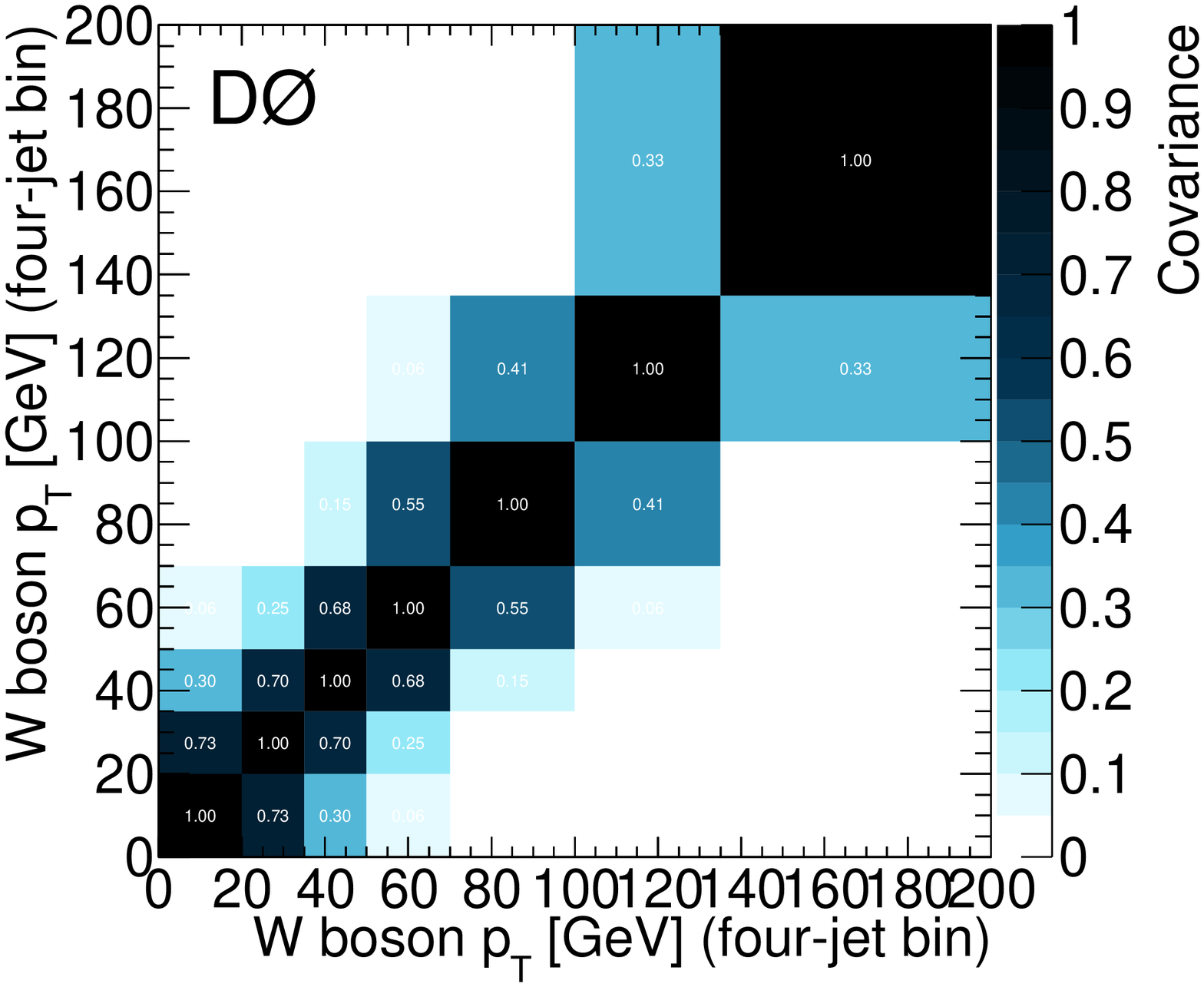}
    \caption{Normalized inverse covariance matrices for unfolded $W$ boson transverse momentum distributions.
      \label{fig:normInvCov_Wpt}
    }
  \end{center}
\end{figure}

\begin{figure}[htbp]
  \begin{center}
    \includegraphics[width=0.35\textwidth]{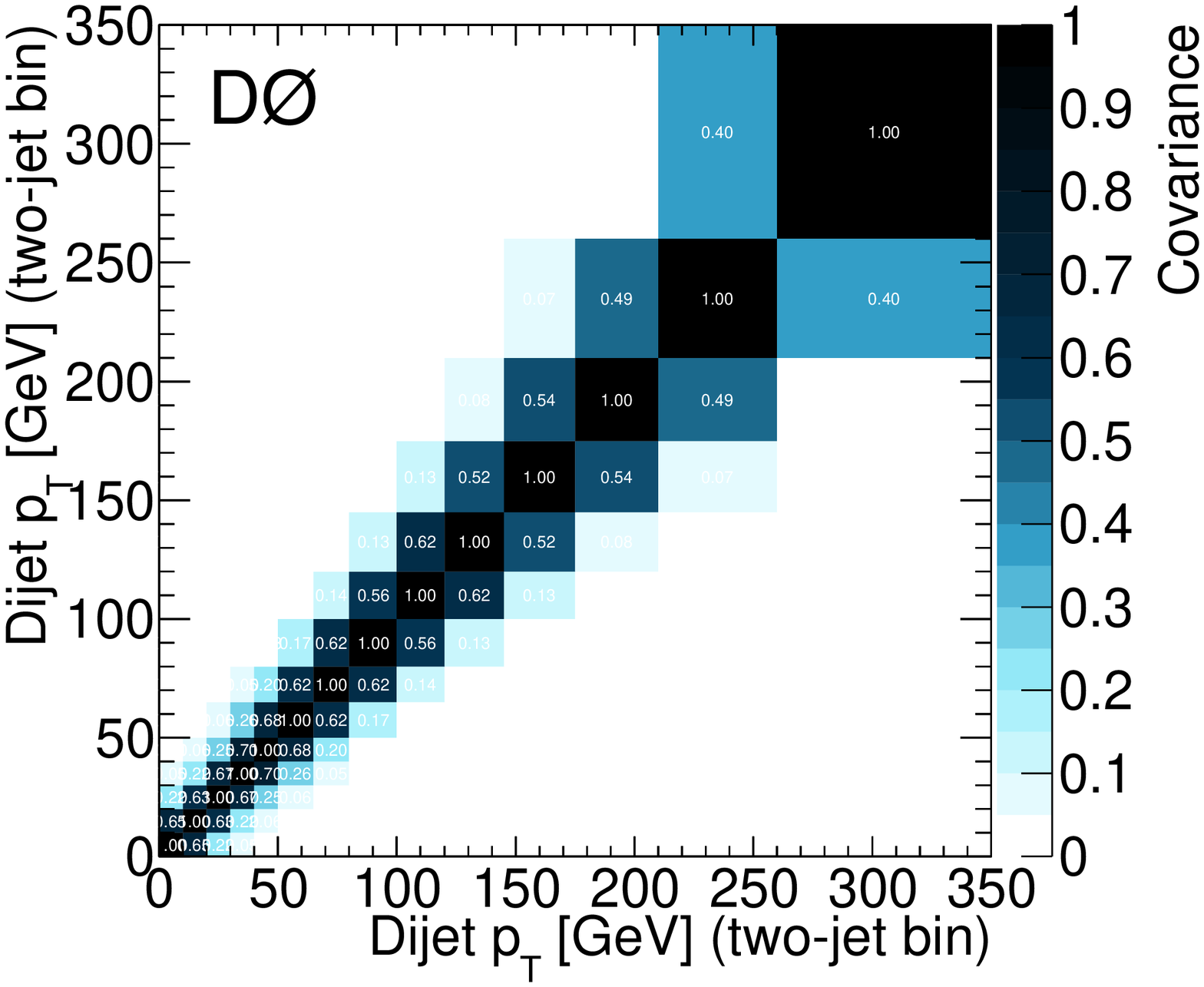}
    \includegraphics[width=0.35\textwidth]{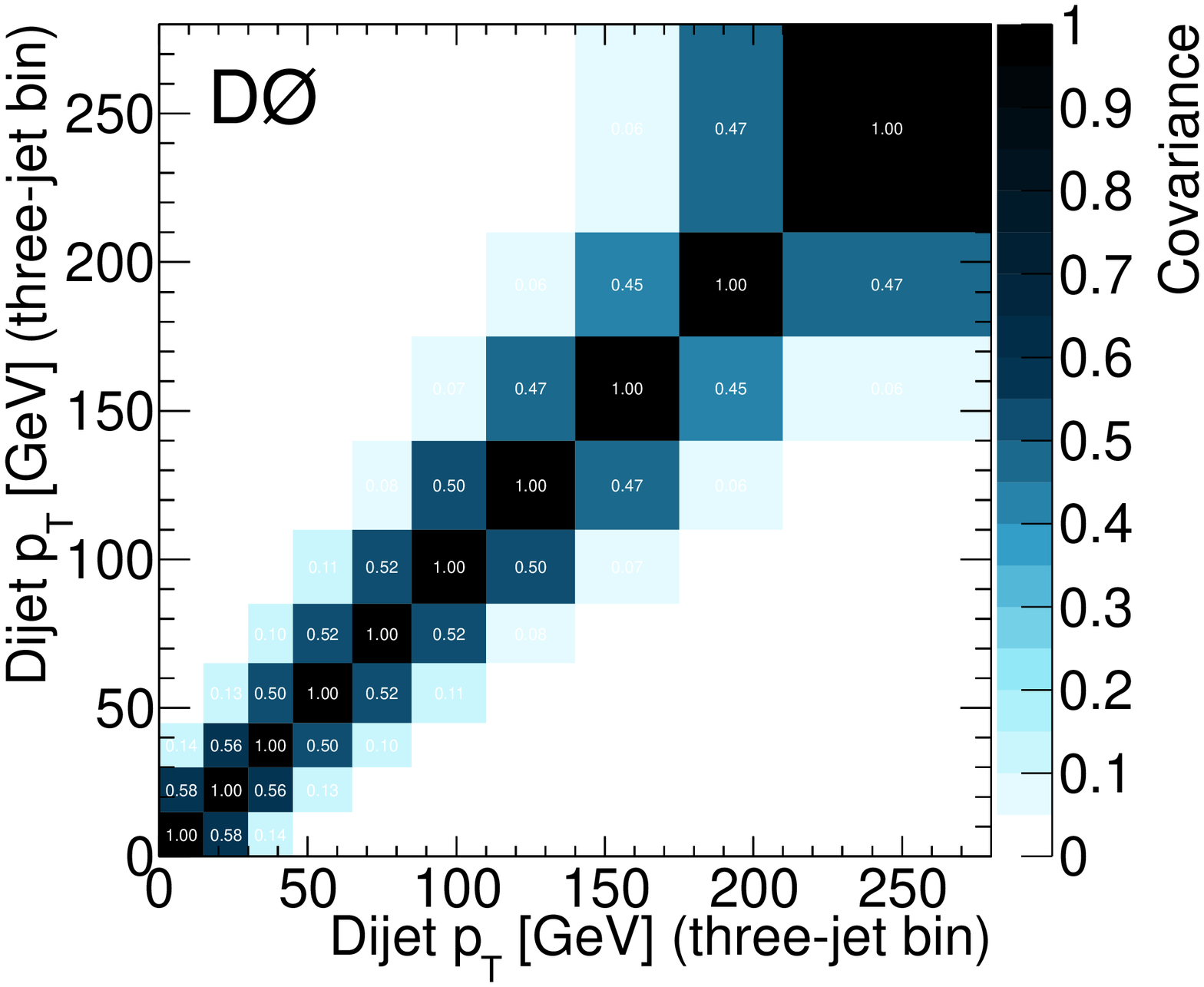}
    \includegraphics[width=0.35\textwidth]{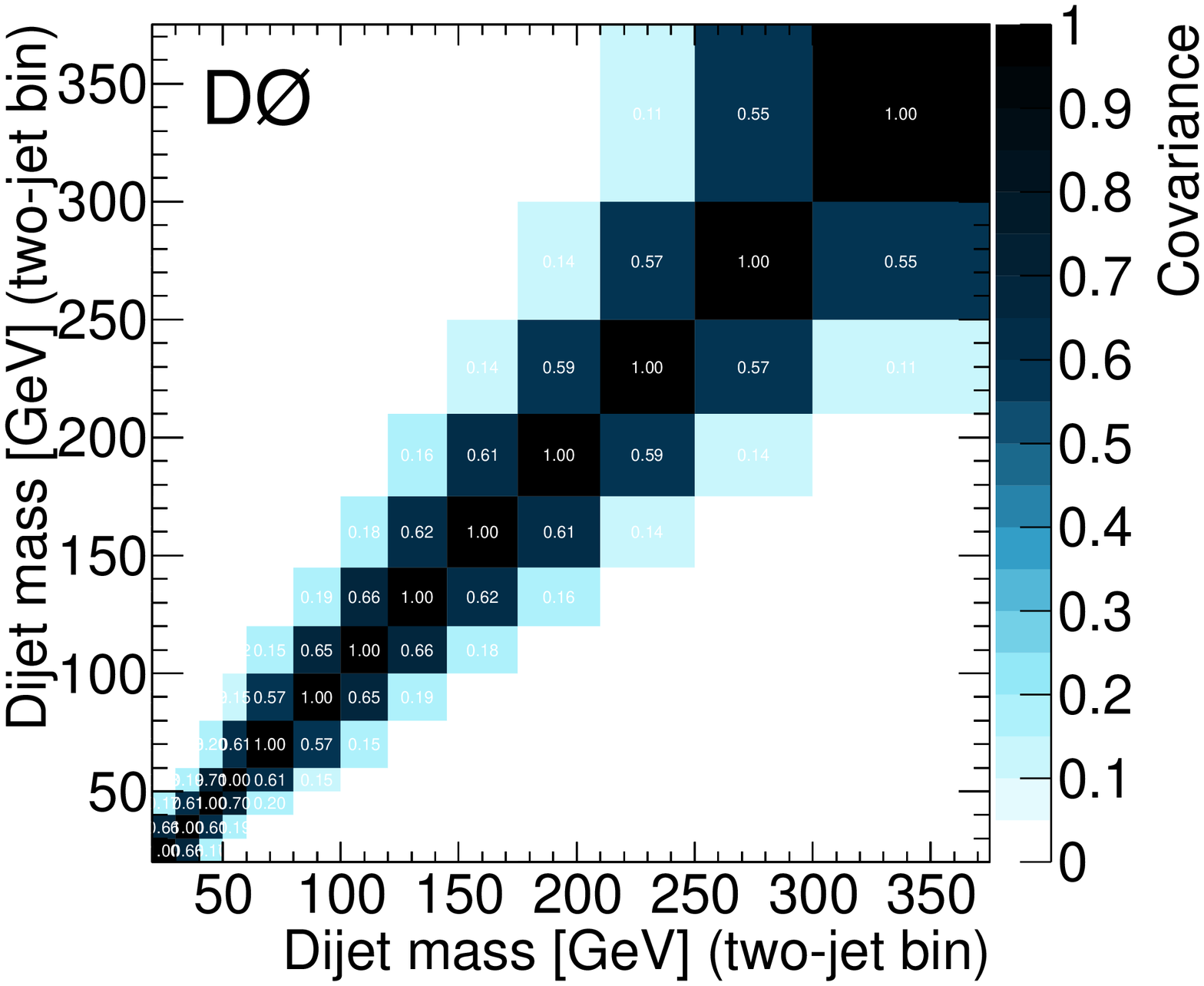}
    \includegraphics[width=0.35\textwidth]{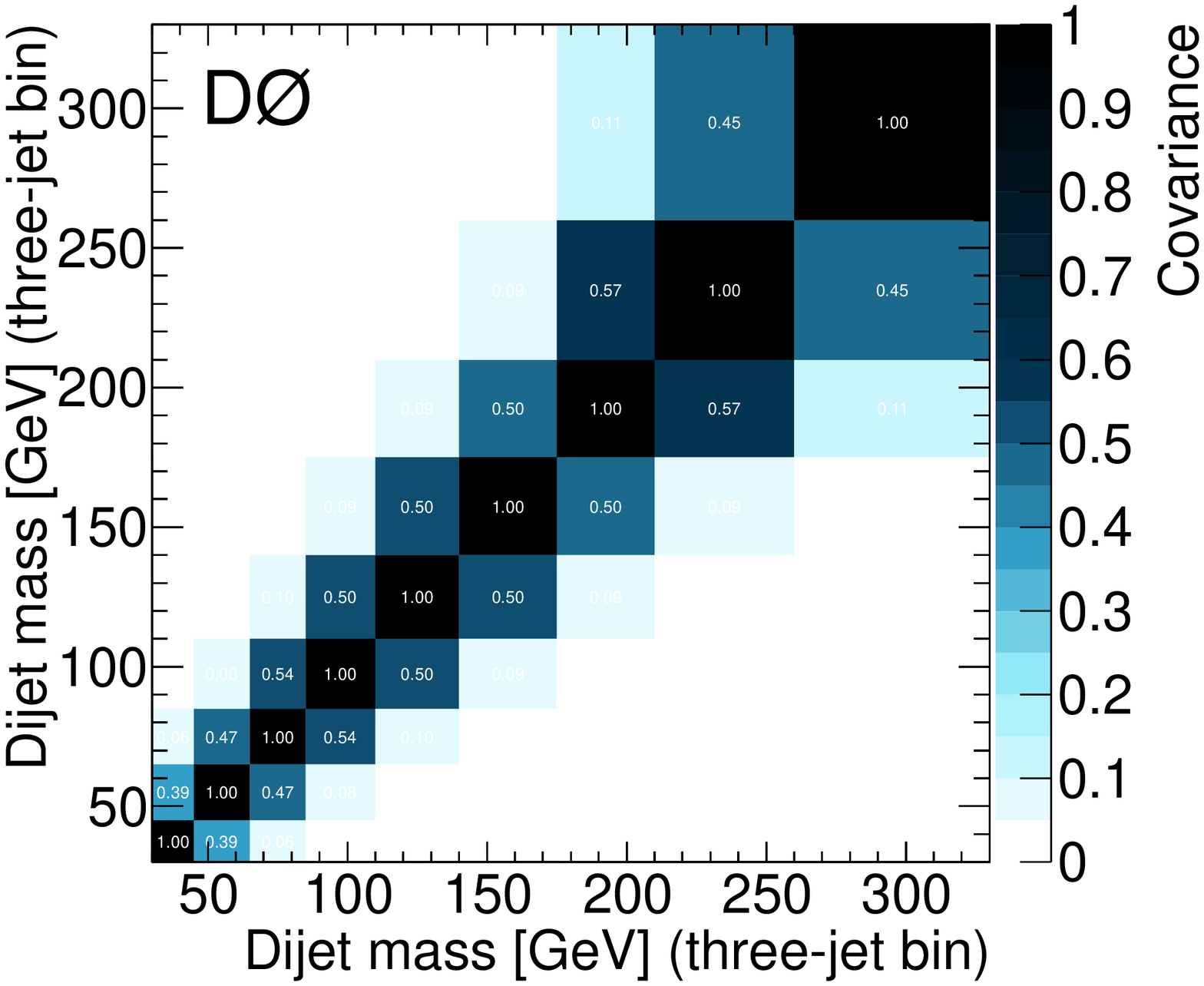}
    \caption{Normalized inverse covariance matrices for unfolded dijet $p_T$ and invariant mass distributions.
      \label{fig:normInvCov_dijetptmass}
    }
  \end{center}
\end{figure}

\begin{figure}[htbp]
  \begin{center}
    \includegraphics[width=0.35\textwidth]{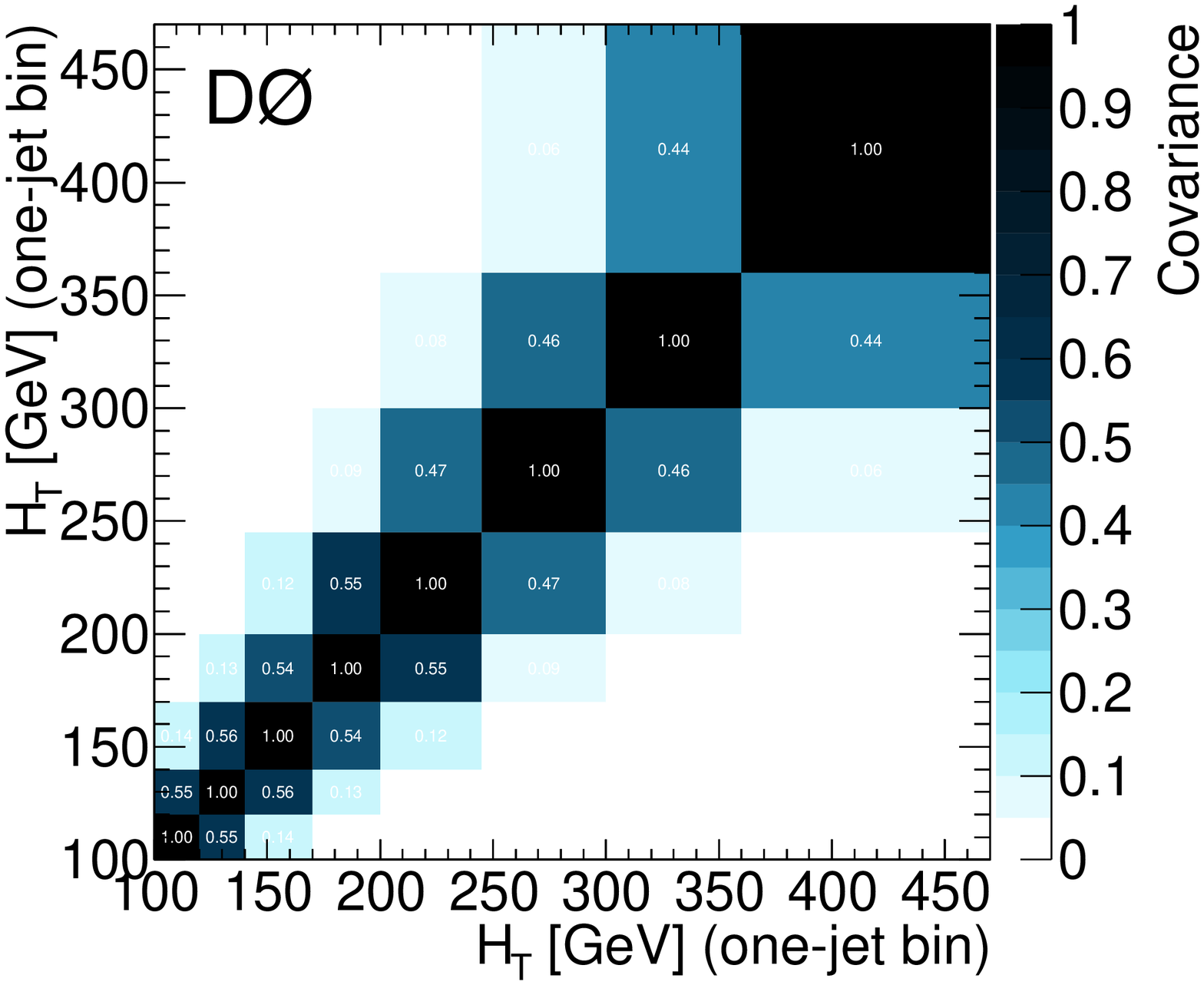}
    \includegraphics[width=0.35\textwidth]{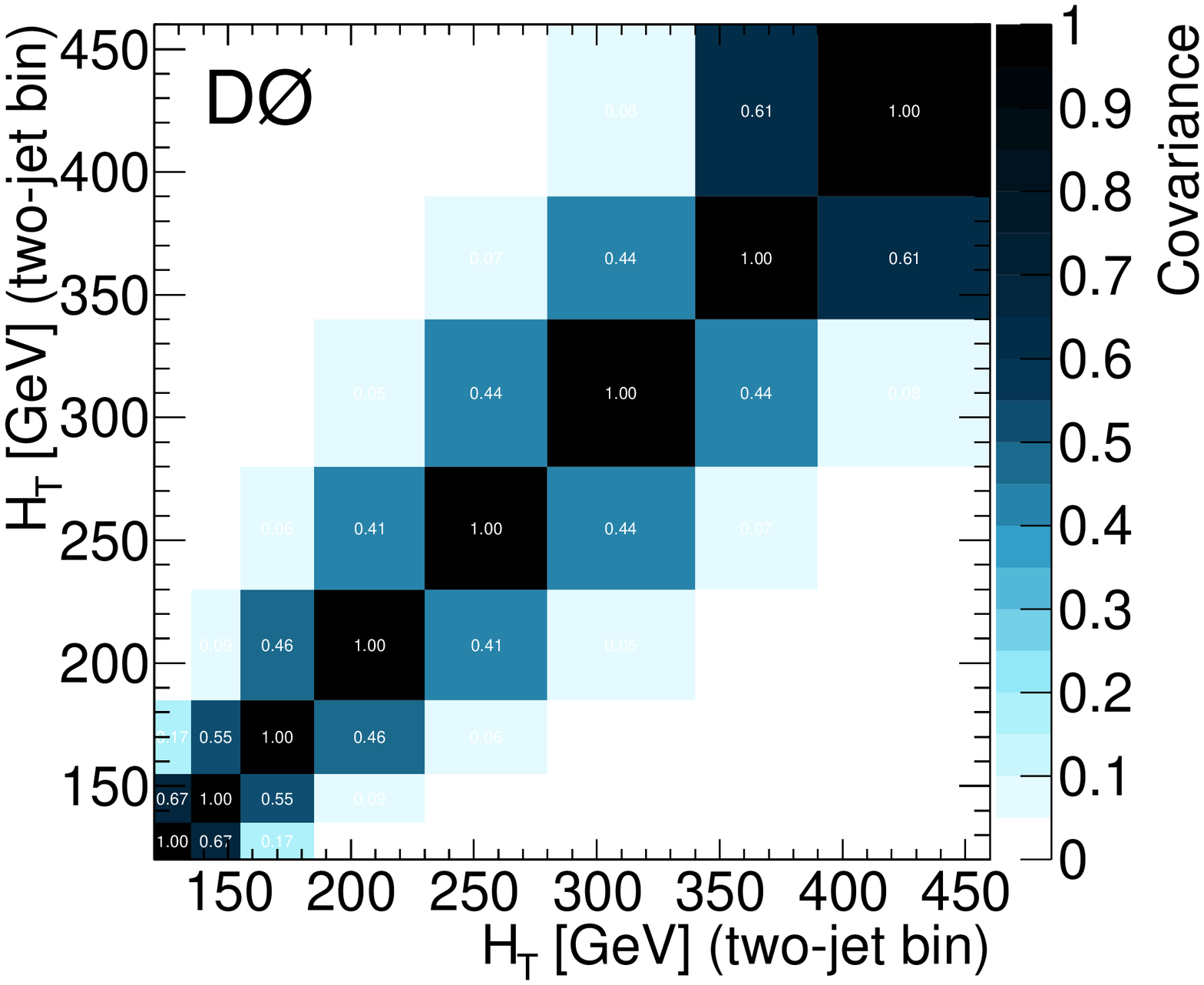}
    \includegraphics[width=0.35\textwidth]{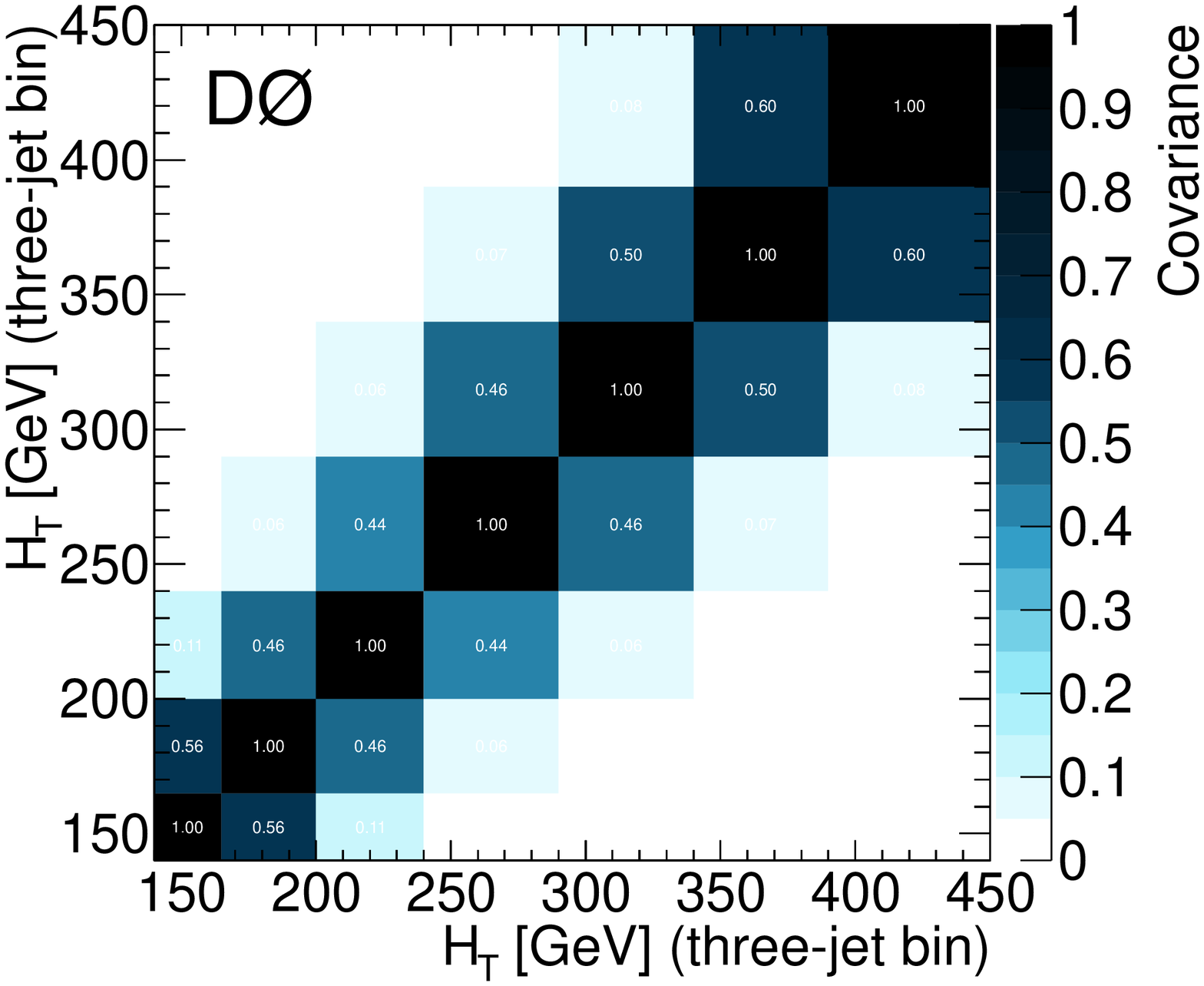}
    \includegraphics[width=0.35\textwidth]{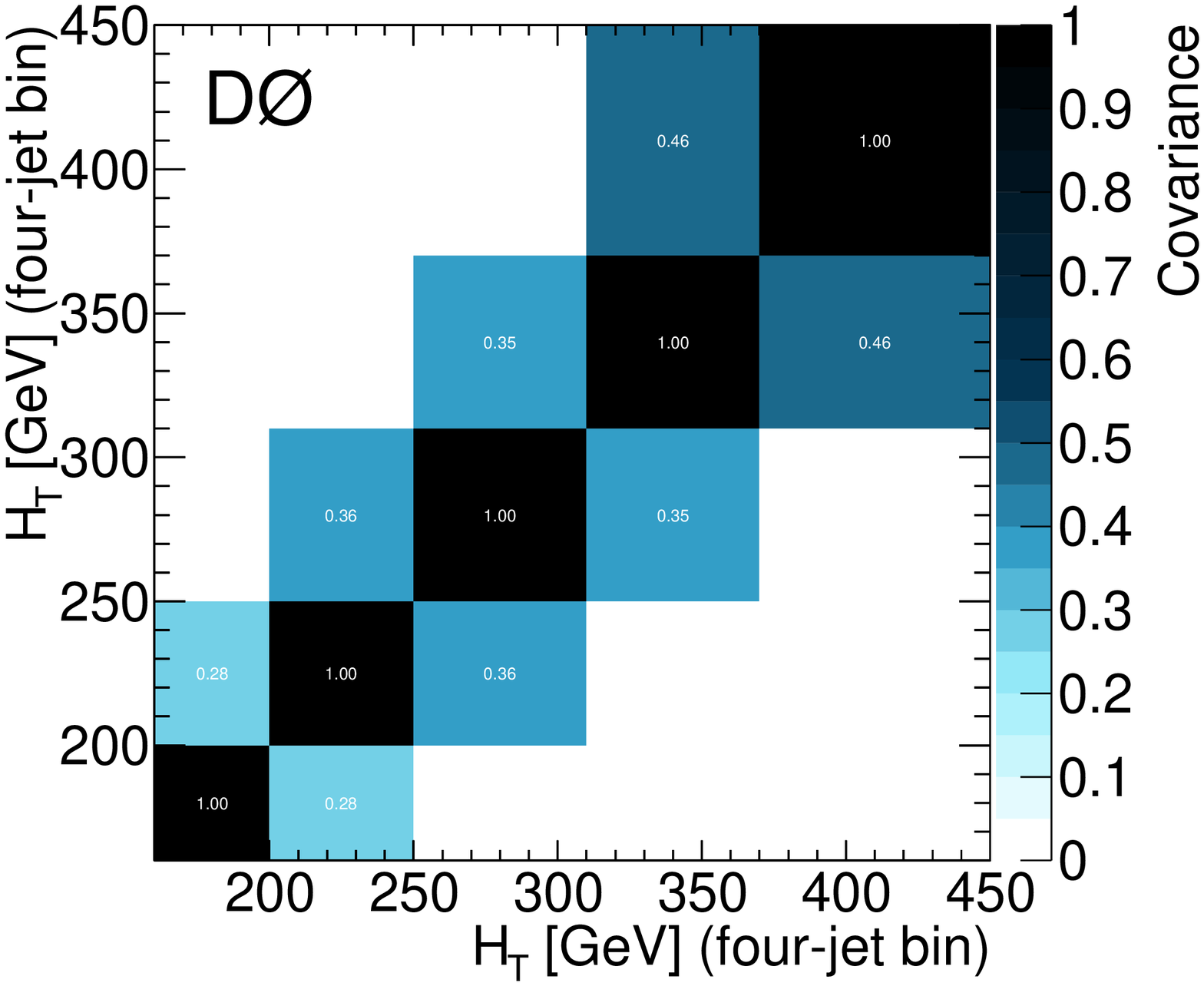}
    \caption{Normalized inverse covariance matrices for unfolded $H_T$ (scalar sum of the transverse energies of the $W$ boson and all jets) distributions.
      \label{fig:normInvCov_jetHT}
    }
  \end{center}
\end{figure}

\begin{figure}[htbp]
  \begin{center}
    \includegraphics[width=0.35\textwidth]{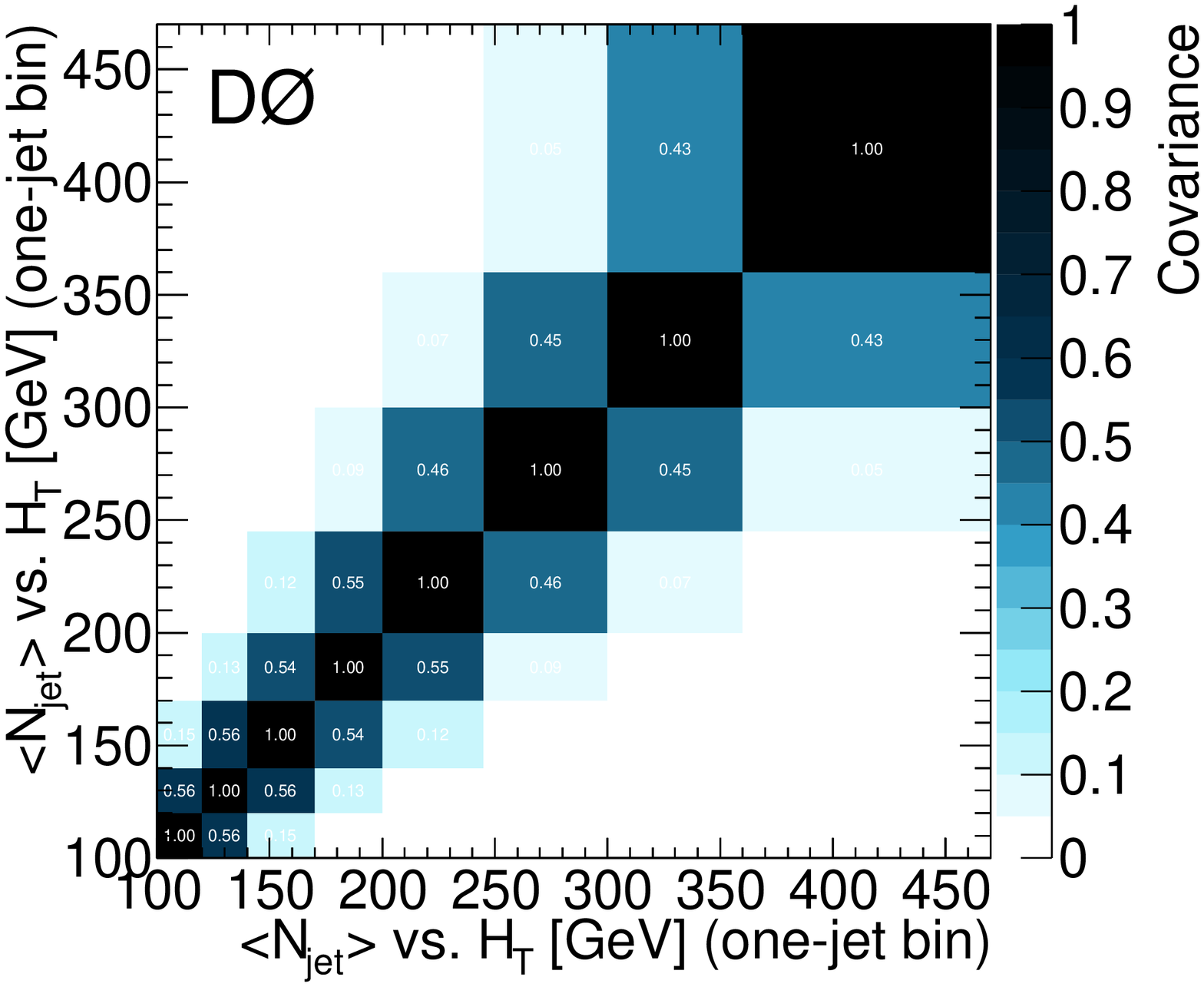}
    \includegraphics[width=0.35\textwidth]{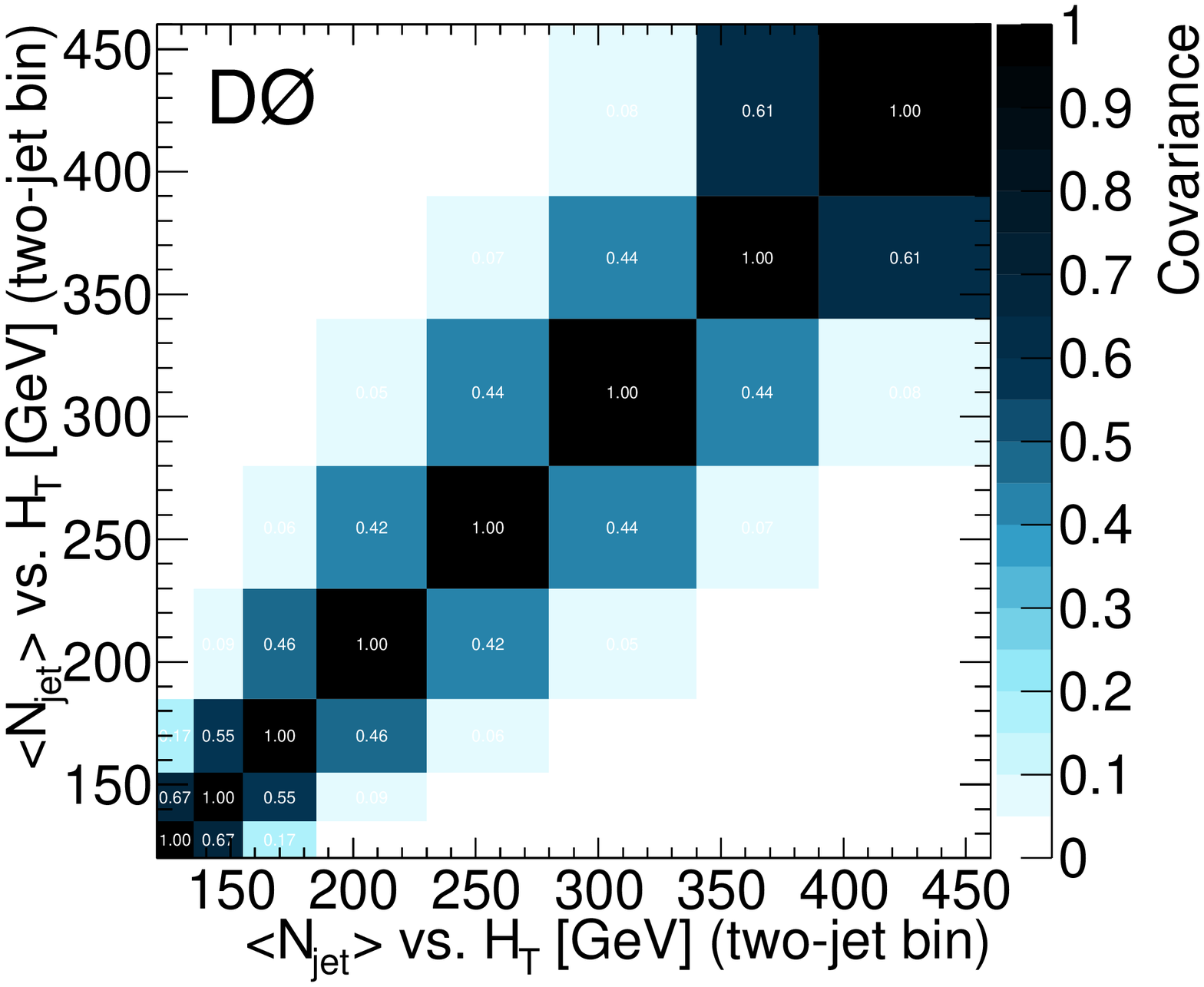}
    \includegraphics[width=0.35\textwidth]{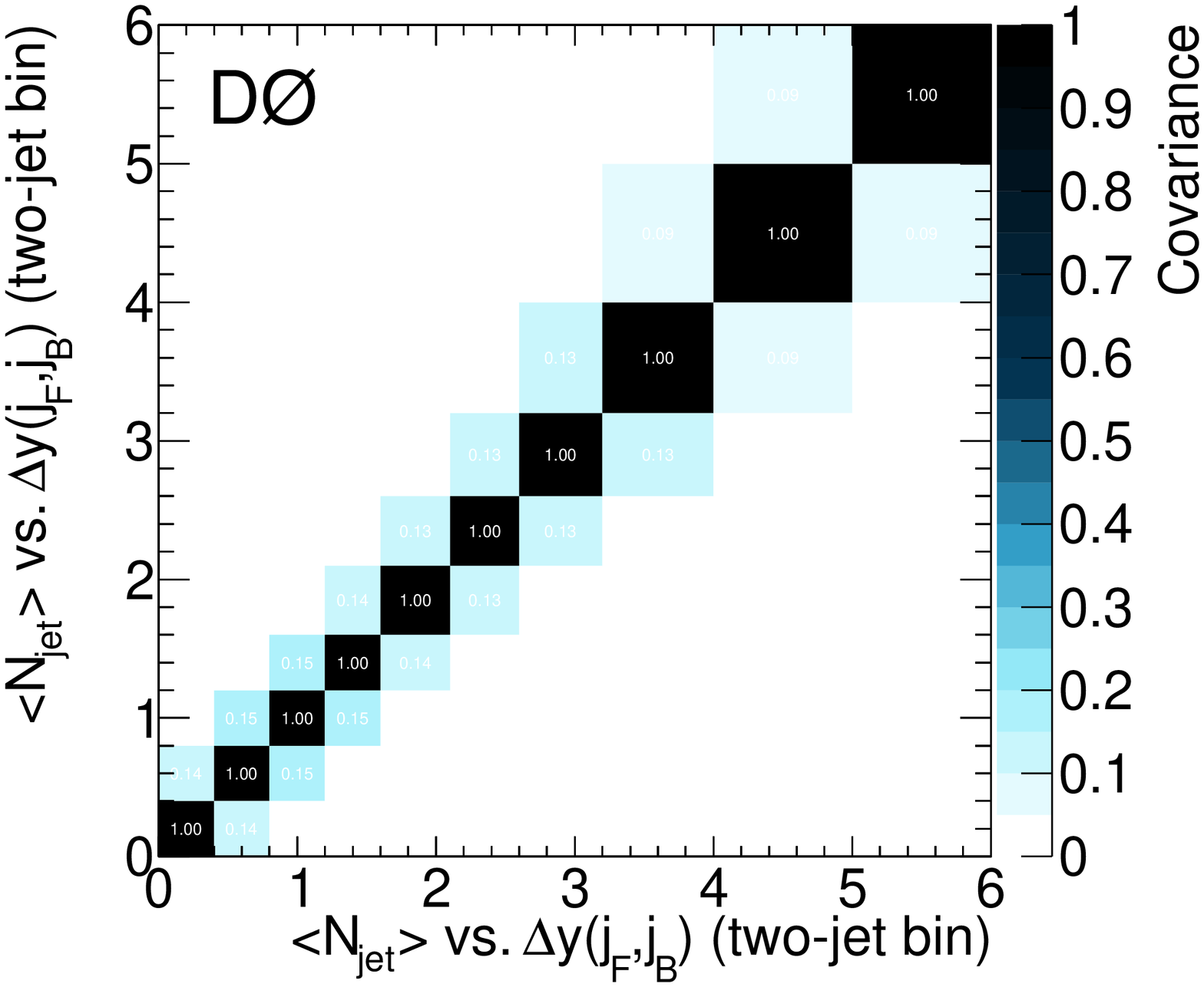}
    \includegraphics[width=0.35\textwidth]{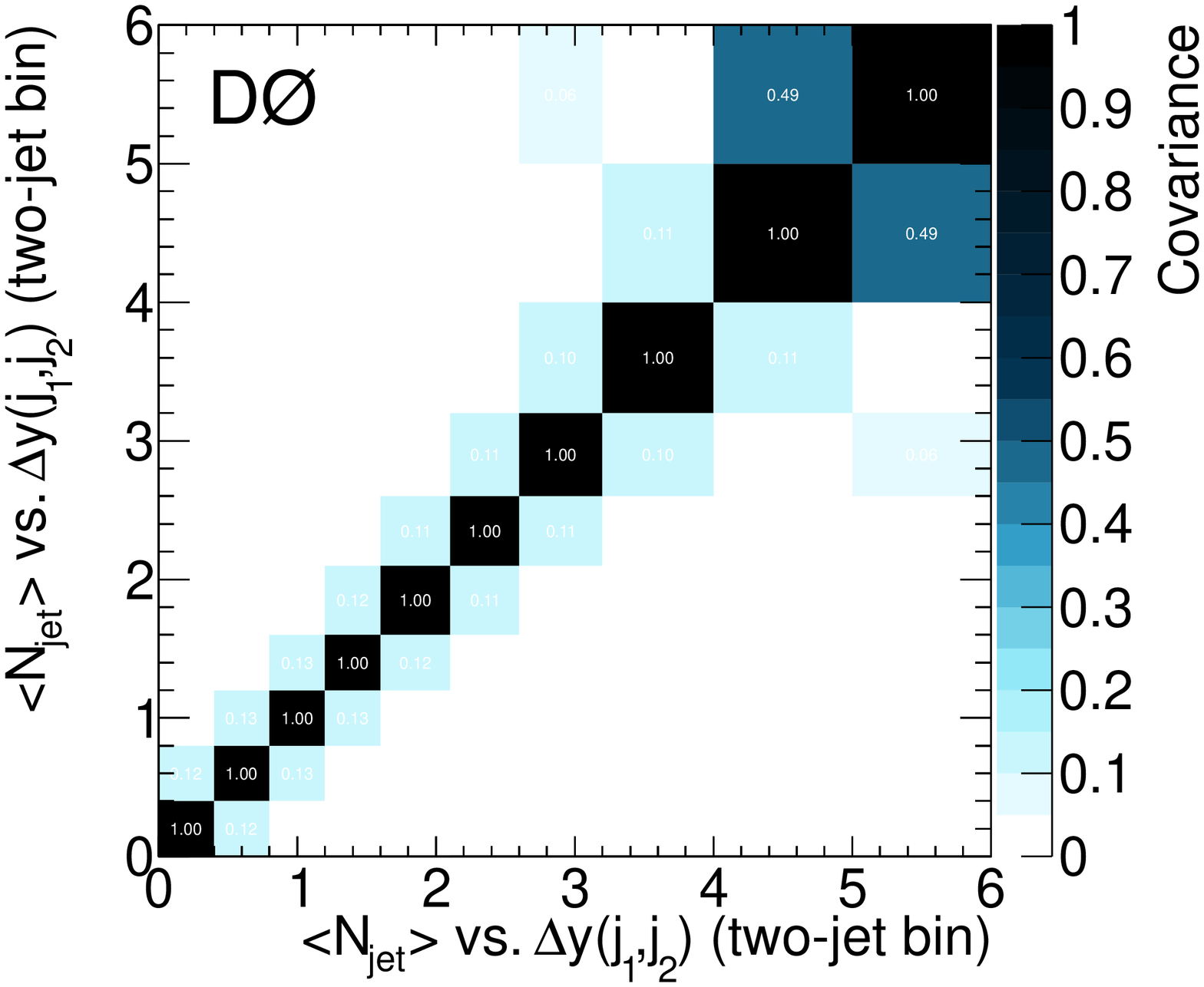}
    \caption{Normalized inverse covariance matrices for the unfolded mean number of jets as a function of $H_T$ in the inclusive one and two jet multiplicity bins
      and as a function of the dijet rapidity separations for the two highest $p_T$ or two most rapidity-separated jets.
      \label{fig:normInvCov_njet}
    }
  \end{center}
\end{figure}

\newpage
\clearpage

\begin{sidewaystable}[htbp]
\caption{\label{tab:invCov-jet1eta}Normalized inverse covariance matrix for the leading jet rapidity distribution.}
\tiny
\begin{ruledtabular}

\end{ruledtabular}
%\end{sidewaystable}

   \vspace{2\baselineskip}

%\begin{sidewaystable}[htbp]
\caption{\label{tab:invCov-jet2eta}Normalized inverse covariance matrix for the second jet rapidity distribution.}
\tiny
\begin{ruledtabular}
%
\end{ruledtabular}
\end{sidewaystable}

\begin{table}[htbp]
\caption{\label{tab:invCov-jet3eta}Normalized inverse covariance matrix for the third jet rapidity distribution.}
\tiny
\begin{ruledtabular}
%
\end{ruledtabular}
\end{table}

\begin{table}[htbp]
\caption{\label{tab:invCov-jet4eta}Normalized inverse covariance matrix for the fourth jet rapidity distribution.}
\tiny
\begin{ruledtabular}
%
\end{ruledtabular}
\end{table}

\begin{table}[htbp]
\caption{\label{tab:invCov-lepton1pt}Normalized inverse covariance matrix for the lepton $p_T$ [one-jet] distribution.}
\tiny
\begin{ruledtabular}
%
\end{ruledtabular}
\end{table}

\begin{table}[htbp]
\caption{\label{tab:invCov-lepton2pt}Normalized inverse covariance matrix for the lepton $p_T$ [two-jet] distribution.}
\tiny
\begin{ruledtabular}
%
\end{ruledtabular}
\end{table}

\begin{table}[htbp]
\caption{\label{tab:invCov-lepton3pt}Normalized inverse covariance matrix for the lepton $p_T$ [three-jet] distribution.}
\tiny
\begin{ruledtabular}
%
\end{ruledtabular}
\end{table}

\begin{table}[htbp]
\caption{\label{tab:invCov-lepton4pt}Normalized inverse covariance matrix for the lepton $p_T$ [four-jet] distribution.}
\tiny
\begin{ruledtabular}
%
\end{ruledtabular}
\end{table}

\begin{table}[htbp]
\caption{\label{tab:invCov-lepton1eta}Normalized inverse covariance matrix for the lepton pseudorapidity [one-jet] distribution.}
\tiny
\begin{ruledtabular}
%
\end{ruledtabular}
\end{table}

\begin{table}[htbp]
\caption{\label{tab:invCov-lepton2eta}Normalized inverse covariance matrix for the lepton pseudorapidity [two-jet] distribution.}
\tiny
\begin{ruledtabular}
%
\end{ruledtabular}
\end{table}

\begin{table}[htbp]
\caption{\label{tab:invCov-lepton3eta}Normalized inverse covariance matrix for the lepton pseudorapidity [three-jet] distribution.}
\tiny
\begin{ruledtabular}
%
\end{ruledtabular}
\end{table}

\begin{table}[htbp]
\caption{\label{tab:invCov-lepton4eta}Normalized inverse covariance matrix for the lepton pseudorapidity [four-jet] distribution.}
\tiny
\begin{ruledtabular}
%
\end{ruledtabular}
\end{table}

\begin{table}[htbp]
\caption{\label{tab:invCov-jet2DeltaRap}Normalized inverse covariance matrix for the $\Delta y_{12}$ [two-jet] distribution.}
\tiny
\begin{ruledtabular}
%
\end{ruledtabular}
\end{table}

\begin{table}[htbp]
\caption{\label{tab:invCov-jet2DeltaRapFB}Normalized inverse covariance matrix for the $\Delta y_{FB}$ [two-jet] distribution.}
\tiny
\begin{ruledtabular}
%
\end{ruledtabular}
\end{table}

\begin{table}[htbp]
\caption{\label{tab:invCov-jet3DeltaRap_3jet}Normalized inverse covariance matrix for the $\Delta y_{12}$ [three-jet] distribution.}
\tiny
\begin{ruledtabular}
%
\end{ruledtabular}
\end{table}

\begin{table}[htbp]
\caption{\label{tab:invCov-jet3DeltaRapFB}Normalized inverse covariance matrix for the $\Delta y_{FB}$ [three-jet] distribution.}
\tiny
\begin{ruledtabular}
%
\end{ruledtabular}
\end{table}

\begin{table}[htbp]
\caption{\label{tab:invCov-jet2dPhi}Normalized inverse covariance matrix for the $\Delta\phi_{12}$ distribution.}
\tiny
\begin{ruledtabular}
%
\end{ruledtabular}
\end{table}

\begin{table}[htbp]
\caption{\label{tab:invCov-jet2dPhiFB}Normalized inverse covariance matrix for the $\Delta\phi_{FB}$ distribution.}
\tiny
\begin{ruledtabular}
%
\end{ruledtabular}
\end{table}

\clearpage
\newpage

\begin{table}[htbp]
\caption{\label{tab:invCov-jet2dR}Normalized inverse covariance matrix for the $\Delta R_{jj}$ distribution.}
\tiny
\begin{ruledtabular}
%
\end{ruledtabular}
\end{table}

\begin{table}[htbp]
\caption{\label{tab:invCov-jet3DeltaRap13}Normalized inverse covariance matrix for the $\Delta y_{13}$ distribution.}
\tiny
\begin{ruledtabular}
%
\end{ruledtabular}
\end{table}

\begin{table}[htbp]
\caption{\label{tab:invCov-jet3DeltaRap23}Normalized inverse covariance matrix for the $\Delta y_{23}$ distribution.}
\tiny
\begin{ruledtabular}
%
\end{ruledtabular}
\end{table}

\clearpage
\newpage

\begin{table}[htbp]
\caption{\label{tab:invCov-W1pt}Normalized inverse covariance matrix for the $W$ boson $p_T$ [one-jet] distribution.}
\tiny
\begin{ruledtabular}
%
\end{ruledtabular}
\end{table}

\begin{table}[htbp]
\caption{\label{tab:invCov-W2pt}Normalized inverse covariance matrix for the $W$ boson $p_T$ [two-jet] distribution.}
\tiny
\begin{ruledtabular}
%
\end{ruledtabular}
\end{table}

\begin{table}[htbp]
\caption{\label{tab:invCov-W3pt}Normalized inverse covariance matrix for the $W$ boson $p_T$ [three-jet] distribution.}
\tiny
\begin{ruledtabular}
%
\end{ruledtabular}
\end{table}

\begin{table}[htbp]
\caption{\label{tab:invCov-W4pt}Normalized inverse covariance matrix for the $W$ boson $p_T$ [four-jet] distribution.}
\tiny
\begin{ruledtabular}
%
\end{ruledtabular}
\end{table}

\begin{sidewaystable}[htbp]
\caption{\label{tab:invCov-dijet2pt}Normalized inverse covariance matrix for the dijet $p_T$ [two-jet] distribution.}
\tiny
\begin{ruledtabular}
%
\end{ruledtabular}
%\end{sidewaystable}

   \vspace{2\baselineskip}

%\begin{sidewaystable}[htbp]
\caption{\label{tab:invCov-dijet3pt}Normalized inverse covariance matrix for the dijet $p_T$ [three-jet] distribution.}
\tiny
\begin{ruledtabular}
%
\end{ruledtabular}
\end{sidewaystable}

\begin{sidewaystable}[htbp]
\caption{\label{tab:invCov-dijet2mass}Normalized inverse covariance matrix for the dijet invariant mass [two-jet] distribution.}
\tiny
\begin{ruledtabular}
%
\end{ruledtabular}
%\end{sidewaystable}

   \vspace{2\baselineskip}

%\begin{sidewaystable}[htbp]
\caption{\label{tab:invCov-dijet3mass}Normalized inverse covariance matrix for the dijet invariant mass [three-jet] distribution.}
\tiny
\begin{ruledtabular}
%
\end{ruledtabular}
\end{sidewaystable}

\begin{table}[htbp]
\caption{\label{tab:invCov-jet1HT}Normalized inverse covariance matrix for the $H_T$ [one-jet] distribution.}
\tiny
\begin{ruledtabular}
%
\end{ruledtabular}
\end{table}

\begin{table}[htbp]
\caption{\label{tab:invCov-jet2HT}Normalized inverse covariance matrix for the $H_T$ [two-jet] distribution.}
\tiny
\begin{ruledtabular}
%
\end{ruledtabular}
\end{table}

\begin{table}[htbp]
\caption{\label{tab:invCov-jet3HT}Normalized inverse covariance matrix for the $H_T$ [three-jet] distribution.}
\tiny
\begin{ruledtabular}
%
\end{ruledtabular}
\end{table}

\begin{table}[htbp]
\caption{\label{tab:invCov-jet4HT}Normalized inverse covariance matrix for the $H_T$ [four-jet] distribution.}
\tiny
\begin{ruledtabular}
%
\end{ruledtabular}
\end{table}

\clearpage
\newpage

\section{Systematic Uncertainty Correlations On Unfolded Distributions}

\begin{center}
{\bf To appear as an Electronic Physics Auxiliary Publication (EPAPS)}
\end{center}

In this appendix, we provide the bin-to-bin correlations of the systematic uncertainties for each 
measurement made in the main text of the paper, in tabular format.
The full correlation matrix is obtained by summing the covariance matrices for each source of systematic uncertainty.

\begin{sidewaystable}[htbp]
\caption{\label{tab:systCorr-jet1eta}Systematic correlation matrix for the leading jet rapidity distribution.}
\tiny
\begin{ruledtabular}

\end{ruledtabular}
%\end{sidewaystable}

   \vspace{2\baselineskip}

%\begin{sidewaystable}[htbp]
\caption{\label{tab:systCorr-jet2eta}Systematic correlation matrix for the second jet rapidity distribution.}
\tiny
\begin{ruledtabular}
%
\end{ruledtabular}
\end{sidewaystable}

\begin{table}[htbp]
\caption{\label{tab:systCorr-jet3eta}Systematic correlation matrix for the third jet rapidity distribution.}
\tiny
\begin{ruledtabular}
%
\end{ruledtabular}
\end{table}

\begin{table}[htbp]
\caption{\label{tab:systCorr-jet4eta}Systematic correlation matrix for the fourth jet rapidity distribution.}
\tiny
\begin{ruledtabular}
%
\end{ruledtabular}
\end{table}

\begin{table}[htbp]
\caption{\label{tab:systCorr-lepton1pt}Systematic correlation matrix for the lepton $p_T$ [one-jet] distribution.}
\tiny
\begin{ruledtabular}
%
\end{ruledtabular}
\end{table}

\begin{table}[htbp]
\caption{\label{tab:systCorr-lepton2pt}Systematic correlation matrix for the lepton $p_T$ [two-jet] distribution.}
\tiny
\begin{ruledtabular}
%
\end{ruledtabular}
\end{table}

\begin{table}[htbp]
\caption{\label{tab:systCorr-lepton3pt}Systematic correlation matrix for the lepton $p_T$ [three-jet] distribution.}
\tiny
\begin{ruledtabular}
%
\end{ruledtabular}
\end{table}

\begin{table}[htbp]
\caption{\label{tab:systCorr-lepton4pt}Systematic correlation matrix for the lepton $p_T$ [four-jet] distribution.}
\tiny
\begin{ruledtabular}
%
\end{ruledtabular}
\end{table}

\begin{table}[htbp]
\caption{\label{tab:systCorr-lepton1eta}Systematic correlation matrix for the lepton pseudorapidity [one-jet] distribution.}
\tiny
\begin{ruledtabular}
%
\end{ruledtabular}
\end{table}

\begin{table}[htbp]
\caption{\label{tab:systCorr-lepton2eta}Systematic correlation matrix for the lepton pseudorapidity [two-jet] distribution.}
\tiny
\begin{ruledtabular}
%
\end{ruledtabular}
\end{table}

\begin{table}[htbp]
\caption{\label{tab:systCorr-lepton3eta}Systematic correlation matrix for the lepton pseudorapidity [three-jet] distribution.}
\tiny
\begin{ruledtabular}
%
\end{ruledtabular}
\end{table}

\begin{table}[htbp]
\caption{\label{tab:systCorr-lepton4eta}Systematic correlation matrix for the lepton pseudorapidity [four-jet] distribution.}
\tiny
\begin{ruledtabular}
%
\end{ruledtabular}
\end{table}

\begin{table}[htbp]
\caption{\label{tab:systCorr-jet2DeltaRap}Systematic correlation matrix for the $\Delta y_{12}$ [two-jet] distribution.}
\tiny
\begin{ruledtabular}
%
\end{ruledtabular}
\end{table}

\begin{table}[htbp]
\caption{\label{tab:systCorr-jet2DeltaRapFB}Systematic correlation matrix for the $\Delta y_{FB}$ [two-jet] distribution.}
\tiny
\begin{ruledtabular}
%
\end{ruledtabular}
\end{table}

\begin{table}[htbp]
\caption{\label{tab:systCorr-jet3DeltaRap_3jet}Systematic correlation matrix for the $\Delta y_{12}$ [three-jet] distribution.}
\tiny
\begin{ruledtabular}
%
\end{ruledtabular}
\end{table}

\begin{table}[htbp]
\caption{\label{tab:systCorr-jet3DeltaRapFB}Systematic correlation matrix for the $\Delta y_{FB}$ [three-jet] distribution.}
\tiny
\begin{ruledtabular}
%
\end{ruledtabular}
\end{table}

\begin{table}[htbp]
\caption{\label{tab:systCorr-jet2dPhi}Systematic correlation matrix for the $\Delta\phi_{12}$ distribution.}
\tiny
\begin{ruledtabular}
%
\end{ruledtabular}
\end{table}

\begin{table}[htbp]
\caption{\label{tab:systCorr-jet2dPhiFB}Systematic correlation matrix for the $\Delta\phi_{FB}$ distribution.}
\tiny
\begin{ruledtabular}
%
\end{ruledtabular}
\end{table}

\newpage
\clearpage

\begin{table}[htbp]
\caption{\label{tab:systCorr-jet2dR}Systematic correlation matrix for the $\Delta R_{jj}$ distribution.}
\tiny
\begin{ruledtabular}
%
\end{ruledtabular}
\end{table}

\begin{table}[htbp]
\caption{\label{tab:systCorr-jet3DeltaRap13}Systematic correlation matrix for the $\Delta y_{13}$ distribution.}
\tiny
\begin{ruledtabular}
%
\end{ruledtabular}
\end{table}

\begin{table}[htbp]
\caption{\label{tab:systCorr-jet3DeltaRap23}Systematic correlation matrix for the $\Delta y_{23}$ distribution.}
\tiny
\begin{ruledtabular}
%
\end{ruledtabular}
\end{table}

\newpage
\clearpage

\begin{table}[htbp]
\caption{\label{tab:systCorr-W1pt}Systematic correlation matrix for the $W$ boson $p_T$ [one-jet] distribution.}
\tiny
\begin{ruledtabular}
%
\end{ruledtabular}
\end{table}

\begin{table}[htbp]
\caption{\label{tab:systCorr-W2pt}Systematic correlation matrix for the $W$ boson $p_T$ [two-jet] distribution.}
\tiny
\begin{ruledtabular}
%
\end{ruledtabular}
\end{table}

\begin{table}[htbp]
\caption{\label{tab:systCorr-W3pt}Systematic correlation matrix for the $W$ boson $p_T$ [three-jet] distribution.}
\tiny
\begin{ruledtabular}
%
\end{ruledtabular}
\end{table}

\begin{table}[htbp]
\caption{\label{tab:systCorr-W4pt}Systematic correlation matrix for the $W$ boson $p_T$ [four-jet] distribution.}
\tiny
\begin{ruledtabular}
%
\end{ruledtabular}
\end{table}

\begin{sidewaystable}[htbp]
\caption{\label{tab:systCorr-dijet2pt}Systematic correlation matrix for the dijet $p_T$ [two-jet] distribution.}
\tiny
\begin{ruledtabular}
%
\end{ruledtabular}
%\end{sidewaystable}

   \vspace{2\baselineskip}

%\begin{sidewaystable}[htbp]
\caption{\label{tab:systCorr-dijet3pt}Systematic correlation matrix for the dijet $p_T$ [three-jet] distribution.}
\tiny
\begin{ruledtabular}
%
\end{ruledtabular}
\end{sidewaystable}

\begin{sidewaystable}[htbp]
\caption{\label{tab:systCorr-dijet2mass}Systematic correlation matrix for the dijet invariant mass [two-jet] distribution.}
\tiny
\begin{ruledtabular}
%
\end{ruledtabular}
%\end{sidewaystable}

   \vspace{2\baselineskip}

%\begin{sidewaystable}[htbp]
\caption{\label{tab:systCorr-dijet3mass}Systematic correlation matrix for the dijet invariant mass [three-jet] distribution.}
\tiny
\begin{ruledtabular}
%
\end{ruledtabular}
\end{sidewaystable}

\begin{table}[htbp]
\caption{\label{tab:systCorr-jet1HT}Systematic correlation matrix for the $H_T$ [one-jet] distribution.}
\tiny
\begin{ruledtabular}
%
\end{ruledtabular}
\end{table}

\begin{table}[htbp]
\caption{\label{tab:systCorr-jet2HT}Systematic correlation matrix for the $H_T$ [two-jet] distribution.}
\tiny
\begin{ruledtabular}
%
\end{ruledtabular}
\end{table}

\begin{table}[htbp]
\caption{\label{tab:systCorr-jet3HT}Systematic correlation matrix for the $H_T$ [three-jet] distribution.}
\tiny
\begin{ruledtabular}
%
\end{ruledtabular}
\end{table}

\begin{table}[htbp]
\caption{\label{tab:systCorr-jet4HT}Systematic correlation matrix for the $H_T$ [four-jet] distribution.}
\tiny
\begin{ruledtabular}
%
\end{ruledtabular}
\end{table}


\begin{thebibliography}{99}
\bibitem{definitions}
  Rapidity is defined as $y$ = $-\ln[(E + p_Z )/(E - p_Z )]$, where $E$ is the energy and $p_Z$ is the momentum component along the proton beam direction.
  $\phi$ is defined as the azimuthal angle in the plane transverse to the proton beam direction.
\end{thebibliography}
\end{document}